\documentclass[11pt]{article}

\usepackage{array}	
\newcolumntype{P}[1]{>{\centering\arraybackslash}p{#1}}		
\newcolumntype{M}[1]{>{\centering\arraybackslash}m{#1}}
\newcommand*\mystrut[1]{\vrule width0pt height0pt depth#1\relax}

\usepackage{tikz}
\usetikzlibrary{patterns}
\usepackage{bm}
\usepackage{empheq}
\usepackage{cases}
\usepackage{multirow}
\usepackage{booktabs}
\usepackage{wrapfig}
\usepackage{soul}
\usepackage{xcolor}

\usetikzlibrary{decorations.pathreplacing,angles,quotes}

\usepackage{listings}
\usepackage[colorlinks=false, backref]{hyperref}

\usepackage{amsmath,amsthm,amsfonts,amssymb,amscd}
\usepackage{float}
\usepackage{graphicx,subfigure}
\graphicspath {{figures/}}
\usepackage[font=footnotesize,labelfont=bf]{caption}

\usepackage[mathscr]{euscript}
\usepackage{stmaryrd}
\usepackage{microtype}
\usepackage{booktabs}
\usepackage{cleveref}
\usepackage{bookmark}
\usepackage{mathrsfs}

\usepackage{emptypage}
\usepackage{tikz}
\usepackage{subfigure}
\usepackage[utf8]{inputenc}
\usepackage[T1]{fontenc}
\usepackage{mathtools}
\usepackage{colonequals}
\usepackage{bbm}

\DeclareMathAlphabet{\mathpzc}{OT1}{pzc}{m}{it}

\theoremstyle{plain}

\theoremstyle{definition}

\newtheorem{remark}{\scshape Remark}

\def\Cw{\overline{C}_w}

\def\p{\partial}

\usepackage{fancyhdr}
\title{{\bf A space-time smooth artificial viscosity method with wavelet noise indicator and shock 
collision scheme, Part 1: the 1-$D$ case.}}
\author{
 {\small {\bf Raaghav Ramani}}
 \vspace{-.05 in}
\\{\footnotesize Department of Mathematics}
\vspace{-.05 in}
\\{\footnotesize University of California}
\vspace{-.05 in}
\\{\footnotesize Davis, CA 95616 USA}
\vspace{-.05 in}
\\{\footnotesize  {\it rramani@math.ucdavis.edu}}\and
 {\small {\bf Jon Reisner}}
 \vspace{-.05 in}
\\{\footnotesize Los Alamos National Lab}
\vspace{-.05 in}
\\{\footnotesize XCP-4 MSF605}
\vspace{-.05 in}
\\{\footnotesize Los Alamos, NM 87544}
\vspace{-.05 in}
\\{\footnotesize {\it reisner@lanl.gov}}
\and
 {\small {\bf Steve Shkoller}}
 \vspace{-.05 in}
\\{\footnotesize Department of Mathematics}
\vspace{-.05 in}
\\{\footnotesize University of California}
\vspace{-.05 in}
\\{\footnotesize Davis, CA 95616 USA}
\vspace{-.05 in}
\\{\footnotesize  {\it shkoller@math.ucdavis.edu}}
}
\date{\today}

\begin{document}

\maketitle

\begin{abstract}
In this first part of two papers, 
we extend the $C$-method developed in \cite{ReSeSh2012} for adding localized, space-time smooth artificial viscosity to nonlinear systems 
of conservation laws that propagate shock waves, rarefaction waves, and contact discontinuities in one space dimension. 
For gas dynamics, the $C$-method 
couples the Euler equations to 
a scalar reaction-diffusion equation, whose solution $C$ serves as a space-time smooth artificial viscosity indicator.

The purpose of this paper is the development of a high-order numerical algorithm for shock-wall collision and bounce-back.   
Specifically, we generalize the original $C$-method by adding a new collision indicator, which naturally activates during shock-wall
collision. Additionally, 
we implement a new high-frequency wavelet-based noise detector together with an efficient and localized noise removal algorithm.    
 To test the 
methodology, we use 
a highly simplified WENO-based discretization scheme.  We show that our scheme improves the order of accuracy of our WENO algorithm,  handles extremely strong discontinuities 
(ranging up to nine orders of magnitude), allows for shock collision and bounce back, and removes high frequency noise. The causes of the well-known
``wall heating'' phenomenon are discussed, and we demonstrate that this particular pathology can be 
effectively treated in the framework of the $C$-method.
   This method is generalized to two space dimensions in the second part of this work \cite{RaReSh2018b}.
\end{abstract}

{\small
\tableofcontents}

\section{Introduction}\label{sec:intro}

This is the first in a two-part series of papers, in which we develop a high-order numerical algorithm to simulate compressible fluid flow
with shock waves and contact discontinuities, as well as shock-wall collision and bounce-back.  In the second part of this series \cite{RaReSh2018b}, we treat problems in two space dimensions. In this first part, we begin the development for one-dimensional flows.

The initial-value problem
for a nonlinear system of conservation laws in one space dimension is given as  
\begin{subequations}\label{nonlinear-conservation-law}
\begin{align}
\partial_t {\bm u}(x,t) + \partial_x F({\bm u}(x,t)) &= 0, \\
{\bm u}(x,t=0) &= {\bm u}_0(x), 
\end{align}
\end{subequations}
where  ${\bm u}(x,t)$ denotes a vector of conserved quantities, $x$ denotes the space coordinate,  and $t$ denotes the time coordinate.
 Many different physical phenomena can be modeled by \eqref{nonlinear-conservation-law},
including gas dynamics, described by the compressible  Euler equations, which shall be the focus of this paper.

It is well known that 
solutions of \eqref{nonlinear-conservation-law} can develop finite-time discontinuities, even for smooth initial
data $\bm{u}_0$. In this case, the discontinuities are propagated according to the \emph{Rankine-Hugoniot}
conditions (see \S\ref{subsec:review}). Consequently, it is important to develop robust numerical schemes that 
can approximate discontinuous solutions. This is a nontrivial task, since approximations to discontinuous 
solutions usually result in the occurrence of small-scale oscillations, or Gibbs-phenomenon;
however, a variety of high-order discretization schemes and techniques have been developed to combat this issue
and produce non-oscillatory  solutions. 
In the case of  1-D gas dynamics,  the construction of non-oscillatory, higher-order,  numerical algorithms such as ENO by Harten, Engquist,  Osher \& Chakravarthy \cite{Harten1987231} and Shu \& Osher \cite{Shu1988439}, \cite{OsherShu1989};  WENO by Liu, Osher, \& Chan \cite{LiOsCh1994} and
Jiang \& Shu \cite{JiangShu1996}; MUSCL by Van Leer \cite{VanLeer1979101},  Colella \cite{Colella1985104}, and  Huynh \cite{Huynh19951565}; or PPM by Colella \& Woodward \cite{CoWo1984}  requires 
carefully chosen {\it reconstruction} and {\it numerical flux}.

Such numerical methods evolve cell-averaged quantities;  to calculate an accurate approximation of the flux at cell-interfaces, these schemes  reconstruct $k$th-order ($k\ge 2$) polynomial
approximations of the solution (and hence the flux) from the computed cell-averages, and thus provide $k$th-order accuracy away from discontinuities.   See, for example, the convergence plots
of Greenough \& Rider \cite{GreenoughRider2004} and  Liska \& Wendroff \cite{Liska2003995}.  Given a polynomial representation of the solution, a strategy is chosen to compute the
most accurate cell-interface flux, and this is achieved by a variety of algorithms.
Centered numerical fluxes, such as Lax-Friedrichs, add dissipation as a mechanism to preserve stability and monotonicity. On the other hand, {\it characteristic-type} upwinding based upon exact (Godunov) or approximate (Roe, Osher, HLL, HLLC) Riemann solvers,
 which preserve monotonicity without adding too much dissipation,  tend to be rather complex and PDE-specific;  moreover,  for strong shocks, other techniques may be
 required  to dampen post-shock oscillations or to yield entropy-satisfying approximations (see Quirk \cite{Quirk1994555}).
 Again, we refer the reader to the papers  \cite{GreenoughRider2004}, \cite{Liska2003995} or Colella \& Woodward \cite{Colella1984115} for a thorough overview, as well as a comparison of the effectiveness of a variety of competitive schemes.
 
Majda \& Osher {\cite{MaOs1977}} have shown that \emph{any} numerical scheme for a
 problem with discontinuities
 will suffer from a formal loss of accuracy near the discontinuity. Nonetheless, the use of high-order 
 schemes is imperative for the resolution of finer structures in smooth regions of the flow. Formally 
 high-order WENO schemes (as well as other high-order methods) maintain high-order accuracy in  regions away from shocks, but 
 are only first-order accurate at the discontinuity.
 
 In order to 
 ascertain the performance of a method, it is essential to conduct numerical tests for a range of problems 
 with different features of varying complexity. These tests are made precise by calculating error norms 
 of the computed solution
 relative to either an exact solution (if available), or a highly resolved solution which may be regarded 
 as the exact solution. Proposed numerical algorithms should demonstrate small error norms and 
 close to optimal convergence for a range of test problems. However, due to the fact that different tests 
 can exhibit very different phenomena and features, it is not so surprising that there are a number of 
 situations in which anomalous behavior of solutions is observed, which results in large errors and 
 poor rates of convergence. Examples of such errors are wall-heating, 
 the carbuncle phenomenon, long wavelength instabilities in slow-moving shocks, and non entropy-satisfying 
 ``expansion shocks'' (see Quirk {\cite{Quirk1994555}} for further details). 
 
In this paper, we continue the development of the $C$-method {\cite{ReSeSh2012}}, a nonlinear artificial viscosity
 modification of the Euler equations of gas dynamics, whose numerical discretization by  a simple WENO-type (or even
 central differencing) scheme can stabilize the type of instabilities noted above.
As proven in {\cite{ReSeSh2012}},  weak solutions of the $C$-method modification of the Euler equations 
converge to the unique (entropy) solutions of the Euler equations as the artificial viscosity parameter tends to zero.
Herein, we present 
 numerical error analysis and order of accuracy studies for a number of classical shock tube experiments;  we
 show that  a highly simplified WENO discretization of the $C$-method yields highly accurate solutions displaying close to optimal rates of convergence.

For instance, we show that for the Sod problem, our simple WENO-type discretization of the $C$-method yields
 smaller errors and faster rates of convergence in the $L^1$, $L^2$, and $L^ \infty$ norms as compared to the 
 same WENO discretization of the unmodified Euler equations.
 In particular, for the difficult problem of shock-wall collision 
(to be introduced in \S{\ref{intro:shock-collision}} and developed in \S{\ref{sec:wallvisc}}) for the 
Sod problem on a grid with 801 cells, we show that the $L^1$ error with the $C$-method
is 35\% of the error without the $C$-method. Moreover, the order of convergence of solutions is 
approximately 0.95, which is close to optimal and more than twice the 
order of convergence when the $C$-method is not employed. 
Similar conclusions hold also for the extremely difficult LeBlanc problem, for which 
we show that the use of the  $C$-method produces  $L^1$ errors that are approximately four times smaller 
 prior to shock-wall collision, and approximately three times smaller post shock-wall collision.

Our quantitative analysis, together with the qualitative observations we make via plot comparison, lead us 
to conclude that the use of a simple discretization of the $C$-method provides a flexible, highly accurate scheme that produces solutions 
with close to optimal rates of convergence for a variety of problems with different features.


\subsection{Using artificial viscosity with conservation laws}
Artificial viscosity is an effective method for 
the  numerical stabilization of shock waves in gas dynamics; the simplest such regularization of  \eqref{nonlinear-conservation-law} 
replaces the right-hand side with the linear second-order operator (see, for instance, \mbox{\cite{Landshoff1955,Wilkins1980,DiPerna1983}})
\begin{equation}\label{linear-viscosity}
\beta \, \Delta x \, \partial_{xx} {\bm u}(x,t)\,,
\end{equation}
where $\beta=O(1)$ is a constant, and $\Delta x $ denotes a small asymptotic parameter that, when the term
\eqref{linear-viscosity} is numerically discretized, represents the grid spacing.

 For each such $\beta >0$, solutions to the regularized conservation law smooth the shock across a 
 small region of width proportional to $\Delta x$, and simultaneously prevent small-scale oscillations from corrupting  sound waves in numerical simulations; nevertheless, 
the uniform application of  diffusion given by \eqref{linear-viscosity} ensures only first-order accuracy  of numerical schemes and overly diffuses wave 
amplitudes and speeds.

In  \cite{vNRi1950}, Von Neumann and Richtmeyer replaced the
uniform linear viscosity \eqref{linear-viscosity} with a nonlinear term given by
\begin{equation}\label{nonlinear-artificial-viscosity}
\beta \, (\Delta x)^2 \partial_x \left( |\partial_x u|\, \partial_x {\bm u} \right)\,,
\end{equation}
which we shall refer to as \emph{classical artificial viscosity}. 
Here,  $u(x,t)$ represents, in the case of the Euler equations of gas dynamics, the velocity of the
fluid. 
The use of the localizing coefficient $|\partial_x u|$ in
\eqref{nonlinear-artificial-viscosity} 
concentrates the addition of viscosity to the narrow intervals  containing shocks,
while 
maintaining high-order accuracy in  regions away from the shock, wherein the solution is smooth.   See also Margolin
\cite{Margolin2018} and Mattsson \& Rider \cite{MaRi2015}  for a description of the origin
and the interpretation of artificial viscosity as a physical phenomenon.

It is now well-known \cite{Lapidus1967154, GeMaDa1966} that classical
artificial viscosity corrects for the over-dissipation of the linear viscosity \eqref{linear-viscosity}, and allows for 
the implementation of numerical methods that are both non-oscillatory at shocks, as well as high-order accurate
in smooth regions. 
On the other hand, the fact that the localizing coefficient $|\partial_x u|$ itself 
becomes highly irregular in regions containing 
shocks often results in the failure of such schemes to suppress spurious oscillations.
This inadequacy of classical artificial viscosity may be observed with the highly singular phenomenon of 
shock-wave wall collision. In this case, large amplitude, high frequency, non-physical oscillations appear in the 
solution post-collision behind the shock curve, and the rough nature of $|\partial_x u|$ in both space and time 
means that the classical artificial viscosity is often unable to remove such oscillations.   

This suggests that a space-time smoothed variant of the localizing coefficient $|\partial_x u|$ might allow for 
a less oscillatory, more accurate solution profile. 
We propose the use of the $C$-method as a 
means of producing such a localizing coefficient. 
A similar method is employed by Cabot \& Cook 
\mbox{\cite{CaCo2004,CaCo2004b}}, who use a 
high-wavenumber indicator together with a Gaussian filter to produce such a function, though we note that 
the produced function is only spatially regularized, and not temporally. See also the work of  
Barter \& Darmofal {\cite{BaDa2010}}, who utilize a PDE-based approach to smooth the localizing function.
As we shall explain below, the function $C(x,t)$ will play the role of $|\partial_x u|$; not only
will it be a space-time smooth approximation, but it will moreover be an envelope for $|\partial_x u|$, maintaining its highly localized 
properties while retaining a certain memory of the behavior of the shock wave.
%

\subsection{Stabilizing shock collision}\label{intro:shock-collision}
In the context of fixed-grid, explicit,  finite-difference schemes, shock-wall collision and bounce-back
leads to egregious oscillatory behavior. This is primarily due to the fact that the shock-wall collision causes an 
immediate change in the sign of the shock speed $\dot{\sigma}(t)$, leading to a discontinuity in 
$\dot{\sigma}(t)$. Consequently, shock-wall collision is a highly singular phenomenon that requires explicit
stabilization. In \S\ref{sec:wallvisc}, we introduce a simple modification of the $C$-method, which we call the wall $C$-method, 
that implements a space-time smooth stabilization for shock-wall collision. 
This method is then applied to various test cases in \S\ref{sec:simulations}, with the computed solutions showing excellent agreement with the exact solution post shock-wall collision. 
Error analysis and convergence tests show that the wall $C$-method produces solutions with much 
smaller errors, even
for the difficult LeBlanc shock tube problem. 

\subsection{High-frequency noise}
The occurrence of high-frequency, often small amplitude, spurious oscillations (or \emph{noise}) is a common
 issue in numerical schemes. One cause of this noise is related to the stability (CFL) condition for explicit
 time-integration methods. A simple method for suppressing such noise is the use of the linear viscosity 
 \eqref{linear-viscosity}, though, as explained above, this often results in the degradation of the solution 
 in regions without noise.
  An alternative is to first decompose the solution using a basis of orthogonal \emph{wavelets}, then 
 truncate the decomposition so as to remove the high frequency components (which correspond with noise), 
 though this may be very computationally expensive and, moreover, 
 requires the use of a fully orthogonal basis of 
 wavelets. In \S\ref{sec:noiseind}, 
 we introduce a hybridized version of the two above methods, wherein wavelets are used 
 to accurately locate high frequency noise, and then a linear viscosity is used, via a localized heat equation
 solver, to remove this noise. This noise detection and removal algorithm is very simple 
 to implement, and is applied to a number of test problems in \S\ref{sec:simulations}. 
 Error analysis shows that the algorithm improves the accuracy of the solution while retaining the order of 
 convergence; in particular, the algorithm is able to suppress high-frequency noise while preserving the 
 amplitude of lower frequency (physical) sinusoidal waves for the Osher-Shu problem. 

\subsection{Outline of the paper}

In \S\ref{subsec:compEuler}, we introduce the compressible Euler equations, the corresponding flux, and the Rankine-Hugoniot jump
conditions. In \S\ref{subsec:review},  we review the original $C$-method, as introduced in  \cite{ReSeSh2012}. 
Then in \S\ref{sec:wallvisc}, we discuss  the problem of shock-wave wall collision, and introduce 
a novel generalization of the $C$-method, which  relies on a new artificial {\it wall viscosity} mechanism
 that suppresses post shock-collision oscillations.  We then introduce our WENO-$C$-$W$ scheme as a discretized version of our
 new $C$-method for shock-wall collision.
In \S\ref{sec:noiseind}, we present a wavelet based \emph{noise indicator}, that locates regions of 
noise containing high frequency oscillations on the discretized domain. A noise removal algorithm, based on a localized
solution of
the heat equation, is then used to remove high frequency oscillations.   We then describe our WENO-$C$-$W$-$N$ algorithm which
adds the noise indicator and noise removal scheme to our WENO-$C$-$W$ method.   Finally, in \S\ref{sec:simulations}, we demonstrate
the efficacy of our method for a number of classical shock tube problems, including the Sod, LeBlanc, Peak, and Osher-Shu tests.   We show
numerical results and order-of-accuracy studies, and in the process, we explain the cause and solution to the wall-heating problem.
In Appendix \ref{sec:appendix},  we describe two WENO schemes that we use for comparison purposes: the first couples WENO with classical
artificial viscosity, while the second couples WENO with Noh's artificial viscosity operator, designed specifically for the 
case of shock-wall collision and the wall-heating phenomenon.

\section{The compressible Euler equations and the original $C$-method} 

\subsection{The conservation laws of gas dynamics}
\label{subsec:compEuler}

The  compressible Euler equations on a 1-$D$ spatial interval $ x_1 \leq x  \leq x_M$, 
 and a time interval $0 \le t \le T$  
are written in vector-form as the following coupled system of nonlinear conservation laws:
 \begin{subequations}
\label{subeq:consLaw}
\begin{alignat}{2}
\partial_t {\bm u}(x,t)+ \partial_x {\bm F}({\bm u}(x,t))  = \bm{0},& && \ \ \  x_1< x< x_M \,,   t > 0, \label{eqn:consLawEvolution} \\
{\bm u}(x,0)  = {\bm u}_0(x),& && \ \ \ x_1 \le x \le x_M \,,   t = 0, \label{eqn:consLawIC}
\end{alignat}
\end{subequations}
where the $3$-vector ${\bm u}(x,t)$ and {\it flux function} ${\bm F}({\bm u}(x,t))$ are defined, respectively, as 
$$
{\bm u} = \left ( \begin{array}{c} \rho \\ \rho u \\ E \end{array} \right ) \quad \text{and} \quad {\bm F}({\bm u}) = \left ( \begin{array}{c} \rho u \\ \rho u^2 + p \\ u \left ( E + p \right ) \end{array} \right )\,,  
$$
 while the given initial data for the problem is
$$
{\bm u}_0(x) = \left (\begin{array}{c} \rho_0(x) \\  (\rho u)_0(x) \\ E_0(x) \end{array}\right )  \,.
$$
The {\it{conservative variables}} $\rho, \ \rho u$, and $E$ denote 
the {\it density},  {\it momentum}, and {\it energy} of a compressible gas, 
while the variable $u$ represents the {\it{velocity field}}.
The variable $p$ denotes the {\it pressure} function, and according to the
ideal gas law  is given by
\begin{equation}\label{eq-of-state}
p = (\gamma - 1) \left ( E - \frac{1}{2} \rho u^2 \right )\,,
\end{equation}
where $ \gamma $ is  the adiabatic constant.    
Equations (\ref{subeq:consLaw}) represent the conservation
of mass, linear momentum, and energy in the evolution of a compressible gas.   

The total energy per unit volume $E$ is the sum of 
kinetic energy and the potential energy, 
\begin{equation}\label{energy_identity}
E= \underbrace{\mystrut{2.5ex} \frac{1}{2}\rho u^2}_{\text{kinetic}} + \underbrace{\mystrut{2.5ex} \frac{p}{\gamma -1}}_{\text{potential}} \,.
\end{equation} 
We also define the {\it{specific internal energy per unit mass}} of the system $e$, defined as 
\begin{equation}\label{defn-internal-energy}
e = \frac{p}{(\gamma-1)\rho}\,, 
\end{equation}
so that the total energy of the system may be written as the sum of the kinetic energy and the internal energy
per unit volume $\rho e$, 
$$
E=\frac{1}{2}\rho u^2 + \rho e\,.
$$
The gradient of the flux ${\bm F}({\bm u})$ is given  by
$$
\mathrm{D}{\bm F}({\bm u}) =
\left[
\begin{array}{ccc}
0 & 1 & 0 \\[0.5em]
\frac{1}{2} (\gamma -3) u^2 & (3-\gamma) u & \gamma -1 \\[0.5em]
- \gamma \frac{u E }{\rho} + (\gamma -1) u^3 &
          \frac{ \gamma E}{\rho} + \frac{3}{2}(1- \gamma ) u^2 & \gamma u
\end{array}\right]
$$
with eigenvalues
\begin{subequations}
\label{subeq:maxWaveSpeed}
\begin{equation*}
  \lambda_1 = u + c\,, \ \lambda_2 = u\,,  \ \lambda_3 = u - c \,,
\end{equation*}
\end{subequations}
where $c =\sqrt{\gamma p /\rho }$ denotes the sound speed
(see, for example, Toro \cite{Toro2009}).   These eigenvalues determine the wave speeds. Since the behavior of the various wave patterns is greatly influenced by the speed of propagation, we define
the {\it maximum wave speed} $S(\bm{u})$ as
\begin{equation}\label{wave-speed}
S(\bm{u}) = [S({\bm u})](t) = \max_{i =1,2,3} \max_{x} \left \{ |\lambda_i(x,t)| \right \} \,.
\end{equation}

We are interested in solutions $\bm{u}$ with discontinuous wave profiles, 
such as those with shock waves and contact discontinuities.       
The  Rankine-Hugoniot  (R-H) conditions
determine the speed $\dot \sigma = \dot \sigma(t)$ of the moving shock or contact discontinuity, 
and represent conservation of mass, linear momentum and energy across the 
discontinuity (see, for example, \cite{Leveque2002}).  For a shock wave, the R-H condition is given by the relation
$$
F(\bm{u} _l) - F(\bm{u} _r) = \dot \sigma ( \bm{u}_l - \bm{u}_r)
$$
where the subscript $l$ denotes the state to the left of the discontinuity, and the 
subscript $r$ denotes the state to the right of
the discontinuity.  This means that the following three {\it jump conditions} must hold:
\begin{subequations}\label{RHconditions}
\begin{align}
\left( \rho_l u_l \right) - \left(\rho_r u_r \right) & = \dot \sigma (\rho_l - \rho_r) \label{RH1} \\
\left(\rho_l u_l^2 + p_l \right) - 
\left( \rho_r u_r^2 + p_r \right) & = \dot \sigma \left( \left( \rho u\right)_l - \left(\rho u\right)_r \right) \label{RH2} \\
\left(u_l (E_l + p_l)  \right) - \left( u_r (E_r + p_r)  \right) &= \dot \sigma (E_l -E_r) \,. \label{RH3}
\end{align} 
\end{subequations}
Uniqueness for weak solutions that have jump discontinuities in general  does not hold, 
unless entropy conditions are satisfied
(see the discussion in \S2.9.4 in \cite{ReSeSh2012}).    However,  solutions obtained in the limit of zero viscosity 
are known to satisfy the entropy condition
and are hence unique.  We refer the reader to \cite{ReSeSh2012} for a discussion of the 
convergence of $C$-method solutions as $ \Delta x \to 0$.
%

\subsection{A review of the original $C$-method}\label{subsec:review}

We now briefly review the $C$-method from  \cite{ReSeSh2012}, which is a spacetime smooth version of classical artificial viscosity
 with a  \emph{compression switch}:
\begin{equation*}
\beta (\Delta x)^2 \, \partial_x \Big( \mathbbm{1}_{(-\infty,0)} (\partial_x u) \, |\partial_x u| \,\partial_x u \Big)\,,
\end{equation*}
where the compression switch $\mathbbm{1}_{(-\infty,0)} (\partial_x u)$ ensures that artificial viscosity is only activiated
during  compression, and not in regions of expansion where there are no shocks. 


The localizing function $C(x,t)$ is given as the solution to the
scalar reaction-diffusion equation\footnote{We note that this scalar reaction-diffusion equation 
is not Galilean invariant. In the current implementation, the $C$-method is viewed purely as a numerical 
tool, whereas the function $C$ may very well be viewed as an important physical quantity, in which case 
the $C$-equation itself should be preserved under Galilean transformations. This can be accomplished by 
the addition of an advection term to the current $C$-equation. We have checked for some 1-$D$ examples 
that the addition of such a term has little effect on the demonstrated success of the $C$-method.}
\begin{equation*}
\partial_t C + \frac{S(\bm{u})}{\Delta x} C - S(\bm{u}) \Delta x \, \partial_{xx} C = \frac{S(\bm{u})}{\Delta x} \, G\,,
\end{equation*}
where the forcing $G$ is 
\begin{equation}\label{C-forcing}
G \equiv G(x,t) = \mathbbm{1}_{(-\infty,0)} (\partial_x u) \, \frac{ | \partial_x u | }{ \max_{x} | \partial_x u |}\,,
\end{equation}
and $S(\bm{u})$ is the maximum wave speed \eqref{wave-speed}. 
The $C$-method artificial viscosity term is then given by 
\begin{equation*}
\tilde{\beta} (\Delta x)^2 \, \partial_x \left( C \, \partial_x u \right)\,, \text{ where } \tilde{\beta} = \frac{\max_{x} | \partial_x u |}{\max_{x} C} \, \beta\,,
\end{equation*}
and the compressible Euler equations coupled with the $C$-method are written as the following Euler-$C$
system:
\begin{subequations}
\label{C-method-1d}
\begin{align} 
\p_t \rho + \p_x (\rho u) & = 0\,, \\
\p_t (\rho u) + \p_x (\rho u^2 + p) & =  \tilde{\beta} (\Delta x)^2 \, \partial_x \left( \rho C \, \partial_x u \right)\,, \\
\p_t E + \p_x (u(E+p)) &=  \tilde{\beta} (\Delta x)^2 \, \partial_x \left( \rho C \, \partial_x (E/\rho) \right) \,, \\
\p_t C + \frac{S(\bm{u})}{\Delta x} C - S(\bm{u}) \Delta x \, \partial_{xx} C &= \frac{S(\bm{u})}{\Delta x} \, G(\partial_x u) \,.
\end{align} 
\end{subequations}
Solutions of the Euler-$C$ equations \eqref{C-method-1d} converge to solutions of the Euler equations \eqref{subeq:consLaw} as $\beta \to 0$ (see
Section 2.9 in \cite{ReSeSh2012} for a proof).  
As was demonstrated in \cite{ReSeSh2012}, a simple WENO-type numerical
discretization of the Euler-$C$ equations \eqref{C-method-1d}   (as
will be described in \S\ref{sec:wallvisc})  is an effective high-order scheme which compares  favorably  to the best state-of-the-art algorithms
for the classical shock-tube experiments of  
Sod, Osher-Shu,  Woodward-Colella, and 
LeBlanc.
 In particular,  this simple WENO-type discretization of the  $C$-method is able to remove the large
overshoot in the LeBlanc contact discontinuity for the internal energy function (see \cite{ReSeSh2012}), whereas the other state-of-the-art schemes were not able
to do so.

Herein, we  generalize the $C$-method to allow for
shock-wave wall collision and bounce-back, and  introduce a wavelet-based \emph{noise indicator} algorithm  that locates high-frequency
noise;  a heat equation-based local solver will be used for noise removal.
We shall also explain the well-known problem of wall-heating (see, for example, \cite{Rider2000,Noh1987}).

\section{A new  $C$-method for shock-wall collision}\label{sec:wallvisc}
We now consider the highly singular problem of shock-wall collision and bounce-back, and specifically, the removal of spurious post collision
oscillations.

\subsection{The Euler-$C$-$W$ system}
As a generalization to the Euler-$C$ system \eqref{C-method-1d}, we consider the following coupled Euler-$C$-$W$ system:
\begin{subequations}\label{EulerC}
\begin{align}
\partial_t \rho + \partial_x (\rho u) &=0, \label{EulerC-density}\\ 
\partial_t (\rho u) + \partial_x (\rho u^2 + p) &=  \partial_x \left(  \mathcal{B} ^{(u)}(t) \, \rho \, C \, \partial_x u \right), \label{EulerC-momentum}\\
\partial_t E + \partial_x (u(E + p)) &=  \partial_x \left( \mathcal{B} ^{(E)}(t) \, \rho \, C \, \partial_x (E / \rho) \right) \,, \label{EulerC-energy}  \\
\partial_t C + \frac{S(\bm{u})}{\varepsilon \Delta x} C - \kappa \Delta x \cdot S(\bm{u}) \partial_{xx} C &
 = \frac{S(\bm{u})}{\varepsilon \Delta x} G  \,,  \label{C-Sod} \\
 \partial_t C_w + \frac{S(\bm{u})}{\varepsilon_w \Delta x} C_w - \kappa_w \Delta x \cdot S(\bm{u}) \partial_{xx} C_w
 & = \frac{S(\bm{u})}{\varepsilon_{w} \Delta x} G \,, \label{Cwall-sod}
\end{align}
\end{subequations}
where
\begin{subequations}
\label{artificial_visc}
\begin{align} 
\mathcal{B} ^{(u )}(t) & = (\Delta x)^2 \cdot \frac{\max_{x} | \partial_x u| }{\max_{x} C} \left( \beta^u   + \beta^{u }_w \cdot \overline{C}_w(t)\right)\,, \\
\mathcal{B} ^{(E )}(t) & = (\Delta x)^2 \cdot \frac{ \max_{x} | \partial_x u| }{ \max_{x} C}
\left(\beta^{E} + \beta^{E}_w \cdot \overline{C}_w(t)\right) \,,
\end{align} 
\end{subequations}
and where
the smooth and localized bump function $\Cw(t)$ is defined as 
\begin{equation}\label{wall-ind-fn}
\overline{C}_w(t) = \frac{C_w(x_M,t)}{\max_{x} C_w(x,t)},
\end{equation}
and $x_M$ denotes the right boundary, where the shock-wall collision and bounce-back is 
assumed (for simplicity) to occur.
Furthermore, $S(\bm{u})$ is the maximum wave-speed \eqref{wave-speed}, and $G = G(x,t)$ is the forcing to the $C$-equation, defined by \eqref{C-forcing}. 
The indicator function $  \mathbbm{1}_{(-\infty,0)}(\partial_x u) $ is
the {\it compression switch}, in which $G$ is non-zero only if $\partial_x u < 0$.   For convenience, we list all of the
parameters and variables associated with the system \eqref{EulerC} in Table \ref{table:sod}.   We note that due to the presence of the
compression switch in the definition of $G$, we can instead define $G(x,t) = \mathbbm{1}_{(-\infty,0)} (\partial_x \rho) \cdot \frac{|\partial_x \rho|}{\max_{x} | \partial_x \rho|} $ and obtain identical results.\footnote{Indeed, this will be our strategy for the 2-$D$ $C$-method that we indroduce
in \cite{RaReSh2018b}.}

We shall explain the use of this new $C_w(x,t)$ function and the localized time-function $\overline{C}_w(t)$ below, when we present
the results of numerical experiments of shock-wall collision.

We remark that the artificial viscosity terms on the right-hand side of the momentum equation \eqref{EulerC-momentum} and  the
energy equation \eqref{EulerC-energy} ensure that the total energy is conserved; in particular, the solution $E(x,t)$ of \eqref{EulerC-energy} 
continues to obey the identity \eqref{energy_identity}.  For simplicity, we consider the case of periodic boundary conditions.  On the one
hand,  integration of
the energy equation \eqref{EulerC-energy} over the spatial domain $[x_1,x_M]$ shows that
$\frac{\mathrm{d}}{\mathrm{d}t} \int_{x_1}^{x_M} E\, \mathrm{d}x = 0$. 
On the other hand,  multiplying the momentum \eqref{EulerC-momentum} by $u$, integrating over the domain  $[x_1,x_M]$, utilizing the
conservation of mass
\eqref{EulerC-density} together with the pressure identity \eqref{eq-of-state}, and the energy equation\eqref{EulerC-energy}, we find that
$$\frac{\mathrm{d}}{\mathrm{d}t} \int_{x_1}^{x_M} \left( \frac{1}{2} \rho u^2 + \frac{p}{\gamma -1} \right) \, \mathrm{d}x = 0 \,.$$
 This shows that the velocity $u$ and pressure $p$ adjust accordingly to maintain the relation \eqref{energy_identity}, and that our
 modified Euler-$C$-$W$ system conserves total energy.

{
\begin{table}[H]
\centering
{\small
\begin{tabular}{|M{4cm} | M{6cm}|} 
 \hline
 Parameter / Variable & Description \\ [0.0em] 
 \hline \hline 
$\beta^{u}$, $\beta^E$ & artificial viscosity coefficients for the momentum 
 and energy, respectively. \\[0.5em] 
\hline
$\beta^{u}_w$, 
 $\beta^E_w$ &  wall viscosity coefficients for the
 momentum and energy, respectively. \\[0.5em] 
 \hline 
 $S(\bm{u})(t)$ & maximum wave speed $\max_{x} \left( \max \left\{ \, | u(x,t) | , | u(x,t) \pm c | \,\right\} \right) $. \\[0.5em] 
 \hline 
 $\varepsilon$, $\varepsilon_w$ & parameters controlling support 
 of $C$ and $C_w$, respectively. \\[0.5em]
 \hline
 $\kappa$, $\kappa_w$ & parameters controlling smoothness 
 of $C$ and $C_w$, respectively. \\[0.5em]
 \hline
 $\overline{C}_w(t)$ &  smooth and localized bump function. \\[0.5em]
 \hline
\end{tabular}}
\caption{Relevant parameters and variables for the Euler-$C$-$W$  system \eqref{EulerC}.}
\label{table:sod}
\end{table}}

\subsection{Boundary conditions for the Euler-$C$-$W$  system}
We consider two types of boundary conditions on the interval $x_1 \leq x \leq x_M$.  
 For many of the test problems, 
we employ the so-called {\it{reflective}} or {\it{solid wall}} boundary 
conditions at $x=x_1$ and $x=x_M$ and $t\ge 0$:
\begin{equation}
\label{var-bcs}
\partial_x \rho (x,t) = 0 \,, \ \ 
\rho u (x,t) =0 \,, \ \
\partial_x E (x,t) =0 \,, \ \  \partial_x C(x,t) = 0 \,, \ \ \partial_x C_w(x,t) = 0 \,.
\end{equation} 
Alternatively, we shall sometimes use the  {\it{free flow}} boundary conditions: 
\begin{equation} 
\label{var-bcs-alternate}
\partial_x \rho (x,t) = 0 \,, \ \ 
\partial_x\left(\rho u\right) (x,t) =0 \,, \ \
\partial_x E (x,t) =0 \,, \ \  \partial_x C(x,t) = 0 \,, \ \ \partial_x C_w(x,t) = 0 \,.
\end{equation}

\subsection{The WENO-$C$-$W$  algorithm}\label{sec-weno-reconstruction-procedure}
\subsubsection{Discretization of the Euler-$C$-$W$ system}
We now describe the simple WENO-based space discretization scheme used for the Euler-$C$-$W$ system \eqref{EulerC}.
We use a  formally fifth-order WENO reconstruction procedure together with upwinding, 
based on the sign of the velocity at the cell edges. We stress that the WENO-type discretization we 
use is highly simplified, and is not meant to be representative of the class of full WENO solvers. 
However, we note that, for certain problems, our simplified WENO-type discretization produces 
solutions with similar errors and convergence rates to those produced using a standard WENO scheme 
(see \S{\ref{sec:comparison-with-other-schemes}}).

The spatial domain $x_1\leq x \leq x_M$ is subdivided into $M$ equally sized cells of width $\Delta x$, where 
the left-most and right-most cells are centered on the left and right boundaries, respectively. We 
denote the cell centers by $x_i$ for $i =1,\ldots,M$, and the cell edges with the fractional index
$$
x_{i+\frac{1}{2}} = \frac{ x_i + x_{i+1}}{2} , \text{ for } i=1,\ldots,M-1 \,.
$$

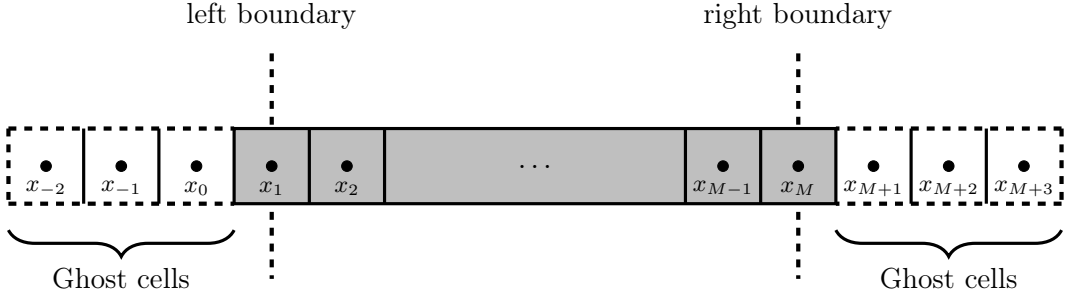
\begin{figure}\label{fig:cells}
\centering
\begin{tikzpicture}
\draw[ultra thick,dashed] (-3.5,3) -- (-3.5,0);
\draw[ultra thick,dashed] (3.5,3) -- (3.5,0);
\filldraw [lightgray] (-4,2) --(4,2) -- (4,1) -- (-4,1);
\draw[very thick] (-4,2) -- (4,2) -- (4,1)--(-4,1)--cycle;
\draw[ultra thick,dashed] (-7,2) -- (-4,2);
\draw[ultra thick,dashed] (7,2) -- (4,2);
\draw[ultra thick,dashed] (-7,1) -- (-4,1);
\draw[ultra thick,dashed] (7,1) -- (4,1);
\draw[ultra thick,dashed] (-7,2) -- (-7,1);
\draw[ultra thick,dashed] (7,2) -- (7,1);
\draw[very thick] (-5,2) -- (-5,1);
\draw[very thick] (-6,2) -- (-6,1);
\draw[very thick] (5,2) -- (5,1);
\draw[very thick] (6,2) -- (6,1);
\draw[very thick] (3,2) -- (3,1);
\draw[very thick] (-3,2) -- (-3,1);
\draw[very thick] (2,2) -- (2,1);
\draw[very thick] (-2,2) -- (-2,1);
\filldraw [black] (-6.5,1.5) circle (2pt);
\filldraw [black] (6.5,1.5) circle (2pt);
\filldraw [black] (-3.5,1.5) circle (2pt);
\filldraw [black] (3.5,1.5) circle (2pt);
\filldraw [black] (5.5,1.5) circle (2pt);
\filldraw [black] (-5.5,1.5) circle (2pt);
\filldraw [black] (4.5,1.5) circle (2pt);
\filldraw [black] (-4.5,1.5) circle (2pt);
\filldraw [black] (2.5,1.5) circle (2pt);
\filldraw [black] (-2.5,1.5) circle (2pt);
\node at (-6.5,1.2) {\footnotesize $x_{-2}$};
\node at (-5.5,1.2) {\footnotesize $x_{-1}$};
\node at (-4.5,1.2) {\footnotesize $x_0$};
\node at (-3.5,1.2) {\footnotesize $x_1$};
\node at (-2.5,1.2) {\footnotesize $x_2$};
\node at (2.5,1.2) {\footnotesize $x_{M-1}$};
\node at (3.5,1.2) {\footnotesize $x_M$};
\node at (4.5,1.2) {\footnotesize $x_{M+1}$};
\node at (5.5,1.2) {\footnotesize $x_{M+2}$};
\node at (6.5,1.2) {\footnotesize $x_{M+3}$};
\node at (0,1.5) {$\ldots$};
\node at (-3.5,3.5) {left boundary};
\node at (3.5,3.5) {right boundary};
\draw[very thick,decoration={brace,amplitude=10pt,mirror,raise=10pt},decorate]
  (-7,1) -- node[below=20pt] {Ghost cells} (-4,1);
  \draw[very thick,decoration={brace,amplitude=10pt,mirror,raise=10pt},decorate]
  (4,1) -- node[below=20pt] {Ghost cells} (7,1);
\end{tikzpicture}
\caption{The grid, together with ghost cells,  for the WENO-$C$-$W$ algorithm.}
\end{figure}

Any quantity evaluated at a cell center $x_i$ shall be denoted by $w_i$, and a quantity evaluated 
at a cell edge  $x_{i+\frac{1}{2}}$ is denoted by $w_{i+\frac{1}{2}}$.
Given a vector $w_i$ corresponding to cell-center values, and vectors $z_{i-\frac{1}{2}}$ and 
$z_{i+\frac{1}{2}}$ corresponding to cell edge  values, we define the $j^{\text{th}}$ component  
by
$$
\left[ \operatorname{WENO}(w_i,z_{i \pm \frac{1}{2}}) \right]_j = \frac{1}{\Delta x} \left( \tilde{w}_{j+\frac{1}{2}} z_{j+\frac{1}{2}} -  \tilde{w}_{j-\frac{1}{2}} z_{j-\frac{1}{2}} \right), 
$$
where the cell-edge values $\tilde{w}_{j+\frac{1}{2}}$ are calculated using a standard fifth-order 
WENO reconstruction procedure (see \cite{JiangShu1996}, \cite{Shu2003}) 
with upwinding based on the sign of $z_{j+\frac{1}{2}}$. 

Then,  defining the vectors $\bm{u} = [\rho,\rho u,E]^T$ and 
$\bm{C} = [C,C_w]^{T}$, we now construct the 
operators $\mathcal{A}_{\operatorname{WENO}}$ and $\mathcal{B}_{\operatorname{WENO}}$ as
\begin{equation}\label{A-weno}
\left[ \mathcal{A}_{\operatorname{WENO}}(\bm{u}_i,\bm{C}_i) \right]  = 
	\begin{bmatrix}
         \left[ \operatorname{WENO}\left( \rho_i, \hat{u}_{i \pm \frac{1}{2}} \right) \right]_i  \\[1.5em]
         \left[ \operatorname{WENO}\left( (\rho u )_i, \hat{u}_{i \pm \frac{1}{2}} \right) \right]_i + \tilde{\partial_4} p_i - \mathcal{B}^{(u)}(t) \cdot \frac{ \tilde{\partial_{C}} \left(u_{i+\frac{1}{2}}\right) - \tilde{\partial_{C}} \left( u_{i-\frac{1}{2}} \right) }{\Delta x} \\[1.5em]
         \left[ \operatorname{WENO}\left( (E + p )_i, \hat{u}_{i \pm \frac{1}{2}} \right) \right]_i - \mathcal{B}^{(E)}(t) \cdot \frac{ \tilde{\partial_{C}} \left( (E/\rho)_{i+\frac{1}{2}} \right) - \tilde{\partial_{C}} \left( (E/\rho)_{i-\frac{1}{2}} \right) }{\Delta x}
        \end{bmatrix} 
\end{equation}
and 
\begin{equation}\label{B-weno}
\left[ \mathcal{B}_{\operatorname{WENO}}(\bm{u}_i,\bm{C}_i) \right]  = 
	\begin{bmatrix}
	\frac{S(\bm{u}_i)}{\varepsilon \Delta x} \left\{ C_i - G_i \right\} + \frac{\tilde{\partial_S} C_{i+\frac{1}{2}} - \tilde{\partial_S} C_{i-\frac{1}{2}}}{\Delta x} \\[1.5em]
	\frac{S(\bm{u}_i)}{\varepsilon_{w} \Delta x} \left\{ \left[C_w\right]_i - G_i \right\} + \frac{\tilde{\partial_S} \left[ C_w\right]_{i+\frac{1}{2}} - \tilde{\partial_S} \left[ C_w\right]_{i-\frac{1}{2}}}{\Delta x}
         \end{bmatrix}. 
\end{equation}

Here, we have used the notation $\tilde{\partial_4} p_i$ to denote the fourth-order central 
difference approximation for the derivative of the pressure at the cell center $x_i$:
$$
\tilde{\partial_4} p_i = \frac{p_{i-2} - 8p_{i-1} + 8p_{i+1}-p_{i+2}}{12 \cdot \Delta x}.
$$
The cell-edge velocities $\hat{u}_{i \pm \frac{1}{2}}$ used for upwinding 
are calculated using a fourth-order averaging:
$$
\hat{u}_{i-\frac{1}{2}} = \frac{-u_{i-2} + 7u_{i-1} + 7u_i - u_{i+1}}{12} \,.
$$
We have also used the notation $\tilde{\partial_C} \left( w_{i+\frac{1}{2}} \right)$ and $\tilde{\partial_S} C_{i+\frac{1}{2}}$ to denote
\begin{align*}
\tilde{\partial_C} \left( w_{i+\frac{1}{2}} \right) &= \rho_{i+\frac{1}{2}}  C_{i+\frac{1}{2}} \tilde{\partial} w_{i+\frac{1}{2}}, \\
\tilde{\partial_S} C_{i+\frac{1}{2}} &= \kappa \Delta x \,  S(\bm{u}_i) \, \tilde{\partial}C_{i+\frac{1}{2}},
\end{align*}
respectively. Here, the notation $z_{i+\frac{1}{2}}$ denotes a quantity calculated at the 
cell edge $x_{i+\frac{1}{2}}$ using the standard averaging
$$
z_{i+\frac{1}{2}} = \frac{z_i + z_{i+1}}{2},
$$
while the quantity $\tilde{\partial} w_{i+\frac{1}{2}}$ denotes the central difference 
approximation for $\partial_x w$ at the cell edge $x_{i+\frac{1}{2}}$, 
$$
\tilde{\partial} w_{i+\frac{1}{2}} = \frac{w_{i+1}-w_i}{\Delta x}. 
$$

Now, given $\bm{u}^n$ at a time $t = t_n = n \Delta t$, we evolve the solution as follows:
\begin{subequations}\label{Euler-semi-discrete}
\begin{align}
\bm{{u}}_i^{n+1} &= \text{RK}\left( \bm{{u}}_i^{n}, \mathcal{A}_{\operatorname{WENO}}(\bm{{u}}_i^{n},\bm{{C}}_i^n) \right) \,, \\
\bm{{C}}_i^{n+1} &=\text{RK}\left( \bm{{C}}_i^{n} ,\mathcal{B}_{\operatorname{WENO}}(\bm{{u}}_i^{n},\bm{{C}}_i^n) \right)  \,,
\end{align}
\end{subequations}
where RK denotes the explicit fourth-order Runge-Kutta  time-integration method.

\subsubsection{Discretization of boundary conditions and ghost node values} 
Boundary conditions for the functions $C$ and $C_w$ are imposed through the assigning of the so-called \textit{ghost node} values. More precisely, the ghost node values for the functions $C$ and $C_w$ are prescribed via an even extension:
\begin{equation}\label{C-ghost-node}
C_{1-k} = C_{1+k}  \quad  \text{and}   \quad C_{M+k} = C_{M-k},
\end{equation}
for $k=1,\ldots,M_g$, where $M_g$ is the number of ghost nodes. For our (formally) fifth-order 
WENO scheme, $M_g = 3$.

The associated boundary conditions for the conservative variables are also imposed via the {ghost node} 
conditions. For the Dirichlet boundary condition, an odd extension is used, while for the  Neumann boundary condition, an even extension is used. More precisely, suppose that we wish to impose the free-flow boundary 
conditions \eqref{var-bcs-alternate}. This is done by choosing the ghost node values as
\begin{subequations}\label{ghost-node-alternate}
\begin{alignat}{3}
\rho_{1-k} &= \rho_{1+k} \quad &&\text{and} & \quad \rho_{M+k} &= \rho_{M-k},\\
\rho u_{1-k} &= \rho u_{1+k} \quad &&\text{and} & \quad \rho u_{M+k} &= \rho u_{M-k},\\
E_{1-k} &= E_{1+k} \quad &&\text{and} & \quad E_{M+k} &= E_{M-k},
\end{alignat} 
\end{subequations}
for $k=1,\ldots,M_g$.

The solid wall boundary conditions \eqref{var-bcs} are imposed
by replacing the even extension of the momentum in \eqref{ghost-node-alternate} with the odd extension 
of the momentum: 
\begin{subequations}\label{ghost-node}
\begin{alignat}{3}
\rho_{1-k} &= \rho_{1+k} \quad &&\text{and} & \quad \rho_{M+k} &= \rho_{M-k},\\
\rho u_{1-k} &= -\rho u_{1+k} \quad &&\text{and} & \quad \rho u_{M+k} &= -\rho u_{M-k},\\
E_{1-k} &= E_{1+k} \quad &&\text{and} & \quad E_{M+k} &= E_{M-k},
\end{alignat} 
\end{subequations}
for $k=1,\ldots,M_g$. Again, it is easy to verify that the density $\rho$ and the energy $E$ satisfy 
the homogenous Neumann boundary condition in \eqref{var-bcs}. To verify that 
the momentum satisfies the homogenous Dirichlet boundary condition, we 
need to use the momentum equation in the semi-discrete form \eqref{Euler-semi-discrete}. Suppose that at time-step $n$, the velocity at the boundaries vanishes: 
$u_{M}^n = u_1^n = 0$.
For simplicity, we restrict to the right boundary in cell $x_M$. 
The even extensions of $\rho$ and $C$ and the 
odd extension of $u$ mean that the diffusion term on the right-hand side of the momentum equation vanishes 
since 
\begin{align*}
\tilde{\partial_{C}} \left( u_{M+\frac{1}{2}} \right) &= \rho_{M+\frac{1}{2}} \cdot C_{M+\frac{1}{2}} \cdot \frac{\left(u_{M+1}-u_M\right)}{\Delta x} \\
&=  \rho_{M-\frac{1}{2}} \cdot C_{M-\frac{1}{2}} \cdot \frac{\left(-u_{M-1}+u_M\right)}{\Delta x} 
=\tilde{\partial_{C}} \left( u_{M-\frac{1}{2}} \right).
\end{align*}
Moreover, since the pressure $p$ is evenly extended, the derivative at the 
boundaries $\tilde{\partial_4} p_M$ and $\tilde{\partial_4} p_1 $, also vanishes. One can also check 
that the derivative of the flux term at the boundary vanishes:
$
\left[ \operatorname{WENO}\left( (\rho u)_i, \hat{u}_{i\pm\frac{1}{2}} \right) \right]_M = 0$.
This means that $\partial_t (\rho u) = 0$ at the boundaries, so that momentum satisfies
$\rho u = 0$ at the boundaries for $t\ge 0$, 
provided that the initial momentum vanishes on the boundaries.

\subsection{Using WENO-$C$-$W$ for the Sod shock-wall collision problem}
The reflection of a shock wave from a fixed wall was first considered in \cite{courant1999supersonic}  
from a theoretical viewpoint (see also \cite{Alpher1954,meyer1957}).  
Further investigations in 
\cite{Noh1987,DONAT199642} were done primarily in the context of the 
wall-heating phenomenon (to be discussed below).
The reflection of a shock-wave from a non-rigid boundary was considered in 
\cite{mazor1992head,Igra1992}, wherein an artificial viscosity method was utilized to stabilize the solution. 

\begin{figure}[H]
\centering
\subfigure[$t=0.20$]{\label{fig:sod-before-collision1}\includegraphics[width=75mm]{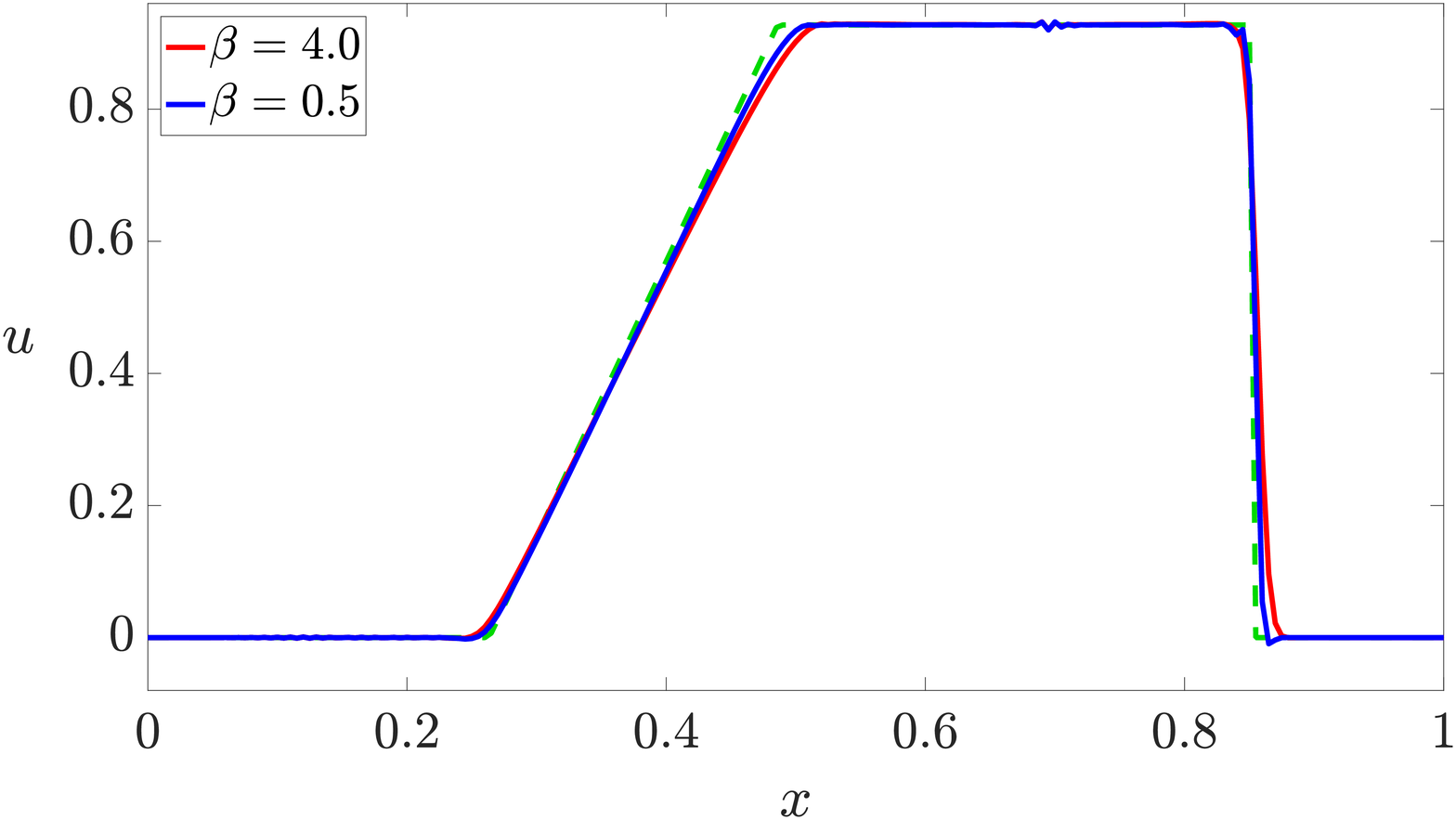}}
\subfigure[$t=0.36$]{\label{fig:sod-after-collision1}\includegraphics[width=75mm]{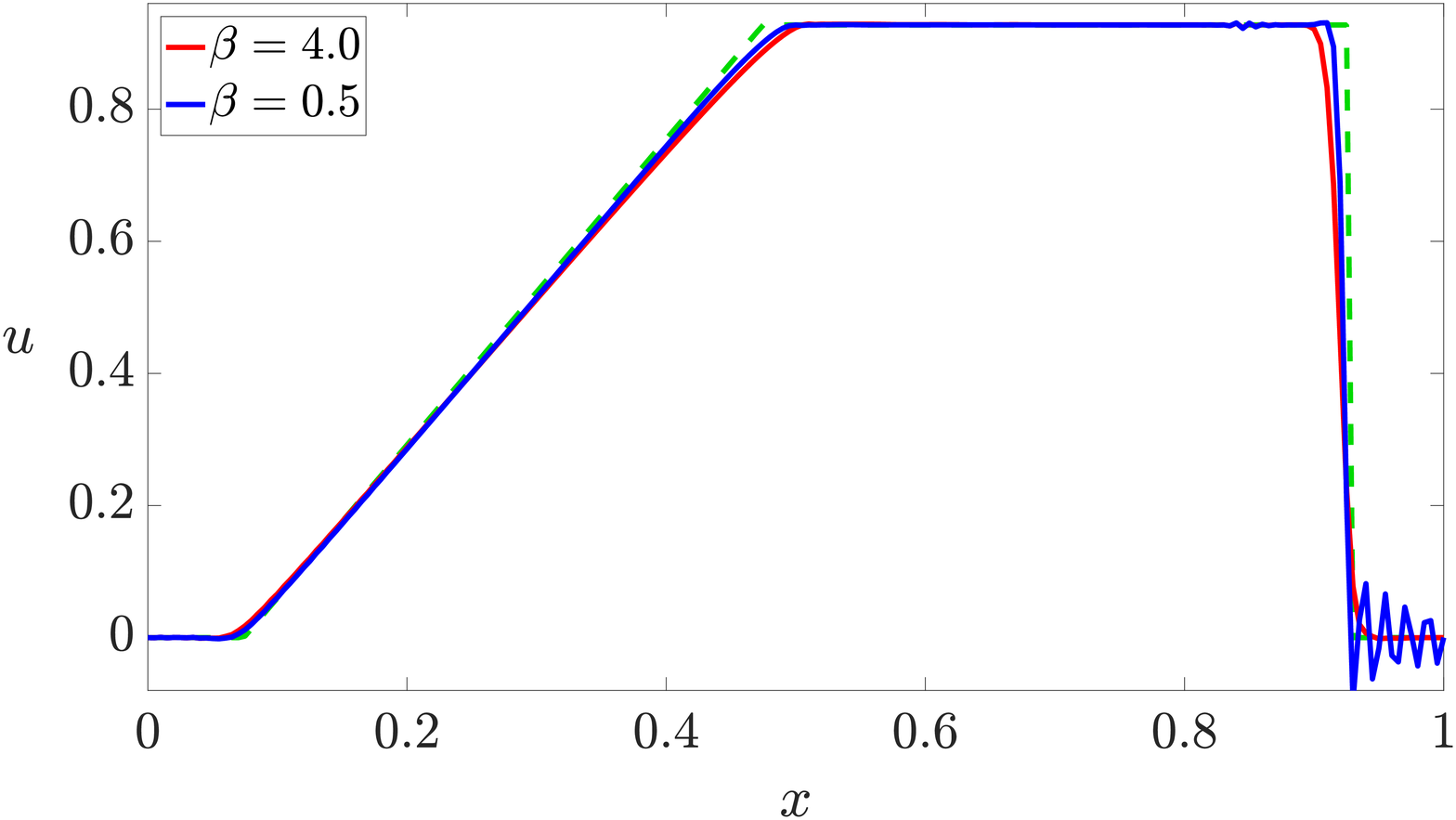}}
\caption{The velocity profile for the Sod shock tube problem, calculated using our 
WENO-$C$ scheme with 201 cells. The blue and red curves are the velocity profiles and the dashed 
green curve is the exact solution.}
\end{figure}

As a motivating example, we first consider the classical Sod shock tube experiment. This is a Riemann problem
on the domain $0 \leq x \leq 1$, with initial data given by 
\begin{equation}\label{sod_initialdata}
\begin{bmatrix}
\rho_0 \\ (\rho u)_0 \\ E_0 
\end{bmatrix}
=
\begin{bmatrix}
1 \\ 0 \\ 2.5
\end{bmatrix}
\mathbbm{1}_{[0,\frac{1}{2})}(x)
+
\begin{bmatrix}
0.125 \\ 0 \\ 0.25
\end{bmatrix}
\mathbbm{1}_{[\frac{1}{2},1]}(x)  \text{ and } \gamma=1.4\,,
\end{equation}
where $\mathbbm{1}_{[a,b)}(x)$ denotes the indicator function on the interval $a \leq x < b$.
The solution consists of a rarefaction wave, a contact discontinuity, and a shock wave. 
The shock propagating to the right collides with the wall, modeled by the point $x=1$, at time $t \approx 0.28$.   
In Fig.\ref{fig:sod-before-collision1}, we show the success of the WENO-$C$ method
for this problem prior to the collision of the shock wave with the wall at $x=1$;  however, as shown 
in Fig.\ref{fig:sod-after-collision1},  the WENO-$C$ scheme (without the addition of the wall function 
$\overline{C}_w(t)$)
is not sufficient to remove spurious oscillations post shock collision in the case of small $\beta=0.5$. 
On the other hand, by setting
$\beta=4.0$,  the velocity  is mostly free of post shock-wall collision oscillations at $t=0.36$, at the expense of an overly diffused
shock profile prior to shock-wall collision at $t=0.2$.  Moreover, for more difficult problems, such as the
LeBlanc problem considered in \S\ref{subsec:leblanc}, very 
precise choices of the artificial viscosity parameters are
required to maintain stability and correct wave speeds. Consequently, it is difficult to choose $\beta$ such that 
the solutions both pre and post shock-wall collision are accurate and noise-free. 
The use of the wall viscosity will provide a nice solution strategy.

\subsubsection{An explanation of the temporal bump function $\Cw(t)$}

 We now explain the use of the new $C_w(x,t)$ function together with the
temporal bump function $\Cw(t)$.  We shall assume, for simplicity, that the shock wave 
is traveling to the right, so that the shock wave collides with the wall  $x = 1$.   
Thanks to the homogeneous Neumann 
boundary condition $\partial_x C_w=0$ at the wall $x=1$, there is a smooth 
 growth (in time) of the amplitude of $C_w(1,t)$ just prior to shock-wall collision, 
followed by a smooth decrease of amplitude during shock bounce-back.

\begin{figure}[H]
\centering
\subfigure[$t=0.200$]{\label{fig:wallvisc-expln1}\includegraphics[width=75mm]{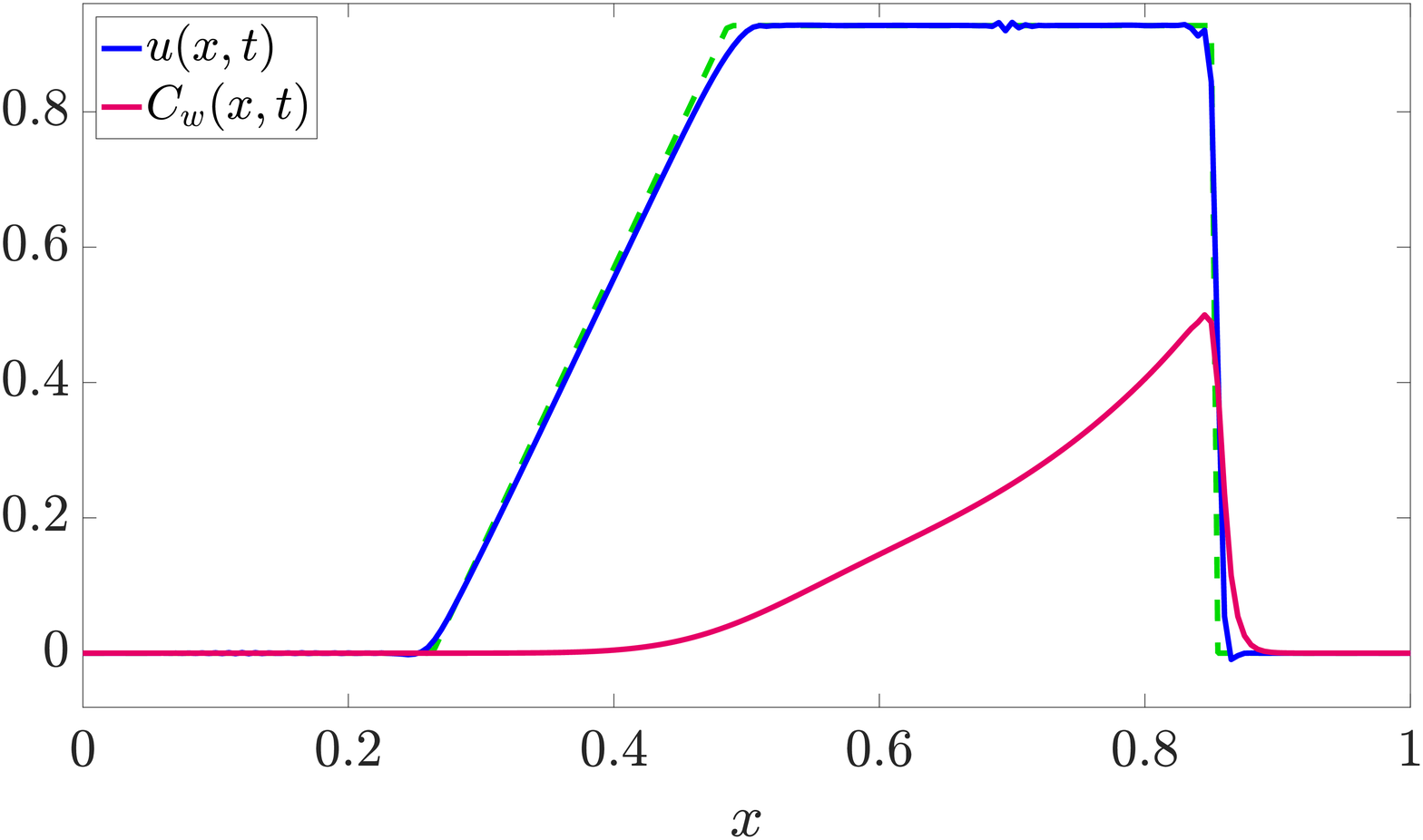}}
\subfigure[$t=0.272$]{\label{fig:wallvisc-expln2}\includegraphics[width=75mm]{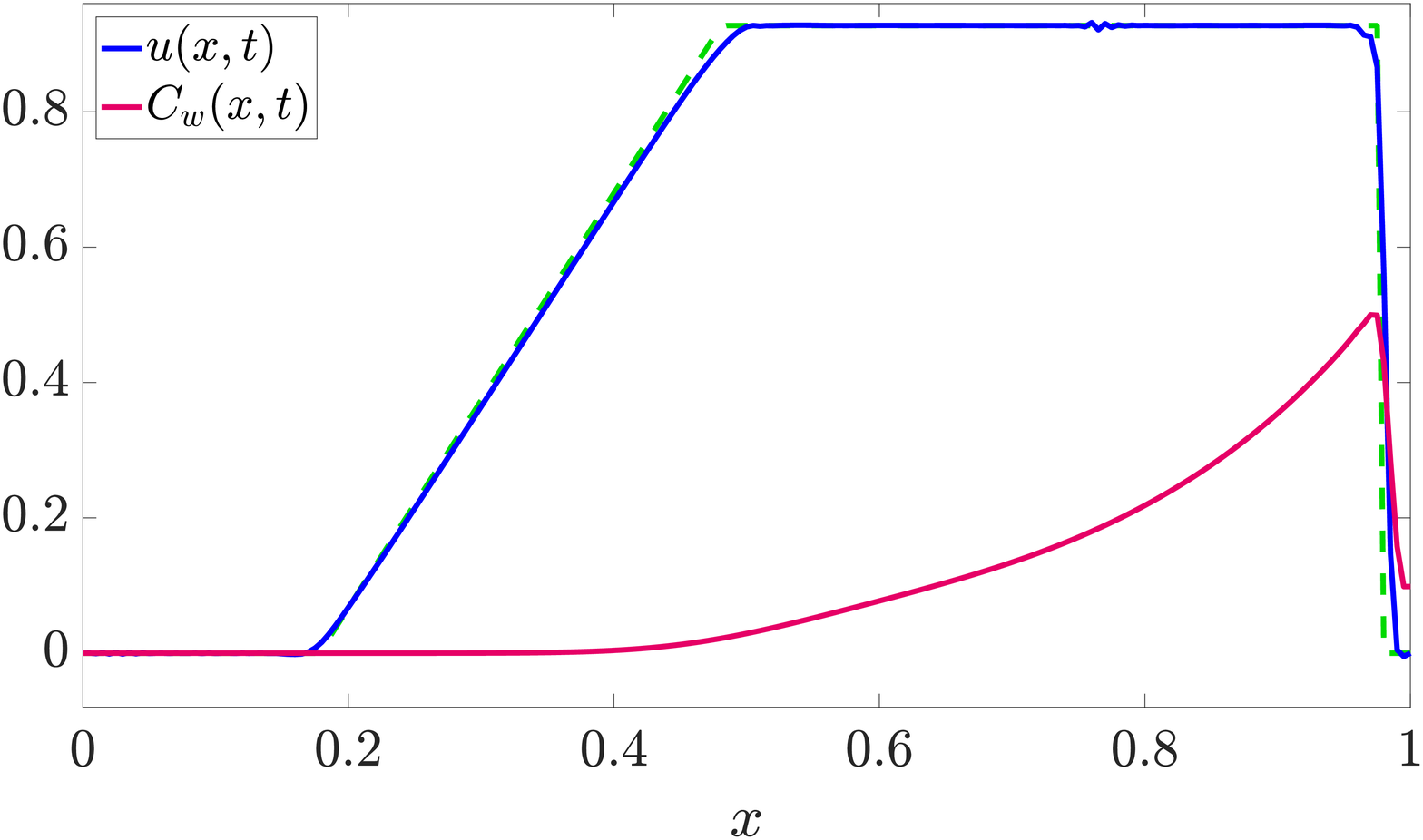}}
\subfigure[$t=0.296$]{\label{fig:wallvisc-expln3}\includegraphics[width=75mm]{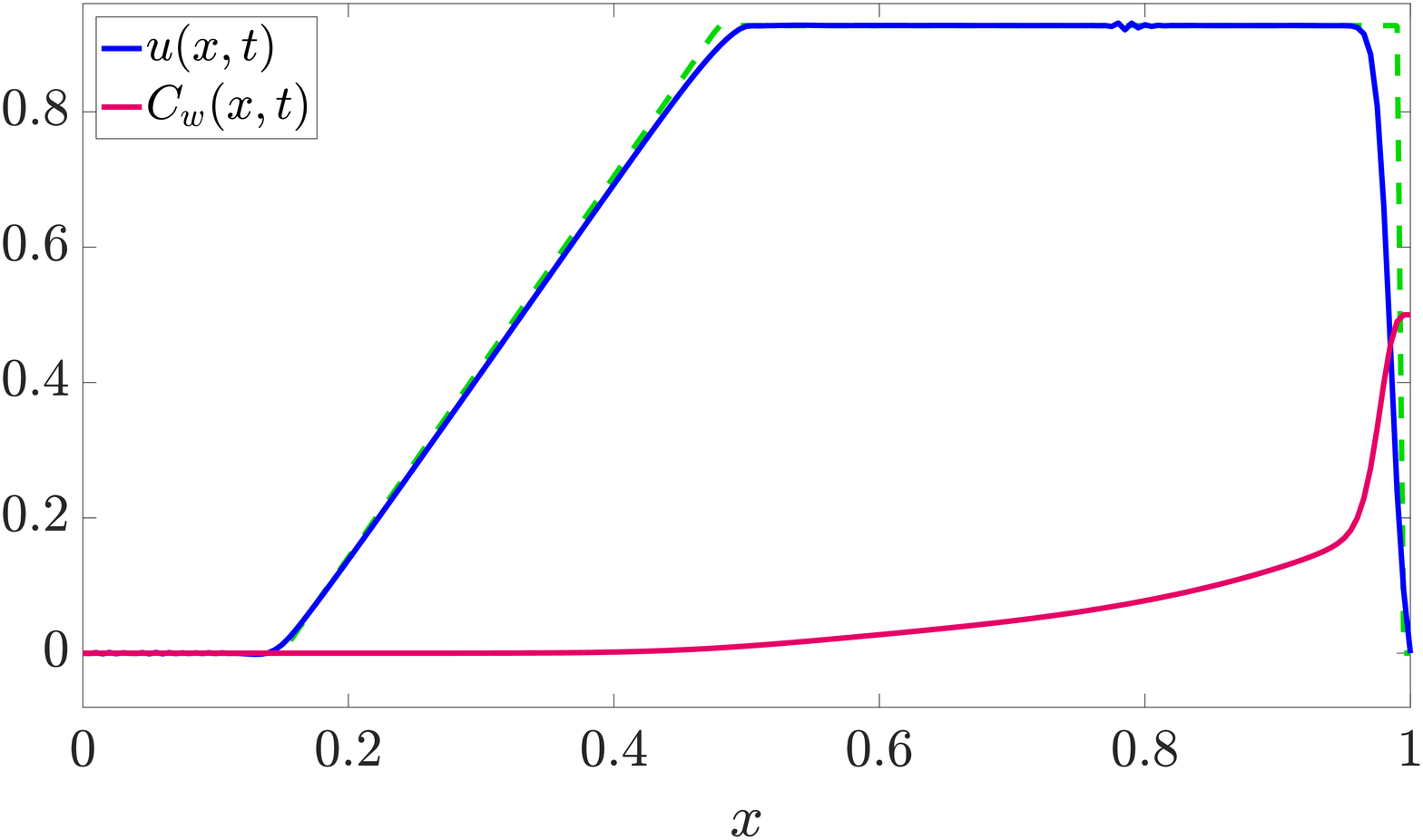}}
\subfigure[$t=0.360$]{\label{fig:wallvisc-expln4}\includegraphics[width=75mm]{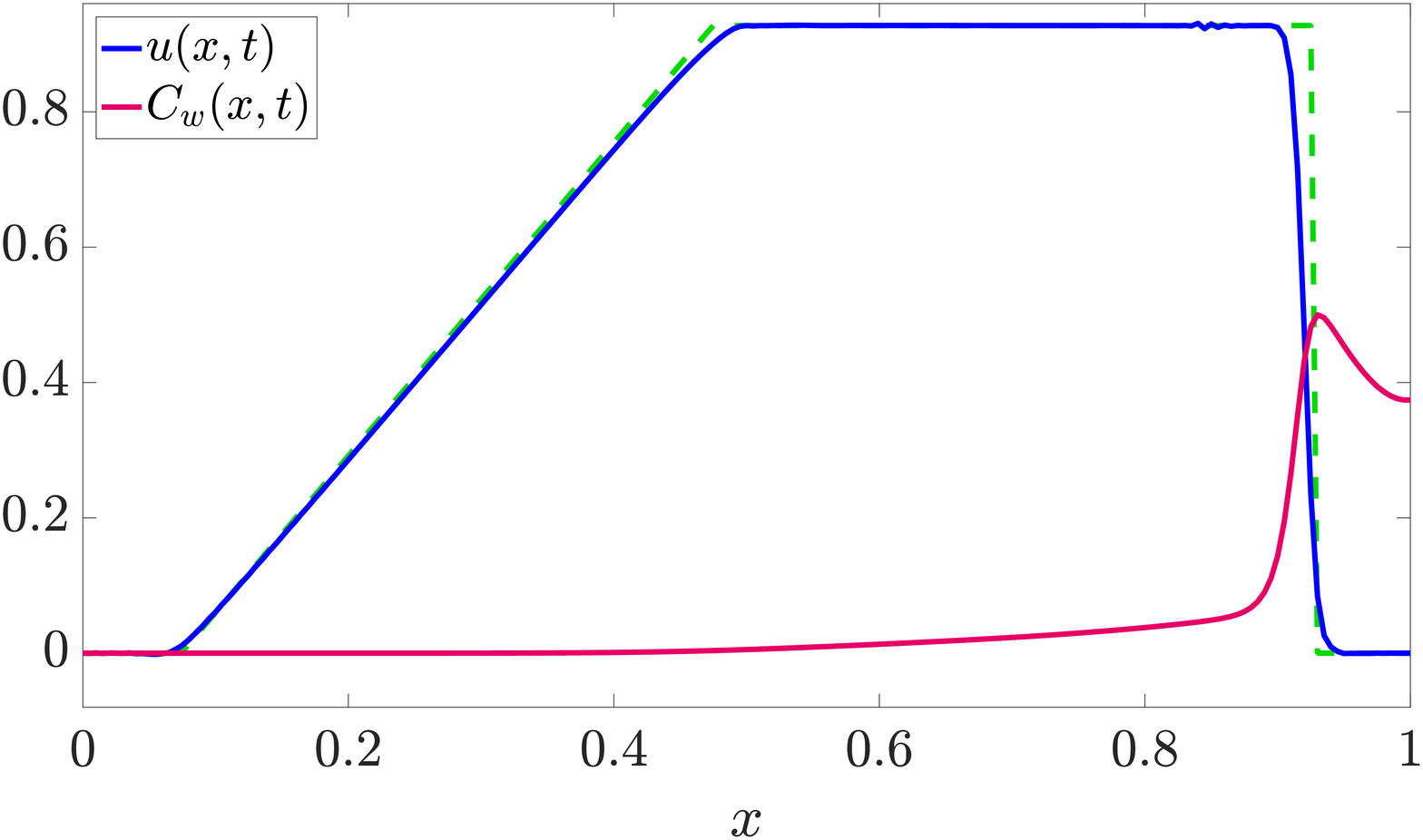}}
\caption{The velocity profile for the Sod shock tube problem, calculated using our 
WENO-$C$-$W$ scheme with 201 cells. The blue curve is the velocity profile and the dashed 
green curve is the exact solution. The red curve is the (normalized and resized) function 
$C_w(x,t)$.}
\label{fig:wallvisc-expln}
\end{figure}

In Fig.\ref{fig:wallvisc-expln}, we illustrate the
WENO-$C$-$W$ scheme as applied to Sod.
While the shock is away from the wall, $C_w(1,t)$ is zero, and 
thus by formula \eqref{wall-ind-fn} so is $\overline{C}_w(t)$; see
the purple curve in Fig.\ref{fig:wallvisc-expln1}.
As the shock approaches the wall (as shown in Fig.\ref{fig:wallvisc-expln2}), the 
Neumann boundary condition for the $C_w$-equation ensures that 
$\overline{C}_w(t)$ increases smoothly, until it reaches a maximum 
when the shock collides with the wall (Fig.\ref{fig:wallvisc-expln3}), before smoothly 
decreasing back to zero as the shock moves away from the wall (Fig.\ref{fig:wallvisc-expln4}). 

\begin{figure}[H]
\centering
\subfigure[$\overline{C}_w(t)$]{\label{fig:wallvisc-expln5}\includegraphics[width=75mm]{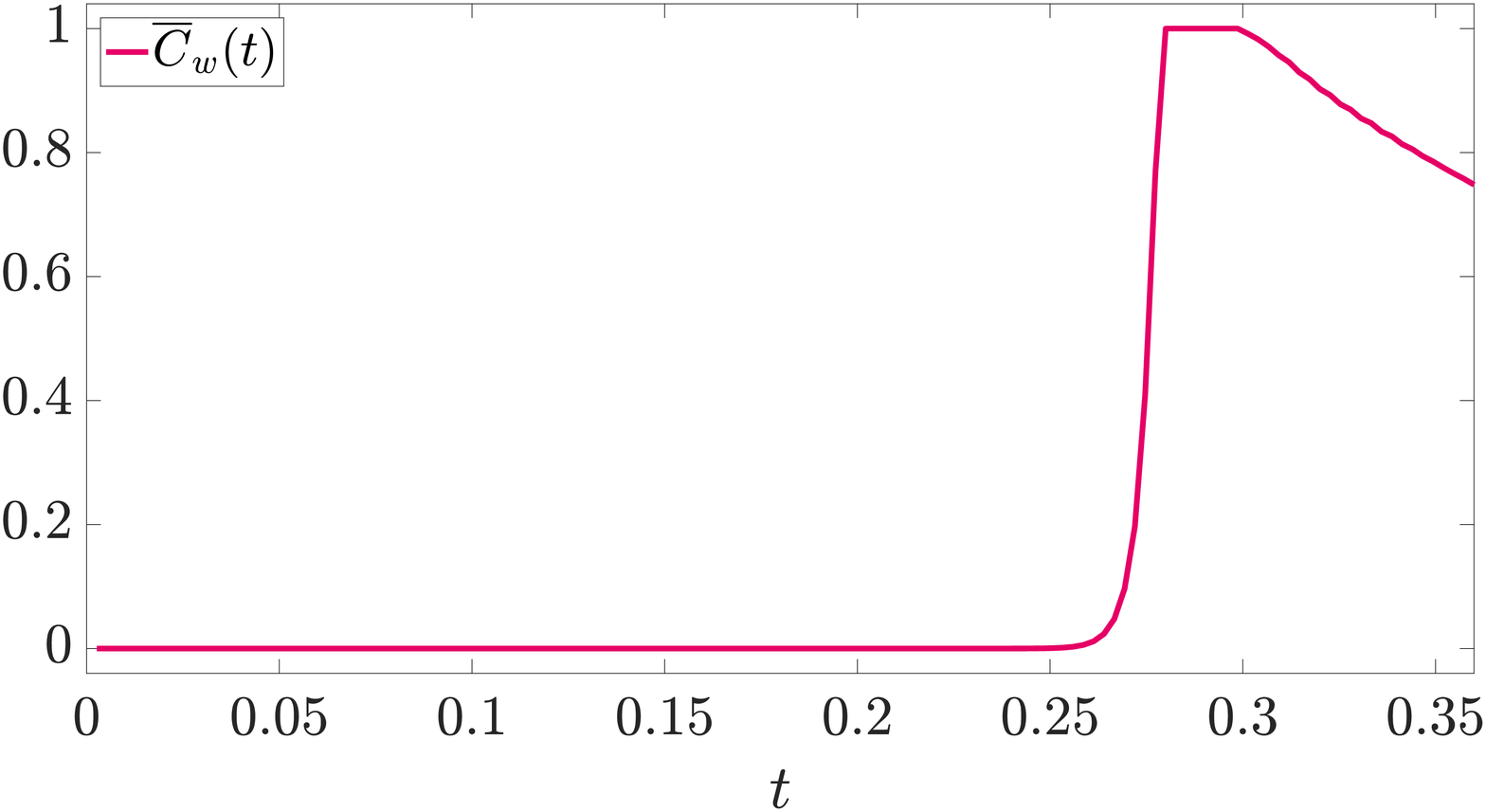}}
\subfigure[zooming in on $\overline{C}_w(t)$ during shock collision]{\label{fig:wallvisc-expln6}\includegraphics[width=75mm]{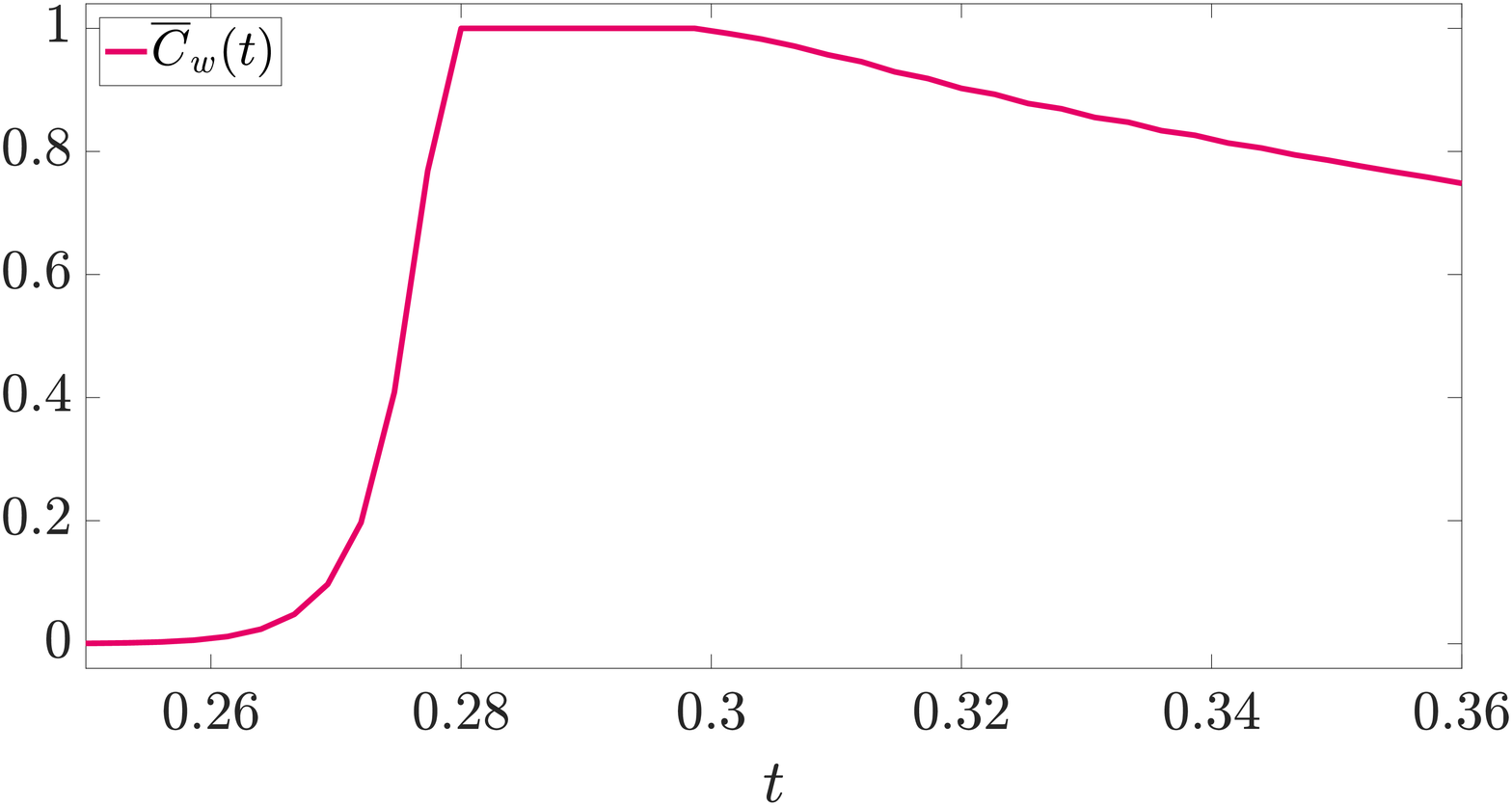}}
\caption{The wall indicator function $\overline{C}_w(t)$ for the 
Sod shock tube problem. The function is zero when the shock is away from the wall, increases 
smoothly as the shock approaches the wall, and reaches a maximum when the shock 
collides with the wall, before decreasing smoothly as the shock moves away from the wall.} 
\label{fig:wallvisc-expln-alt}
\end{figure}

In Fig.\ref{fig:wallvisc-expln-alt}, we plot the graph of $\overline{C}_w(t)$. 
The localized nature of the temporal bump function $\overline{C}_w(t)$ means that the extra viscosity, 
given by $\beta_w$ in \eqref{artificial_visc},
is added only during shock-wall collision and bounce-back; prior to collision, no extra viscosity is added and
the solution is consequently not overly diffused.
In \S \ref{sec:simulations}, we apply the 
WENO-$C$-$W$ scheme to a number of different shock tube problems for shock collision and bounce-back. 

\subsubsection{A generalization of our algorithm to shock-shock collision problems}

We remark here that a shock hitting a wall is simply a special case of shock-shock collision;
 indeed, the 
shock-wall collision problem may be viewed as the collision between two identical shocks but with different signs 
for the shock speed. A simple generalization of the Euler-$C$-$W$ algorithm which  allows for arbitrary shock-shock collision
is obtained by
 redefining the temporal bump function \mbox{\eqref{wall-ind-fn}} with the new function 
\begin{equation}\label{wall-ind-fn-general}
\overline{C}_w(t) = \sum_{i} \frac{C_w(x^*_i,t)}{\max_{x} C_w(x,t)},
\end{equation}
where $x^*_i(t)$ denotes the time-dependent local minima of the function $C_w(x,t)$ and approximates the location of the
shock-shock collision (at the collision time).
The functions $x^*_i(t)$ 
are analogous to the time-independent wall location $x_M$ in the shock-wall collision
problem (where the location of the collision is predetermined).

\begin{figure}[H]
\centering
\subfigure[$t=0.160$]{\label{fig:shock-shock1}\includegraphics[width=75mm]{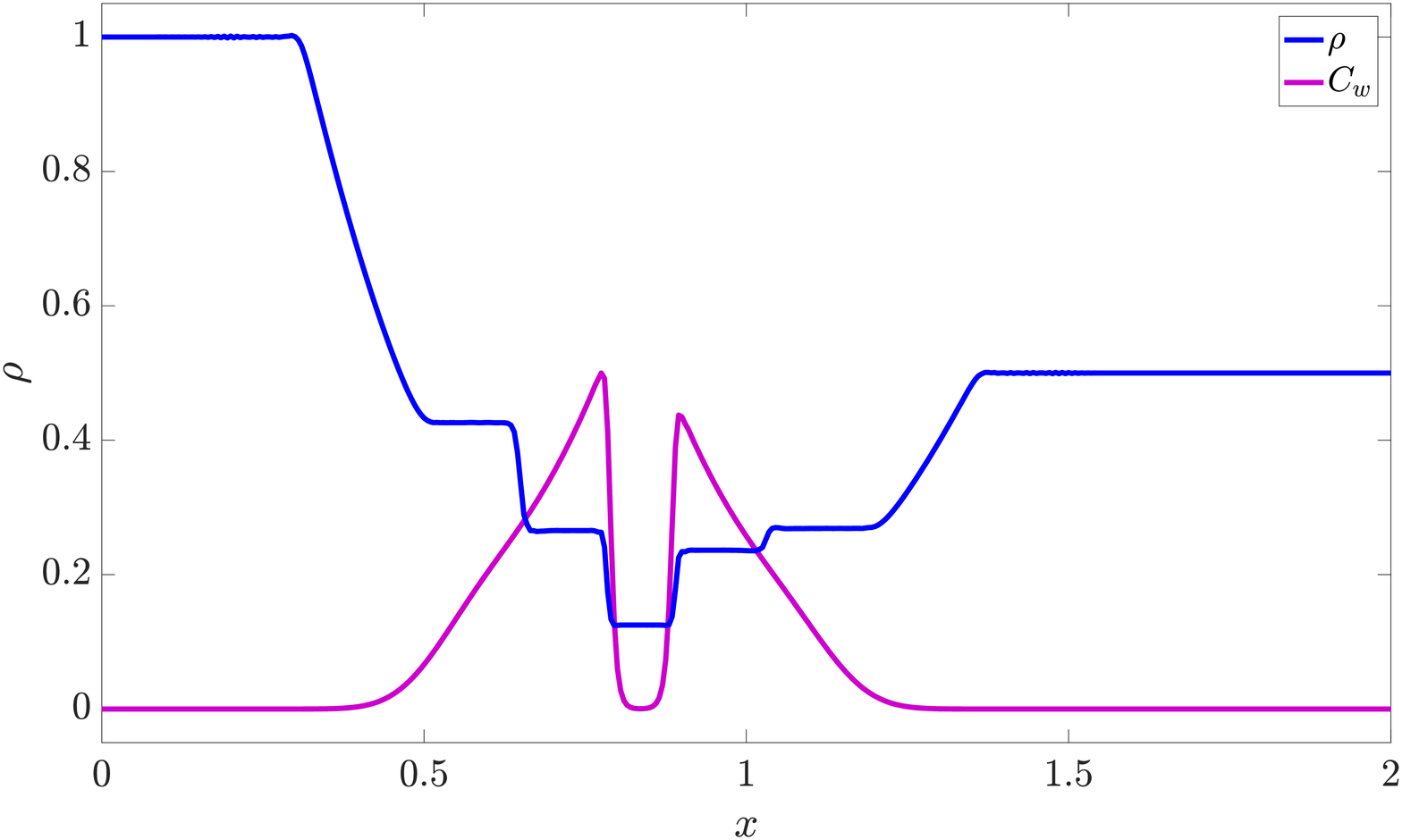}}
\subfigure[$t=0.184$]{\label{fig:shock-shock2}\includegraphics[width=75mm]{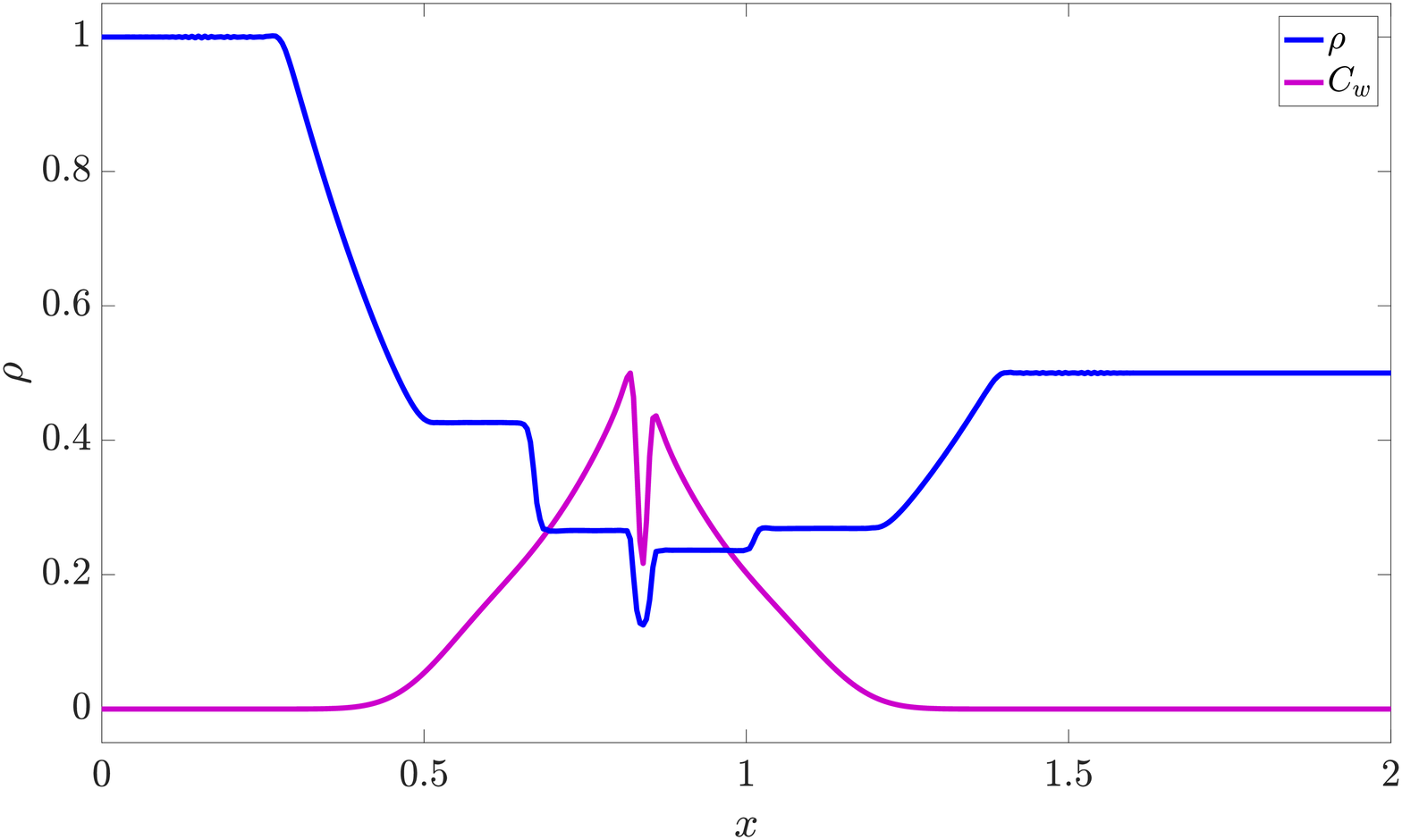}}
\subfigure[$t=0.192$]{\label{fig:shock-shock3}\includegraphics[width=75mm]{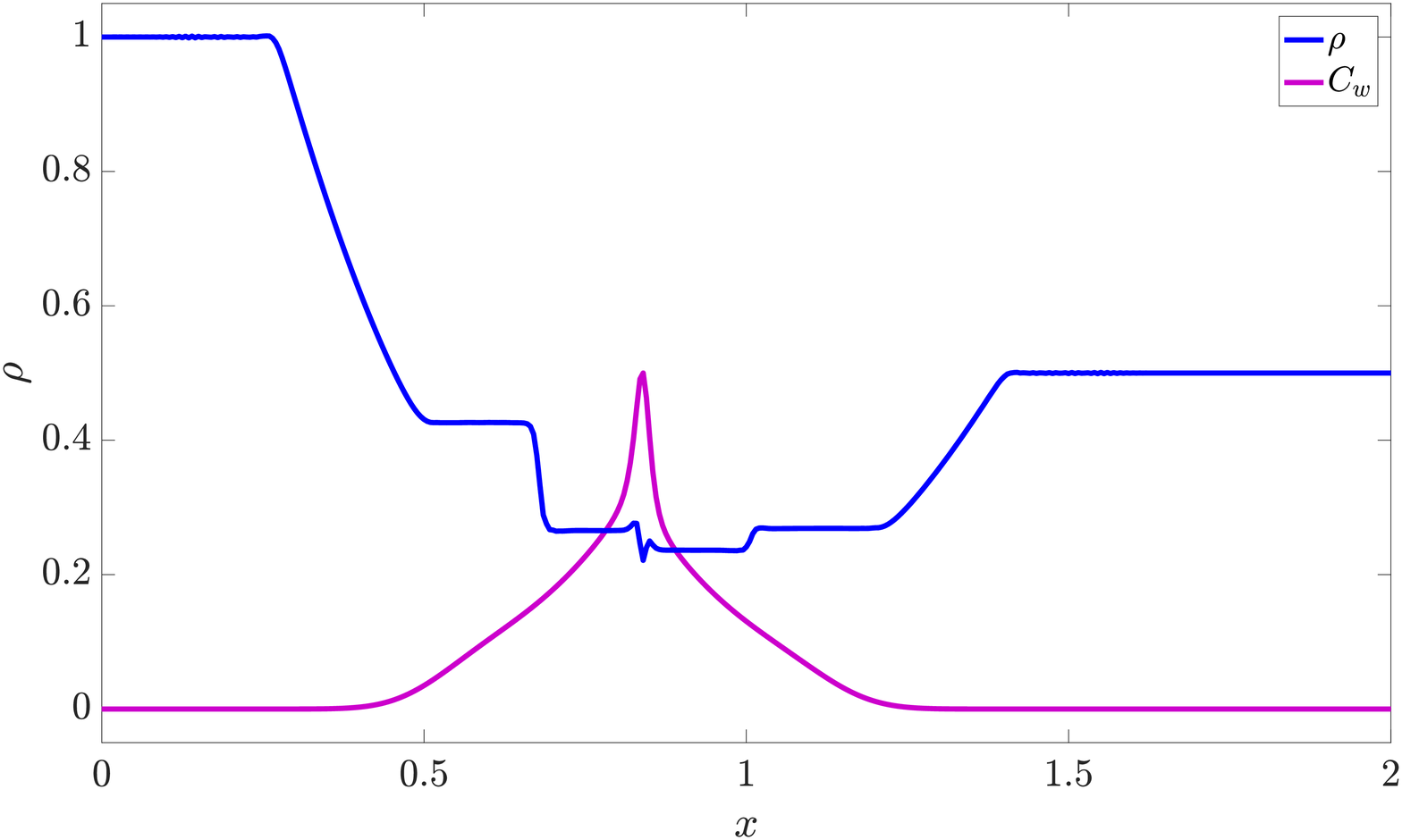}}
\subfigure[$\overline{C}_w(t)$]{\label{fig:shock-shock4}\includegraphics[width=75mm]{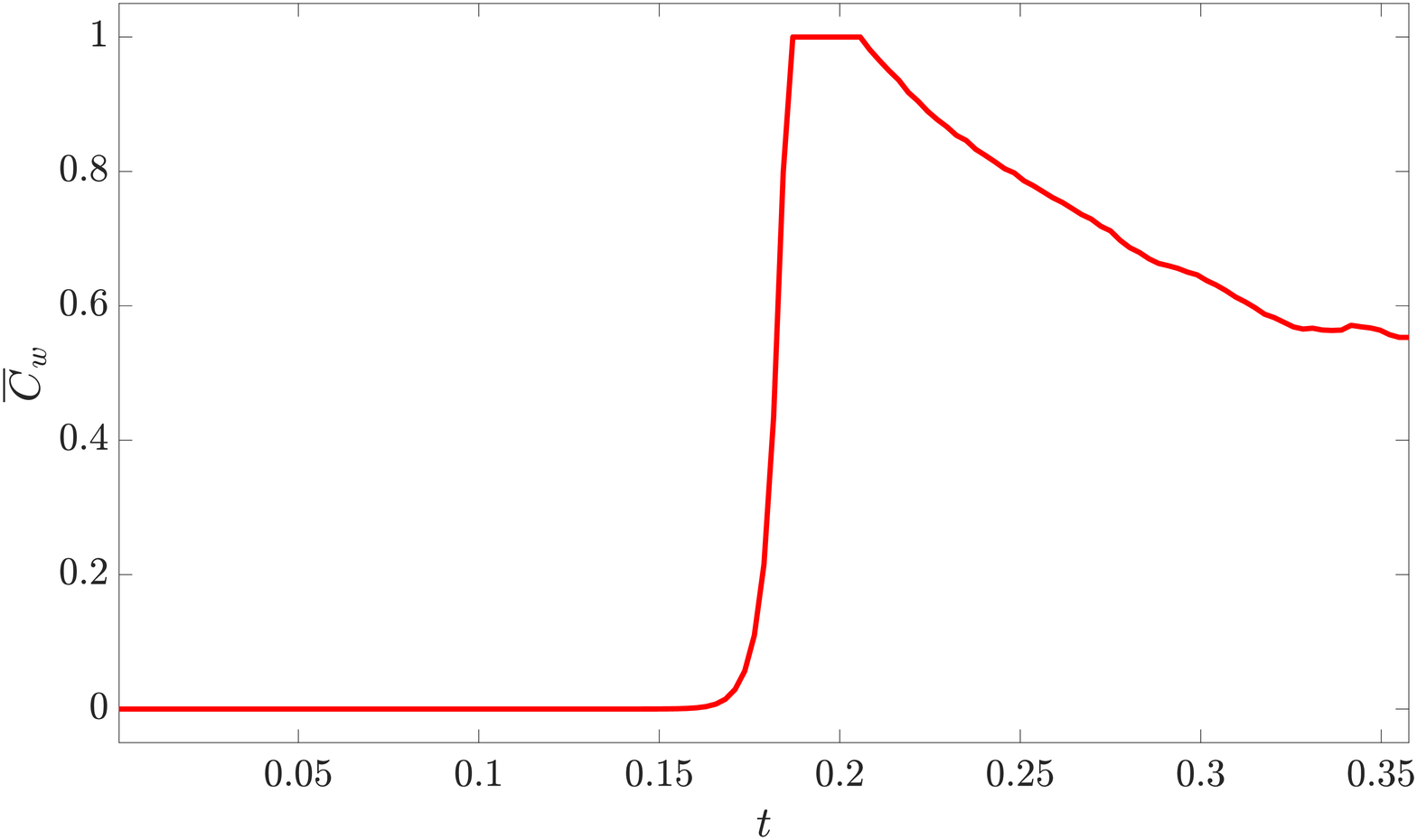}}
\caption{The density profile for a non-identical shock-shock collision problem. 
The blue curve is the density profile and the purple curve in Fig.\ref{fig:shock-shock1}-\ref{fig:shock-shock3} 
is the (normalized and resized) function $C_w(x,t)$. The red curve in Fig.\ref{fig:shock-shock4} is the 
temporal bump function $\overline{C}_w(t)$.}
\label{fig:shock-shock-collision}
\end{figure}

Fig.{\ref{fig:shock-shock-collision}} shows the density function during shock-shock collision.  Also shown, is the temporal bump function $\overline{C}_w$,
which
 naturally increases  as two shock waves approach one another, and provides a natural method for the addition of spacetime smooth
 additional artificial viscosity during the shock-shock collision process.     As can be seen, the two shocks collide at $t =0.192$, at which time
 the function $\overline{C}_w$ achieves its maximum value.    We  examine this problem of shock-shock collision in great detail in {\cite{RaReSh2019}}.

\section{A wavelet-based noise indicator: the WENO-$C$-$W$-$N$ method}\label{sec:noiseind}

\begin{figure}[H]
\centering
\subfigure[velocity profile at $t = 0.20$]{\label{fig:noise-at-contact1}\includegraphics[width=75mm]{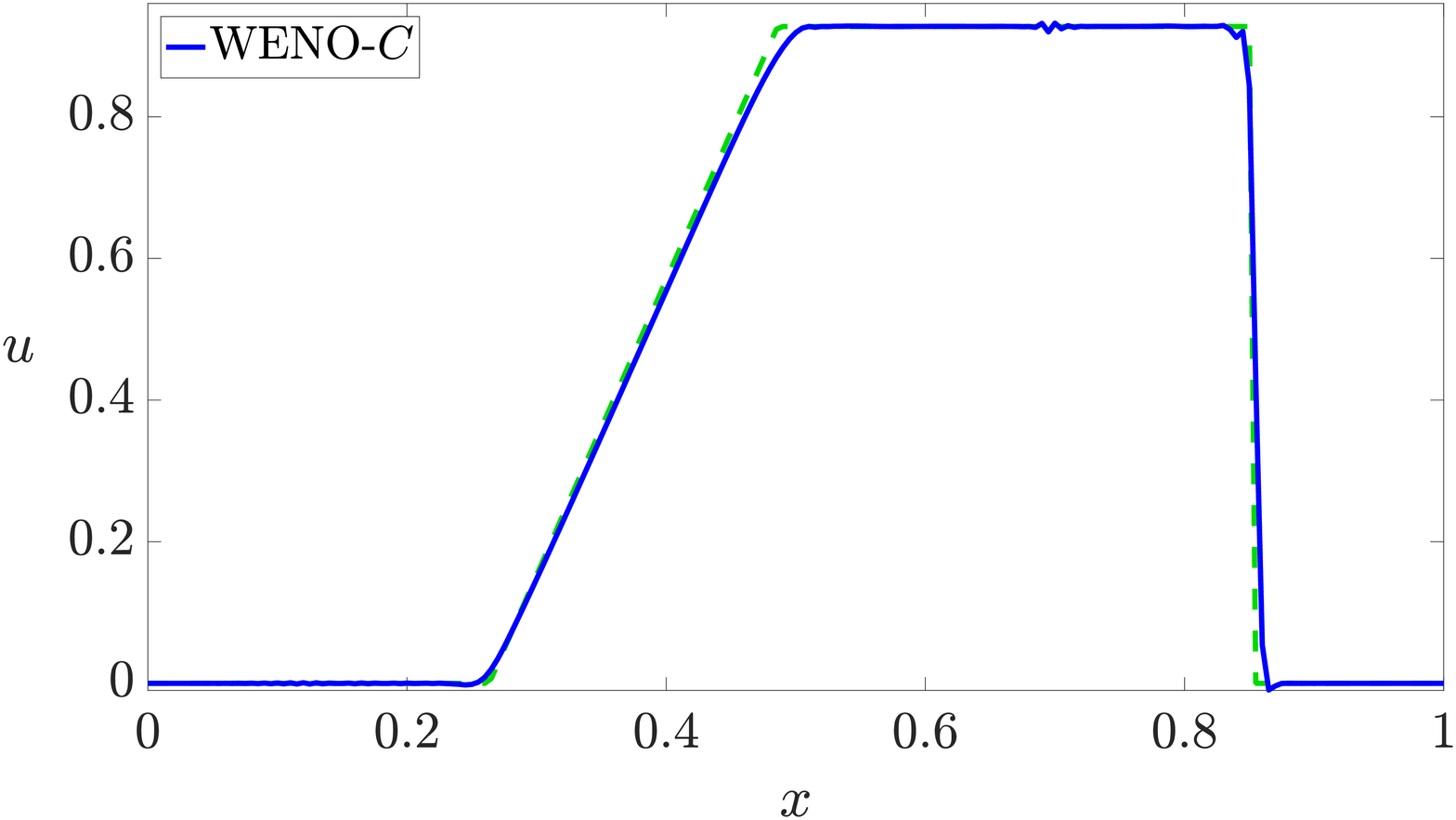}}
\subfigure[zooming in on the noise]{\label{fig:noise-at-contact2}\includegraphics[width=75mm]{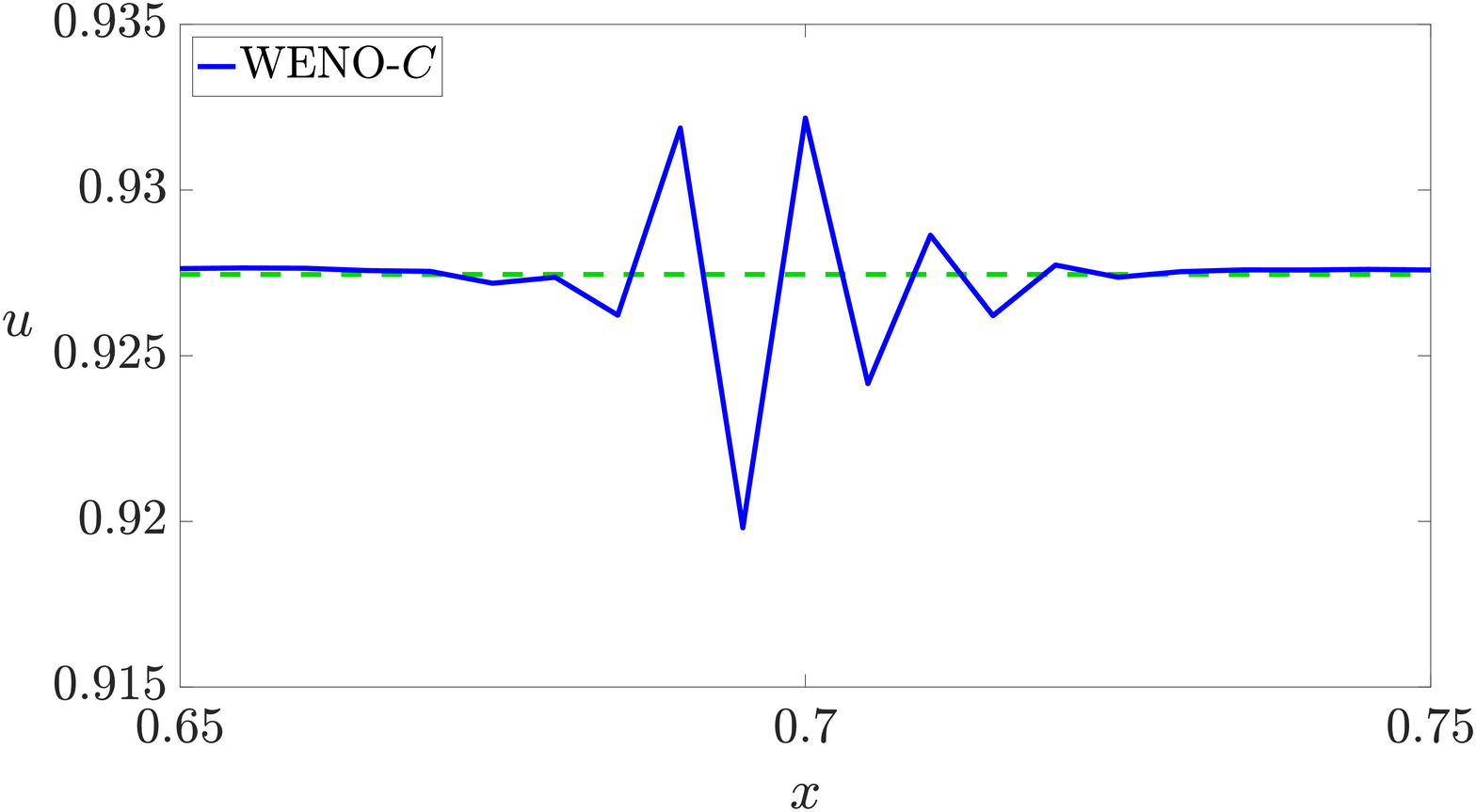}}
\caption{The velocity profile for the Sod shock tube problem, calculated using our WENO-$C$ 
scheme with 201 cells. The blue curve is the velocity profile and the dashed green curve 
is the exact solution. There is noise in the region $x \in [0.65,0.75]$. This is the location 
of the contact discontinuity in the density and momentum profiles.}
\label{fig:noise-at-contact}
\end{figure}

Numerical solutions of gas dynamics often develop 
high-frequency noise. 
 These (often small amplitude) spurious oscillations can occur if the time-step is too large
or because of the smearing of 
contact discontinuities.   Large time-step noise can be seen
with any explicit numerical scheme, while noise in the velocity field at the contact discontinuity is illustrated  
in Fig.\ref{fig:noise-at-contact} for the Sod problem.
 This noise is caused by the slightly different slopes that the momentum and 
density profiles have at the contact discontinuity. 

To deal with the occurrence of spurious noise, we implement a localized {\it{wavelet}}-based noise indicator.
Wavelets were first used in fluid dynamics in the analysis of turbulence by Farge \cite{Farge1992} and Meneveau \cite{Meneveau1991}. They have also been used in the numerical solution of PDE
on adaptive grids (see the review paper 
\cite{SchVas2010}).
 With regards to noise detection and removal, wavelets have
 generally been used in the form of a nonlinear {\it{filter}}, in which
 a noisy function is first decomposed using wavelets, and
 the function is  then 
 {\it{de-noised}} by retaining only the low-frequency components. Such filtering techniques often over-smooth 
 the noisy data, or introduce additional Gibbs-like oscillations \cite{Coifman1995}. 
 
  The main novelty of our approach is the use
 of wavelets only for high-frequency noise detection, while noise removal is achieved by a highly localized heat equation approach.

\subsection{Construction of wavelets}
A wavelet is like a traditional wave (sine or cosine waves), but localized in space i.e. it 
has a finite support. We define a \emph{mother wavelet} 
$\psi(x) = \psi_{0,0}(x)$ that represents the lowest frequency oscillation, and then use a dyadic 
scaling and integral translation to produce wavelets of higher frequencies:
$$
\psi_{r,s}(x) = 2^{r/2}\psi(2^rx-s); \,\, r = 0,1,2,\ldots \text{ and } s = \pm 1, \pm 3, \ldots, \pm (2^r-1). 
$$

Suppose that the spatial domain is given by $x_1 \leq x \leq x_M$. For our purposes, there
 are two key properties that the wavelet family $\{ \psi_{r,s} \}$ needs to satisfy:
\begin{enumerate}
\item Zero mean: 
$$
\int_{x_1}^{x_M} \psi(x) \,\mathrm{d}x = 0. 
$$
Note that due to the dyadic scaling and integral translation, this condition also ensures that
 wavelets of higher frequency have zero mean.
 
\item ``Quasi-orthogonality'' of the form:
$$
\int_{x_1}^{x_M} \psi_{r,s}(x) \cdot \psi_{r,s'}(x) \,\mathrm{d}x = 0, \text{ for } r \geq 0 \text { and } 0 \leq s, s' \leq 2^r - 1. 
$$
That is, each wavelet is orthogonal to every other wavelet of the same frequency. This is to 
ensure that one can locate exactly where each frequency is active. 
\end{enumerate}
%

\begin{figure}[H]
\centering
\scalebox{.5}{\begin{tikzpicture}
\filldraw [black] (-5,0) circle (2pt);
\filldraw [black] (0,0) circle (2pt);
\filldraw [black] (5,0) circle (2pt);
\draw[- >,very thick,dashed] (-7,0) -- (7,0);
\node at (6.7,0.3) {\huge $x$};
\draw[blue,very thick] (-5,0) -- (-2.5,-2) -- (0,4) -- (2.5,-2) -- (5,0);
\node at (-5,0.5) {\huge $x_{2i-1}$};
\node at (0,0.5) {\huge $x_{2i}$};
\node at (5,0.5) {\huge $x_{2i+1}$};
\end{tikzpicture}}
\caption{The highest frequency wavelet $\psi_i$.}
\label{fig:highest-frequency-wavelet}
\end{figure}
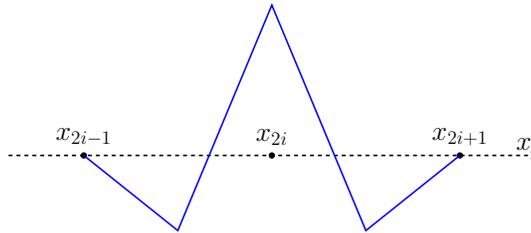

We define our wavelets to take the form shown in Fig.\ref{fig:highest-frequency-wavelet}. 
Since we are only interested in the highest frequency noise, we provide the exact formula for the 
highest frequency wavelet as
\begin{equation}\label{highest-freq-wavelet}
\psi_i (x) = \left\{ \begin{alignedat}{5}
		&-\frac{a}{\Delta x} (x-x_{2i-1}), \quad & \text{if} && \quad x_{2i-1} &\leq x \leq \,\, && x_{2i-\frac{1}{2}} \\[0.5em]
		&+\frac{3a}{\Delta x} (x-x_{2i}) + a, \quad & \text{if} && \quad x_{2i-\frac{1}{2}} &\leq x \leq && x_{2i} \\[0.5em]
		&-\frac{3a}{\Delta x} (x-x_{2i}) + a, \quad & \text{if} && \quad  x_{2i} &\leq x \leq && x_{2i+\frac{1}{2}}  \\[0.5em]
		&+\frac{a}{\Delta x} (x- x_{2i+1}), \quad & \text{if} && \quad  x_{2i+\frac{1}{2}} & \leq x \leq && x_{2i+1}
		\end{alignedat}\right.\vspace{1em}
\end{equation}
for each $i = 1, 2, \ldots, \frac{M-1}{2}$, where the notation $x_{k+\frac{1}{2}}$ denotes the midpoint of 
$x_k$ and $x_{k+1}$. 
It is clear from formula \eqref{highest-freq-wavelet} that  
each $\psi_i$ is supported in the interval $\mathcal{I}_i \coloneqq [x_{2i-1},x_{2i+1}]$. 

The constant $a \coloneqq \sqrt{3/ \Delta x}$ in \eqref{highest-freq-wavelet} 
is a normalization factor to ensure that 
the wavelets have $L^2$ norm equal to 1. Since the highest frequency wavelets have 
disjoint supports, it is obvious that the quasi-orthogonality property is satisfied. 
One can also check that the zero mean property is satisfied.

\subsection{High-frequency noise detection}\label{sec:how-does-noise-ind-work}

Given a discretized spatial domain, 
the highest frequency wavelet is supported over two grid cells and is shown in Fig.\ref{fig:highest-frequency-wavelet}.
There are $\frac{M-1}{2}$ two-cell intervals in the computational domain.   Each two-cell interval is denoted by
$ \mathcal{I} _j$, and there is a highest frequency wavelet $\psi_j(x)$ corresponding to each 
$ \mathcal{I} _j$ for every $j=1,...,\frac{M-1}{2}$.

For a given function $f(x)$, we 
next compute the {\it wavelet coefficients} $ \mathcal{C} _j(f)$ for this function.
For each $j=1,..., \frac{M-1}{2}$,  
$$
\mathcal{C} _j(f) := \langle f, \,\psi_j \rangle_{L^2} = \int_{ \mathcal{I} _j} f(x) \, \psi_j(x) \,\mathrm{d}x  \text{ for } j=1,..., \frac{M-1}{2}\,.
$$

Given the cell-center values $f(x_{2j-1}), f(x_{2j}), f(x_{2j+1})$, we can approximate the given function $f(x)$ on 
the interval $\mathcal{I}_j= [x_{2j-1}, x_{2j+1}]$ by a piecewise linear function $\tilde{f}(x)$; in particular, 
we define $\tilde{f}(x)$ by linear interpolation of the cell-center values 
of $f(x)$. We then approximate the wavelet coefficients by $\mathcal{C}_j(f) \approx \mathcal{C}_j(\tilde f)$,  and can compute
\begin{equation}\label{l2-inner-product}
\mathcal{C}_j(\tilde f) = \langle \tilde f, \psi_j \rangle_{L^2} = -\sqrt{\frac{\Delta x}{48}} \cdot  \Big[f(x_{2j+1}) - 2f(x_{2j}) + f(x_{2j-1}) \Big] \,.
\end{equation}
Notice that the right-hand side of \eqref{l2-inner-product} is proportional to the second-order central difference 
approximation to $f''(x_{2i})$. Also, note that if 
$f(x_{2i}) = \frac{1}{2} \left(f(x_{2i+1}) + f(x_{2i-1}) \right)$, i.e., if the 
function $\tilde{f}$ is linear on $\mathcal{I}_i$, then the associated wavelet coefficient is 
zero. This is crucial in ensuring that only the \emph{highest} frequency noise is detected. 

The magnitude of the wavelet coefficients grows with the amplitude of the 
high-frequency oscillations. For example, consider the case that $f(x)$ is
a hat function over the interval $\mathcal{I}_j= [x_{2j-1}, x_{2j+1}]$ and that
$f(x_{2j-1j}) = f(x_{2j+1}) = 0$. Then the amplitude of the oscillation is given by the magnitude at the peak of the hat, $f(x_{2j})$, and
$|\mathcal{C}_j(f)|$ is proportional 
to $f(x_{2j})$. Consequently, $| \mathcal{C}_j(f) | $ grows linearly with the amplitude of the oscillation. 

On the other hand, suppose that we have a lower frequency oscillation, given by a hat function 
that spans 4 cells, say the intervals $\mathcal{I}_j$ and $\mathcal{I}_{j+1}$. In each of these 
intervals, the oscillating function is linear, so that the associated wavelet coefficients $\mathcal{C}_j(f)$ 
and $\mathcal{C}_{j+1}(f)$ are equal to zero. This illustrates the fact that the highest frequency 
wavelets detect \emph{only} the highest frequency noise. 

\subsection{Noise detection in the presence of a shock wave}\label{sec:noise-detection}
\begin{figure}[H]
\centering
\subfigure[shock is ``in phase'' with the wavelet]{\label{fig:wavelet-zero}\includegraphics[width=75mm]{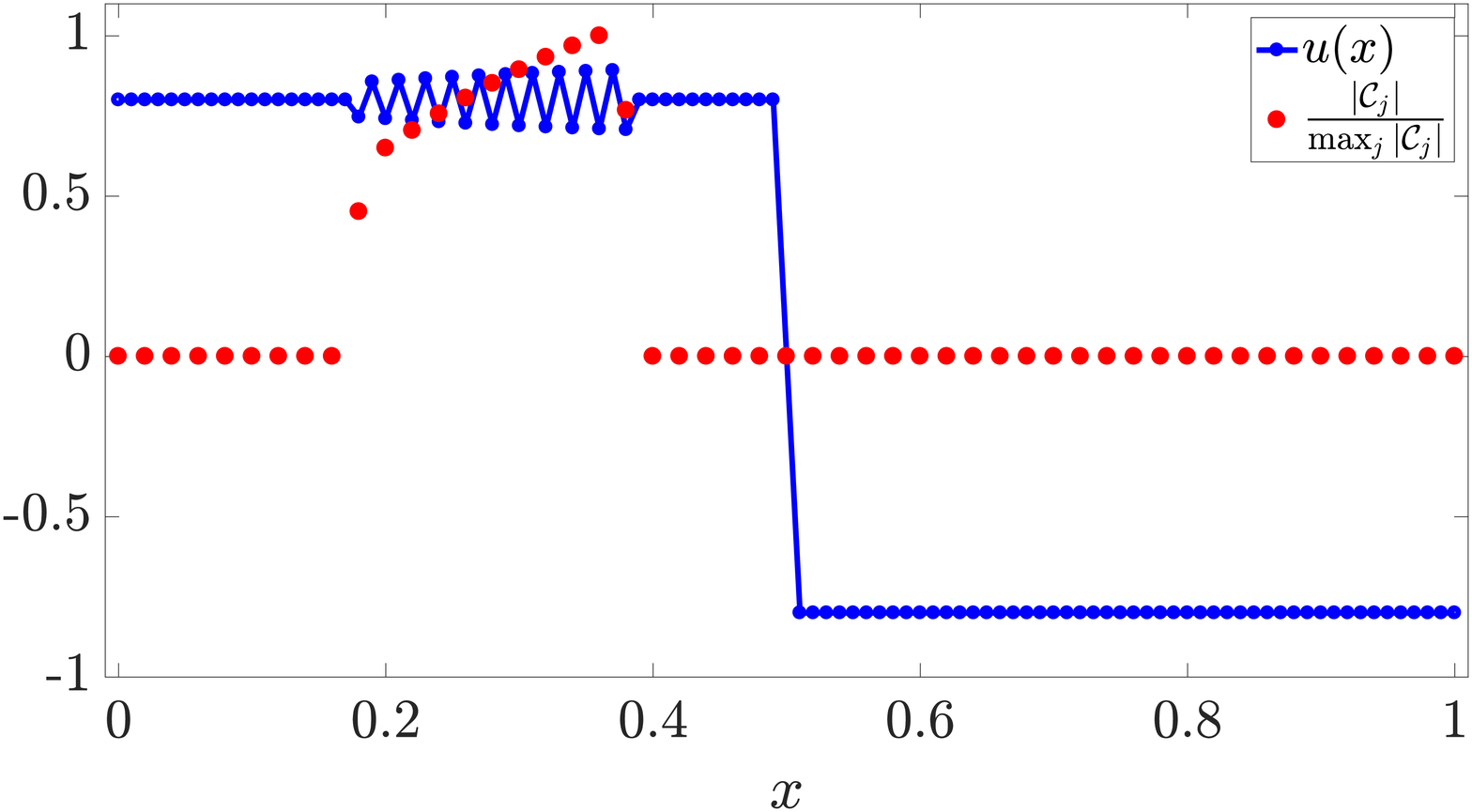}}
\subfigure[shock is ``out of phase'' with the wavelet]{\label{fig:wavelet-large}\includegraphics[width=75mm]{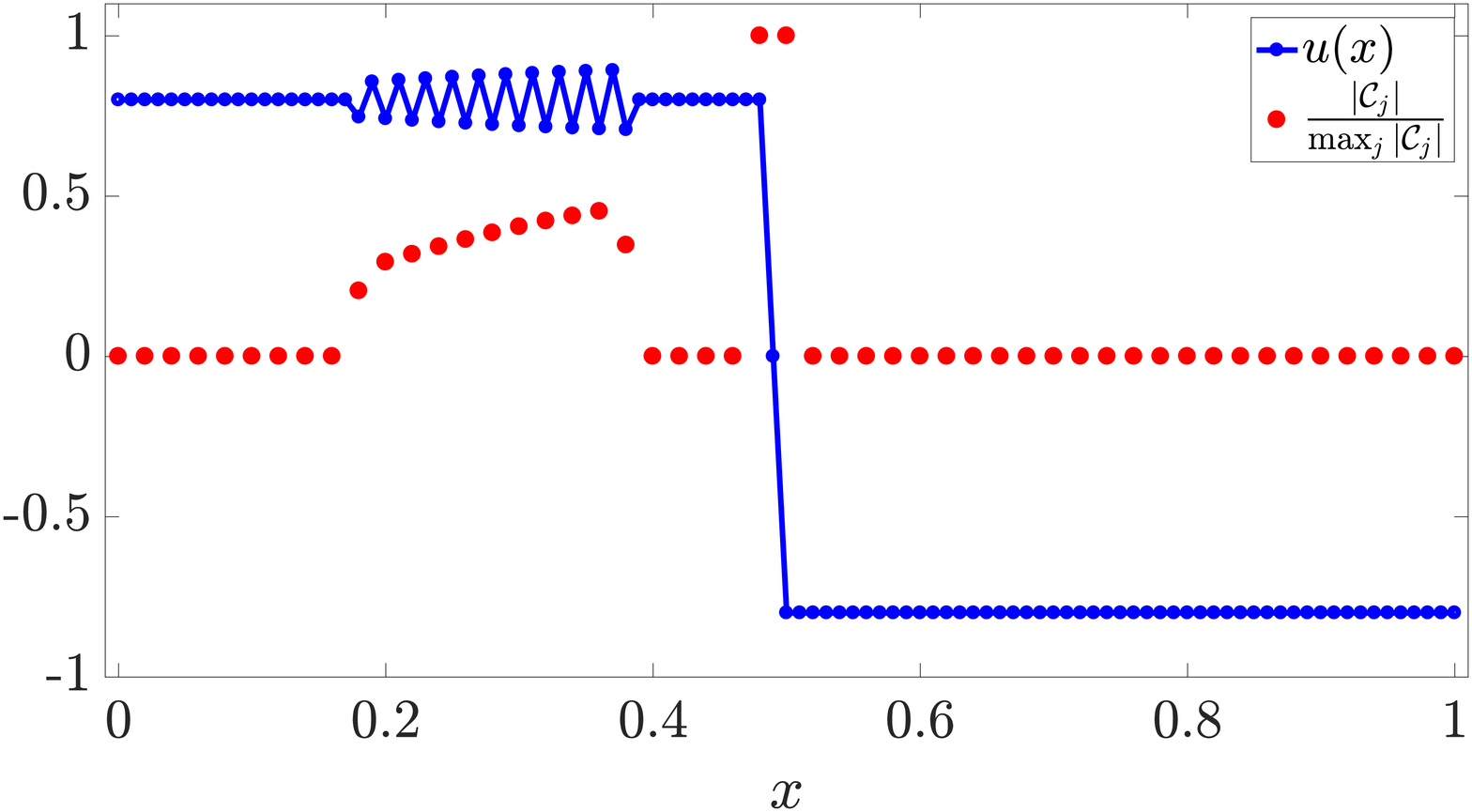}}
\caption{Wavelet coefficients at the shock curve compared with wavelet coefficients in regions 
where there is noise. The blue curve is $u(x)$, and the red dots indicate the relative magnitude 
of the associated wavelet coefficient $\mathcal{C}_j(u)$. 
The wave profile in Fig.\ref{fig:wavelet-large} is identical 
to that in Fig.\ref{fig:wavelet-zero}, but shifted by 
one cell to the left.}
\label{fig:wavelet-coeff-expln}
\end{figure}

We next examine
 the noise detection algorithm,  applied to a function $u(x)$ which has a shock discontinuity.
For $j=1,...,(M-1)/2$, we again compute the wavelet coefficients  $\mathcal{C}_j(u)$ for each two-cell interval $ \mathcal{I} _j$
according to \eqref{l2-inner-product}.  Suppose the shock discontinuity occurs spans the two-cell interval $ \mathcal{I} _j$; then,
on $ \mathcal{I} _j$ the shock curve is essentially linear and $C_j(u)=0$, but if the shock is out of phase by one cell with $ \mathcal{I} _j$,
then the wavelet coefficient  $C_j(u)$ can be large (see 
Fig.\ref{fig:wavelet-coeff-expln}).

In order to avoid over-diffusion at the shock, we prevent noise detection near shock discontinuities.
This is achieved by noting that the function $C(x,t)$ attains a local maximum for points $x$ along the shock curve. 
Consequently, we locate the local maximums of $C(x,t)$, by 
 finding the cells for which $\partial_x C = 0$ and $\partial_{xx} C < 0$. We then deactivate the noise detection
 in the cells surrounding the shock curve.  
  
Having deactivated the noise indicator  near the discontinuity, the largest wavelet coefficients are 
now those where the high-frequency oscillations exist. We may then define the \emph{noise detector function} 
$\mathbbm{1}_{\operatorname{noise}}(x)$ as follows: for each $j=1,..., \frac{M-1}{2}$ and  
$x \in \mathcal{I} _j$, 
we set $\mathbbm{1}_{\operatorname{noise}}(x)=1$ if $ |\mathcal{C} _j(u)| > \mathcal{C} _{\text{ref}} >0$ and set
$\mathbbm{1}_{\operatorname{noise}}(x)=0$ otherwise.

The constant $\mathcal{C}_{\operatorname{ref}}$ is obtained by computing the wavelet coefficient of a standard hat function
on the interval 
$[-\Delta x, +\Delta x]$ with amplitude $\delta h$:
\begin{equation}\label{cref}
\mathcal{C}_{\operatorname{ref}} = \delta h \cdot \sqrt{\frac{\Delta x}{12}}\,.  
\end{equation} 

\subsection{Noise removal algorithm}\label{sec:noise-removal}
Having described the noise detection algorithm, we next propose an efficient scheme for 
removing noise from a given function $u(x)$ by
solving a localized heat equation over the collection of intervals $ \mathcal{I} _j$ where high-frequency noise has been detected.

The union of all noisy intervals $ \mathcal{I} _j$ consists of  $K$ connected intervals $V_1,..., V_K$.   For each set $V_k$, $k=1,...,K$,
we define the set $W_k$ by affixing one cell on the left and one cell on the right.
We then solve a localized 
heat equation for the ``de-noised'' solution $w(x,\tau)$ in each of the domains $W_k$:
\begin{subequations}\label{localized-heat-equation}
\begin{alignat}{2}
\partial_\tau {w}(x,\tau) &= \eta \cdot \partial_{xx} {w}(x,\tau), &&\text { for } x \in  W_k \text{ and }  \tau \geq 0 \,, \\
 {w}(x,0) &= u(x), &&\text{ for } x \in W_k \,, \\
 {w}(x,\tau) &= u(x), &&\text{ for } x \in \partial W_k \,,
\end{alignat}
\end{subequations}
where $0< \eta \ll 1$ is a small constant, which we refer to as the noise removal viscosity.   The function
 $w(x,\tau) = u(x)$ for $x \in \left( \bigcup_{k=1}^K W_k \right)^\mathcal{C}$ and $\tau \geq 0$. 
We remark that the time $\tau$ is a ``fictitious'' time, introduced for the 
diffusion mechanism.    Equation (\ref{localized-heat-equation}b) is the initial condition over the intervals where noise has been 
detected, and (\ref{localized-heat-equation}c) is a Dirichlet boundary condition ensuring that $w(x,\tau)$ continuously transitions to $u(x)$.


We use an explicit scheme to solve \eqref{localized-heat-equation}, and in practice, one time-step is sufficient to remove noise.
If an explicit time-stepping scheme is used to solve \eqref{localized-heat-equation}, it is
not necessary to construct the domains $W_k$. Instead, one can simply use the 
noise indicator function $\mathbbm{1}_{\operatorname{noise}}(x)$, and solve a modified heat equation:
\begin{subequations}
\begin{alignat}{2}
\partial_\tau {w}(x,\tau) &= \eta \cdot \mathbbm{1}_{\operatorname{noise}}(x) \cdot \partial_{xx} {w}(x,\tau), \quad &&\text {for } x_1<x<x_M\text{ and }  \tau \geq 0, \label{heat-equation-local1}\\
 {w}(x,0) &= u(x), &&\text{for } x_1<x<x_M, \\
 {w}(x,\tau) &= u(x), &&\text{for } x=x_1 \text{ and } x=x_M.
\end{alignat}
\end{subequations}

The utilization of an explicit scheme results in the stability constraint 
$\eta \Delta \tau/ (\Delta x)^2 < 1/2$. 
However, in practice, we have found that much smaller values 
$\eta \Delta \tau /(\Delta x)^2 \ll 1/2$ are sufficient to dampen spurious noise. We also remark that 
the use of a single time-step means that the noise removal provided by the localized heat equation 
can be viewed as a 
\emph{filtering} process, in which noise is removed through a local averaging. Consequently, the 
averaging provided by the Laplacian term on the right-hand side of {\eqref{heat-equation-local1}}, 
namely $(w_{i+1}-2w_i+w_{i-1})/2$, can be replaced by other local averages, such as that provided 
by Gaussian filtering {\cite{CaCo2004}}. 

However, we wish to stress that there is a distinction between the operation of 
\emph{smoothing} a noisy function and the noise removal process we have outlined. While, of course, 
removing high-frequency noise does indeed smooth the function, because we remove highly localized 
(in space and time) packets of oscillations, the procedure is quite different to more traditional smoothing 
algorithms, in which one uses truncation of frequencies in Fourier space or the analogous hyperviscosity 
operators in physical space. As such, it is difficult to obtain analytically the truncation error by means of 
a Taylor expansion, but it is possible to measure the error improvement by virtue of convergence
studies comparing the algorithm with and without the noise removal algorithm activated. We provide results 
of such studies in \S{\ref{sec:simulations}}.

\subsection{The WENO-$C$-$W$-$N$ algorithm}\label{sec-weno-algorithm}

We now describe how we implement the above noise indicator algorithm for the 
Euler equations. The algorithm proceeds in two stages; in 
the first stage, we use the WENO-$C$-$W$ scheme described in \S\ref{sec-weno-reconstruction-procedure} 
to solve for an intermediary 
solution $\bm{\tilde{u}} = \left[ \tilde{\rho}, \tilde{\rho u}, \tilde{E} \right]^T$; in the second stage, we 
feed this intermediary solution $\bm{\tilde{u}}$ into the noise indicator algorithm to de-noise the solution.  The two-stage process is now
described.
\begin{enumerate}
\item An intermediary solution $\bm{\tilde{u}}$ is obtained as
\begin{align*} 
\bm{\tilde{u}}_i &= \text{RK}\left( \bm{{u}}_i^{n}, \mathcal{A}_{\operatorname{WENO}}(\bm{{u}}_i^{n},\bm{{C}}_i^n) \right) \,, \\
\bm{{C}}_i^{n+1} &=\text{RK}\left( \bm{{C}}_i^{n} ,\mathcal{B}_{\operatorname{WENO}}(\bm{{u}}_i^{n},\bm{{C}}_i^n) \right)  \,.
\end{align*} 
\item The  intermediate velocity $\tilde{u}_i $ is then 
de-noised  using the procedure described 
in \S \ref{sec:noise-detection} and \S\ref{sec:noise-removal},  
producing the noise-free velocity $u(x,t_{n+1})$.  The
 updated solution $\bm{u}(x,t_{n+1})$ is then obtained as
$$
\renewcommand\arraystretch{1.5}
\bm{u}(x,t_{n+1}) \equiv 
\left(
\rho(x,t_{n+1}) , \rho u (x,t_{n+1}) , E(x,t_{n+1}) \right)
\coloneqq 
\left(
\tilde{\rho}(x) \,,  \tilde{\rho}(x) \cdot u(x,t_{n+1}) \,, \tilde{E} (x) \right) \,.
$$
\end{enumerate}

\begin{remark}\label{remark-noise} 
Implementation of our noise removal scheme
has been motivated by the high-frequency oscillations of the velocity field that occur exacly
at the contact discontinuity (see, for example,  Fig.{\ref{fig:noise-at-contact}}). 
We note that once the noise indicator function 
$\mathbbm{1}_{\mathrm{noise}}(x)$ is computed, 
there are many possible choices for the noise removal portion of the algorithm.  For example,
while our implemented algorithm only removes high-frequency oscillations form the velocity,
we could also remove such oscillations from $\rho$ and $E$, and we could in place of the velocity field,
instead remove oscillations from the momentum $\rho u$.   We have found that any of these
choices produce the same 
 relative errors in the one-dimensional test problems considered herein.   Moreover, as we demonstrate in 
\S{\ref{sec:simulations}},  the removal of high-frequency oscillations from $u$ alone,  is sufficient to remove noise from
the density and energy as well.
A more detailed examination of various noise removal algorithms (as well as those more ideally suited for parallelization) is made in {\cite{RaReSh2019}}.
\end{remark}


\section{Numerical simulations of classical shock tube experiments}\label{sec:simulations}
In this section, we show results of the discretized $C$-method for a variety of classical shock tube experiments.  For some of the
problems, we will compare against WENO-based classical artificial viscosity schemes and Noh schemes.  See Appendix 
\ref{sec:appendix} and Table \ref{table:schemes} for a description of all of the numerical methods
employed herein.

{
\begin{table}[H]
\centering
{\small
\begin{tabular}{|M{4cm} | M{6cm}|} 
 \hline
 Parameter / Variable & Description \\ [0.0em] 
 \hline \hline 
$\beta^{u}$, $\beta^E$ & artificial viscosity coefficients for the momentum 
 and energy, respectively. \\[0.5em] 
\hline
$\beta^{u}_w$, 
 $\beta^E_w$ &  wall viscosity coefficients for the
 momentum and energy, respectively. \\[0.5em] 
 \hline 
 $\delta h$, $\eta$ & amplitude of noise and noise removal viscosity, respectively. \\[0.5em] 
 \hline 
 $\varepsilon$, $\varepsilon_w$ & parameters controlling support 
 of $C$ and $C_w$, respectively. \\[0.5em]
 \hline
 $\kappa$, $\kappa_w$ & parameters controlling smoothness 
 of $C$ and $C_w$, respectively. \\[0.5em]
 \hline
\end{tabular}}
\caption{Relevant parameters and variables used in the numerical tests.}
\label{table:parameters}
\end{table}}

As with any artificial viscosity scheme, parameters must be chosen for the problem under consideration. 
Before presenting our numerical results,  we consider this issue for
the $C$-method, whose parameters are 
listed in Table {\ref{table:parameters}}. The artificial viscosity parameters $\beta$ are chosen in the following 
manner: we set $\beta^E=0$, choose $\beta^u$, $\beta^u_w$ large enough so that 
post-shock oscillations are removed both pre and post shock-wall collision, then choose $\beta^E_w$ large 
enough so that the wall-heating phenomenon (discussed later in \S{\ref{sec-sod-shock-tube}}) does not occur.
A similar philosophy is applied to choice of parameters for the noise detection and removal algorithm:  first, we determine
the amplitude of highest-frequency oscillation $\delta h$ and  then choose the artificial viscosity parameter $\eta$ large enough 
to diffuse the noise.

The parameters $\varepsilon$, $\varepsilon_w$, $\kappa$, and $\kappa_w$ are $O(1)$ constants.  
Setting a larger value for $\varepsilon$ or $\varepsilon_w$ serves to increase the support of the corresponding
$C$-function, while increasing the value of $\kappa$ or $\kappa_w$ produces smoother $C$-functions. 
For certain problems, smoothing the $C$-variables by using a larger $\kappa$ further minimizes noise that
occurs in the solution.

In Appendix {\ref{sec:appendix2}} we demonstrate the accuracy of the $C$-method when the 
values of the
parameters $\varepsilon$, $\varepsilon_w$, $\kappa$, and $\kappa_w$ are fixed values for the 
different test problems. It is shown that the differences between the solutions computed using the 
 optimized parameter sets we use for the 
problems in this section and the fixed-parameter sets we use for the runs in Appendix {\ref{sec:appendix2}} are 
minimal, and that the fixed choice of parameters can be used for general problems. However, we wish to 
emphasize that one of the strengths of the $C$-method is its 
flexibility  to optimize  parameters for specific features associated with particular data. 
 
The error analysis and convergence studies we perform for the numerical experiments considered in the 
following sections use the $L^1$, $L^2$, and 
$L^ \infty$ norms. Given two functions $f(x)$ and $g(x)$ defined on the computational grid with $M$ cells, 
these error norms are defined by
\begin{subequations}
\begin{alignat}{2}
&\lvert \lvert f- g\rvert \rvert_{L^1}  &&=  \frac{1}{M} \sum_{i=1}^M | f(x_i) - g(x_i) |\,, \label{error-formula} \\
&\lvert \lvert f- g\rvert \rvert_{L^2}  &&=  \sqrt{\frac{1}{M} \sum_{i=1}^M | f(x_i) - g(x_i) |^2} \,, \label{error-L2}\\
&\lvert \lvert f- g\rvert \rvert_{L^ \infty} &&= \max_{i=1,\ldots,M}  | f(x_i) - g(x_i) |\,. \label{error-Linf}
\end{alignat}
\end{subequations}
As stated  by Greenough \& Rider {\cite{GreenoughRider2004}},  the 
$L^1$ and $L^2$ norms provide a global view of the errors in the computed solution, whereas the 
$L^ \infty$ norm highlights local errors, such as the undershoot or overshoot that occurs at a discontinuity.
Thus, these three norms together provide a precise \emph{quantitative} description of the errors of numerical 
solutions, and complement the \emph{qualitative} evidence we provide through the visualization of 
numerical simulations.

\subsection{Linear advection}\label{sec-linear-advection}
We begin by considering a linear advection problem to demonstrate the high-order convergence of the 
base WENO scheme. The domain is $0 \leq x \leq 1$, the adiabatic constant is $\gamma=1.4$, the 
initial data is
$$
\begin{bmatrix}
\rho_0 \\ (\rho u)_0 \\ E_0 
\end{bmatrix}
=
\begin{bmatrix}
1 + 0.5 \sin(2 \pi x) \\ 1 + 0.5 \sin(2 \pi x) \\ 0.5 + 0.25 \sin(2 \pi x) + \frac{1}{\gamma -1}
\end{bmatrix}\,,
$$
and we employ periodic boundary conditions. In the exact solution, 
the velocity and pressure remain a 
constant value of 1, while the initial density field is advected by the velocity, so that the density at time $t$ 
satisfies $\rho(x,t) = \rho_0(x-t)$. We employ our simplified WENO scheme on grids with 51, 101, 201,
 and 401 cells.  Each simulation is run with a CFL number of approximately 0.6, and the final time is $t=1.0$. 

In Table {\ref{table:linear-convergence}}, 
we list the $L^1$ error of the computed density minus the exact solution; as expected, the solutions 
converge with almost fifth-order accuracy.

\begin{table}[H]
\centering
\renewcommand{\arraystretch}{1.0}
\scalebox{0.8}{
\begin{tabular}{|lc|cccc|}
\toprule
\midrule
\multirow{2}{*}{\textbf{Scheme}} &  & \multicolumn{4}{c|}{\textbf{Cells}}\\

{}  & & 51   & 101    & 201   & 401\\
\midrule
\multirow{2}{*}{WENO} & Error & 
$7.298 \times 10^{-6}$  & $2.318 \times 10^{-7}$  & $7.526 \times 10^{-9}$ & $2.654 \times 10^{-10}$\\
				    & Order & -- & 4.977   & 4.945  & 4.826\\
\midrule
\bottomrule
\end{tabular}}
\caption{$L^1$ error of the computed density minus the exact solution and convergence for the 
linear advection problem.}
\label{table:linear-convergence}
\end{table}

\subsection{The Sod shock tube problem}\label{sec-sod-shock-tube}

The data for the Sod shock tube is given in \eqref{sod_initialdata}, with the
exact solution given by  a shock wave, a rarefaction wave, and a contact discontinuity. 
To simulate the shock-wave wall collision, we employ reflective boundary conditions \eqref{var-bcs}.
In the tests below, we employ our WENO-$C$-$W$ scheme with 201 cells. 

\subsubsection{The wall-heating problem in the Sod shock tube}\label{sod-bounce-back}
We first demonstrate the well-known wall-heating problem (\cite{Noh1987, Rider2000}) in which an anomalous slope appears in
the density and internal energy upon shock bounce-back.   We shall then explain what the root cause of this problem is, and its solution.

\begin{figure}[H]
\centering
\subfigure[$t=0.20$: pre shock-wall collision]{\label{fig:sod-before-collision3}\includegraphics[width=75mm]{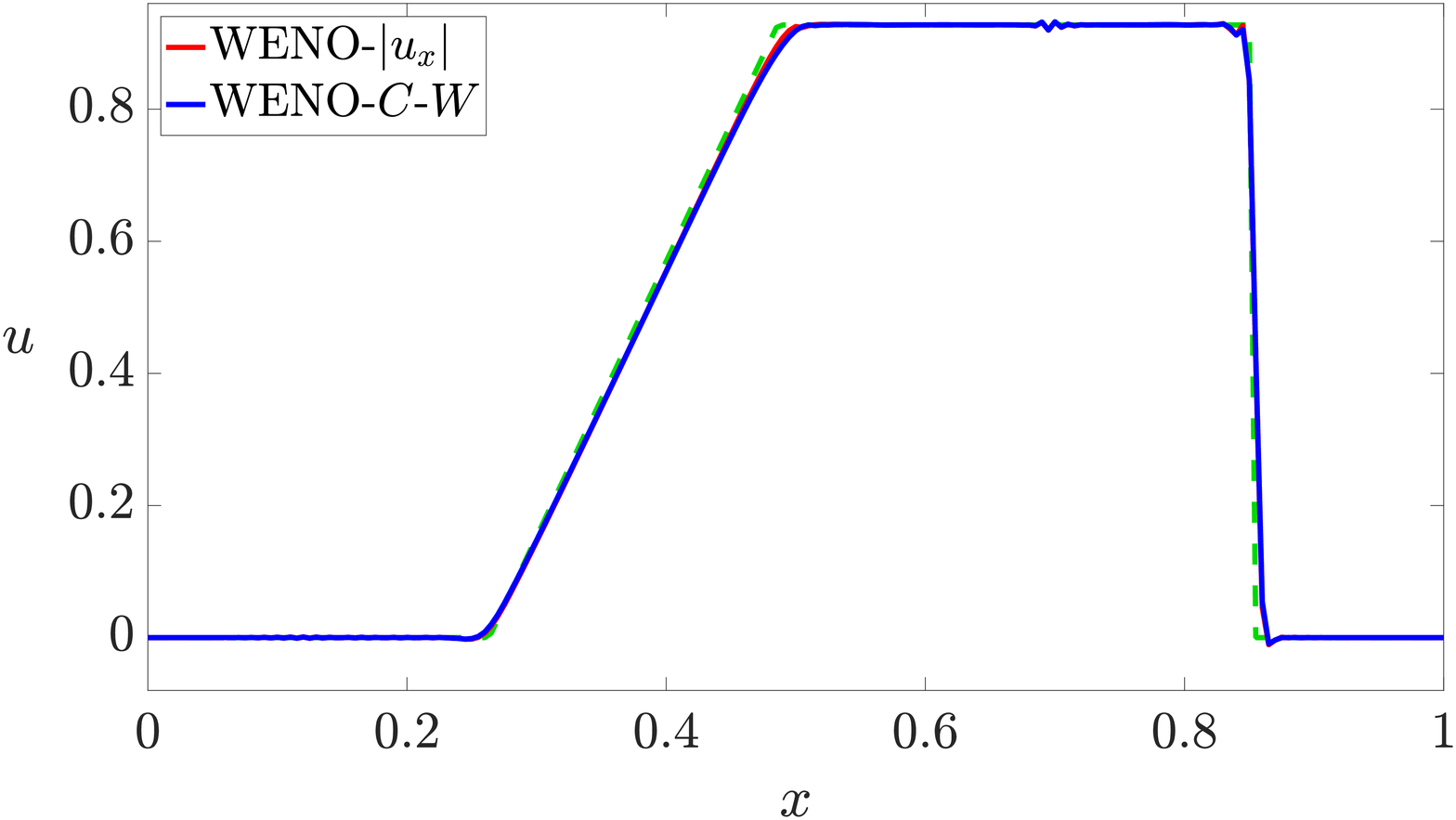}}
\subfigure[$t=0.36$: post shock-wall collision]{\label{fig:sod-after-collision3}\includegraphics[width=75mm]{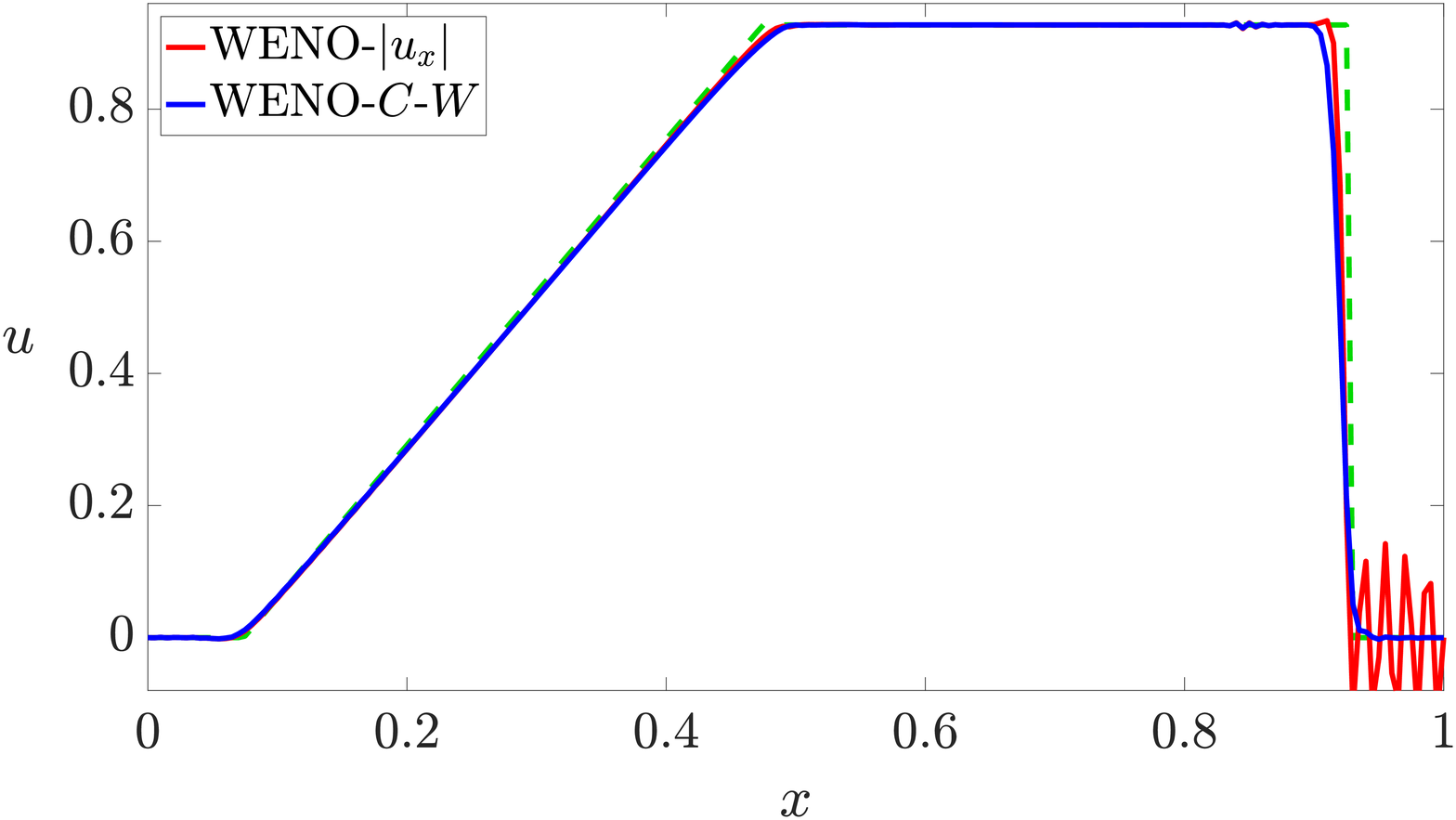}}
\caption{The velocity profile for the Sod shock tube problem, with the wall viscosity activated for the 
momentum equation. The dashed green curve is the exact solution.}
\label{fig:sod-collision3}
\end{figure}

We begin by choosing the parameters
in equations \eqref{EulerC}  and  \eqref{artificial_visc} as 
\begin{alignat*}{4}
\beta^u&=0.5, \qquad  \beta^E&=0.0, \qquad   \beta^u_w &= 3.0, \qquad  \beta^E_w &= 0.0 \\
\varepsilon&=1.0, \qquad  \kappa&=5.0, \qquad  \varepsilon_w &=50.0, \qquad  \kappa_w &=1.0 \,.
\end{alignat*}
The resulting solutions for the 
velocity before and after the shock-wall collision are shown in Fig.\ref{fig:sod-collision3}. Before 
the shock collision with the wall (Fig.\ref{fig:sod-before-collision3}), the solution maintains a sharp shock 
front. After the shock collision (Fig.\ref{fig:sod-after-collision3}), the high-frequency oscillations behind the 
shock wave are damped-out for sufficiently large $\beta^u_w>0$, while maintaining a sharp front. 

\begin{figure}[H]
\centering
\subfigure[$t=0.36$: post shock-wall collision]{\label{fig:sod-after-collision-rho1}\includegraphics[width=75mm]{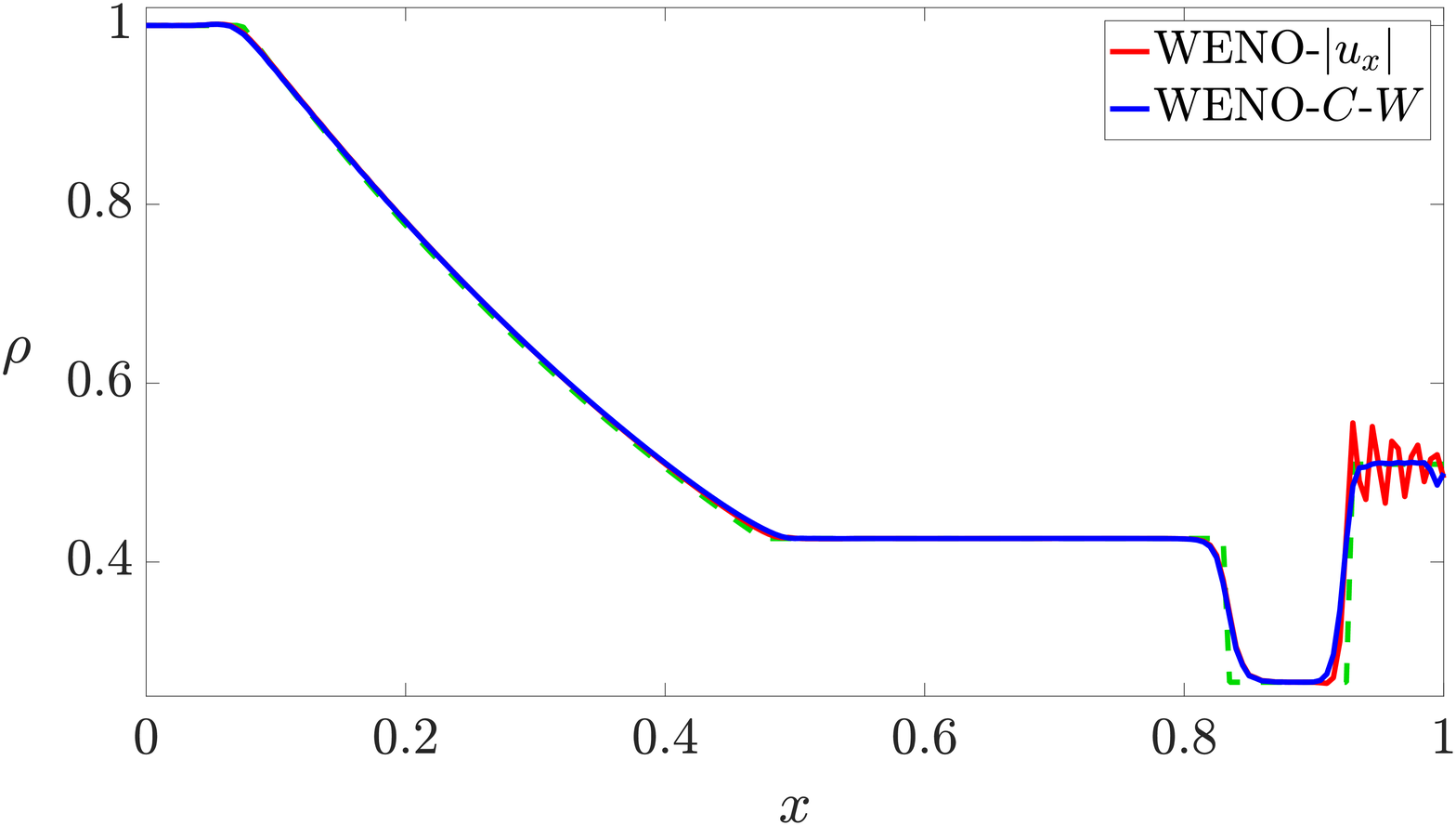}}
\subfigure[$t = 0.36$: zooming in on the undershoot]{\label{fig:sod-after-collision-zoom-rho1}\includegraphics[width=75mm]{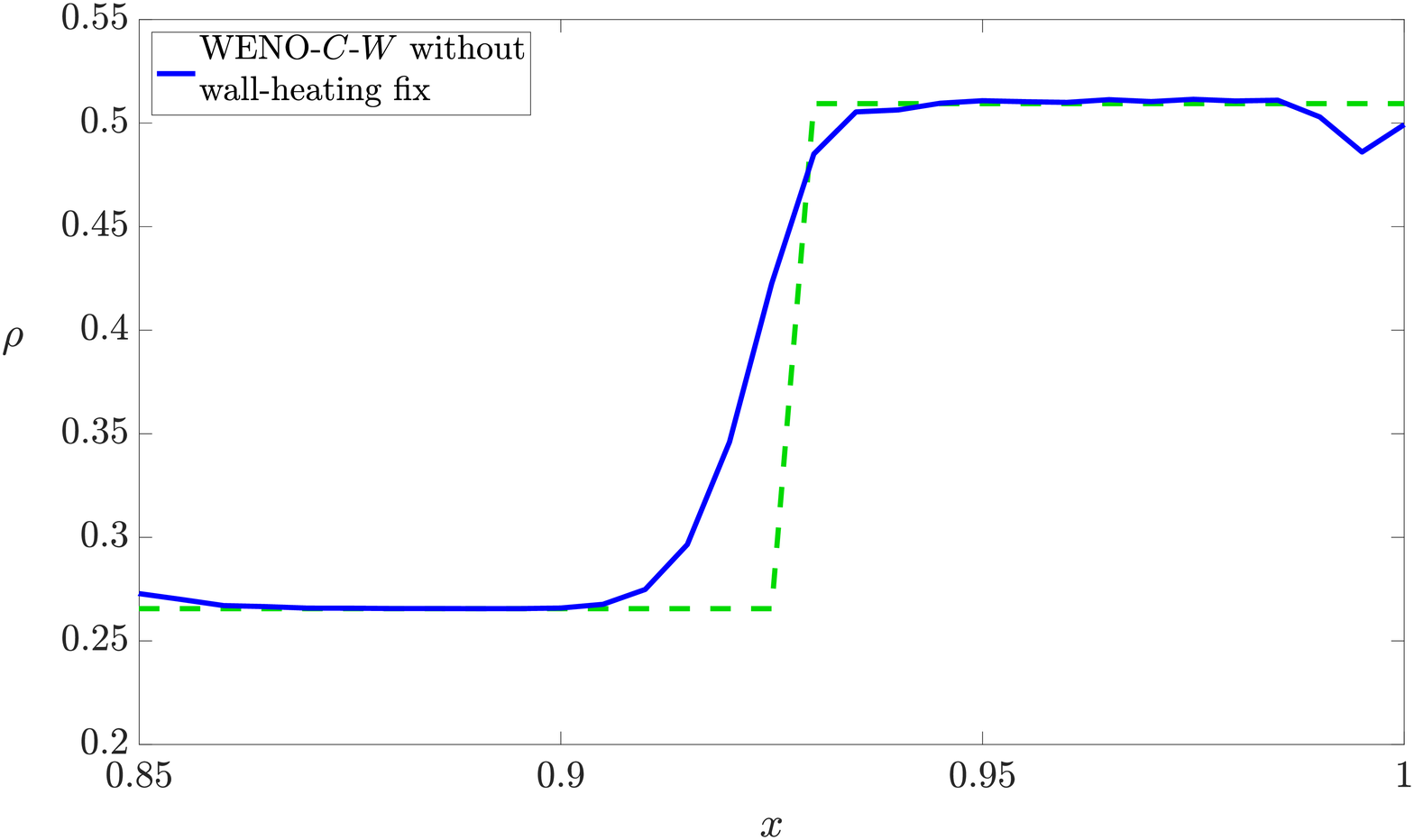}}
\caption{The density profile for the Sod shock tube problem, calculated with the wall viscosity activated 
for the momentum equation. The dashed green curve is the exact solution.}
\label{fig:sod-collision-rho1}
\end{figure}

While  post-collision oscillations in the density profile are suppressed,  Fig.\ref{fig:sod-collision-rho1} shows the presence of the
anomalous density slope $\partial_x\rho(1,t)$ at the wall (which should be zero). 
This incorrect slope is termed {\it{wall heating}} because the undershoot in the density results in 
an overshoot in the 
 internal energy \eqref{defn-internal-energy} (and hence temperature) at the wall.   Noh \cite{Noh1987} suggested
that wall heating would occur in {\it{any}} artificial viscosity scheme, and is in fact built into the 
exact solutions of the difference equations of the artificial viscosity method.  Menikoff \cite{Menikoff1993} 
argues that wall-heating is caused by the {\it{smearing}} of the shock curve that occurs with any artificial 
viscosity scheme, and is thus unavoidable.    Rider \cite{Rider2000} 
argues that incorrect wave speeds result in too much or too little dissipation.   


In fact, it appears that the wall-heating error 
is the result of the misalignment of the {\it gradient of fluxes} for the density, momentum and 
energy equations, which in turn is caused by a slight difference in the speed of the shock fronts for the 
density, momentum and energy.   
We define the {\it{forcing terms}}
\begin{align*}
\mathcal{H}(\rho) &= -\partial_x ( \rho u ), \\
\mathcal{H}(\rho u ) &= -\partial_x (\rho u^2 + p) + \partial_x \left( \mathcal{B}^{(u)}(t)\,\rho\,C\,\partial_x u \right), \\
\mathcal{H}( E ) &= -\partial_x (u(E+p)) + \partial_x \left( \mathcal{B}^{(E)}(t)\,\rho\,C\,\partial_x (E/\rho) \right). 
\end{align*}

\begin{figure}[H]
\centering
\subfigure[$t=0.2725$: pre-collision]{\label{fig:bad-flux-compare1}\includegraphics[width=75mm]{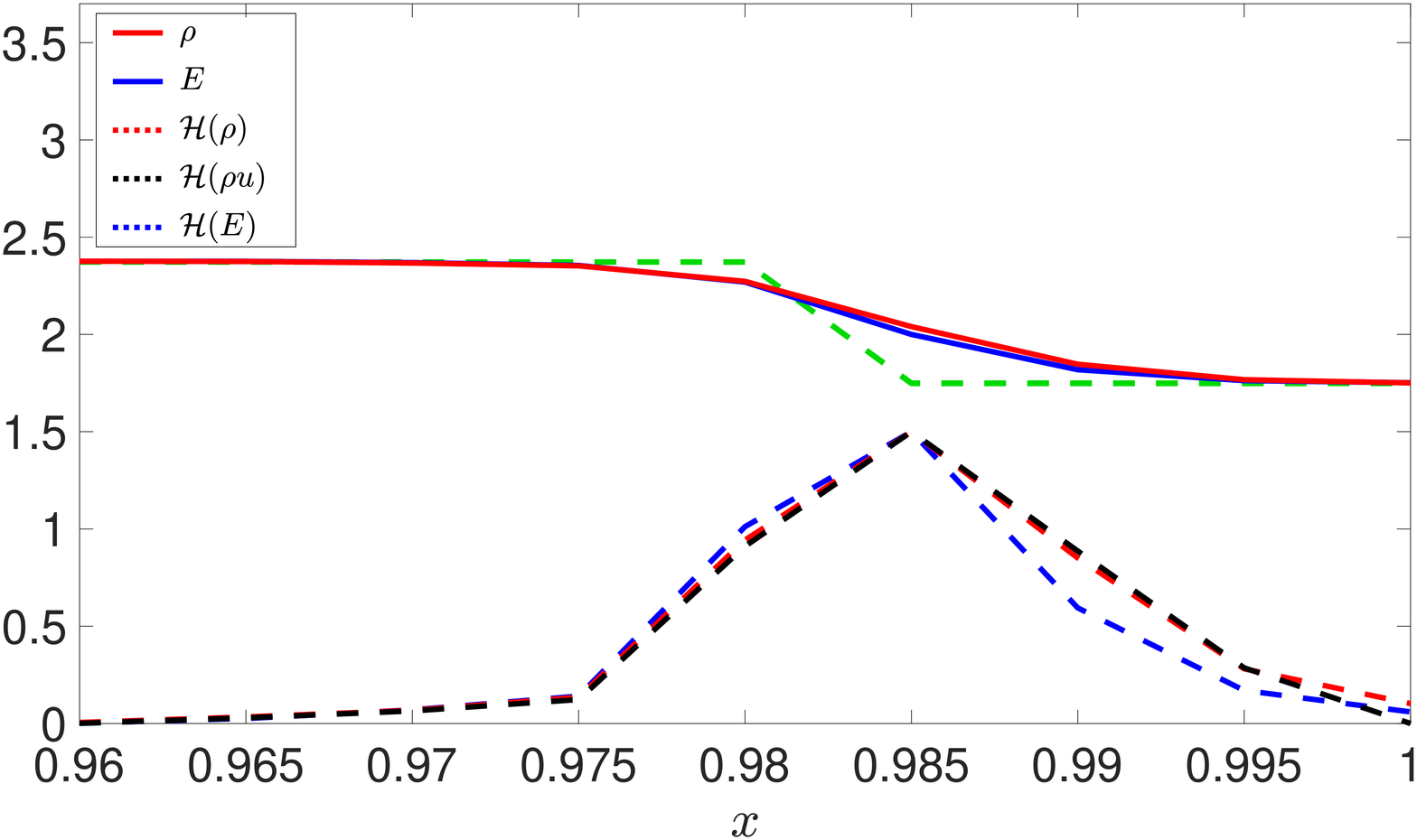}}
\subfigure[$t=0.2925$: post-collision]{\label{fig:bad-flux-compare2}\includegraphics[width=75mm]{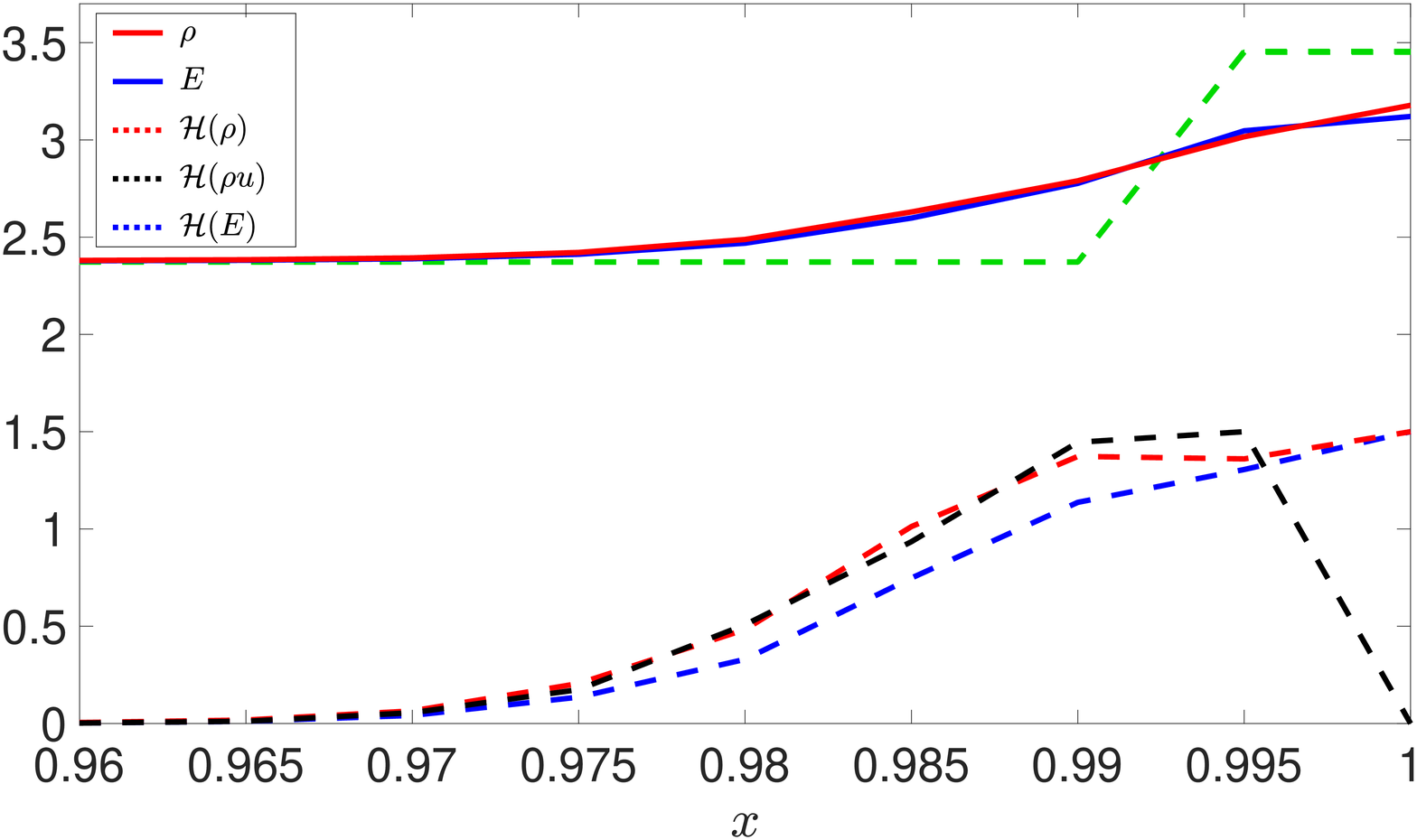}}
\subfigure[$t=0.3000$: post-collision]{\label{fig:bad-flux-compare3}\includegraphics[width=75mm]{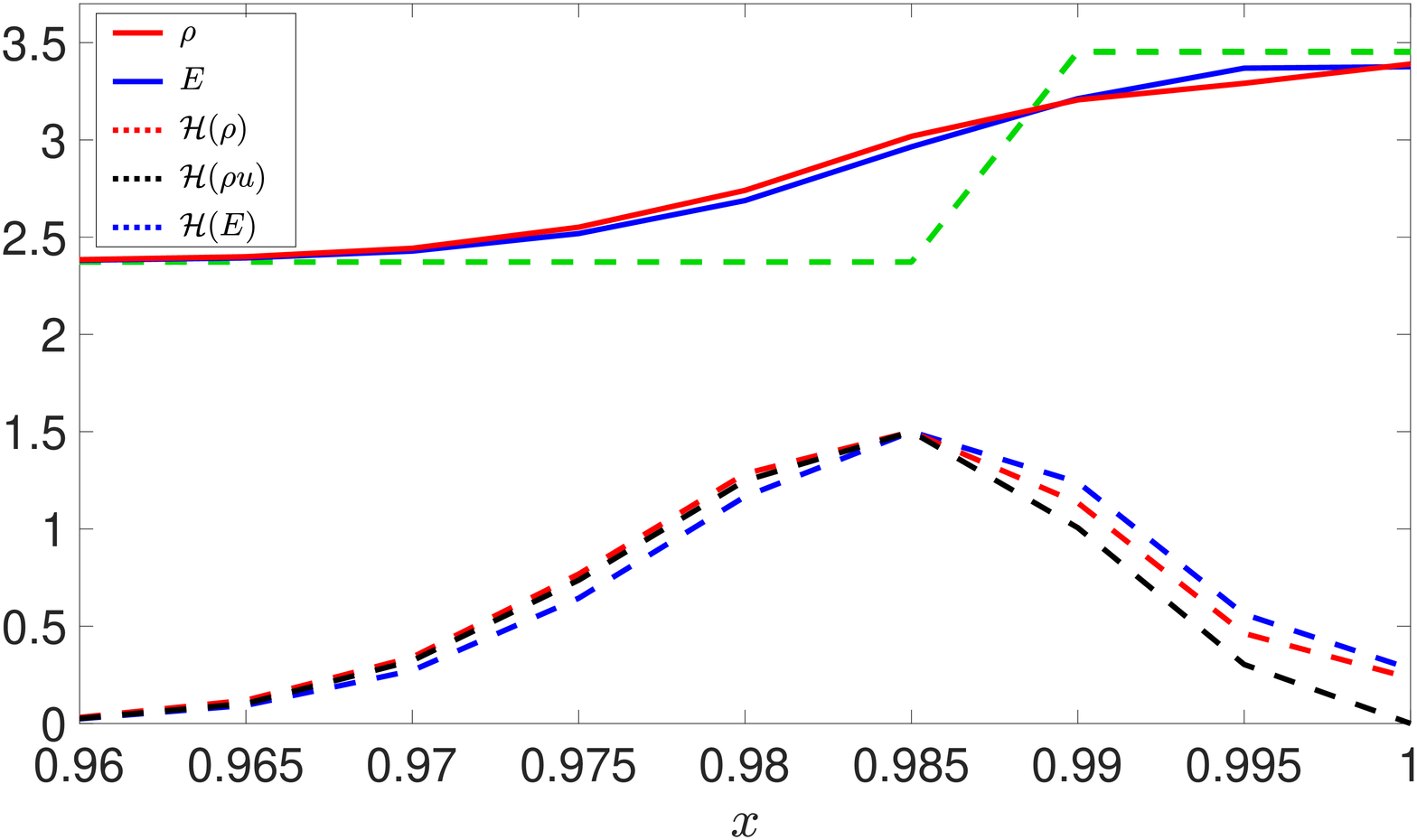}}
\subfigure[$t=0.3050$: post-collision]{\label{fig:bad-flux-compare4}\includegraphics[width=75mm]{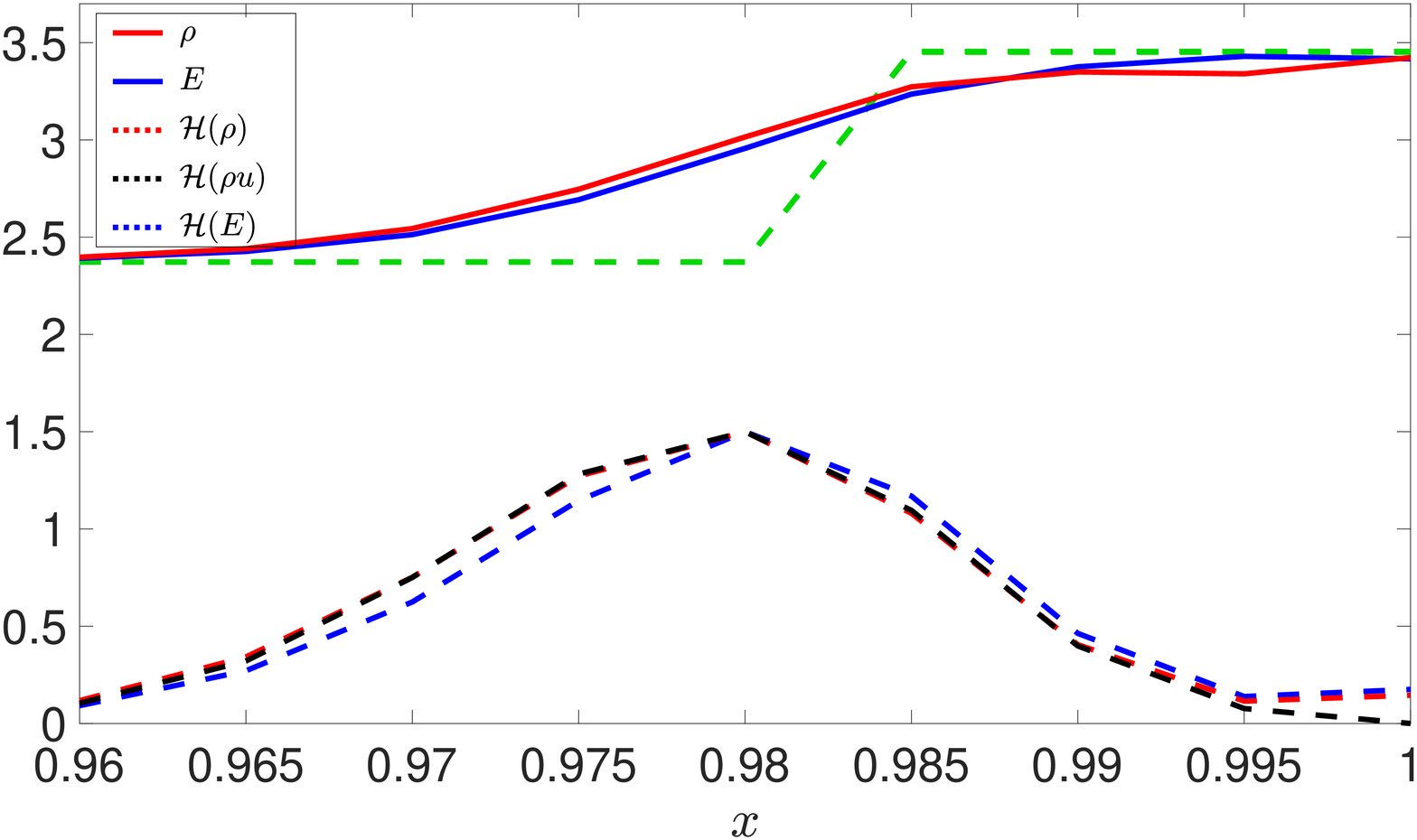}}
\caption{Comparison of the energy and energy forcing term $\mathcal{H}(E)$ 
(blue/blue dashed) with the density, suitably 
resized, and the density forcing term $\mathcal{H}(\rho)$ (red/red dashed) and the 
momentum forcing term $\mathcal{H}(\rho u)$ (black dashed) for the Sod shock tube problem with the wall 
viscosity activated for the momentum equation. The green dashed curve is the exact solution. The figures 
shown are zoomed in at the shock just before or just after the shock front has collided with the wall at $x=1$.}
\label{fig:bad-flux-compare}
\end{figure}

In Fig.\ref{fig:bad-flux-compare}, we compare the energy and density profiles,
 along with the terms $\mathcal{H}(\rho)$, $\mathcal{H}(\rho u )$, and $\mathcal{H}(E)$, 
 all suitably resized\footnote{More precisely, we plot the following: first,
$\frac{3}{2} \frac{\mathcal{H}}{\max_{\Omega}  \mathcal{H}}$ for each of the forcing terms 
$\mathcal{H}(\rho)$, $\mathcal{H}(\rho u)$, and  $\mathcal{H}(E)$; second, the function 
$1.1928+4.4403 \rho$; and finally, the energy $E$.} for ease of comparison, at various times just before or 
after the shock 
fronts have collided with the wall, zoomed-in on the region next to the wall. In 
Fig.\ref{fig:bad-flux-compare1}, the density and energy profiles are very similar, but the forcing terms
are slightly misaligned; it is clear that $\mathcal{H}(E)$ is 
slightly behind both $\mathcal{H}(\rho)$ and $\mathcal{H}(\rho u)$. 
This misalignment causes the solution profiles for the 
energy and density to begin to diverge, as can be seen in Fig.\ref{fig:bad-flux-compare2}. 
Again, there is a misalignment between the forcing terms $\mathcal{H}(E)$ and $\mathcal{H}(\rho)$. 
As the shock moves 
away from the wall in Fig.\ref{fig:bad-flux-compare3} and Fig.\ref{fig:bad-flux-compare4},
the difference between the solution profiles is now clear. Even though the forcing terms
 are now better aligned, the earlier misalignment ensures that the difference between the 
 energy and density profiles is permanent.

\subsubsection{A solution to the wall-heating problem}
The solution to the wall heating problem suggested by 
Noh \cite{Noh1987} is the addition of a heat conduction term to the energy equation. For the WENO-Noh 
scheme we implement in this study, we shall use a heat conduction term of the form\footnote{In equations (2.1)-(2.5) in the paper of Noh {\cite{Noh1987}}, there is, in fact, an additional
term proportional  to $-\rho |\partial_x u|^2 \partial_x u$ 
on the right-hand side of  {\eqref{Noh-heat-conduction}}. We have found that this term is not
 necessary to  remove the wall-heating error, and thus omit it from the Noh scheme we implement in 
this  paper.}
\begin{equation}\label{Noh-heat-conduction}
\partial_x \left( \beta^E_{\operatorname{Noh}}\,  \rho \, |\partial_x u| \,\partial_x e \right), 
\end{equation}
where $e=p/(\gamma-1)\rho = c_v \Theta$ is the   internal energy of the system, proportional to the temperature $\Theta$, with $c_v$ the specific heat capacity at a constant volume. 

We use the following artificial (wall)  viscosity for the 
 energy equation \eqref{EulerC-energy}:
\begin{equation}\label{heat-conduction}
\partial_x \left( \mathcal{B}^{(E)}(t) \, \rho \, C\, \partial_x(E/\rho) \right) \,.
\end{equation}
There are two differences between the terms  \eqref{heat-conduction}  and \eqref{Noh-heat-conduction}: 
 \begin{enumerate}
 \item While \eqref{Noh-heat-conduction} uses the oscillatory  localizing coefficient $|\partial_x u(x,t)|$, we instead use the space-time smooth localizer  $C(x,t)$.
 \item We use $\partial_x (E/\rho)$ in our diffusion operator rather than the function $\partial_x e$.   This  
 difference can be explained as follows:
 equation \eqref{defn-internal-energy} shows that
\begin{equation}\label{twoterms}
\partial_x \left( \mathcal{B}^{(E)}(t) \, \rho \, C\, \partial_x(E/\rho) \right) = 
\partial_x \left( \mathcal{B}^{(E)}(t) \, \rho \, C\, \partial_xe \right) + 
\partial_x \left( \mathcal{B}^{(E)}(t) \, \rho \, C\, u \partial_x u \right)\,.
\end{equation} 
Hence,   \eqref{twoterms} has a similar form to  \eqref{Noh-heat-conduction} (with $C$ replacing $|u_x|$), but with
the additional term $\partial_x \left( \mathcal{B}^{(E)}(t) \, \rho \, C\, u \partial_x u \right)$.   Indeed, the two terms in \eqref{twoterms}
are both proper diffusion operators near shock waves.  This is easy to see: multiplying \eqref{twoterms} by $E$ and integrating by parts then shows that
 $$
 \int_{x_1}^{x_M}  \mathcal{B}^{(E)}(t) \, \rho \, C\, \partial_x e \p_x E \, dx +
  \int_{x_1}^{x_M}  \mathcal{B}^{(E)}(t) \, \rho \, C\, u \,  \partial_x u \p_x E \, dx \,.
 $$
At the shock, $\p_x e$ has the same sign as $\p_x E$ so that  $\int_{x_1}^{x_M}  \mathcal{B}^{(E)}(t) \, \rho \, C\, \partial_x e \p_x E \, dx \ge 0$;
moreover,  in the case of 
a right-traveling shock front,  $\p_x u < 0$, $\p_x E < 0$ and $u > 0$ at the shock, so that 
$  \int_{x_1}^{x_M}  \mathcal{B}^{(E)}(t) \, \rho \, C\, u \,  \partial_x u \p_x E \, dx \ge 0$, while for a
 a left-moving shock, $\p_x u < 0$, $\p_x E > 0$ and $u < 0$, so that once again $  \int_{x_1}^{x_M}  \mathcal{B}^{(E)}(t) \, \rho \, C\, u \,  \partial_x u \p_x E \, dx \ge 0$.
 This then ensures that 
 \eqref{heat-conduction} is a {\it dissipative} operator and that the structure of the artificial viscosity term in  \eqref{heat-conduction} 
 adjusts the Noh-type dissipation $\partial_x \left( \mathcal{B}^{(E)}(t) \, \rho \, C\, \partial_xe \right)$ by the velocity-dependent term
 $\partial_x \left( \mathcal{B}^{(E)}(t) \, \rho \, C\, u \partial_x u \right)$.
 \end{enumerate}
 
In our numerical experiments, presented below, we compare Noh's scheme, called WENO-Noh (see Appendix \ref{sec:appendix}), with our WENO-$C$-$W$ scheme. For
WENO-Noh, we set
$
\beta^{u}_{\operatorname{Noh}} = 15.0$ and $ \beta^{E}_{\operatorname{Noh}} = 10.0$ in \eqref{EulerC-noh}.
These viscosity coefficients were chosen in the following manner: $\beta^{u}_{\operatorname{Noh}}$ was 
first chosen large enough to suppress the post-collision oscillations, and then 
$\beta^{E}_{\operatorname{Noh}}$ was chosen to correct the wall-heating error. 
In our WENO-$C$-$W$ scheme, we set 
$\beta^u = 0.5$, $ \beta^{u}_w = 3.0$, $\beta^{E}=0.0$, and $\beta^{E}_w = 6.0$.

\begin{figure}[H]
\centering
\subfigure[$t=0.20$: velocity, pre-collision]{\label{fig:sod-before-collision4}\includegraphics[width=75mm]{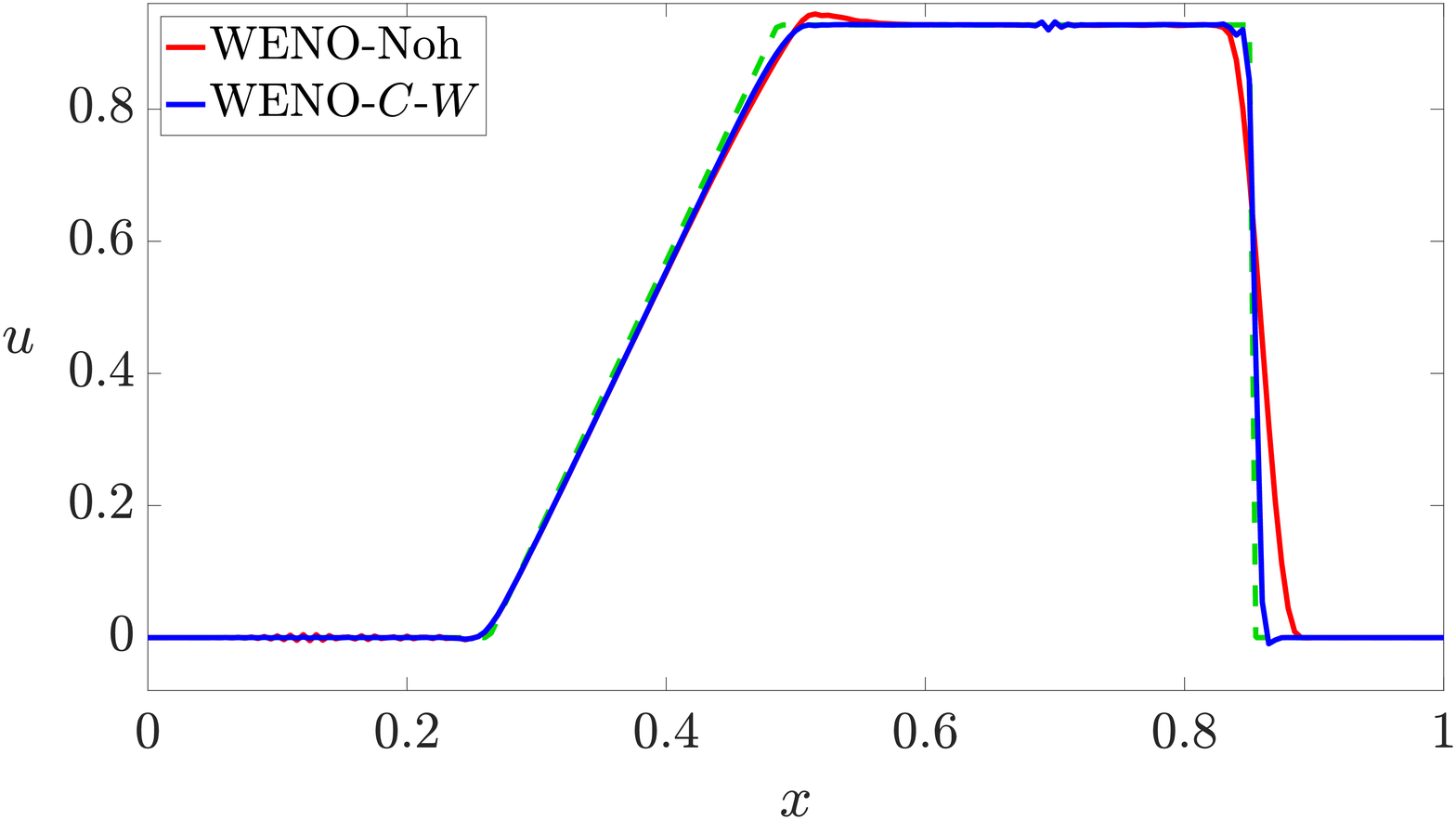}}
\subfigure[$t=0.36$: velocity, post-collision]{\label{fig:sod-after-collision4}\includegraphics[width=75mm]{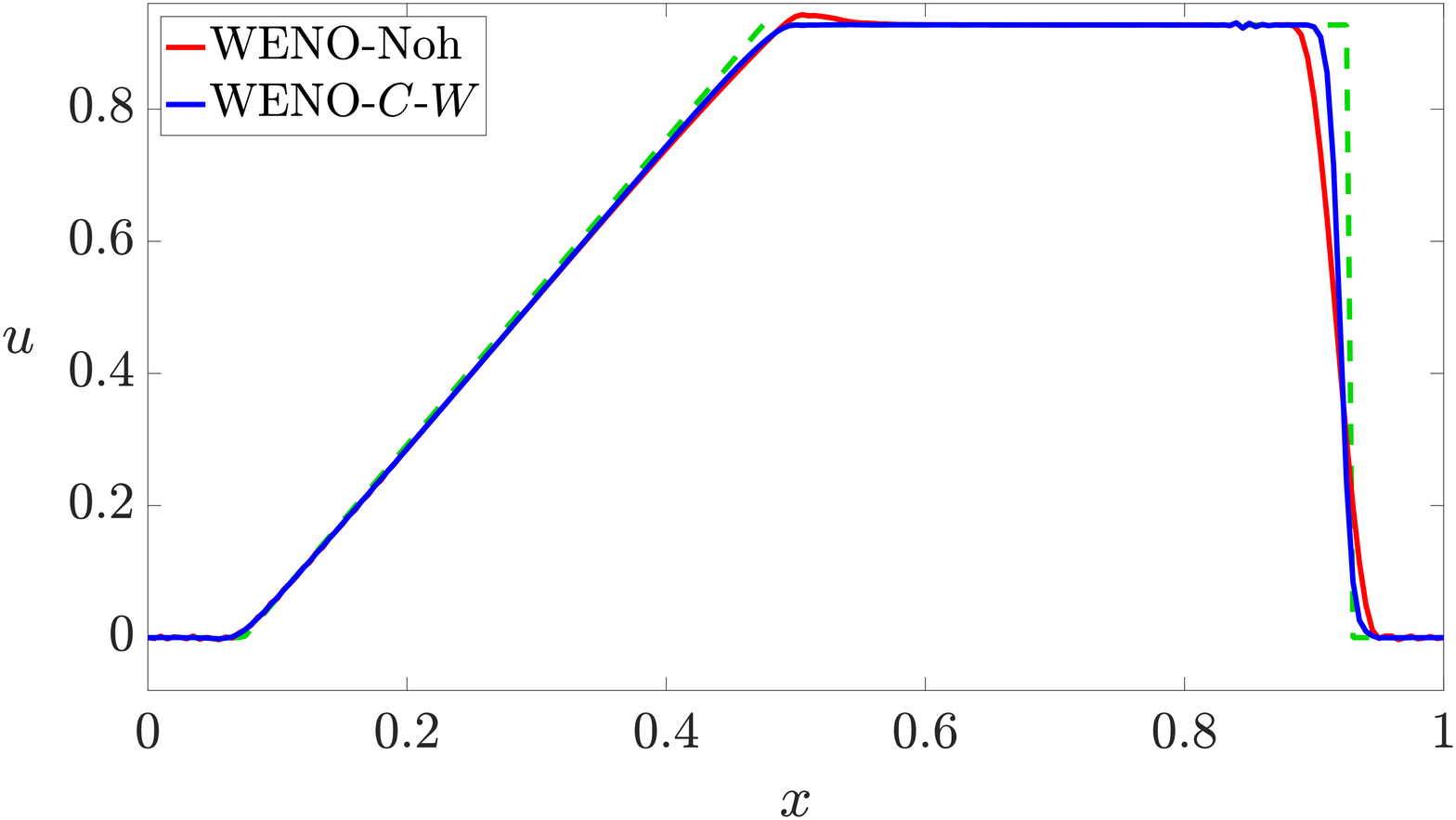}}
\subfigure[$t=0.36$: density, post-collision]{\label{fig:sod-after-collision-rho2}\includegraphics[width=75mm]{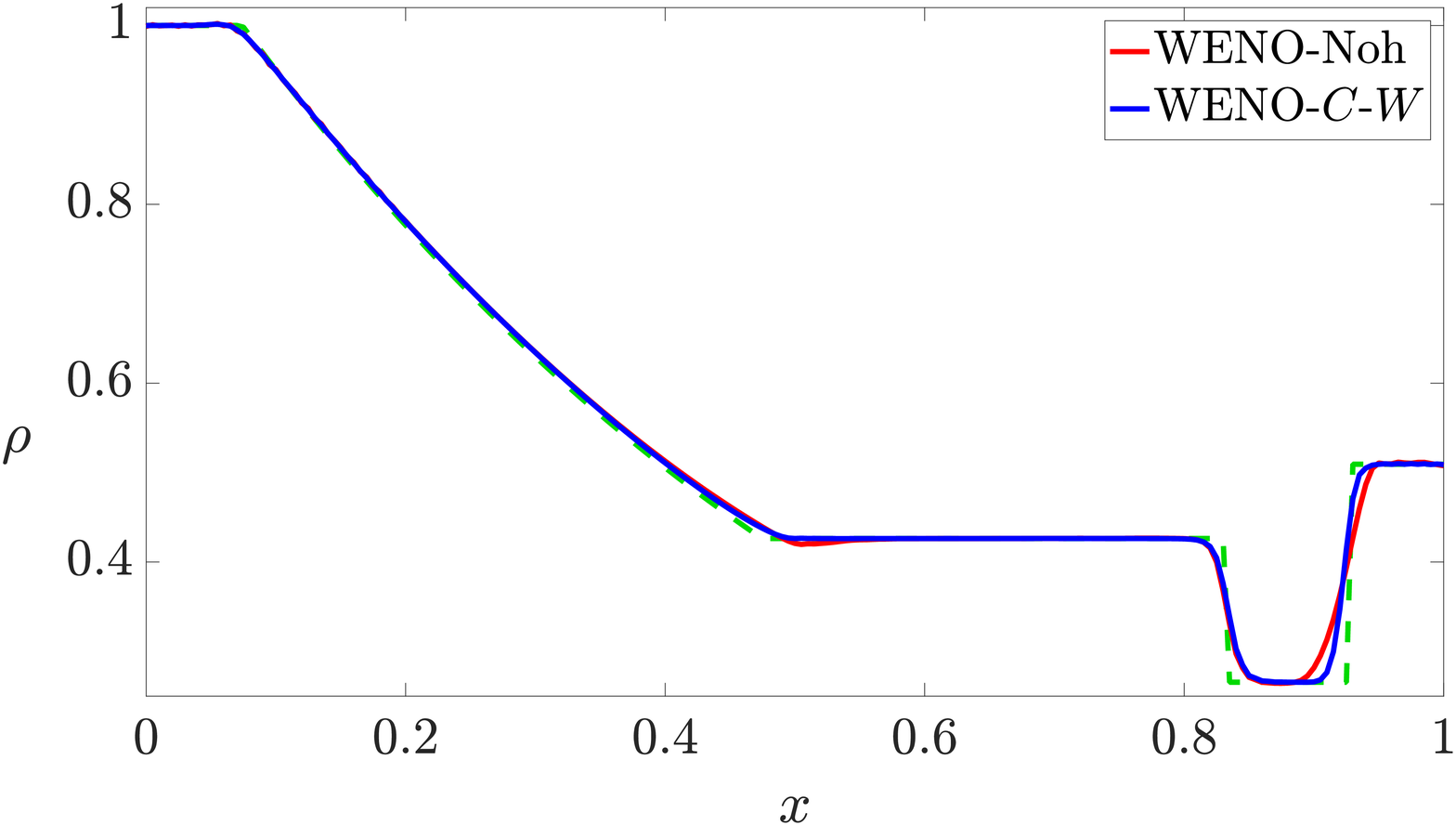}}
\subfigure[$t=0.36$: density, post-collision zoom-in]{\label{fig:sod-after-collision-zoom-rho2}\includegraphics[width=75mm]{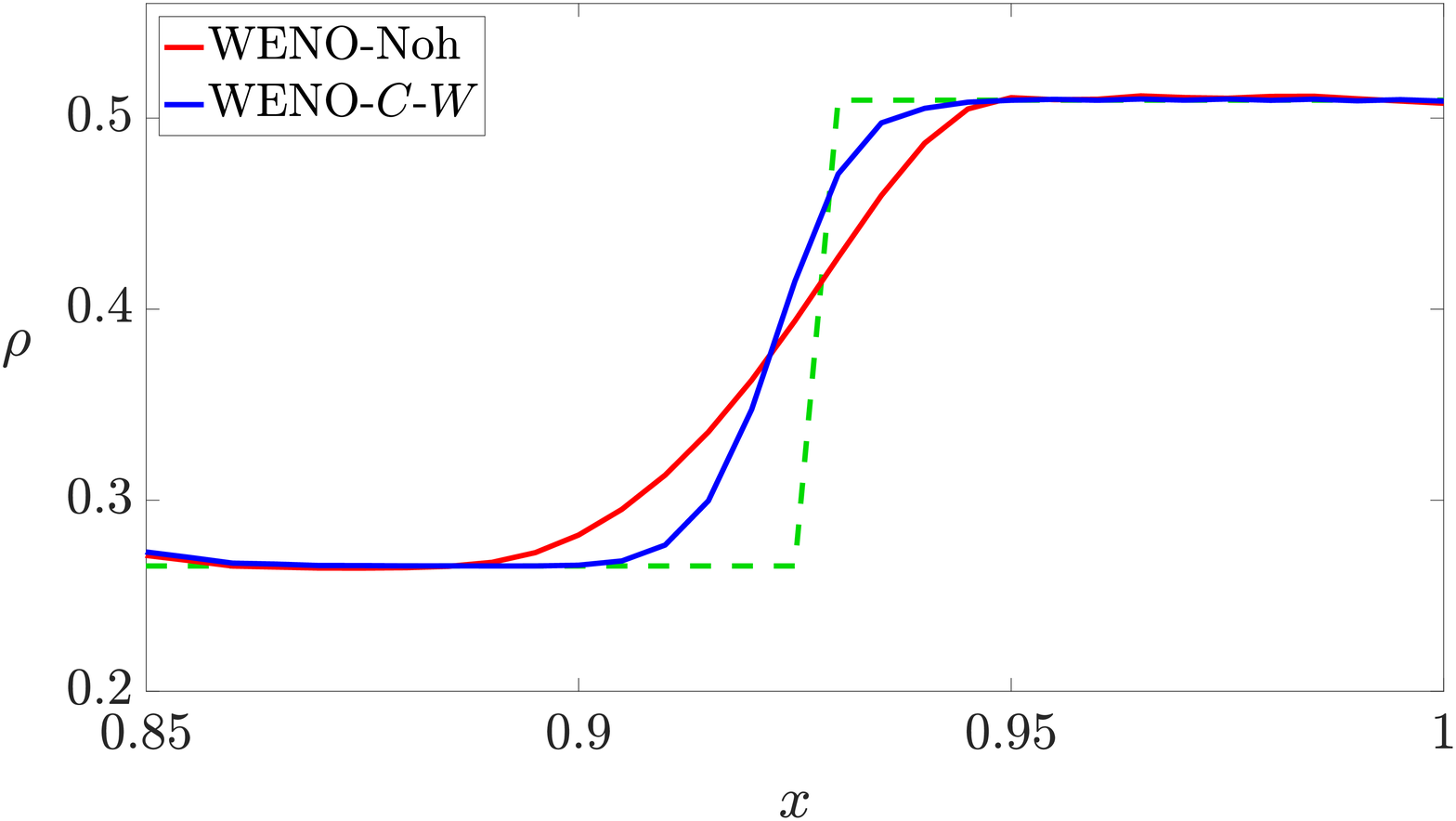}}
\caption{The velocity and density
 profiles for the Sod shock tube problem before  and  after shock-wall collision.}
\label{fig:sod-collision-veleng1}
\end{figure}

In Fig.\ref{fig:sod-collision-veleng1}, we compare the velocity and density
profiles computed with the two schemes above. It is clear that the WENO-$C$-$W$ scheme 
produces a superior solution both before and after the shock-wall collision. The large amount 
of viscosity needed in the WENO-Noh scheme post-collision means that the solution prior to 
shock-wall collision is affected, with a smeared shock curve and overshoot at the top of the 
expansion wave. Moreover, even the relatively large value of $\beta^{u}_{\operatorname{Noh}}$ as 
compared with $\beta^{u} + \beta^{u}_{w}$ is unable to fully suppress the oscillations behind the 
shock curve that occur post-collision. This is due to the smoothness of the localizing coefficient 
$C$ as compared with the 
rough nature of $|\partial_x u|$. 

As is the case with the velocity profile, 
prior to shock collision the WENO-$C$-$W$ scheme produces a superior solution for 
the density profile, with a much 
sharper shock front and more accurate expansion wave. Post-collision, the heat conduction terms
ensure that neither of the methods exhibit the wall heating error. However, there are still small oscillations
present in the solution computed with the WENO-Noh scheme, and the shock front is much more
smeared than that of the solution computed with WENO-$C$-$W$. 

\begin{figure}[H]
\centering
\subfigure[$t=0.2725$: pre-collision]{\label{fig:good-flux-compare1}\includegraphics[width=75mm]{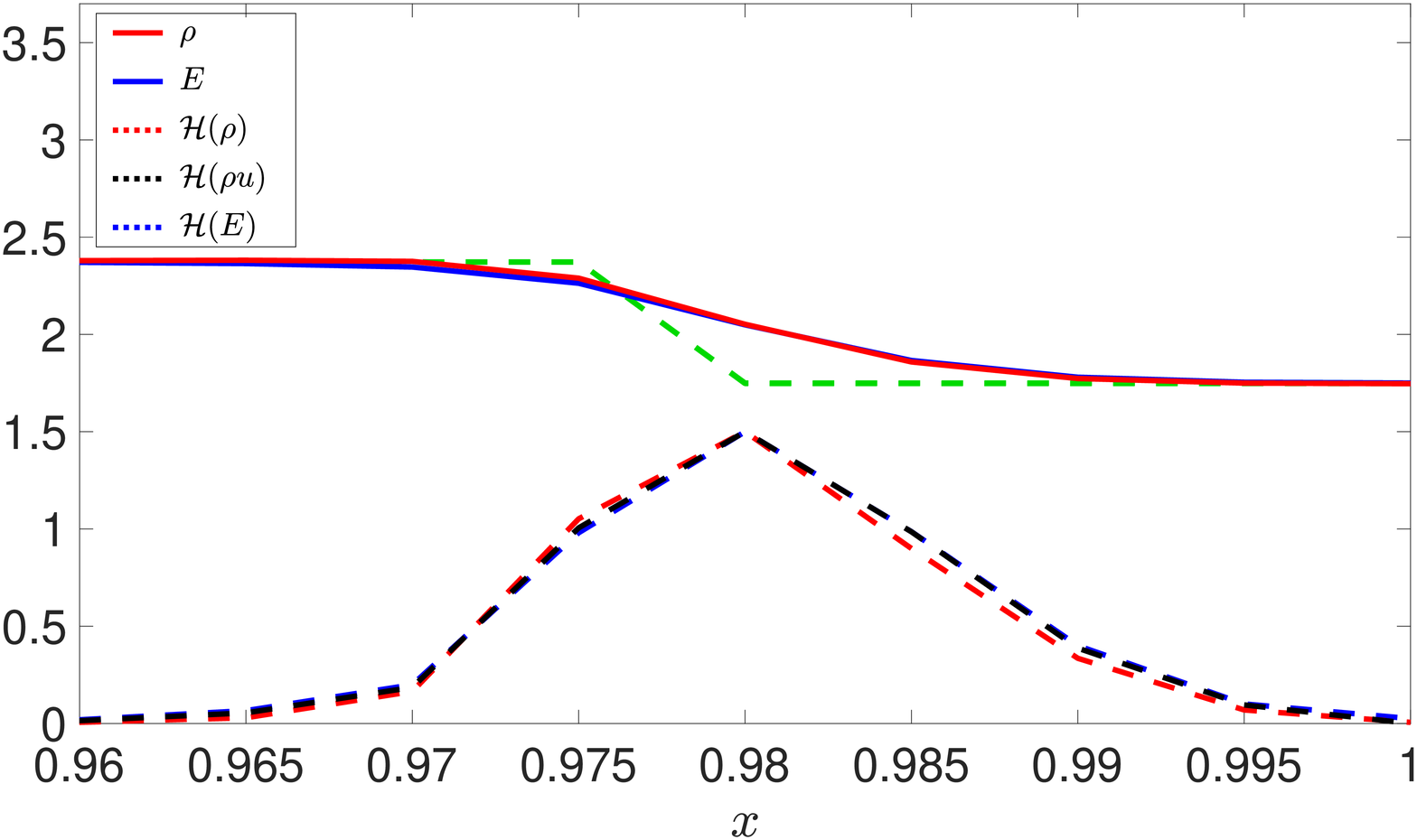}}
\subfigure[$t=0.2925$: post-collision]{\label{fig:good-flux-compare2}\includegraphics[width=75mm]{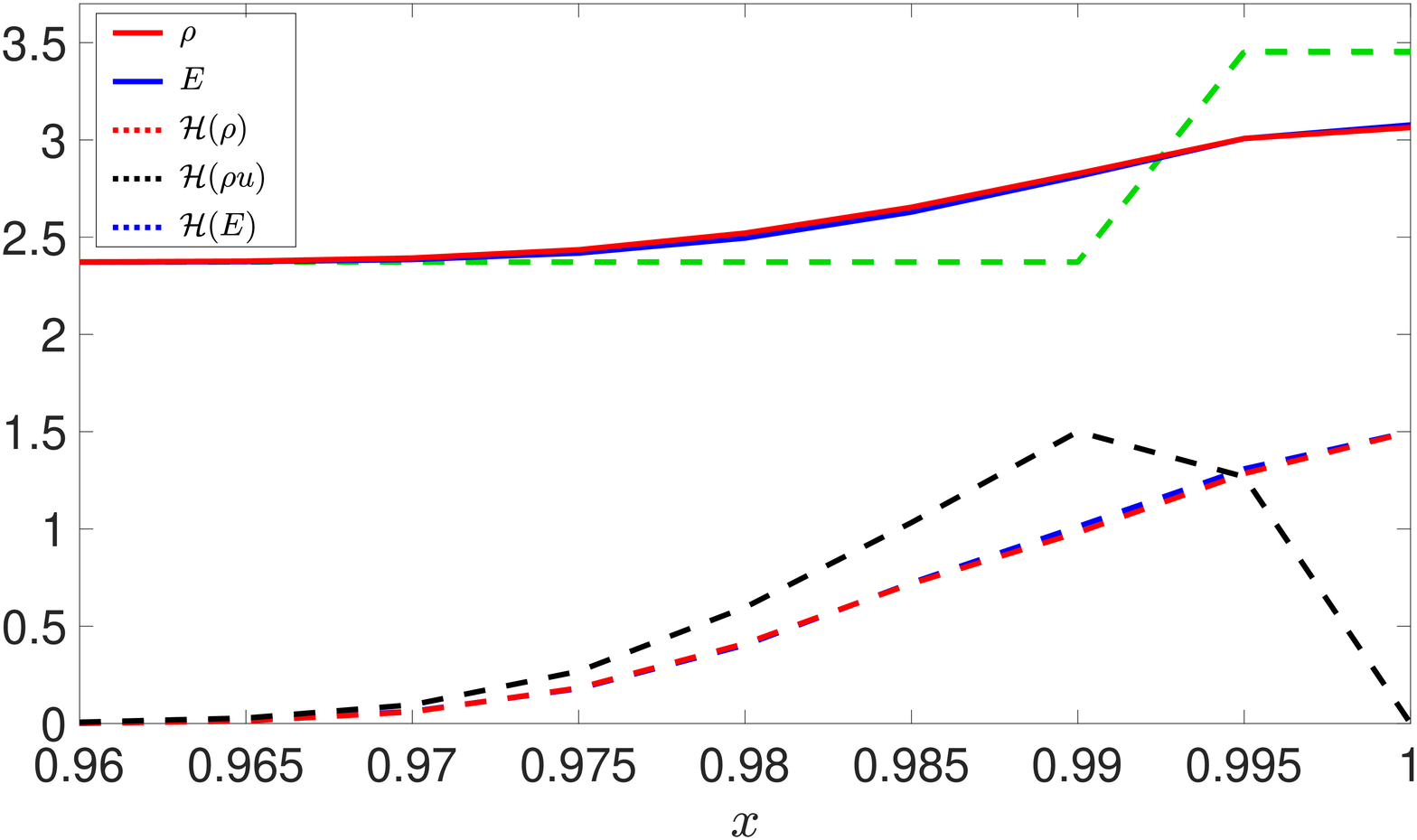}}
\subfigure[$t=0.3000$: post-collision]{\label{fig:good-flux-compare3}\includegraphics[width=75mm]{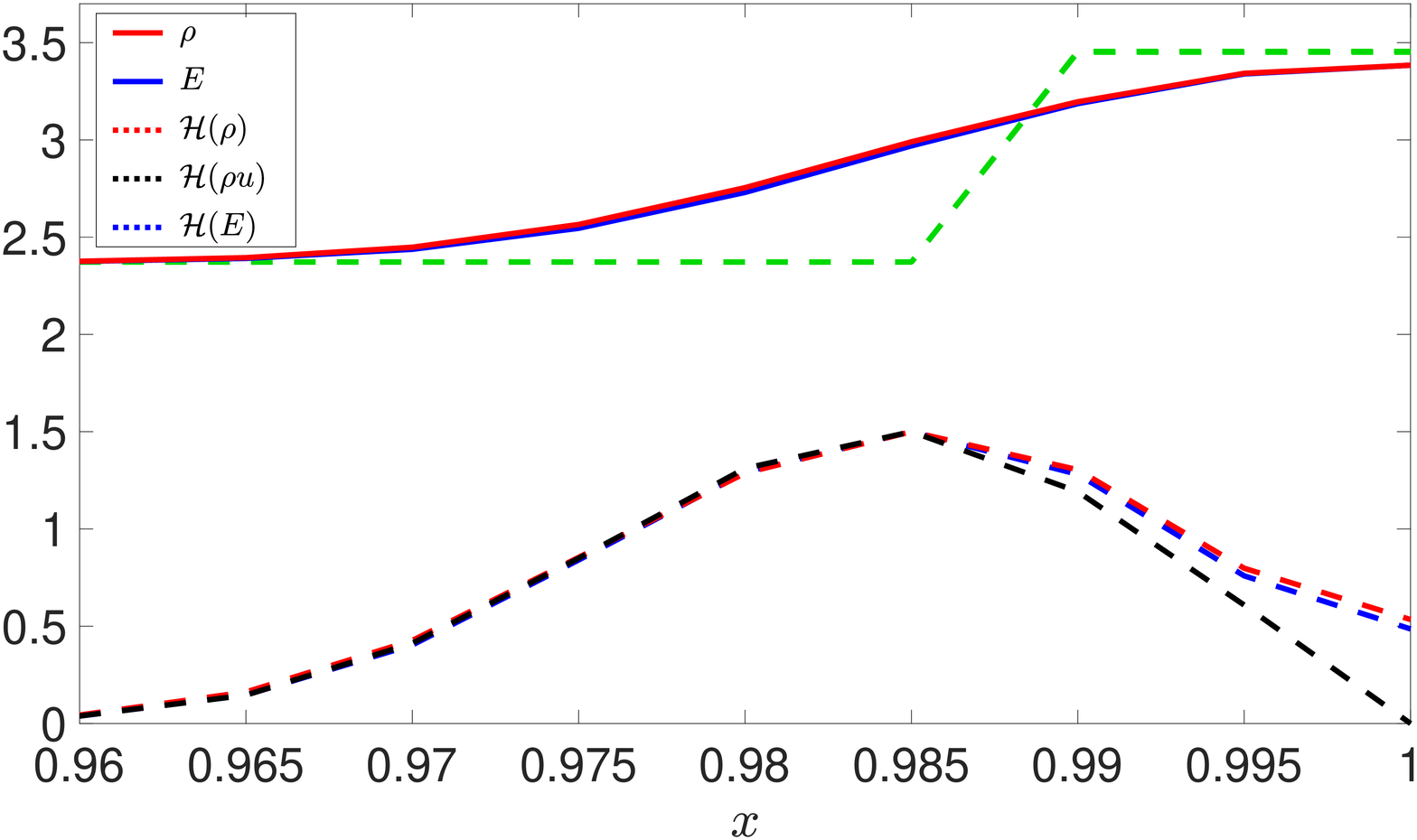}}
\subfigure[$t=0.3050$: post-collision]{\label{fig:good-flux-compare4}\includegraphics[width=75mm]{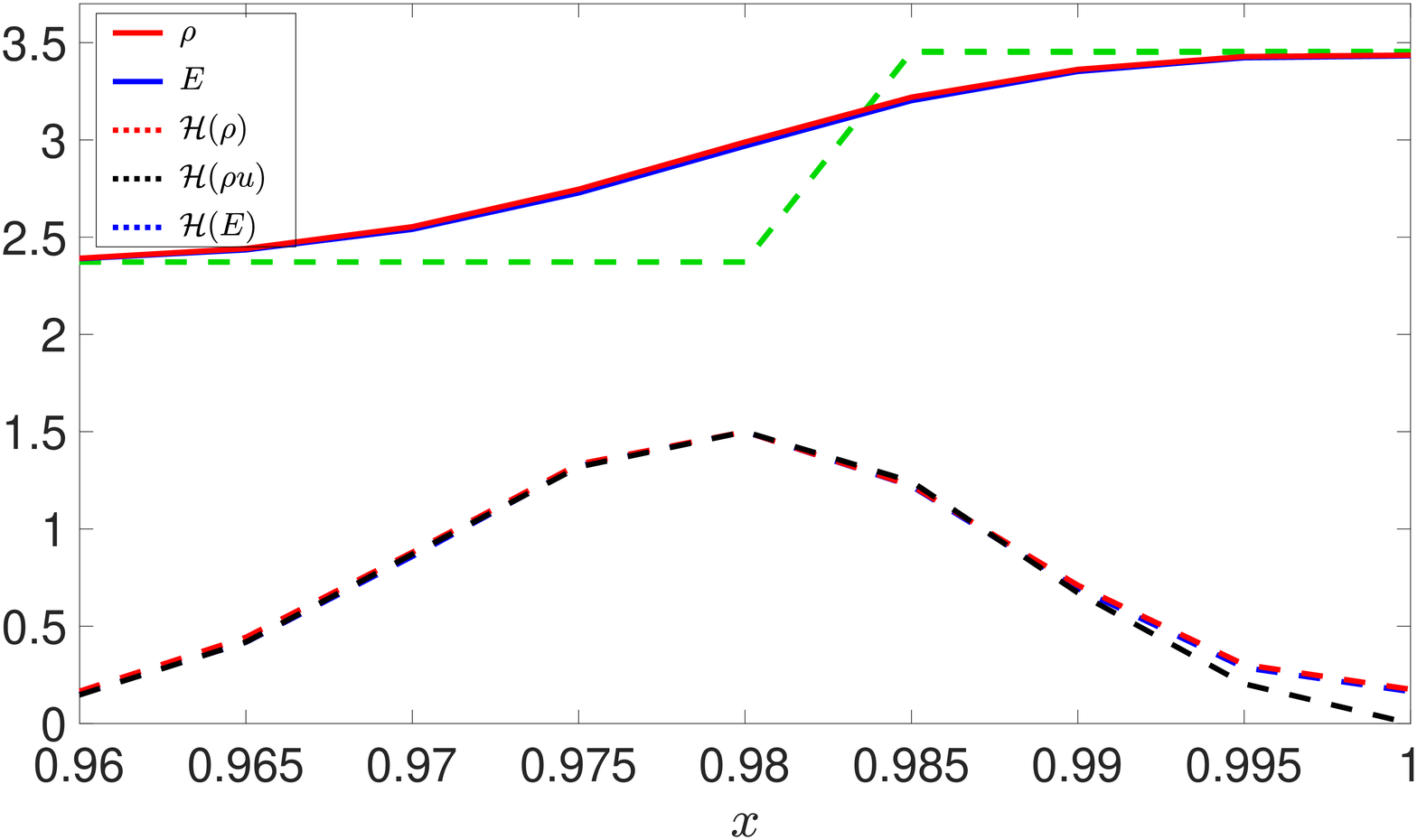}}
\caption{Comparison of the energy and energy forcing term $\mathcal{H}(E)$ 
(blue/blue dashed) with the density, suitably 
resized, and the density forcing term $\mathcal{H}(\rho)$ (red/red dashed) and the 
momentum forcing term $\mathcal{H}(\rho u)$ (black dashed) for the Sod shock tube problem with the wall 
viscosity activated for the momentum and energy equations. 
The green dashed curve is the exact solution. The figures 
shown are zoomed in at the shock just before or just after the shock front has collided with the wall at $x=1$.}
\label{fig:good-flux-compare}
\end{figure}

Finally, comparing Fig.\ref{fig:bad-flux-compare} and Fig.\ref{fig:good-flux-compare}, we 
see that the wall viscosity for the energy equation has properly
aligned the forcing terms $\mathcal{H}(\rho)$, $\mathcal{H}(\rho u)$ and $\mathcal{H}(E)$.    Realignment of the gradient of fluxes removes the
wall-heating problem, created by the smearing of the shock fronts. The artificial viscosity term \eqref{heat-conduction} is responsible for this realignment.

%

\subsubsection{Noise removal with the noise indicator}

We now employ our noise indicator algorithm to the Sod shock tube problem with the aim of removing 
the noise present in the velocity profile at the contact discontinuity.    

\begin{figure}[H]
\centering
\subfigure[$t=0.20$]{\label{fig:sod-noise-indicator1}\includegraphics[width=75mm]{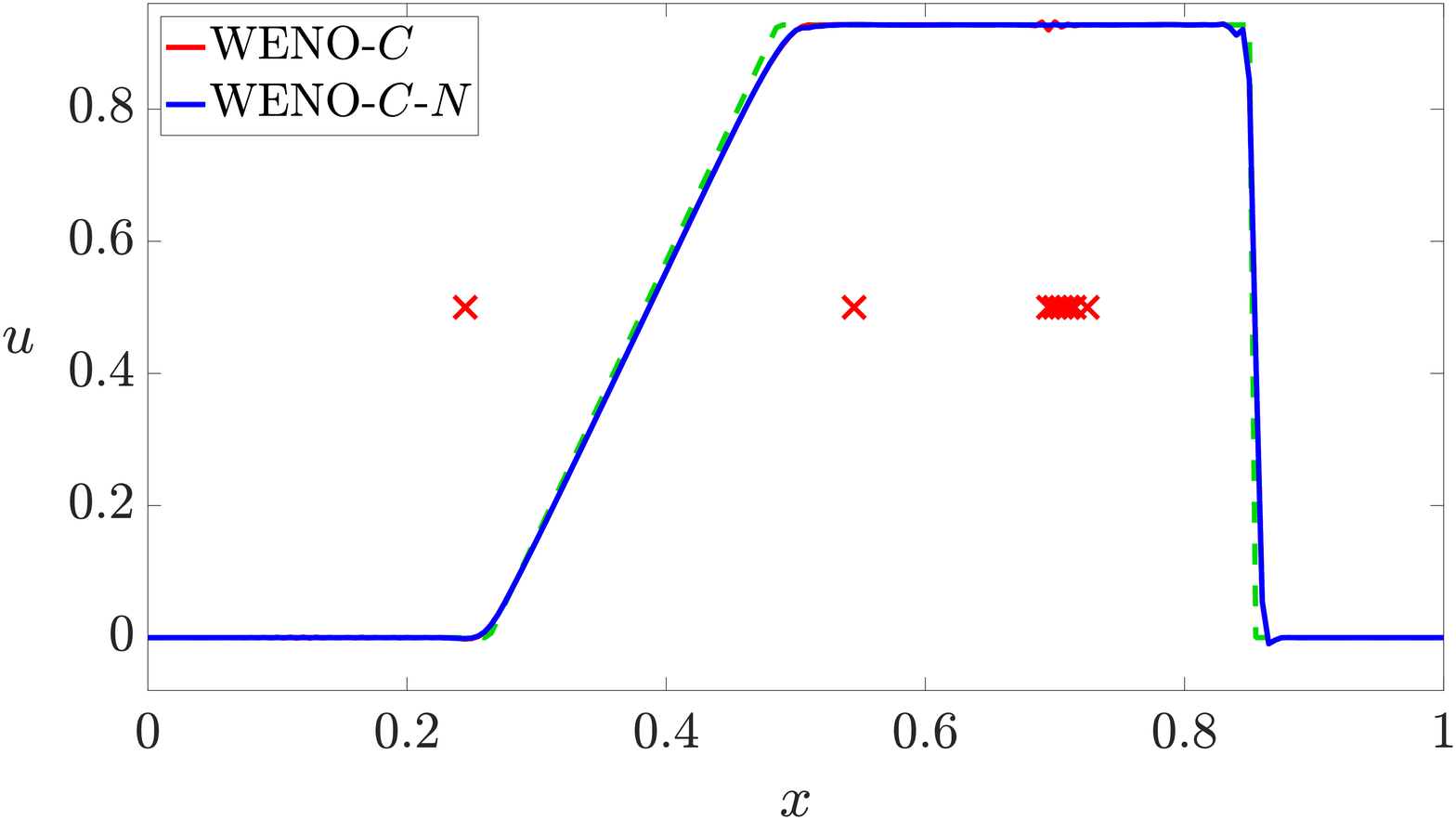}}
\subfigure[$t=0.20$: zooming in on the noise at the contact discontinuity]{\label{fig:sod-noise-indicator2}\includegraphics[width=75mm]{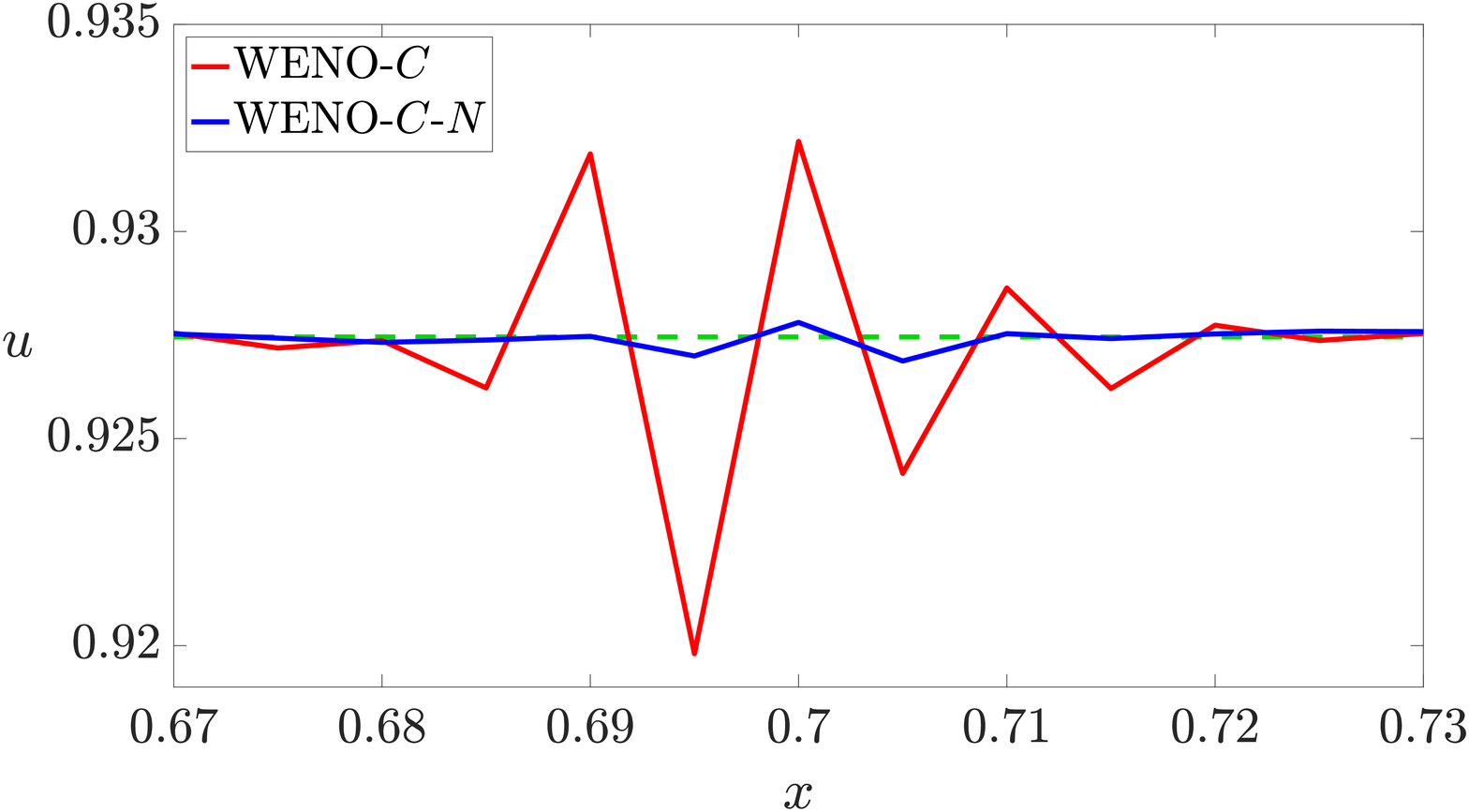}}
\caption{Comparison of the velocity profiles for the Sod shock tube problem computed with 
WENO-$C$ and WENO-$C$-$N$ with 201 cells.
The red crosses in Fig.\ref{fig:sod-noise-indicator1} indicate where the function 
$\mathbbm{1}_{\operatorname{noise}}(x)$ is active.}
\label{fig:sod-noise-indicator}
\end{figure}

In the test below, we employ our WENO-$C$-$N$ scheme with $\eta$ chosen 
such that $\eta \Delta \tau /\Delta x^2 = 0.005$ in the heat equation used for noise removal;  an explicit time-stepping scheme is used
and only one time-step is taken.  For the noise detection algorithm, $C_{\text{ref}}$ in \eqref{cref}  is computed using
$\delta h = 0.0001$.   Fig.\ref{fig:sod-noise-indicator2} shows that the noise indicator removes the spurious
oscillations from the velocity profile. The localized diffusion mechanism ensures that the solution
in other regions is unchanged, and this is demonstrated in Fig.\ref{fig:sod-noise-indicator1}. We note that the 
noise indicator algorithm affects neither the sharpness of the shock front nor the speed of the shock, nor the order of the numerical
method (which we shall show results for below).

 \begin{wrapfigure}{L}{0.4\textwidth}
\includegraphics[scale=0.15]{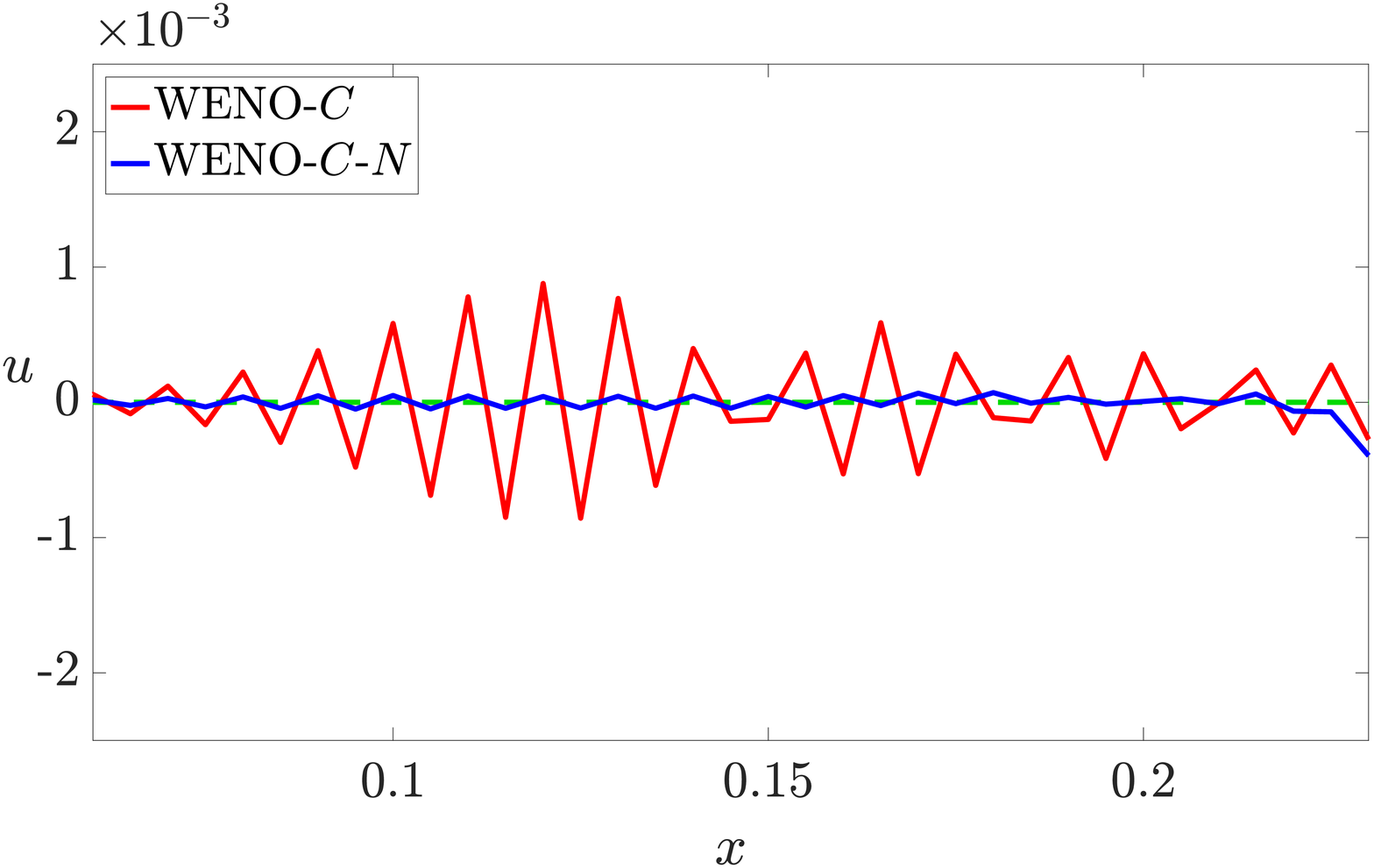}
\end{wrapfigure}
There is also high-frequency noise present to left of the expansion wave (shown in the figure to the left); again, the noise indicator 
detects and removes this noise.

\begin{figure}[H]
\centering
\subfigure[$t=0.36$]{\label{fig:sod-noise-indicator-wall-viscosity1}\includegraphics[width=75mm]{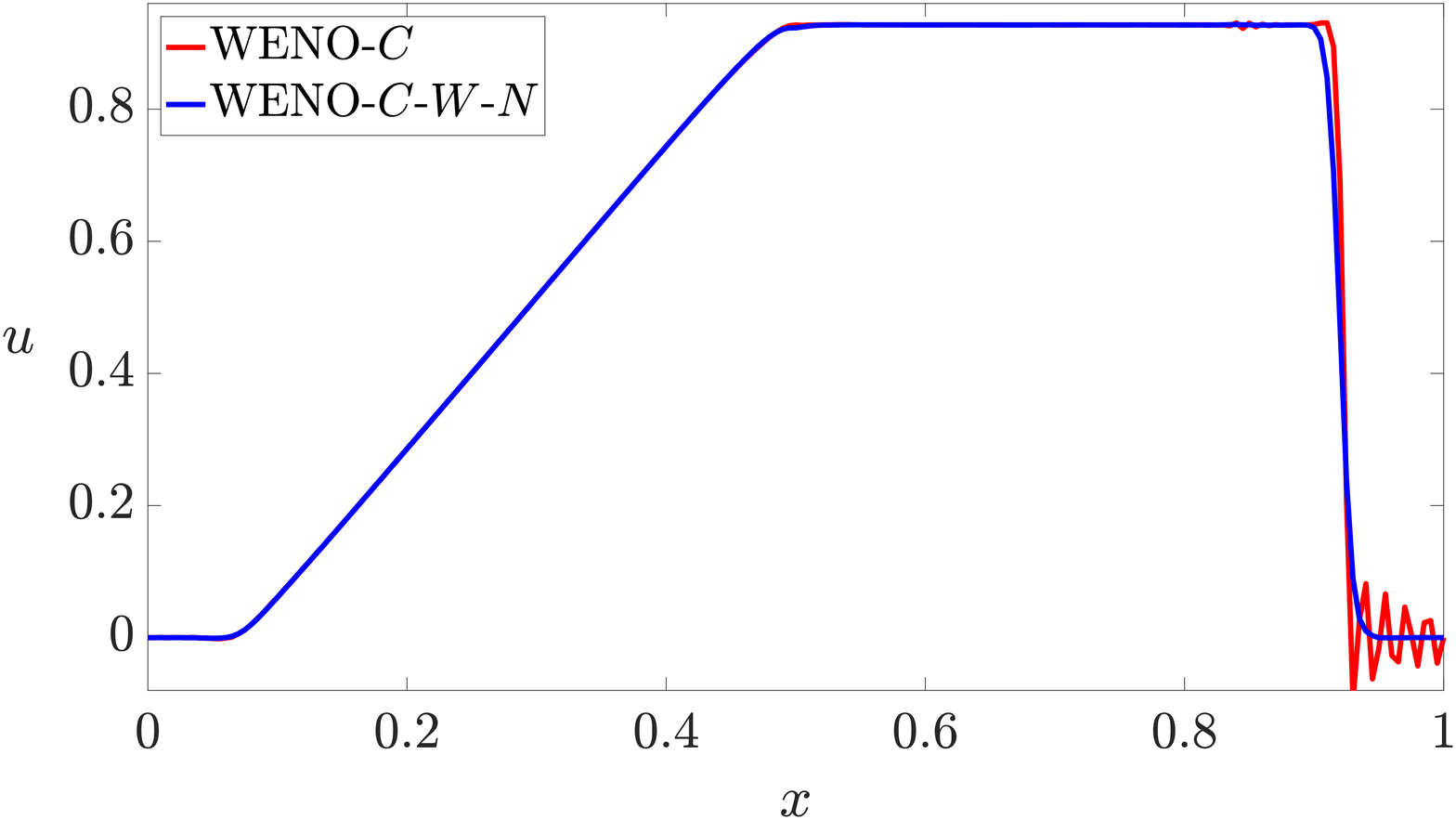}}
\subfigure[$t=0.90$]{\label{fig:sod-noise-indicator-wall-viscosity2}\includegraphics[width=75mm]{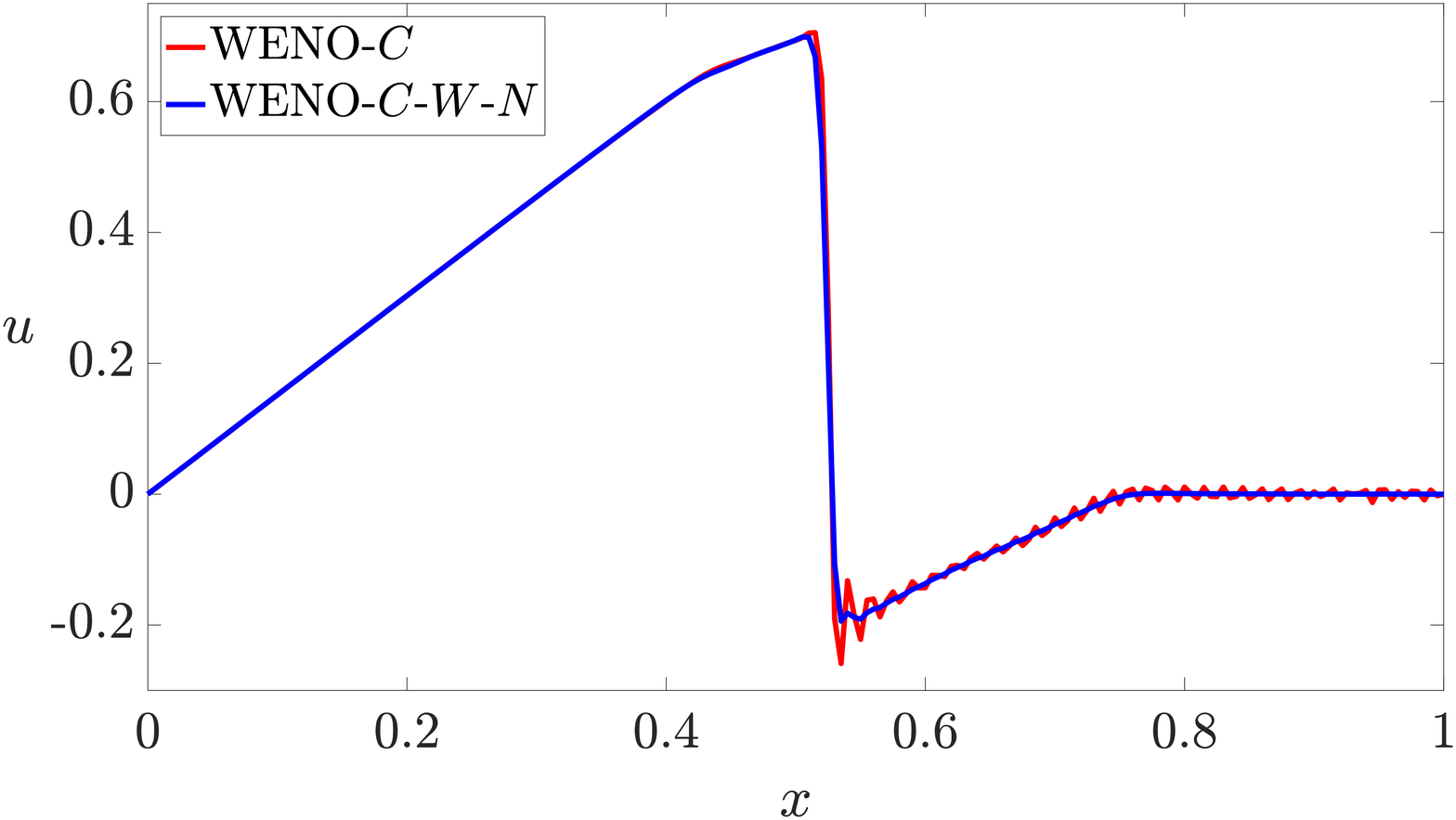}}
\caption{Comparison of the velocity solution profile produced using 
WENO-$C$ and WENO-$C$-$W$-$N$ for the post-collision bounce-back in the Sod shock tube problem with 
201 cells.}
\label{fig:sod-noise-indicator-wall-viscosity}
\end{figure}

In Fig.\ref{fig:sod-noise-indicator-wall-viscosity1}, we show the velocity profile, computed using the 
 WENO-$C$-$W$-$N$  scheme, after the shock-wall collision. Here, all of the post-collision noise is damped
 by the wall viscosity. The WENO-$C$-$W$-$N$ scheme removes spurious oscillations in the solution, 
 while ensuring that a sharp shock front and the correct wave speed are retained, even after multiple 
 shock-wall collisions, as shown in Fig.\ref{fig:sod-noise-indicator-wall-viscosity2}. 
 
\subsubsection{Error analysis and convergence tests}

We now compare the errors of the various numerical schemes given in Table \ref{table:schemes} in Appendix \ref{sec:appendix} applied to the
Sod shock-wall collision and bounce-back problem.
The solutions are computed with all parameters fixed across the different methods,
giving an objective evaluation of each scheme.
 
As advised by Greenough \& Rider {\cite{GreenoughRider2004}}, and in order
to fairly compare our results with those found in the numerics literature, we use the CFL number equal to
0.6.
We have found that the use of a smaller CFL number of 0.3 does not (appreciably)  change the 
conclusions of our numerical tests. For instance, the solutions produced using WENO and 
WENO-$C$-$W$-$N$ show (roughly) the same relative error and order of convergence when CFL=0.3 as 
they do when CFL=0.6. 
However, the presence of the nonlinear artificial viscosity terms in the Euler-$C$ equations, when combined with
an explicit time-integration scheme, place restrictions on the CFL number that would otherwise not be present
in the stand-alone WENO algorithm.  In particular, for the Sod test problem, due to the additional artificial viscosity
present during the shock-wall collision phase, we have found an upper bound on the CFL number to be  $\approx$ 0.7.
While 
our stand-alone WENO scheme  is (formally) stable for 
much larger CFL numbers,  the relative 
error  and the  order of convergence  degrades as the CFL number is increased. Indeed, it is demonstrated
by Greenough \& Rider {\cite{GreenoughRider2004}} that only 75\% of the fifth-order convergence rate of WENO is achieved when 
CFL=1.0, whereas the full fifth-order convergence is achieved for CFL=0.6. Therefore, the use of the smaller 
CFL=0.6 is also necessary for the stand-alone WENO scheme.

\begin{table}[H]
\centering
\renewcommand{\arraystretch}{1.0}
\scalebox{0.8}{
\begin{tabular}{|lc|cccc|}
\toprule
\midrule
\multirow{2}{*}{\textbf{Scheme}} &  & \multicolumn{4}{c|}{\textbf{Cells}}\\

{}  & & 101   & 201    & 401   & 801\\
\midrule
\multirow{2}{*}{WENO} & Error & 
$1.662 \times 10^{-2}$  & $1.772 \times 10^{-2}$  & $1.086 \times 10^{-2}$ & $8.214 \times 10^{-3}$\\
				    & Order & -- & -0.093   & 0.706  & 0.403\\
\midrule
\multirow{2}{*}{WENO-$|u_x|$} & Error & 
$1.534 \times 10^{-2}$ & $1.441 \times 10^{-2}$  & $8.444 \times 10^{-3}$  & $5.864 \times 10^{-3}$\\
				    & Order & -- & 0.090   & 0.771  & 0.526\\
\midrule
\multirow{2}{*}{WENO-Noh} & Error & 
$3.436 \times 10^{-2}$ & $1.799 \times 10^{-2}$  & $9.117 \times 10^{-3}$  & $4.795 \times 10^{-3}$\\
				    & Order & -- & 0.934   & 0.980  & 0.927\\
\midrule
\multirow{2}{*}{WENO-$N$} & Error & 
$1.667 \times 10^{-2}$ & $1.666 \times 10^{-2}$   & $1.064 \times 10^{-2}$  & $7.262 \times 10^{-3}$\\
				    & Order & -- & 0.001   & 0.648  & 0.550\\
\midrule
\multirow{2}{*}{WENO-$C$} & Error & 
$1.520 \times 10^{-2}$ & $1.160 \times 10^{-2}$   & $6.453 \times 10^{-3}$  & $3.927 \times 10^{-3}$\\
				    & Order & -- & 0.390   & 0.846  & 0.717\\
\midrule
\multirow{2}{*}{WENO-$C$-$N$} & Error & 
$1.504 \times 10^{-2}$ & $1.134 \times 10^{-2}$   & $6.412 \times 10^{-3}$  & $3.780 \times 10^{-3}$\\
				    & Order & -- & 0.407   & 0.823  & 0.763\\
\midrule
\multirow{2}{*}{WENO-$C$-$W$} & Error & 
$1.990 \times 10^{-2}$ & $1.151 \times 10^{-2}$   & $5.774 \times 10^{-3}$  & $3.019 \times 10^{-3}$\\
				    & Order & -- & 0.790   & 0.995  & 0.936\\
\midrule
\multirow{2}{*}{WENO-$C$-$W$-$N$} & Error & 
$1.983 \times 10^{-2}$ & $1.146 \times 10^{-2}$   & $5.770 \times 10^{-3}$  & $3.018 \times 10^{-3}$\\
				    & Order & -- & 0.791   & 0.990  & 0.935\\
\midrule
\bottomrule
\end{tabular}}
\caption{Post shock-wall collision ($t=0.36$) $L^1$ error of the computed velocity minus the exact solution and convergence for the 
Sod  problem with shock-wall collision and bounce-back.}
\label{table:sod-error-post}
\end{table}

In Table \ref{table:sod-error-post}, we list the $L^1$ error of the computed velocity minus the exact solution, and study the 
order of convergence for the Sod problem with shock-wall collision and bounce-back.
WENO-$C$ produces solutions that are significantly better than 
those produced with 
WENO-$|u_x|$, which  are significantly better than those solutions produced with  the stand-alone WENO algorithm. 
Note that the use of the noise removal algorithm consistently improves the error bounds, while maintaining the order of accuracy.

On the coarser grids containing 101 or 201 cells, the WENO-$C$-$W$ and WENO-$C$-$W$-$N$ schemes 
produce solutions with slightly larger errors than the solution produced with WENO-$C$-$N$; this is caused 
by the slight smearing of the shock, 
post wall collision. The solutions are, however, {\it{qualitatively}} significantly better, as evidenced by 
Fig.\ref{fig:sod-collision-veleng1} and Fig.\ref{fig:sod-noise-indicator-wall-viscosity} above, as well as 
Fig.\ref{fig:sod-weno-c-comparison} below.
Both WENO-$C$-$W$ and WENO-$C$-$W$-$N$ maintain a relatively high order of accuracy,
whereas the presence of the post-collision noise ensures that both WENO and WENO-$|u_x|$ have 
convergence rates that are irregular and relatively poor.

 \begin{figure}[H]
\centering
\subfigure[$t=0.36$, velocity post-collision]{\label{fig:sod-weno-c-comparison1}\includegraphics[width=75mm]{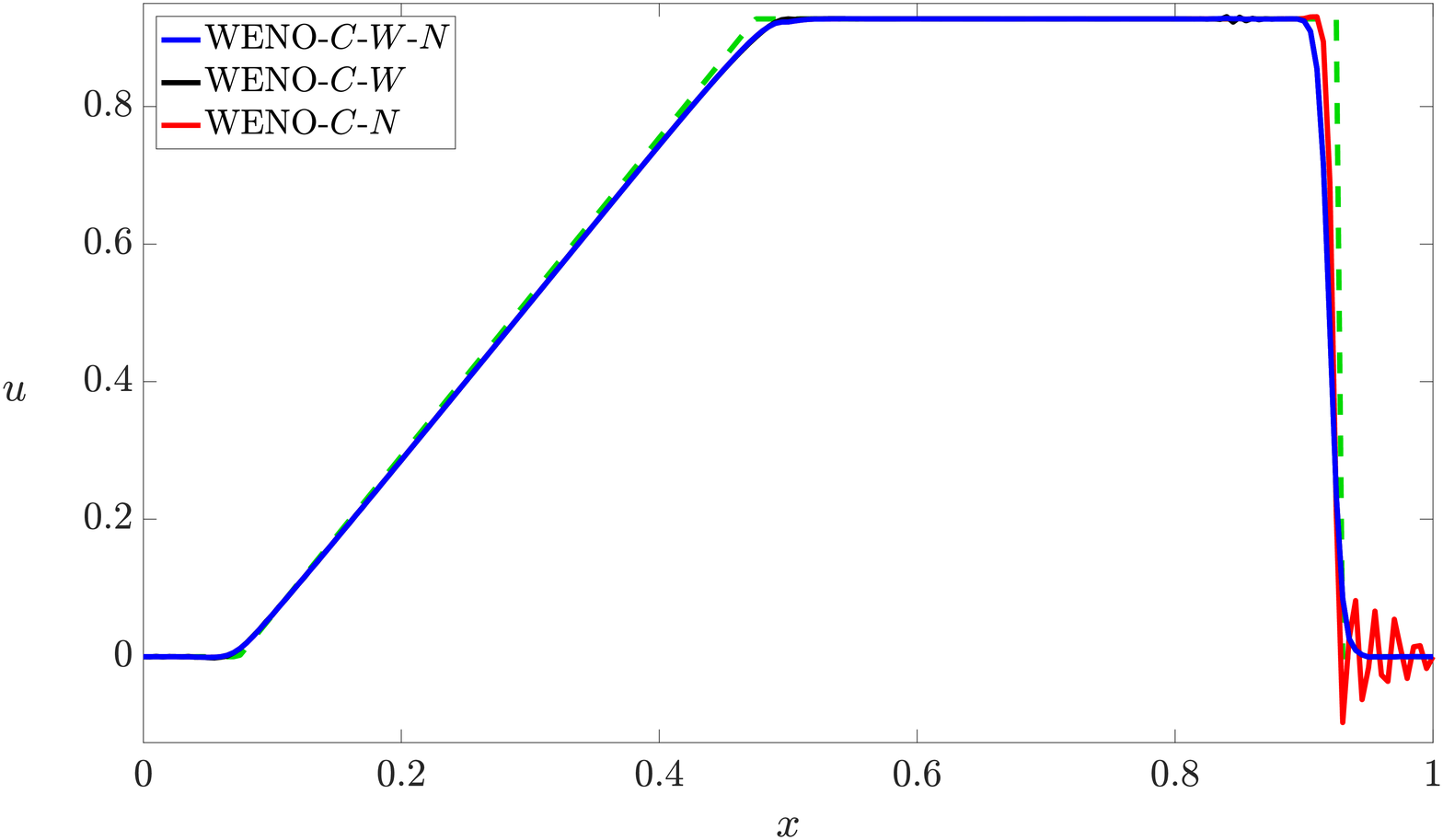}}
\subfigure[$t=0.36$, velocity post-collision zoom-in]{\label{fig:sod-weno-c-comparison2}\includegraphics[width=75mm]{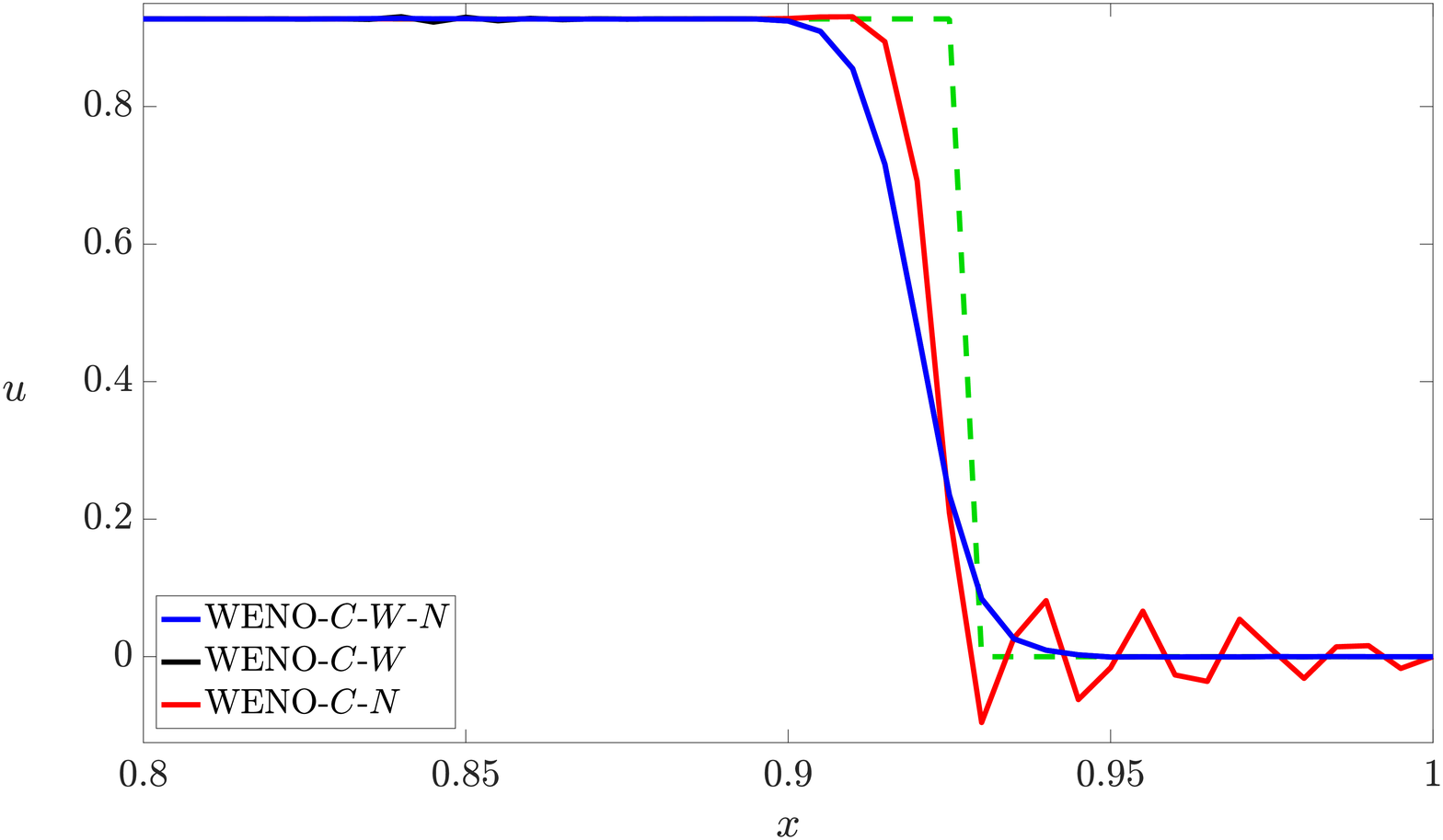}}
\caption{Comparison of the velocity solution profile produced using 
WENO-$C$-$W$-$N$, WENO-$C$-$W$, and WENO-$C$-$N$ for the post-collision bounce-back in the Sod shock tube problem with 
201 cells.}
\label{fig:sod-weno-c-comparison}
\end{figure} 

We remark that our conclusions described above do not change if we replace the $L^1$ norm with either the  $L^2$ or $L^ \infty$ 
norms. We list in Table {\ref{table:sod-error-post-2}} the $L^2$ and $L^ \infty$ error analysis for the 
post shock-wall collision velocity for the Sod problem, where the velocity is computed using 
either WENO-$|u_x|$ or WENO-$C$-$W$-$N$.

For the $L^2$ error analysis, we first note the odd behavior for WENO-$|u_x|$ on the coarser grids with 
101 and 201 cells. The increase in the $L^2$ error despite mesh refinement is caused by the base WENO 
scheme; referring to Table {\ref{table:sod-error-post}}, we see that the $L^1$ error of solutions produced with 
WENO increases as the mesh is refined from 101 to 201 cells. This phenomenon 
is due to the large oscillations that occur post shock-wall collision. The WENO-$C$-$W$-$N$ scheme removes
these oscillations, at the cost of a slight smearing of the shock front; this smearing results in a larger $L^2$ 
error on coarse grids when compared with WENO-$|u_x|$, but smaller $L^2$ errors and better rates of 
convergence as the mesh is refined. 

Table {\ref{table:sod-error-post-2}} shows that the $L^ \infty$ errors for both WENO-$|u_x|$ and 
WENO-$C$-$W$-$N$ grow as the mesh is refined. As noted in {\cite{GreenoughRider2004}}, this is due 
to the localization of the error at the shock. However, we remark that the $L^ \infty$ errors
for WENO-$C$-$W$-$N$ are smaller than the $L^ \infty$ errors for WENO-$|u_x|$ on the grids
with 201, 401, and 801 cells; moreover, these errors grow at a faster rate for WENO-$|u_x|$ than for 
WENO-$C$-$W$-$N$. 

In addition to the 
figures and qualitative evidence provided, the $L^1$, $L^2$, and $L^ \infty$ error studies indicate that 
the $C$-method produces highly accurate solutions with close to optimal rates of convergence for the 
Sod shock-wall collision and bounce-back test.

\begin{table}[H]
\centering
\renewcommand{\arraystretch}{1.0}
\scalebox{0.8}{
\begin{tabular}{|llc|cccc|}
\toprule
\midrule
\multirow{2}{*}{\textbf{Norm}} & \multirow{2}{*}{\textbf{Scheme}} &  & \multicolumn{4}{c|}{\textbf{Cells}}\\

& {}  &   & 101 & 201 & 401 & 801\\
\midrule
 \multirow{4}{*}{\vspace{-1.25em}${L^2}$} & \multirow{2}{*}{WENO-$|u_x|$} & Error & 
 $4.775 \times 10^{-2}$ & $6.068 \times 10^{-2}$ & $4.640 \times 10^{-2}$ & $3.765 \times 10^{-2}$  \\
				  &  & Order & -- & -0.346 & 0.387 & 0.302 \\[1.25em]

& \multirow{2}{*}{WENO-$C$-$W$-$N$} & Error & 
 $6.953 \times 10^{-2}$  & $6.098 \times 10^{-2}$ & $4.423 \times 10^{-2}$ & $3.324 \times 10^{-2}$ \\
				  &  & Order & -- & 0.189 & 0.463 & 0.412  \\[1.25em]
\midrule
 \multirow{4}{*}{\vspace{-1.25em}$L^ \infty$} & \multirow{2}{*}{WENO-$|u_x|$} & Error & 
 $4.262 \times 10^{-1}$  & $7.417 \times 10^{-1}$ & $8.024 \times 10^{-1}$ & $8.925 \times 10^{-1}$ \\
				  &  & Order & -- & -0.799 & -0.113 & -0.154  \\[1.25em]

& \multirow{2}{*}{WENO-$C$-$W$-$N$} & Error & 
 $5.663 \times 10^{-1}$ & $6.926 \times 10^{-1}$ & $7.124 \times 10^{-1}$ & $7.456 \times 10^{-1}$  \\
				  &  & Order & -- & -0.290 & -0.041 & -0.066  \\[1.25em]
\midrule
\bottomrule
\end{tabular}}
\caption{Post shock-wall collision ($t=0.36$) $L^2$ and $L^ \infty$ error of the computed velocity
 minus the exact solution and convergence for the 
Sod  problem with shock-wall collision and bounce-back.} 
\label{table:sod-error-post-2}
\end{table} 


\subsubsection{Comparison with other schemes}\label{sec:comparison-with-other-schemes}

For the purposes of benchmarking our WENO and WENO-$N$ schemes prior to 
shock-wall collision, we present error analysis and convergence rates comparing our simplified WENO scheme
with the scheme utilized by Greenough and Rider in \cite{GreenoughRider2004}. The WENO scheme 
that is presented in \cite{GreenoughRider2004} is 
formally fifth-order accurate in space, with time integration 
done using a total variation diminishing (TVD) third-order Runge-Kutta method. Flux-splitting is accomplished
using a method similar to the Lax-Friedrichs flux-splitting (see
\cite{GreenoughRider2004} for the details). We will refer to this method
as RK3-WENO5.

The error norm utilized in \cite{GreenoughRider2004} is of the form 
$$
\lvert \lvert \rho(\cdot,t) - \rho^*(\cdot,t) \rvert \rvert _{L^1_{GR}} \coloneqq \frac{1}{M} \sum_{i=1}^{M} \frac{| \rho(x_i ,t ) - \rho^*(x_i,t)| }{| \rho^*(x_i,t) |}\,, 
$$
where $\rho$ is the computed density and $\rho^*$ is the exact solution for the density. 
We will refer to this norm as the $L^1_{GR}$ norm. 

In Table \ref{table:sod-error-GR}, we calculate the $L^1_{GR}$ errors for the density for the Sod
shock tube problem computed with WENO and WENO-$N$, and compare them with the corresponding values
in \cite{GreenoughRider2004}. All simulations were run with a CFL number of 0.6. We see that our simplified WENO scheme and noise indicator algorithm compare well with 
the more complicated RK3-WENO5 scheme. Consequently, using our simplified WENO algorithm for the 
purposes of comparison in our error analysis for the artificial viscosity methods presented is justified; that is to 
say, comparing the performance of our artificial viscosity methods with our simplified WENO scheme is 
similar to comparing the performance of our artificial viscosity methods with the more complicated 
(and `industry-standard') RK3-WENO5. 

\begin{table}[H]
\centering
\renewcommand{\arraystretch}{1.0}
\scalebox{0.8}{
\begin{tabular}{|lc|ccc|}
\toprule
\midrule
\multirow{2}{*}{\textbf{Scheme}} &  & \multicolumn{3}{c|}{\textbf{Cells}}\\

{}  & & 101   & 201    & 401  \\
\midrule
\multirow{2}{*}{WENO} & Error & 
$1.57 \times 10^{-2}$  & $7.93 \times 10^{-3}$  & $4.49 \times 10^{-3}$ \\
				    & Order & -- & 0.99   & 0.82  \\
\midrule
\multirow{2}{*}{WENO-N} & Error & 
$1.60 \times 10^{-2}$ & $7.90 \times 10^{-3}$  & $4.37 \times 10^{-3}$ \\
				    & Order & -- & 1.02   & 0.85  \\
\midrule
\multirow{2}{*}{RK3-WENO5} & Error & 
$1.58 \times 10^{-2}$ & $8.24 \times 10^{-3}$  & $4.47 \times 10^{-3}$  \\
				    & Order & -- & 0.93   & 0.88  \\
\midrule
\bottomrule
\end{tabular}}
\caption{Pre shock-wall collision ($t=0.20$) $L^1_{GR}$ error analysis and convergence tests for the 
density for the Sod shock tube problem.}
\label{table:sod-error-GR}
\end{table}

\subsection{The Noh problem}\label{subsec:noh}
As a further example of wall-heating, we next consider the classical 1-$D$ planar Noh problem
\mbox{\cite{Noh1987,Liska2003995}}. The domain of interest is $-0.5 \leq x \leq 0.5$, the 
adiabatic constant is $\gamma = 5/3$, and the initial data is given by
$$
\begin{bmatrix}
\rho_0 \\ (\rho u)_0 \\ E_0 
\end{bmatrix}
=
\begin{bmatrix}
1 \\ 1 \\ 0.5 + \frac{10^{-6}}{\gamma -1}
\end{bmatrix}
\mathbbm{1}_{[-0.5,0)}(x)
+
\begin{bmatrix}
1 \\ -1 \\ 0.5 + \frac{10^{-6}}{\gamma -1}
\end{bmatrix}
\mathbbm{1}_{[0,0.5]}(x)\,. 
$$
The solution for this problem consists of two infinite strength shocks propagating with speed $1/3$ 
outwards from the origin, with a state of constant density and pressure left behind. 

\begin{figure}[H]
\centering
\subfigure[$t=1.0$: density, comparison of WENO-$C$ and WENO]{\label{fig:noh1}\includegraphics[width=75mm]{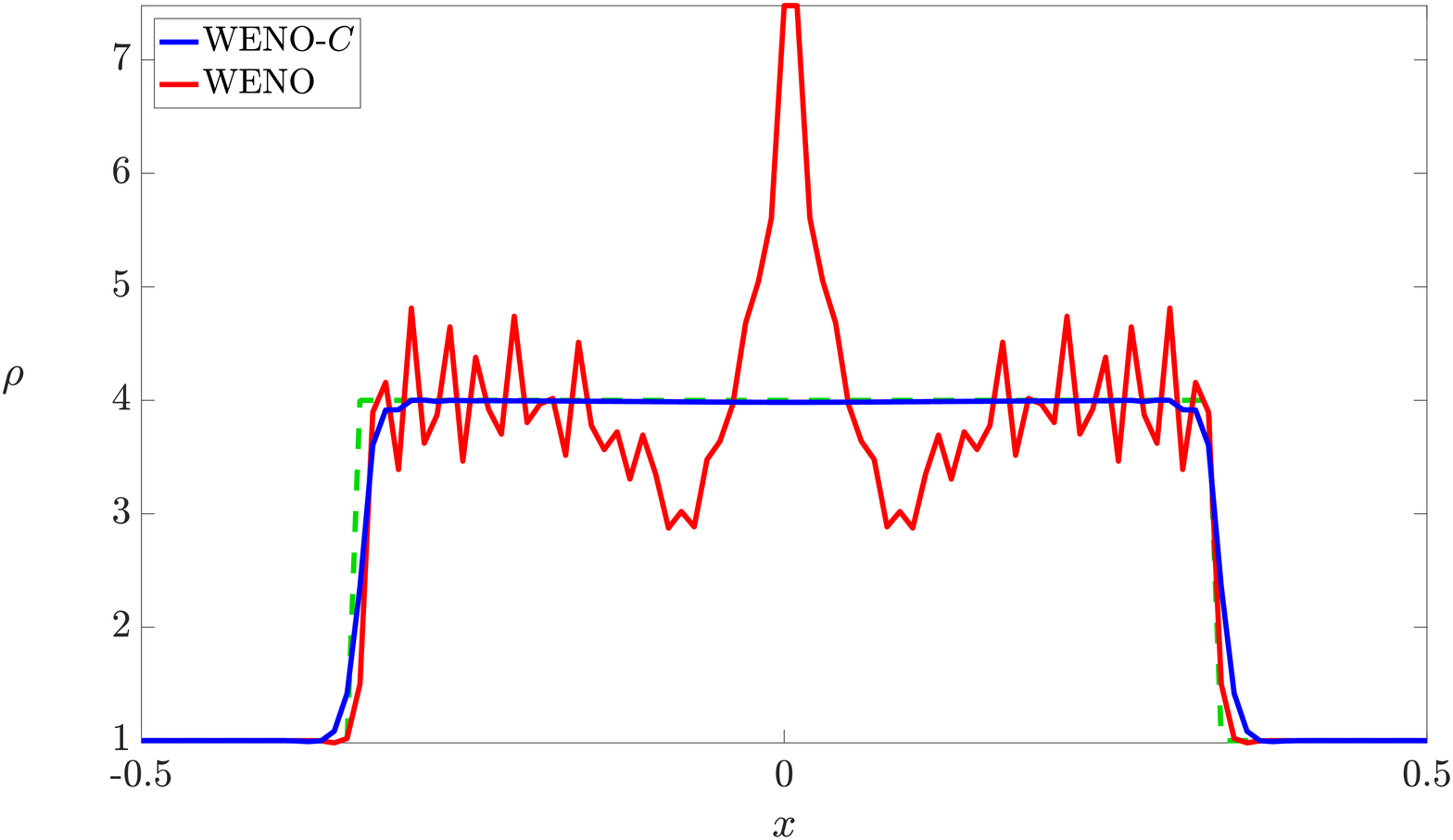}}
\subfigure[$t=1.0$: density, comparison of WENO-$C$ with different grid spacings]{\label{fig:noh2}\includegraphics[width=75mm]{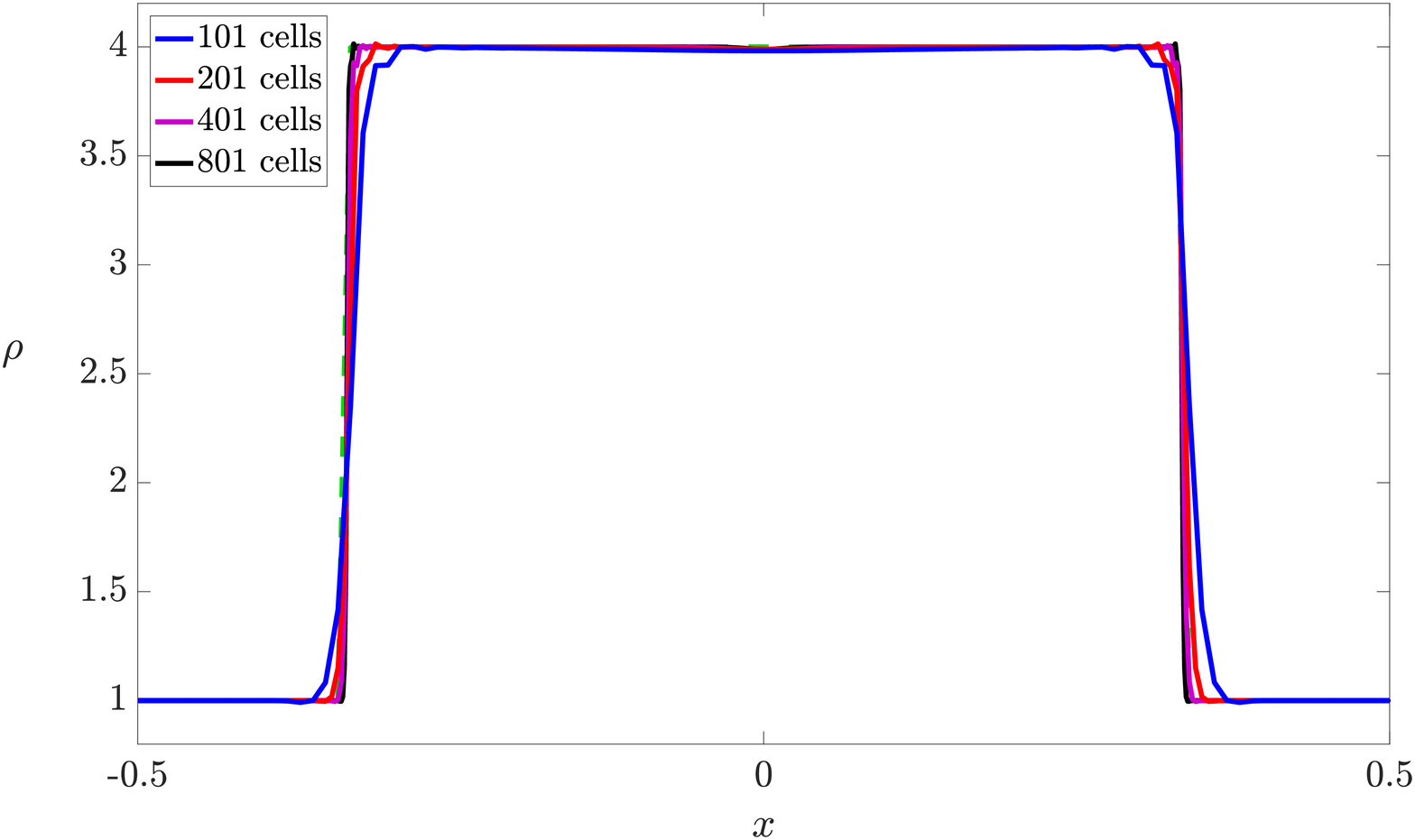}}
\subfigure[$t=1.0$: density zoom-in at the shock, comparison of WENO-$C$ with different grid spacings]{\label{fig:noh3}\includegraphics[width=75mm]{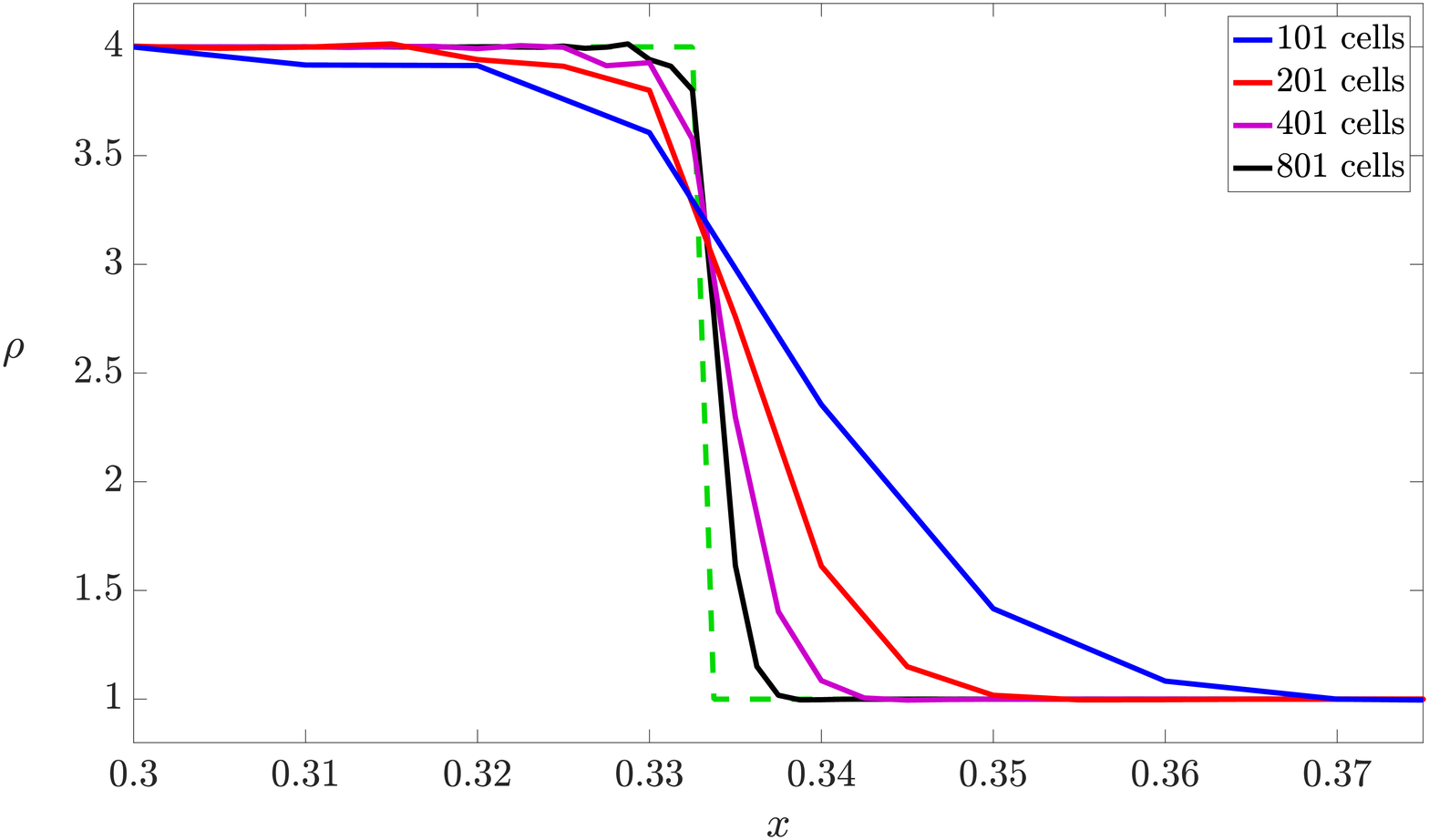}}
\caption{The density profile at time $t = 1.0$ for the Noh problem, with the solution
computed using (a) WENO or (b,c) WENO-$C$. The dashed green curve is the exact solution.}
\label{fig:noh}
\end{figure}

As demonstrated in {\cite{Liska2003995}}, most schemes tend to produce the anomalous wall-heating at 
the center origin. We shall utilize our WENO-$C$ method (i.e. no shock-wall 
collision algorithm) with 101 cells. We choose the relevant parameters as
\begin{gather*}
\beta^u=1.0, \qquad  \beta^E=10.0, \qquad  \varepsilon=50.0, \qquad  \kappa=1.0 \,.
\end{gather*} 
The value of $\beta^u$ is chosen large enough to 
eliminate post-shock oscillations, while $\beta^E$ is chosen to minimize the wall-heating. 
In Fig.{\ref{fig:noh}}, we compare the solutions computed using WENO and WENO-$C$; it is clear that 
WENO-$C$ produces a much more accurate solution, with the post-shock oscillations and wall-heating 
eliminated.

\subsection{The LeBlanc shock tube problem}\label{subsec:leblanc}

\begin{figure}[H]
\centering
\subfigure[$t=6.0$: internal energy]{\label{fig:weno-leblanc1}\includegraphics[width=75mm]{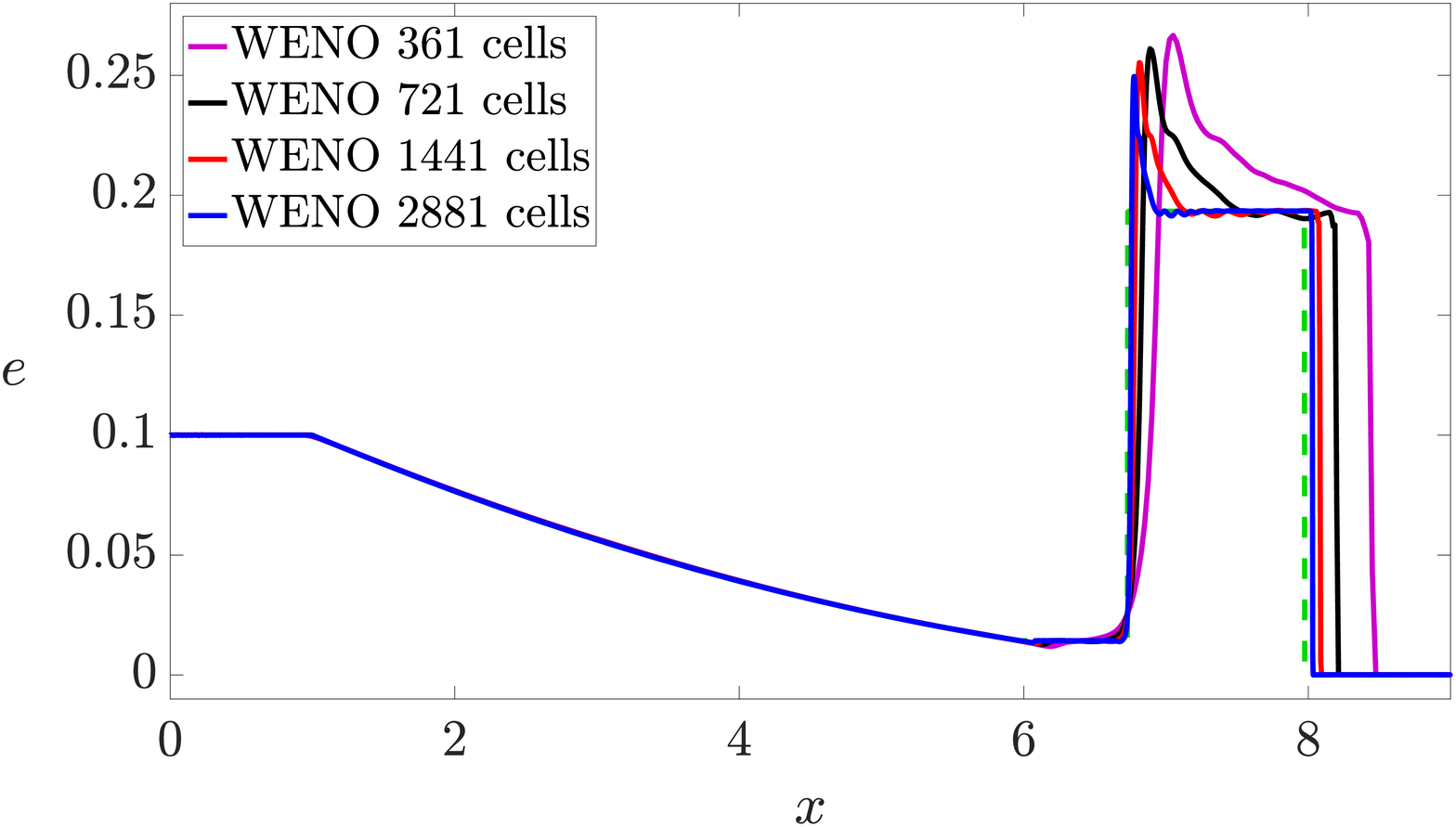}}
\subfigure[$t=6.0$: internal energy, zoomed in]{\label{fig:weno-leblanc2}\includegraphics[width=75mm]{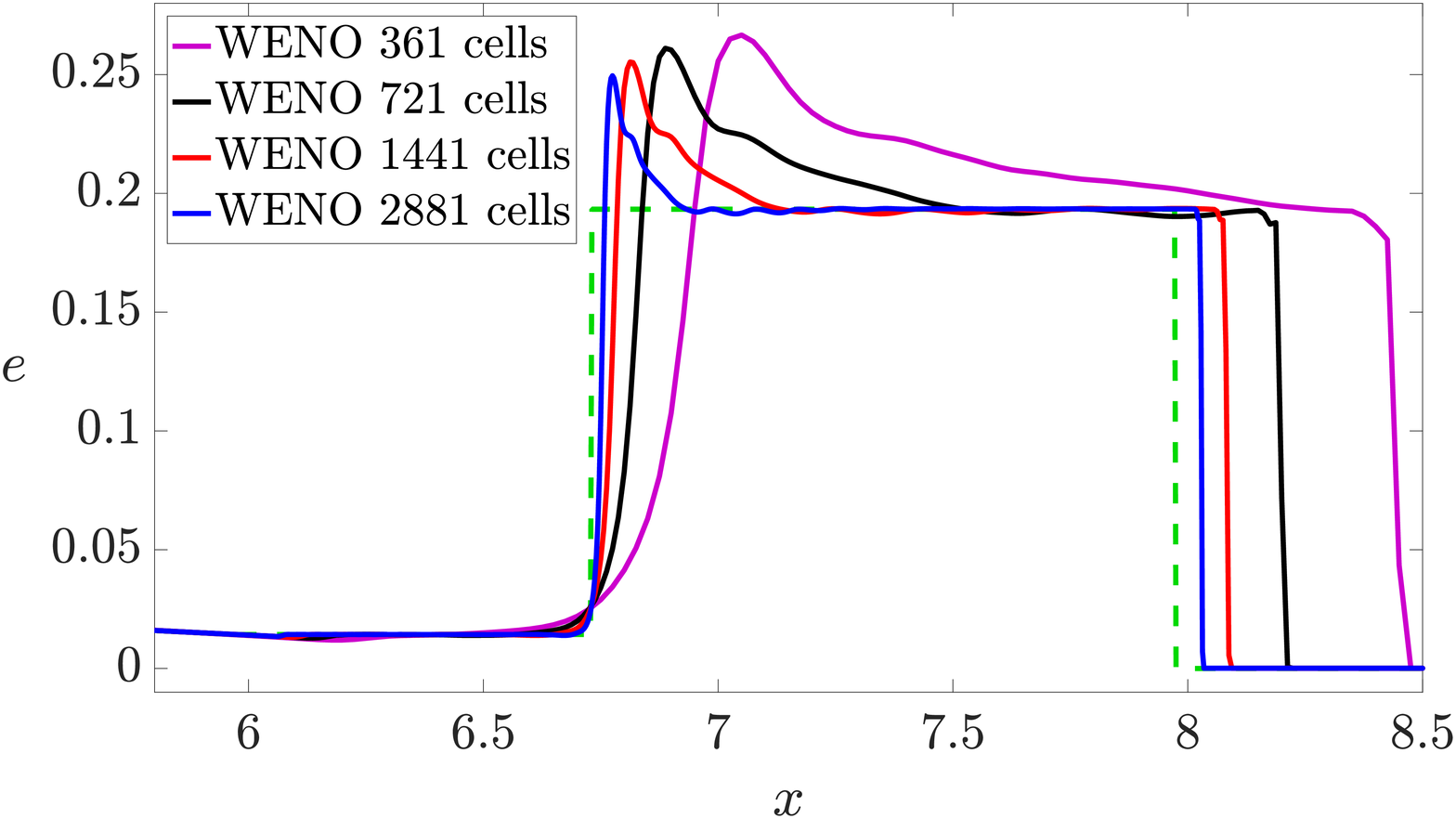}}
\caption{The internal energy profile at time $t = 6.0$ for the LeBlanc shock tube problem, with the solution
computed using WENO. The dashed green curve is the exact solution.}
\label{fig:weno-leblanc}
\end{figure}

We now turn our attention to the LeBlanc shock tube problem. Here, the domain of 
interest is $0 \leq x \leq 9$, the adiabatic constant is $\gamma = \frac{5}{3}$, 
and the initial data is given by
$$
\begin{bmatrix}
\rho_0 \\ (\rho u)_0 \\ E_0 
\end{bmatrix}
=
\begin{bmatrix}
1 \\ 0 \\ 10^{-1}
\end{bmatrix}
\mathbbm{1}_{[0,3)}(x)
+
\begin{bmatrix}
10^{-3} \\ 0 \\ 10^{-9}
\end{bmatrix}
\mathbbm{1}_{[3,9]}(x)\,. 
$$
The large jump in the initial energy $E_0$ produces a very strong shock wave, making the 
LeBlanc shock-tube problem a very difficult test case. Most numerical schemes tend to 
produce large overshoots in the   internal energy
at the contact discontinuity, which results in a loss of accuracy in the shock speed, as shown 
in Fig.\ref{fig:weno-leblanc}. This overshoot in the internal energy is in fact an example of 
wall-heating; a small undershoot in the density and the continuity of the pressure at the contact 
produce this observed overshoot in the internal energy. We refer the reader
to \cite{ReSeSh2012,Liu20098872,Loubere2005105} for further details.

Our strategy is to add an additional diffusion term to the right-hand side 
of the energy equation \eqref{EulerC-energy} that will serve to remove 
the large overshoot in the   internal energy at the contact discontinuity. Specifically, we solve an 
additional $C$-equation for a variable $C^e$ forced by $| \partial_x e |/ \max_{x} | \partial_x e|$.    Thus, 
equation \eqref{EulerC-energy}  is replaced by 
\begin{equation}
\partial_t E + \partial_x (uE + up) =  \partial_x \left( \mathcal{B} ^{(E)}(t) \, \rho \, C \, \partial_x (E/\rho) \right) + \partial_x \left( \mathcal{B} ^{(e)}(t) \, \rho \, C^e \, \partial_x \left( E/\rho \right) \right) \,,
\end{equation}
where the function $C^e$ is computed using
\begin{equation}
\partial_t C^e + \frac{S(\bm{u})}{\varepsilon_e \Delta x} C^e - \kappa_e \Delta x \cdot S(\bm{u}) \partial_{xx} C^e  = \frac{S(\bm{u})}{\varepsilon_e \Delta x} G^e \,.  \label{C-energy-LeBlanc}
\end{equation}
The artificial viscosity coefficients are given by \eqref{artificial_visc} and 
$$
\mathcal{B} ^{(e)}(t) = (\Delta x)^2 \cdot \frac{ \max_{x} | \partial_x u| }{ \max_{x} C^e }
\left(\beta^{e} + \beta^{e}_w \cdot \overline{C}_w(t)\right) \,, 
$$
and $C$, $C_w(x,t)$, and $\Cw(t)$ are defined by
\eqref{C-Sod}, \eqref{Cwall-sod}, and \eqref{wall-ind-fn}, respectively. 
The forcing to the $C^e$ equation \eqref{C-energy-LeBlanc} is 
$$
G^e(x,t) = \mathbbm{1}_{(0,\infty)}(\partial_x u) \frac{ | \partial_x e |}{\max_{x} | \partial_x e |} \,.
$$
Here, the indicator function $  \mathbbm{1}_{(0,\infty)} \, (\partial_x u) $ represents an
 {\it expansion switch}, in which $G^e$ is non-zero only if $\partial_x u > 0$.
 
\subsubsection{Stabilizing shock-wall collision}
To simulate the collision of the shock-wave with the wall, we use solid wall boundary conditions \eqref{var-bcs}.
Motivated by the results for the Sod shock tube problem
 presented in \ref{sod-bounce-back}, we add wall viscosity to the momentum 
and energy equations; we choose the parameters as
\begin{alignat*}{6}
\beta^u&=0.001, \qquad  \beta^E&=0.0, \qquad  \beta^e&=0.4, \qquad \beta^u_w&= 4.0, \qquad  \beta^E_w&= 0.0, \qquad \beta^e_w&=0.0, \\
\varepsilon&=1.25, \qquad  \kappa&=10.0,  \qquad  \varepsilon_e&=1.25, \qquad  \kappa_w&=14.0, \qquad  \varepsilon_w &=50.0, \qquad  \kappa_w&=4 \,.
\end{alignat*}
We employ our WENO-$C$-$W$ scheme with 721 cells. Since this is a more challenging problem than 
the Sod shock tube problem, we use the smaller CFL number of 0.25. 

For the purpose of comparison, 
we also implement the WENO-Noh scheme, with the parameters in \eqref{EulerC-noh} set as
 $\beta^u_{\operatorname{Noh}} = 8.0$ and
$\beta^E_{\operatorname{Noh}} = 9.0$. 
These parameters were chosen with the aim of suppressing post-collision noise while preventing 
the occurrence of the wall heating error. We remark that WENO-Noh failed for CFL=0.25, and 
required the much smaller CFL $\approx 0.045$ to run. 

\begin{figure}[H]
\centering
\subfigure[$t=6.0$: velocity]{\label{fig:leblanc-before-collision1}\includegraphics[width=75mm]{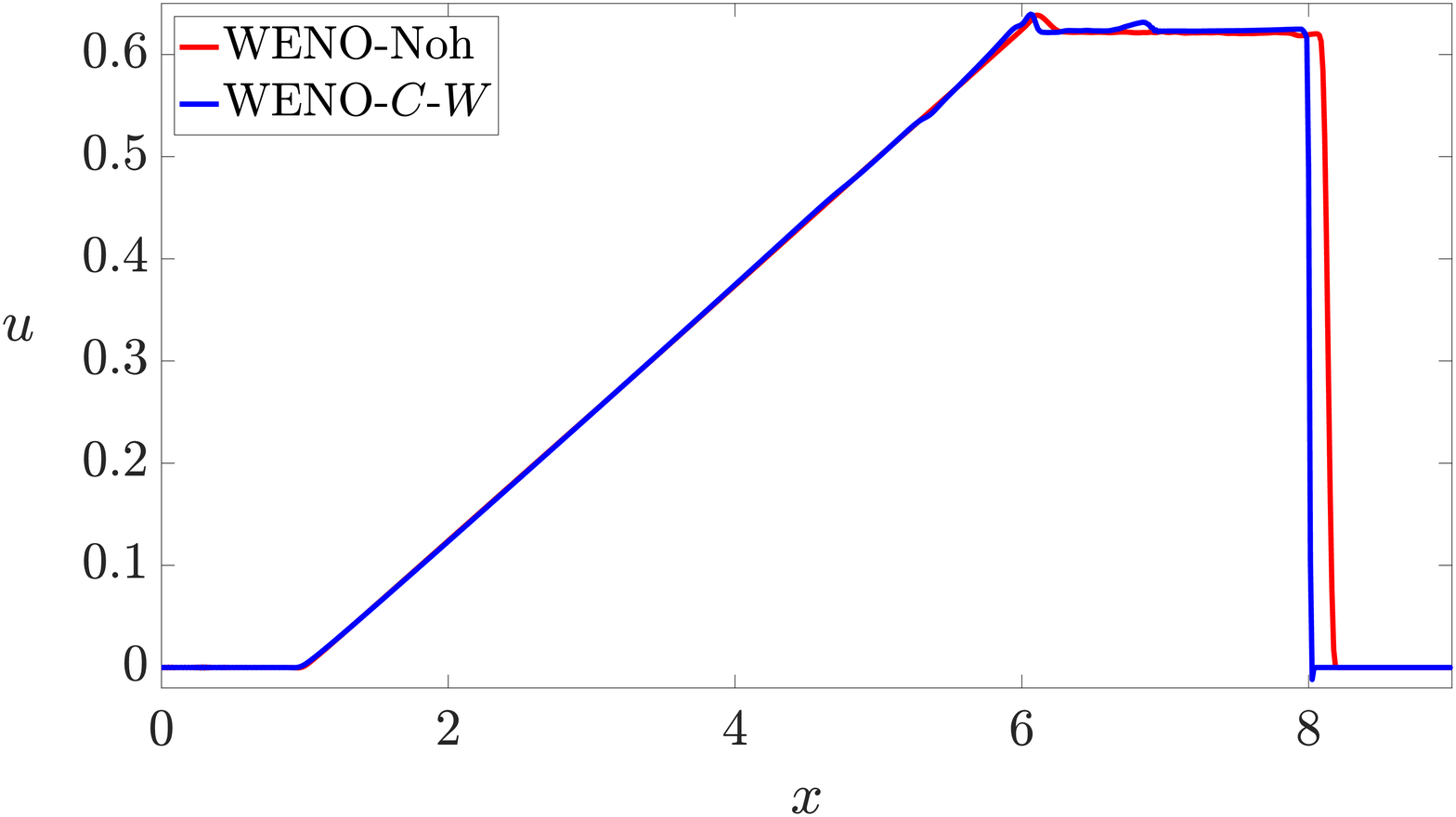}}
\subfigure[$t=6.0$:   internal energy]{\label{fig:leblanc-before-collision-eng1}\includegraphics[width=75mm]{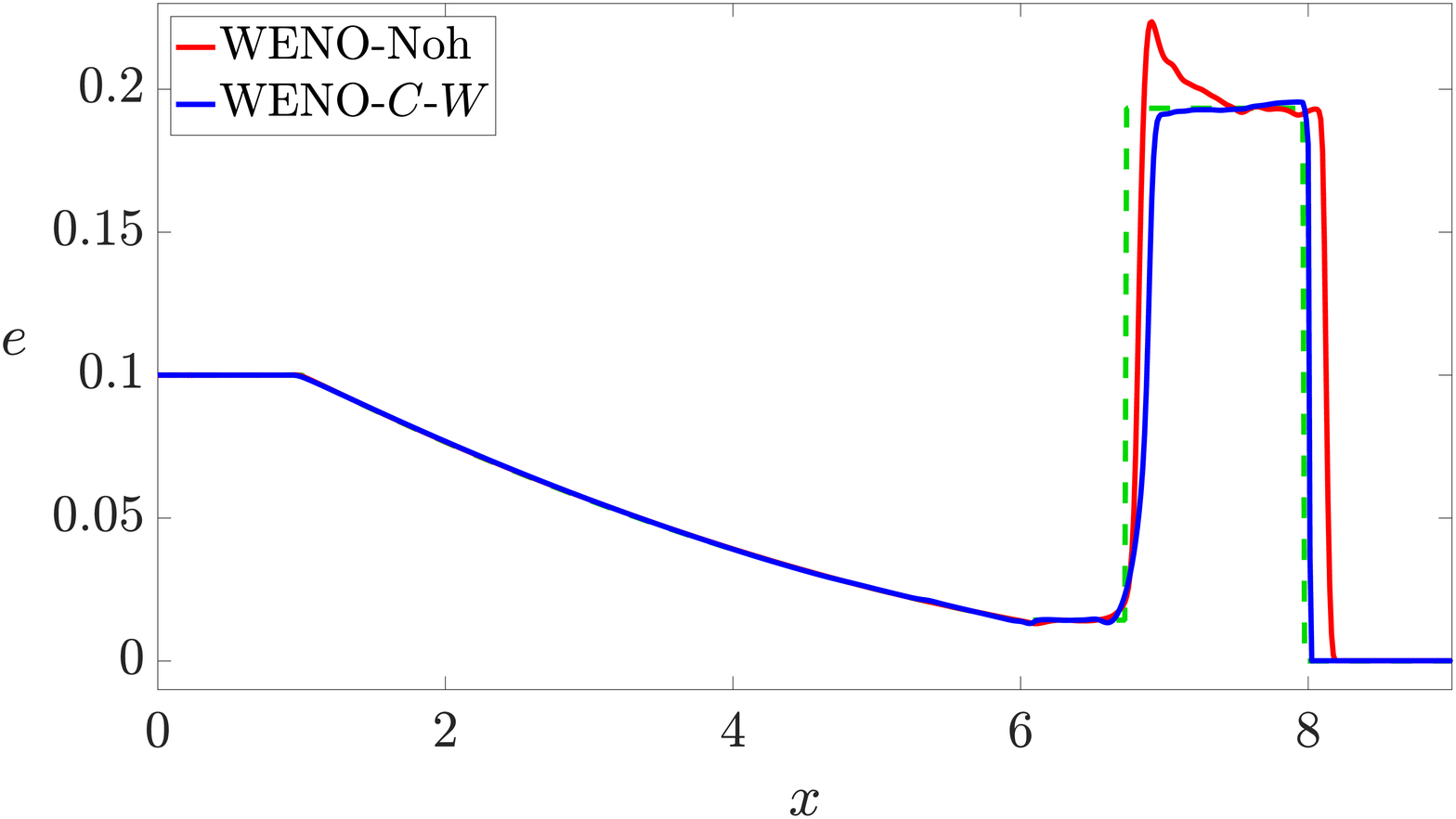}}
\caption{The (a) velocity and (b)   internal energy profiles for the LeBlanc shock tube problem 
before the collision with the wall. The solution is computed with viscosity activated for the 
momentum and energy equations.}
\label{fig:leblanc-before-collision}
\end{figure}

The shock-wave moves to the right and collides with the right wall at time $t \approx 7.2$. 
Prior to shock collision, the WENO-Noh scheme produces a solution with an overshoot in the internal energy 
at the contact discontinuity. This results in an incorrect shock front and wave speed. 
The viscosity for the momentum at the shock and the energy at the 
contact discontinuity in our WENO-$C$-$W$ scheme remove 
post-shock oscillations and the overshoot in the internal 
energy, respectively, as shown in Fig.\ref{fig:leblanc-before-collision}. 

\begin{figure}[H]
\centering
\subfigure[$t=8.0$: velocity]{\label{fig:leblanc-after-collision1}\includegraphics[width=75mm]{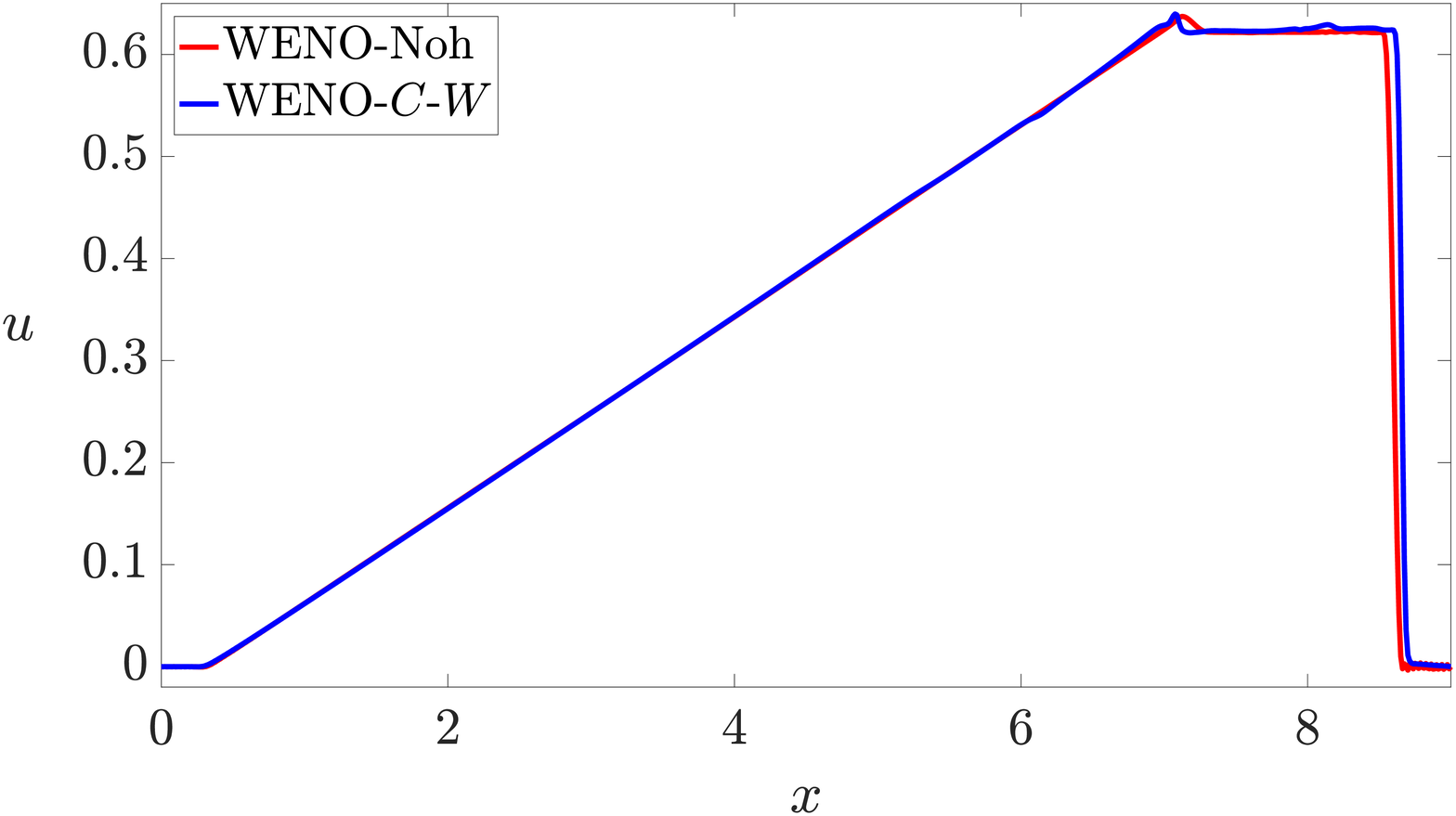}}
\subfigure[$t=8.0$:   internal energy]{\label{fig:leblanc-after-collision-eng1}\includegraphics[width=75mm]{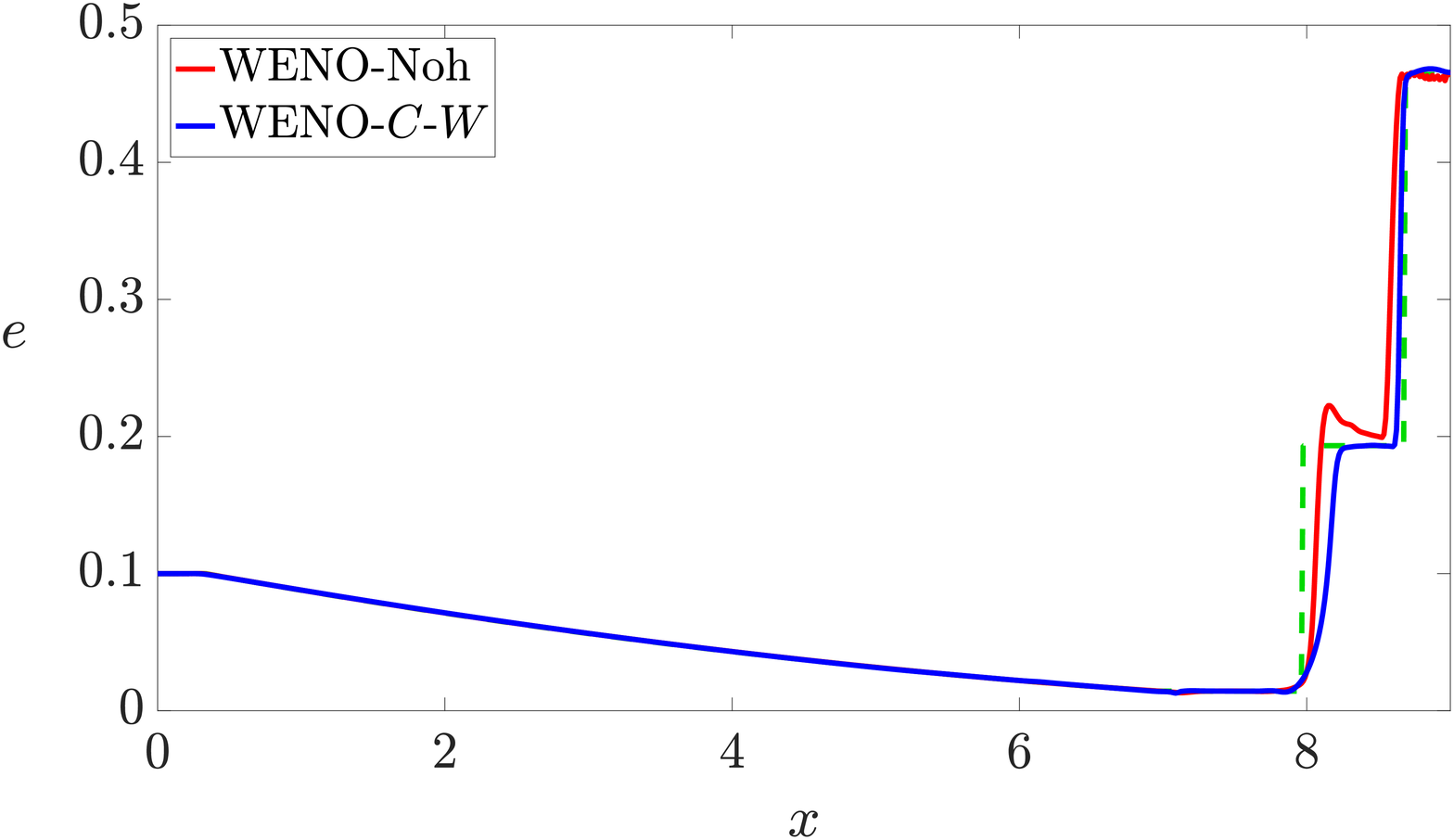}}
\subfigure[$t=8.0$:   internal energy, zoomed in at the wall]{\label{fig:leblanc-after-collision-zoom-inteng}\includegraphics[width=75mm]{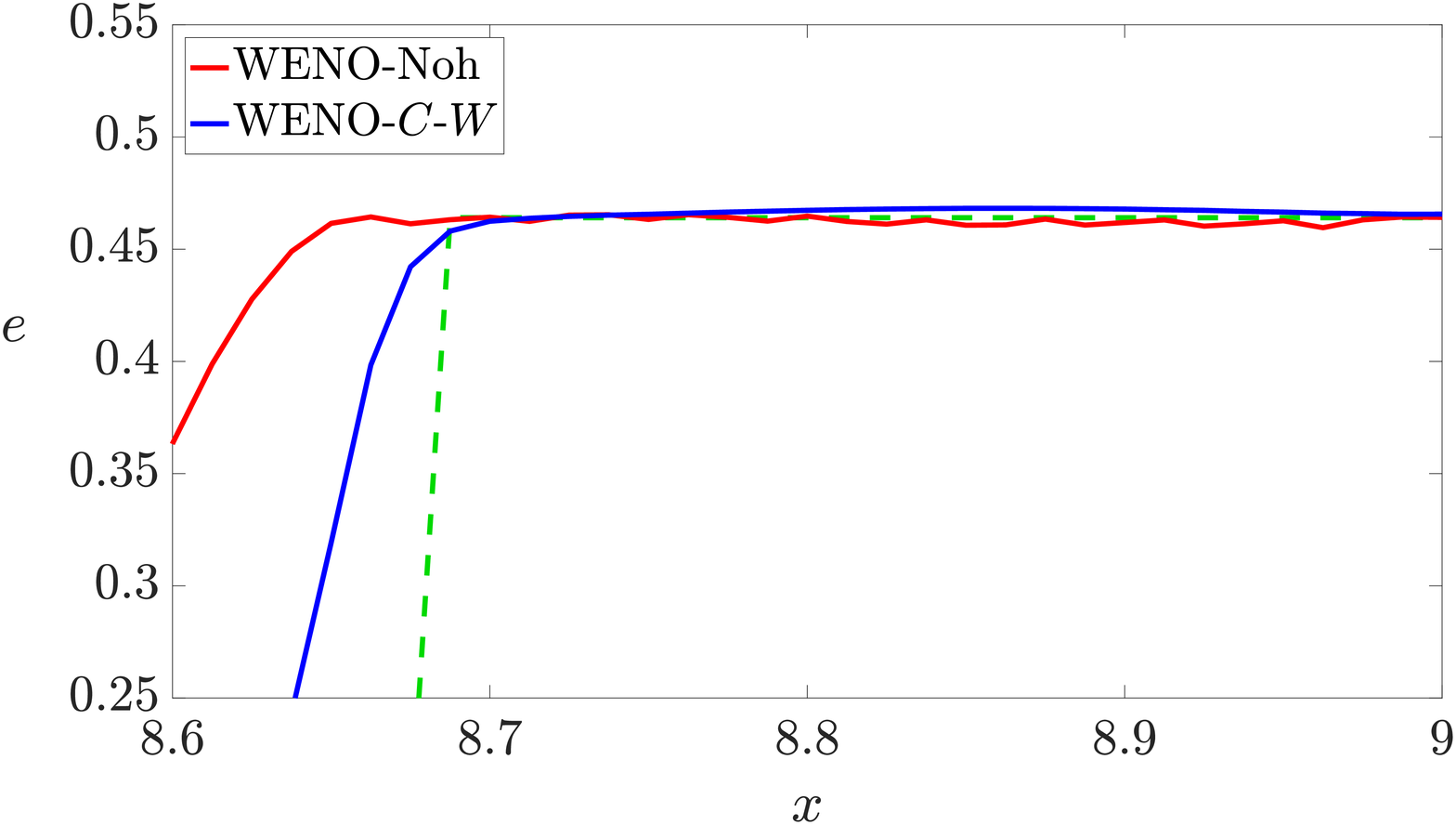}}
\subfigure[$t=8.0$: density, zoomed in at the wall]{\label{fig:leblanc-after-collision-zoom-rho}\includegraphics[width=75mm]{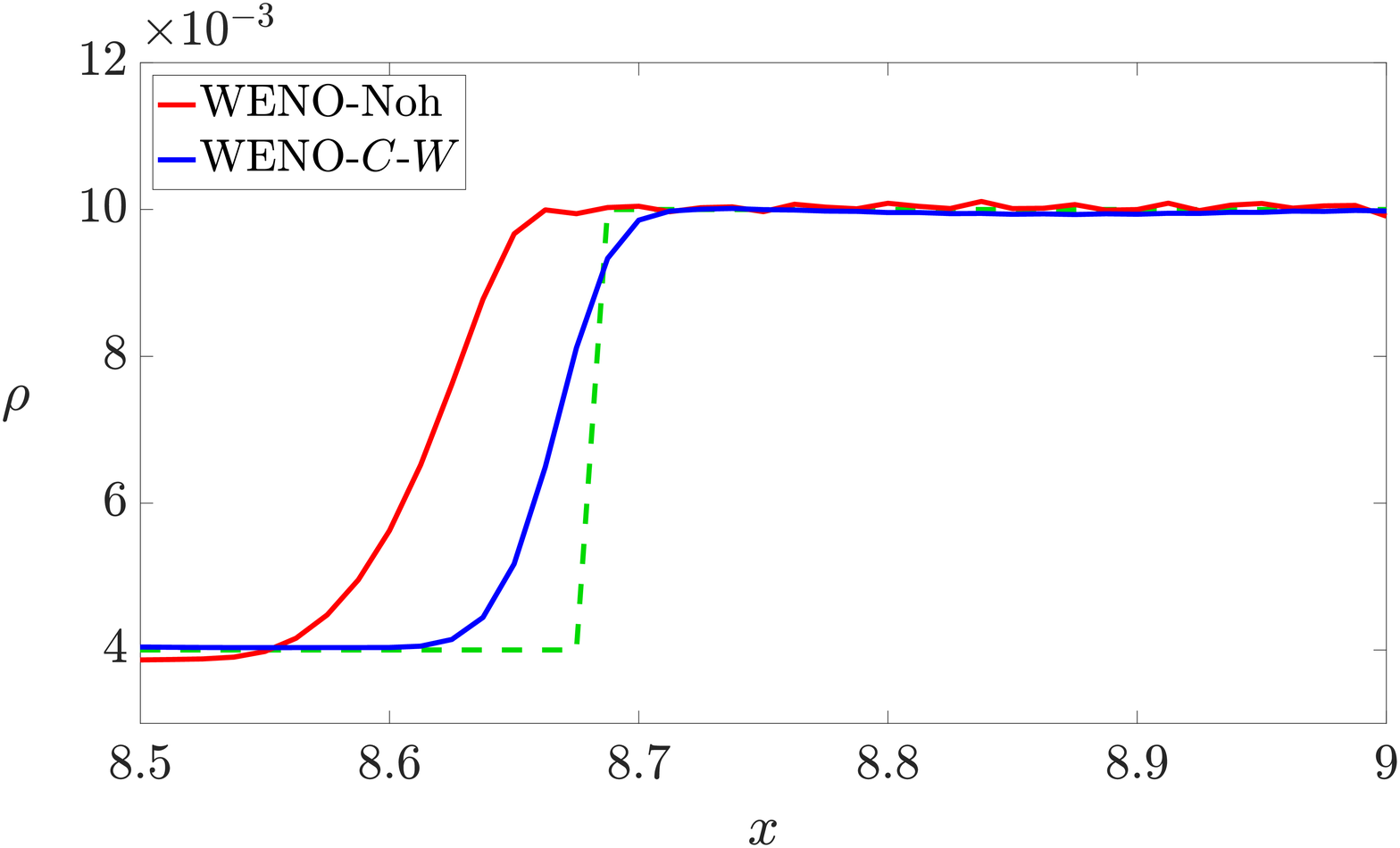}}
\caption{The (a) velocity, (b)   internal energy, (c) zoomed in internal energy and (d) zoomed in density  profiles
for the LeBlanc shock tube problem 
after the collision with the wall. The solution computed with WENO-$C$-$W$ has the wall 
viscosity activated for the momentum and energy equations.}
\label{fig:leblanc-after-collision}
\end{figure}

Post shock-wall collision, the wall viscosity for the momentum and energy equations damp-out the 
oscillations behind the shock, while ensuring that the solution 
maintains a sharp shock front and the 
correct shock speed (see Fig.\ref{fig:leblanc-after-collision}). Moreover, the wall viscosity for 
the energy equation 
prevents the wall heating error from occurring, as shown in 
Fig.\ref{fig:leblanc-after-collision-zoom-rho}. 
Due to the lack of smoothness of the localizing artificial viscosity coefficient 
$|\partial_x u|$, the WENO-Noh scheme is unable to fully suppress all the post-collision oscillations, though 
the heat conduction term in the energy equation prevents the wall heating error from occurring.  
In Fig.\ref{fig:leblanc-after-collision-zoom-inteng}, we zoom
in on the   internal energy profile near the wall; it is evident that the solution computed with
WENO-$C$-$W$ is better than that computed with WENO-Noh, but there is a small error between 
the computed solution and the exact solution. This error occurs because of a very small inaccuracy 
in the density profile, shown in Fig.\ref{fig:leblanc-after-collision-zoom-rho}. Since the 
density is so small here, and since the   internal energy is given by \eqref{defn-internal-energy},
even tiny errors are greatly amplified, making it very difficult to get a completely accurate solution for the
  internal energy.  

\subsubsection{Error analysis and convergence tests}
We now compare the errors of the various numerical schemes
 listed in Table \ref{table:schemes} applied
to the LeBlanc shock tube problem, with the various relevant parameters fixed across the different methods. 
The $L^1$ errors in the velocity are computed using formula \eqref{error-formula}, and are listed
 in Table \ref{table:leblanc-error-pre} (time $t = 6.0$) and 
Table \ref{table:leblanc-error-post} (time $t=8.0$). All of the simulations were run with a CFL number of 
0.25, except for WENO-Noh, which required CFL $\approx 0.045$. 

Prior to shock-wall collision, it is evident from Table \ref{table:leblanc-error-pre} that the $C$-method produces
a solution that is significantly better than those solutions produced without the $C$-method. The $L^1$ errors
for the velocity computed using WENO-$C$ are almost an order of magnitude smaller than the 
$L^1$ errors for the velocity
computed using WENO and WENO-Noh. This is primarily due to the removal of the overshoot
in the internal energy, which results in an accurate shock speed. 

On the other hand, the removal of the odvershoot in the internal energy through the use of $C^e$ results in 
a more smeared contact discontinuity, as shown in Fig.\ref{fig:leblanc-before-collision-eng1}. The
smearing of the contact discontinuity results in a non-physical ``bump'' appearing in the velocity profile, 
as shown in Fig.\ref{fig:leblanc-before-collision1}. Note that this bump does not appear in the velocity
profile computed using WENO-Noh, since in this case the contact discontinuity is sharper, at the expense
of a large overshoot in the internal energy. We suggest, however, that this defect in the solution computed 
using $C^e$ is relatively insignificant when compared against the magnitude of the error in the internal 
energy solutions computed without $C^e$. This is primarily due to the fact that the internal energy error 
results in a highly inaccurate shock speed which, in turn, leads to a highly inaccurate solution, as evidenced 
by Table {\ref{table:leblanc-error-pre}}. On the other hand, the velocity bump error arises from the 
correction of the overshoot in the internal energy, and subsequently the shock speed and location; the latter
two corrections result in a much more accurate solution overall, again demonstrated in Table 
{\ref{table:leblanc-error-pre}}.
Moreover, we note that the bump error decreases with mesh 
refinement approximately four times as fast as the overshoot error, as shown in 
Table \ref{table:leblanc-overshoot-error}. Here, the overshoot/bump error is computed by calculating 
the difference between the value at the peak of the 
overshoot/bump and the value of the exact solution there\footnote{This error is thus a \emph{local}
 $L^ \infty$ error.}.

\begin{table}[H]
\centering
\renewcommand{\arraystretch}{1.25}
\scalebox{0.8}{
\begin{tabular}{|lc|cccc|}
\toprule
\midrule
\multirow{2}{*}{\textbf{Scheme}} &  & \multicolumn{4}{c|}{\textbf{Cells}}\\

{}  & & 361     & 721   & 1441 & 2881\\
\midrule
\multirow{2}{*}{WENO} & Error & 
$3.469 \times 10^{-2}$  & $1.659 \times 10^{-2}$ & $8.010 \times 10^{-3}$ & $4.016 \times 10^{-3}$\\
				    & Order & --  &1.065  & 1.050 & 0.996\\
\midrule
\multirow{2}{*}{WENO-Noh} & Error & 
$2.546 \times 10^{-2}$  & $1.239 \times 10^{-2}$  & $6.001 \times 10^{-3}$ & $3.010 \times 10^{-3}$\\
				    & Order & --  & 1.040 & 1.045  & 0.996\\
\midrule
\multirow{2}{*}{WENO-$N$} & Error & 
$3.468 \times 10^{-2}$  & $1.661 \times 10^{-2}$  & $8.015 \times 10^{-3}$ & $4.022 \times 10^{-3}$\\
				    & Order & --   & 1.062 & 1.051  & 0.995\\
\midrule
\multirow{2}{*}{WENO-$C$} & Error & 
$7.190 \times 10^{-3}$   & $3.959 \times 10^{-3}$  & $2.008 \times 10^{-3}$ & $1.096 \times 10^{-3}$\\
				    & Order & --  & 0.864 & 0.976 & 0.873\\
\midrule
\multirow{2}{*}{WENO-$C$-$N$} & Error & 
$7.169 \times 10^{-3}$   & $3.881 \times 10^{-3}$  & $2.007 \times 10^{-3}$ & $1.113 \times 10^{-3}$\\
				    & Order & --  & 0.885 & 0.951 & 0.851\\
\midrule
\bottomrule
\end{tabular}}
\caption{Pre shock-wall collision ($t = 6.0$) $L^1$ error analysis and convergence tests for the
 velocity for the LeBlanc shock tube problem.}
\label{table:leblanc-error-pre}
\end{table}

\begin{table}[H]
\centering
\renewcommand{\arraystretch}{1.0}
\scalebox{0.8}{
\begin{tabular}{|lc|cccc|}
\toprule
\midrule
\multirow{2}{*}{\textbf{Scheme}} &  & \multicolumn{4}{c|}{\textbf{Cells}}\\
{}  &   & 361    & 721   & 1441 & 2881\\
\midrule
\multirow{2}{*}{WENO} & Overshoot Error & 
 $7.330 \times 10^{-2}$  & $6.780 \times 10^{-2}$ & $6.200 \times 10^{-2}$ & $5.620  \times 10^{-2}$\\
				    & Order & --     &0.113   & 0.129 & 0.142 \\
\midrule
\multirow{2}{*}{WENO-$C$} & Bump Error & 
 $1.500 \times 10^{-2}$  & $9.700 \times 10^{-3}$ & $6.600 \times 10^{-3}$ & $3.900 \times 10^{-3}$\\
				    & Order & --     & 0.629   & 0.556 & 0.759 \\
\midrule
\bottomrule
\end{tabular}}
\caption{Comparison of the overshoot error in the internal energy and the bump error in the velocity
for solutions to the LeBlanc shock tube problem at time $t = 6.0$.}
\label{table:leblanc-overshoot-error}
\end{table}

We note here that Table {\ref{table:leblanc-error-pre}} seems to suggest that the WENO and 
WENO-Noh schemes produce solutions that converge at first-order,
 even though the solutions computed using these schemes are very poor, 
as shown, for example, in Fig.{\ref{fig:weno-leblanc}}. This ``super-convergence'' {\cite{JiangShu1996}} is 
due to large errors on coarser meshes, rather than smaller errors on finer meshes, and is therefore 
superficial. On the other hand, 
the WENO-$C$ and WENO-$C$-$N$ schemes produce solutions with much smaller errors, and suggest close 
to first-order convergence.

Post shock-wall collision, the wall $C$-method produces a highly accurate non-oscillatory solution, while 
ensuring that the wall heating error does not occur. While the WENO-Noh scheme is able to suppress
most of the oscillations, the large amount of viscosity needed due to the lack of smoothness of 
$|\partial_x u|$ results in a shock front that is too smeared, as well as the imposition of a smaller time-step. 
Again, we see that the noise indicator algorithm
serves primarily as an error correction mechanism, removing small-scale high-frequency oscillations from
the solution. 

\begin{table}[H]
\centering
\renewcommand{\arraystretch}{1.0}
\scalebox{0.8}{
\begin{tabular}{|lc|cccc|}
\toprule
\midrule
\multirow{2}{*}{\textbf{Scheme}} &  & \multicolumn{4}{c|}{\textbf{Cells}}\\

{}  &   & 361    & 721   & 1441 & 2881\\
\midrule
\multirow{2}{*}{WENO} & Error & 
 $2.160 \times 10^{-2}$  & $9.832 \times 10^{-3}$ & $5.336 \times 10^{-3}$ & $2.896  \times 10^{-3}$\\
				    & Order & --     &1.136   & 0.882 & 0.882 \\
\midrule
\multirow{2}{*}{WENO-Noh} & Error & 
 $1.528 \times 10^{-2}$  & $6.544 \times 10^{-3}$  & $3.407 \times 10^{-3}$  & $1.668  \times 10^{-3}$ \\
				    & Order & --     & 1.224  & 0.942 & 1.030\\
\midrule
\multirow{2}{*}{WENO-$N$} & Error & 
 $2.141 \times 10^{-2}$   & $9.684 \times 10^{-3}$  & $5.178 \times 10^{-3}$  & $2.793  \times 10^{-3}$ \\
				    & Order & --     & 1.144  & 0.903 & 0.891 \\
\midrule
\multirow{2}{*}{WENO-$C$} & Error & 
 $5.703 \times 10^{-3}$   & $3.486 \times 10^{-3}$  & $2.024 \times 10^{-3}$  & $1.052  \times 10^{-3}$\\
				    & Order & --     & 0.710 & 0.785 & 0.944 \\
\midrule
\multirow{2}{*}{WENO-$C$-$N$} & Error & 
 $5.627 \times 10^{-3}$   & $3.384 \times 10^{-3}$  & $2.001 \times 10^{-3}$  & $1.045  \times 10^{-3}$\\
				    & Order & --    & 0.734 & 0.758 & 0.937  \\
\midrule
\multirow{2}{*}{WENO-$C$-$W$} & Error & 
 $6.077 \times 10^{-3}$   & $3.170 \times 10^{-3}$  & $1.694 \times 10^{-3}$  & $8.257  \times 10^{-4}$ \\
				    & Order & --    & 0.939 & 0.904 & 1.037  \\
\midrule
\multirow{2}{*}{WENO-$C$-$W$-$N$} & Error & 
 $6.064 \times 10^{-3}$   & $3.143 \times 10^{-3}$  & $1.703 \times 10^{-3}$  &  $8.363  \times 10^{-4}$\\
				    & Order & --    &  0.948 & 0.884 & 1.026 \\
\midrule
\bottomrule
\end{tabular}}
\caption{Post shock-wall collision ($t=8.0$) $L^1$ error analysis and convergence tests for the 
velocity for the LeBlanc shock tube problem.}
\label{table:leblanc-error-post}
\end{table} 

\subsection{The Peak shock tube problem}

We next consider the Peak shock tube problem, introduced in \cite{Liska2003995}. The domain of interest is $0.1 \leq x  \leq 0.6$, the adiabatic gas constant is $\gamma = 1.4$, and the 
initial data is given by 
$$
\begin{bmatrix}
\rho_0 \\ (\rho u)_0 \\ E_0 
\end{bmatrix}
=
\begin{bmatrix}
0.1261192 \\ 11.1230540 \\ 1.962323 \times 10^{3}
\end{bmatrix}
\mathbbm{1}_{[0.1,0.5)}(x)
+
\begin{bmatrix}
6.591493 \\ 14.932505 \\ 24.800422
\end{bmatrix}
\mathbbm{1}_{[0.5,0.6]}(x). 
$$
The difficulty in simulating solutions to Peak is due to the fact that the shock wave moves significantly slower than the expansion wave;
moreover, 
the distance between the 
contact discontinuity and the shock is very small, resulting in a sharp and narrow peak in the density. 
Most schemes produce inaccurate velocity profiles with large overshoots and low-frequency noise at the 
expansion wave \cite{Liska2003995,GreenoughRider2004}. 

The stand-alone WENO scheme produces a similarly poor velocity profile, but the $C$-method can be used to produce a good solution.
Since the noise appears in the velocity profile in the 
region with the rarefaction wave, and since the usual $C$-method includes a compression 
switch so that artificial viscosity is active only in regions of compression, we  employ an additional $C$-equation for the rarefaction wave,
whose solution is
 $C^r(x,t)$. We consider the following modification to 
(\ref{EulerC}b):
\begin{align*}
&\partial_t (\rho u) + \partial_x (\rho u^2 + p) =  \partial_x \left( \rho\, \left(\tilde{\beta}^u \, C + \tilde{\beta}^r\,C^r\right) \, \partial_x u \right) \,, \\
&\partial_t C^r + \frac{S(\bm{u})}{\varepsilon \Delta x} C^r - \kappa \Delta x \cdot S(\bm{u}) \partial_{xx} C^r = \frac{S(\bm{u})}{\varepsilon \Delta x} G^r \,, 
\end{align*}
where
$C$ is the solution to \eqref{C-Sod}, and where
$\tilde{\beta}^u = \frac{\max_{x} |\p_x u |}{\max_{x} C} \beta^u $ and $\tilde{\beta}^r = \frac{\max_{x} |\p_x u |}{\max_{x} C^r} \beta^r$,
with
$$
G^r(x,t) = \mathbbm{1}_{(0,+\infty)} (\partial_x u) \cdot \frac{|\partial_x u(x,t)|}{\max_{x} | \partial_x u(x,t)|} \,.
$$
We remark that we have omitted the wall function $\overline C_w$  since we are not simulating the
shock-wall collision for this problem.

\begin{figure}[H]
\centering
\subfigure[$t=0.0039$: velocity]{\label{fig:peak1}\includegraphics[width=75mm]{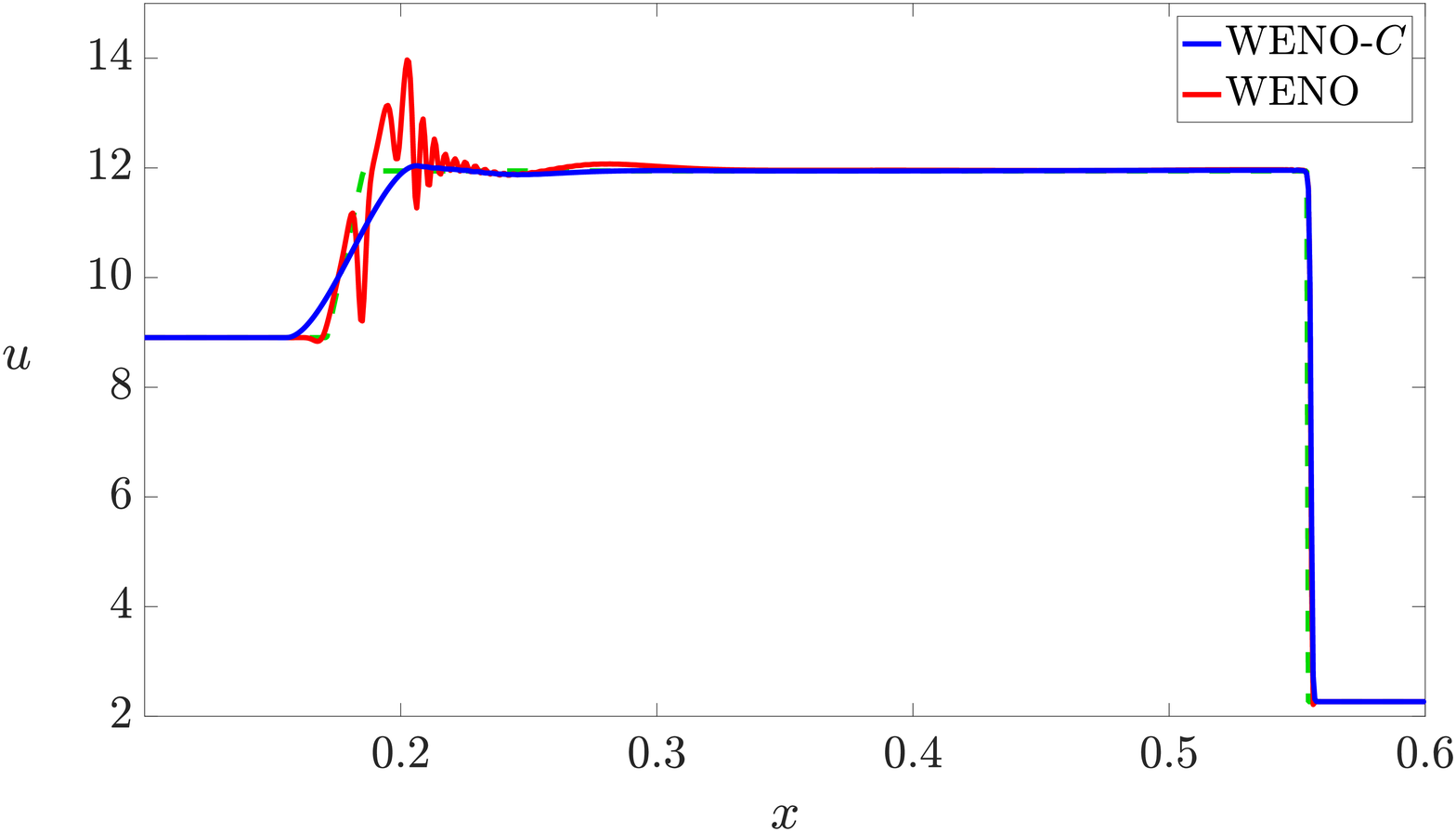}}
\subfigure[$t=0.0039$: velocity, zoomed in]{\label{fig:peak-zoom1}\includegraphics[width=75mm]{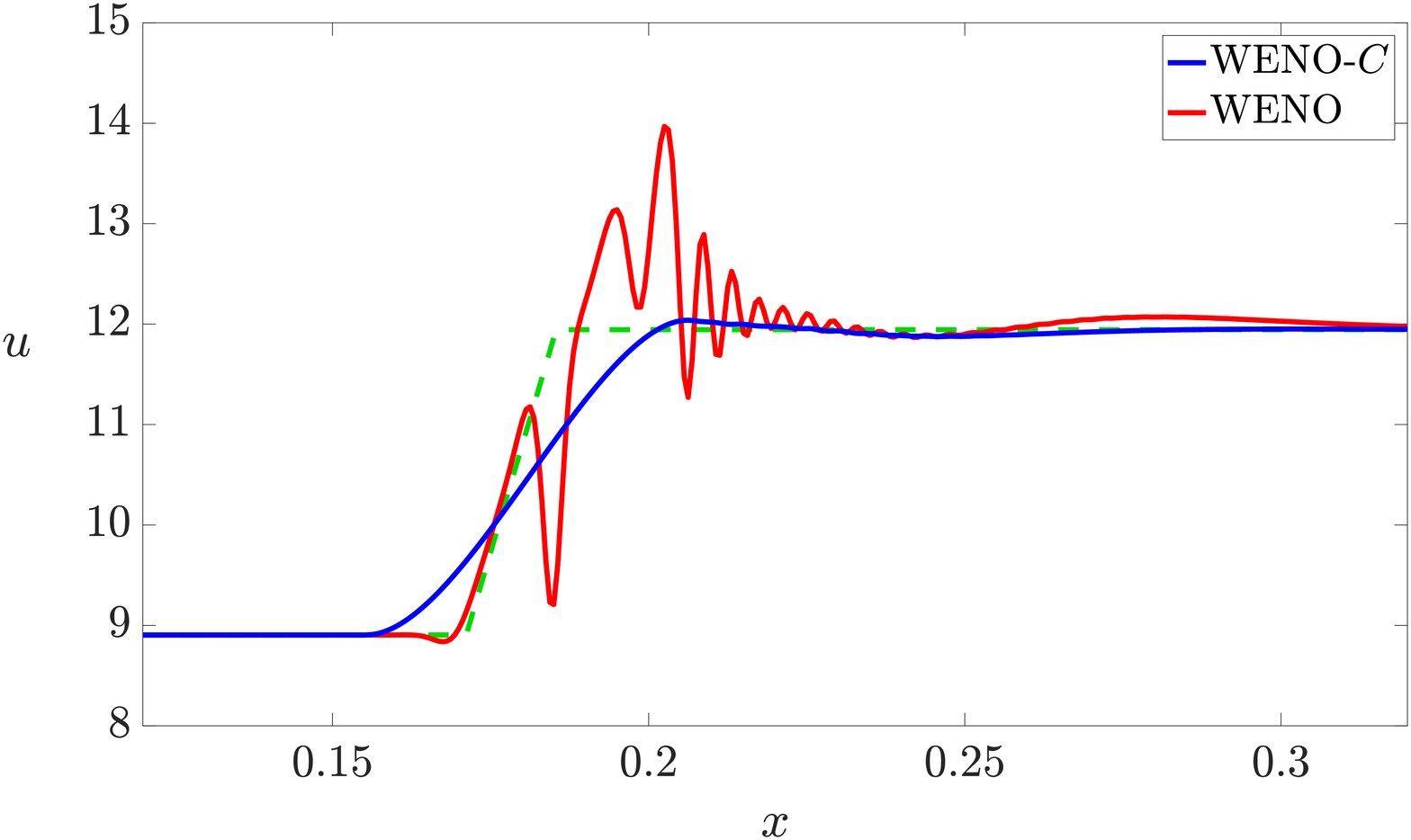}}
\caption{Comparison of WENO and WENO-$C$ for the Peak shock tube problem with 801
cells. The green curve is the exact solution.}
\label{fig:peak}
\end{figure}

WENO and WENO-$C$  (with the above modification) are used on a grid with 801 cells and 
with a time-step $\Delta t \approx 3.55 \times 10^{-6}$, giving CFL=0.6. The final time is $t=0.0039$, and the 
results are shown in Fig.\ref{fig:peak}. The relevant parameters are chosen as
\begin{gather*}
\beta^u=1.0, \qquad  \beta^r=10.0, \qquad \varepsilon=10.0, \qquad  \kappa=40.0,  \qquad \varepsilon_r=1.0, \qquad \kappa_r=20.0 \,.
\end{gather*}
As shown in Fig.\ref{fig:peak}, the extra viscosity provided by $\beta^r$ 
at the rarefaction wave removes the large overshoot and 
low frequency non-physical oscillations that are present in the solution produced with WENO. 

In \cite{Liska2003995}, an error analysis of various schemes applied to the Peak shock tube problem with the
above specifications is provided. We compute the $L^1$ and $L^2$ errors for the computed velocity minus the exact solution, using 
\eqref{error-formula} for the $L^1$ error and 
$$
\lvert \lvert u(\cdot,t) - u^*(\cdot,t) \rvert \rvert_{L^2} = \sqrt{\frac{1}{M} \sum_{i=1}^M | u(x_i,t) - u^*(x_i,t) |^2}\,,
$$
where $M$ is the number of cells used in the simulation and $u^*$ is the exact solution. Following 
\cite{Liska2003995}, we list the errors in percentage form with the ratio in question given by 
$\frac{\lvert \lvert u-u^* \rvert \rvert}{\lvert \lvert u^* \rvert \rvert}$. We also list the smallest error computed 
from all the schemes considered 
in \cite{Liska2003995}; namely,  the error computed from the scheme of Liu and Lax 
\cite{LiuLax1998,LiuLax2003}, which we will refer to as LL. We see in Table \ref{table:peakl} that WENO-$C$ compares very well 
with LL, with the solution producing smaller errors in both the $L^1$ and $L^2$ norms. 

\begin{table}[H]
\centering
\renewcommand{\arraystretch}{1.0}
\scalebox{0.8}{
\begin{tabular}{|llc|c|}
\toprule
\midrule
\multirow{2}{*}{\textbf{Norm}} & \multirow{2}{*}{\textbf{Scheme}} &  & \multicolumn{1}{c|}{\textbf{Cells}}\\

& {}  &   & 801\\
\midrule
 \multirow{6}{*}{\vspace{-1.25em}$\Vert u-u^* \rVert_{L^1}$} & \multirow{2}{*}{WENO} & Error & 
 $1.057 \times 10^{-1}$  \\
				  &  & \% & 1.0 \%  \\[1.25em]

& \multirow{2}{*}{WENO-$C$} & Error & 
 $7.260 \times 10^{-2}$  \\
				  &  & \% & 0.7 \%  \\[1.25em]
& \multirow{2}{*}{LL} & Error & 
 --  \\
				  &  & \% & 0.8 \% \\
\midrule
 \multirow{6}{*}{\vspace{-1.25em}$\lVert u -u^* \rVert_{L^2}$} & \multirow{2}{*}{WENO} & Error & 
 $5.168 \times 10^{-1}$  \\
				  &  & \% & 4.7 \%  \\[1.25em]

& \multirow{2}{*}{WENO-$C$} & Error & 
 $4.684 \times 10^{-1}$  \\
				  &  & \% & 4.3 \%  \\[1.25em]
& \multirow{2}{*}{LL} & Error & 
 --  \\
				  &  & \% & 4.4 \% \\
\midrule
\bottomrule
\end{tabular}}
\caption{$L^1$ and $L^2$ error analysis for the velocity $u$ for the Peak
shock tube problem at time $t=0.0039$.}
\label{table:peakl}
\end{table}

This test demonstrates the flexibility of the $C$-method; although a standard WENO scheme produces an
inaccurate and oscillatory solution, a very simple modification of the $C$-method allows for the suppression
of these oscillations, resulting in a more accurate solution. 

\subsection{The Osher-Shu shock tube problem}

The Osher-Shu shock tube problem, introduced in \cite{OsherShu1989}, simulates a shock front, perturbed by sinusoidal
 fluctuations. 
The computational domain is $-1 \leq x \leq 1$, $\gamma =1.4$, with  initial
data
\begin{equation}\label{osher-shu-initial}
\begin{bmatrix}
\rho_0 \\ (\rho u)_0 \\ E_0 
\end{bmatrix}
=
\begin{bmatrix}
3.857143 \\ 10.14185 \\ 39.1666
\end{bmatrix}
\mathbbm{1}_{[-1,-0.8)}(x)
+
\begin{bmatrix}
1 + 0.2\sin(5 \pi x) \\ 0 \\ 2.5
\end{bmatrix}
\mathbbm{1}_{[-0.8,1]}(x) \,. 
\end{equation}
We employ free-flow boundary 
conditions \eqref{var-bcs-alternate} at the left wall $x=-1$, and solid wall boundary conditions \eqref{var-bcs} at the right wall $x=1$.

\subsubsection{Noise removal with the noise indicator}
In order to test the efficacy of our noise detection and removal algorithm for the Osher-Shu test, we 
perform our numerical simulations using too large a time-step  and hence a numerically unstable 
CFL number, which 
produces spurious high-frequency oscillation behind the shock\footnote{Artificially inflating the CFL number
allows us to model a typical scenario in computational physics in which  a DNS-type simulation requires 
a prohibitively small time-step, and forces simulations that require entering the unstable CFL regime.
Our objective is to demonstrate that this high-frequency instability can be supressed by use
of our localized noise removal algorithm.}. 
Of course, high-frequency oscillations can be created by numerous numerical instabilities,
but an unstable CFL number creates the prototypical oscillation pattern for testing a noise removal scheme.

Our goal is to remove the high-frequency noise from the solution without affecting the low-frequency sinusoidal
oscillations that are the main feature of this test problem. To this end, we first compute a solution using 
WENO with 1025 cells with a time-step $\Delta t = 5.0 \times 10^{-4}$, giving a CFL number of 1.2.

 The relatively large number of 
cells and time-step produce noise with a frequency that is significantly higher than the lower frequency
non-spurious oscillations present in the solution.  The WENO-$N$ scheme is used with 
the reference coefficient $C_{\mathrm{ref}}$ in \eqref{cref} calculated using 
$\delta h = 10^{-3}$. The 
noise removal viscosity $\eta$ is chosen such that $\eta \Delta \tau / \Delta x^2 = 0.25$ and only one time-step is taken in the
heat equation.
Since an 
exact solution is not available for this problem, our ``exact'' solution is computed with 
WENO using 8193 cells and a time-step of $\Delta t = 3.125 \times 10^{-5}$, so that CFL $\approx 0.6$.  

\begin{figure}[H]
\centering
\subfigure[$t=0.36$: velocity]{\label{fig:oshu-noise1}\includegraphics[width=75mm]{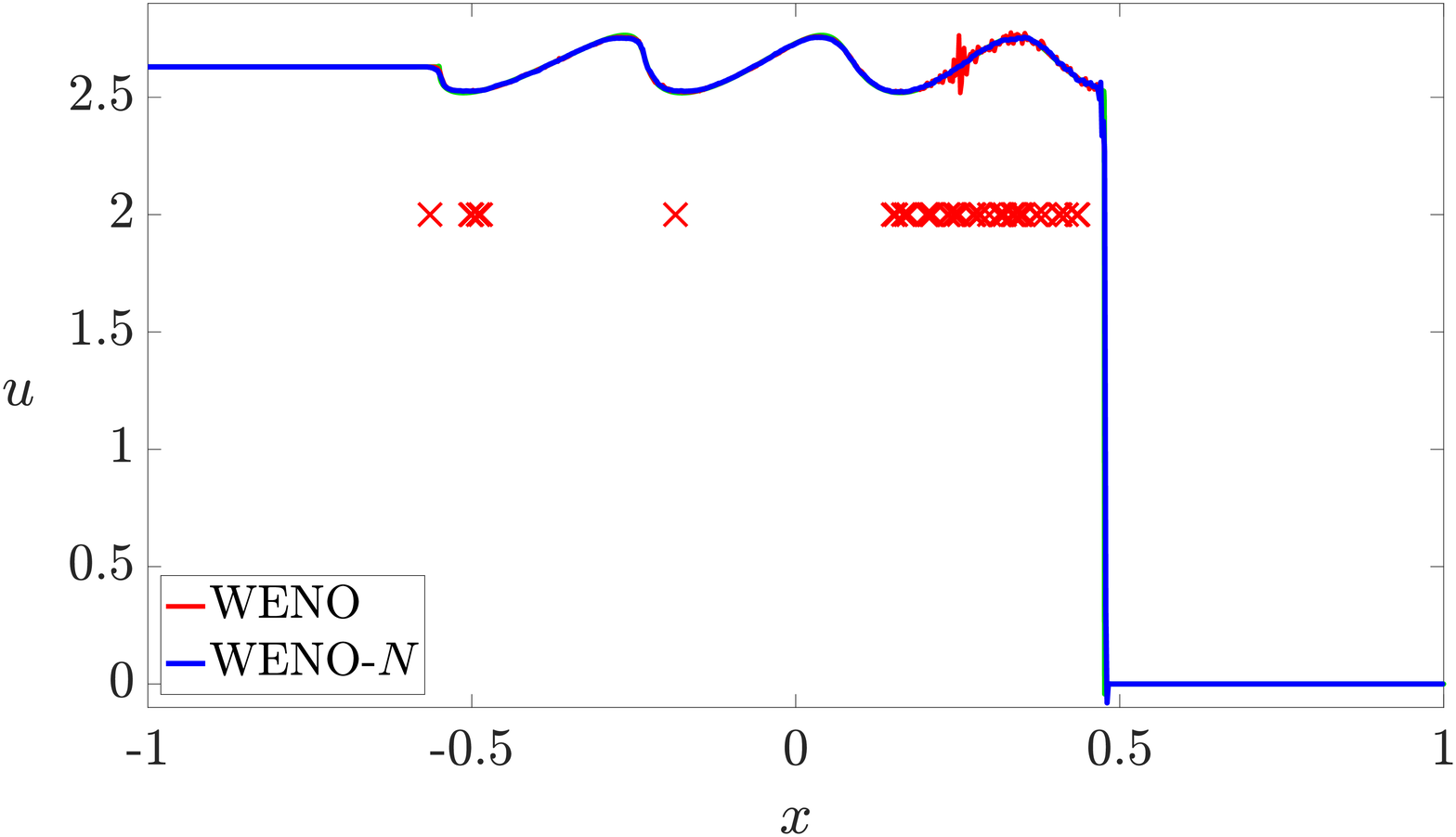}}
\subfigure[$t=0.36$: velocity, zoomed in]{\label{fig:oshu-noise-zoom1}\includegraphics[width=75mm]{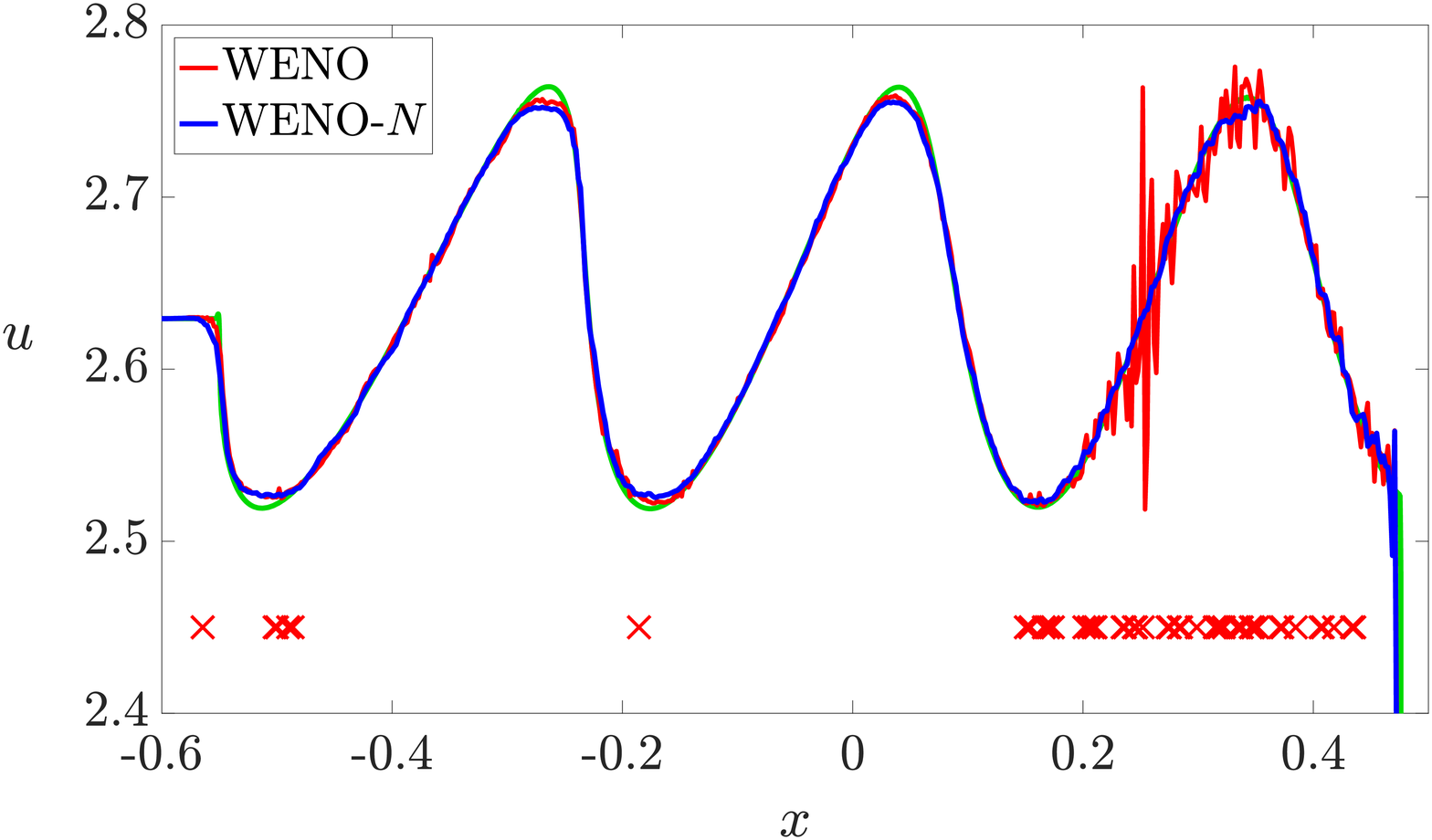}}
\caption{Comparison of WENO and WENO-$N$ for the Osher-Shu problem with 1025
cells. The red crosses indicate where the 
noise indicator function $\mathbbm{1}_{\operatorname{noise}}(x)$ is active. The green curve is 
the ``exact'' solution.}
\label{fig:oshu-noise}
\end{figure}

In Fig.\ref{fig:oshu-noise}, we compare the solutions computed with WENO and WENO-$N$. The 
noise indicator algorithm locates and removes the high-frequency noise present in the solution, without
affecting the sinusoidal oscillations. 
The sharpness of the shock front remains unaffected with the use of the noise indicator, due
to the deactivation of noise detection in a small region surrounding the shock. 

For the purpose of benchmarking our noise detection and removal algorithm, we also conduct 
tests in which we use linear (hyperviscosity) operators 
(see \mbox{\cite{Landshoff1955,Wilkins1980,CaShWh1998,PaPo1988,CaCo2004,CaCo2004b}}) of the form
\begin{equation}\label{hyperviscosity}
(-1)^{r-1} \beta_r (\Delta x)^{2r-1} \frac{\partial^{2r} u}{\partial x^{2r}}
\end{equation}
to remove noise, where $r \geq 1$.
The equations of motion we consider are the Euler equations 
{\eqref{eqn:consLawEvolution}} with the term {\eqref{hyperviscosity}} on the right-hand side of the 
momentum equation. When numerically approximated using our WENO-type discretization, the 
resulting scheme is referred to as the WENO-$\Delta^r u$ scheme. 
We perform numerical tests for the WENO-$\Delta^r u$ scheme with
$r=1, 2, 3$, and set $\beta_1=0.2$, $\beta_2=0.05$, and $\beta_3=0.01$, with these values determined 
\emph{a posteriori} to optimize the resulting solutions.

\begin{figure}[H]
\centering
\subfigure[WENO-$N$]{\label{fig:oshu-convergence1}\includegraphics[width=75mm]{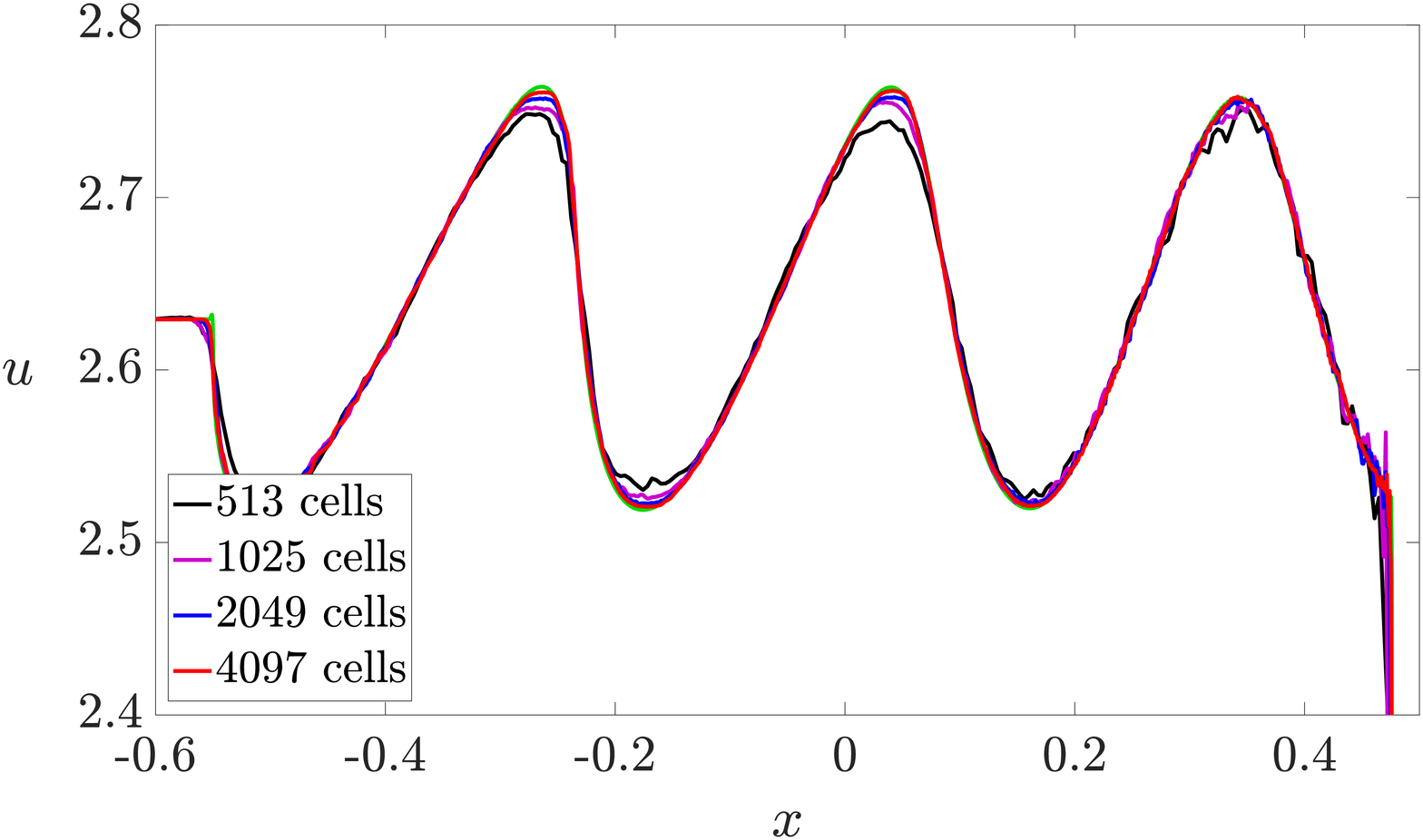}}
\subfigure[WENO-$\Delta u$]{\label{fig:oshu-convergence2}\includegraphics[width=75mm]{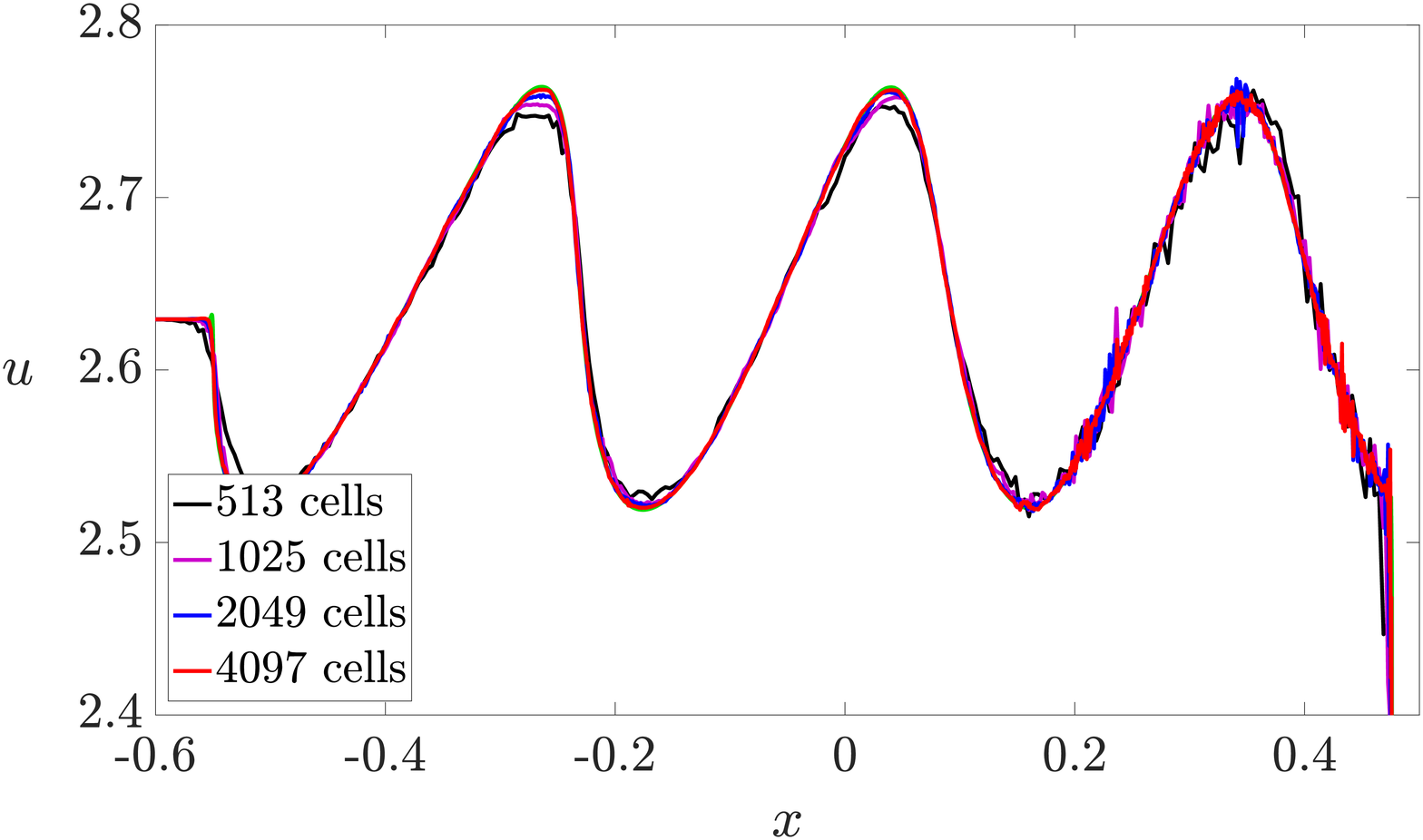}}
\subfigure[WENO-$\Delta^2 u$]{\label{fig:oshu-convergence3}\includegraphics[width=75mm]{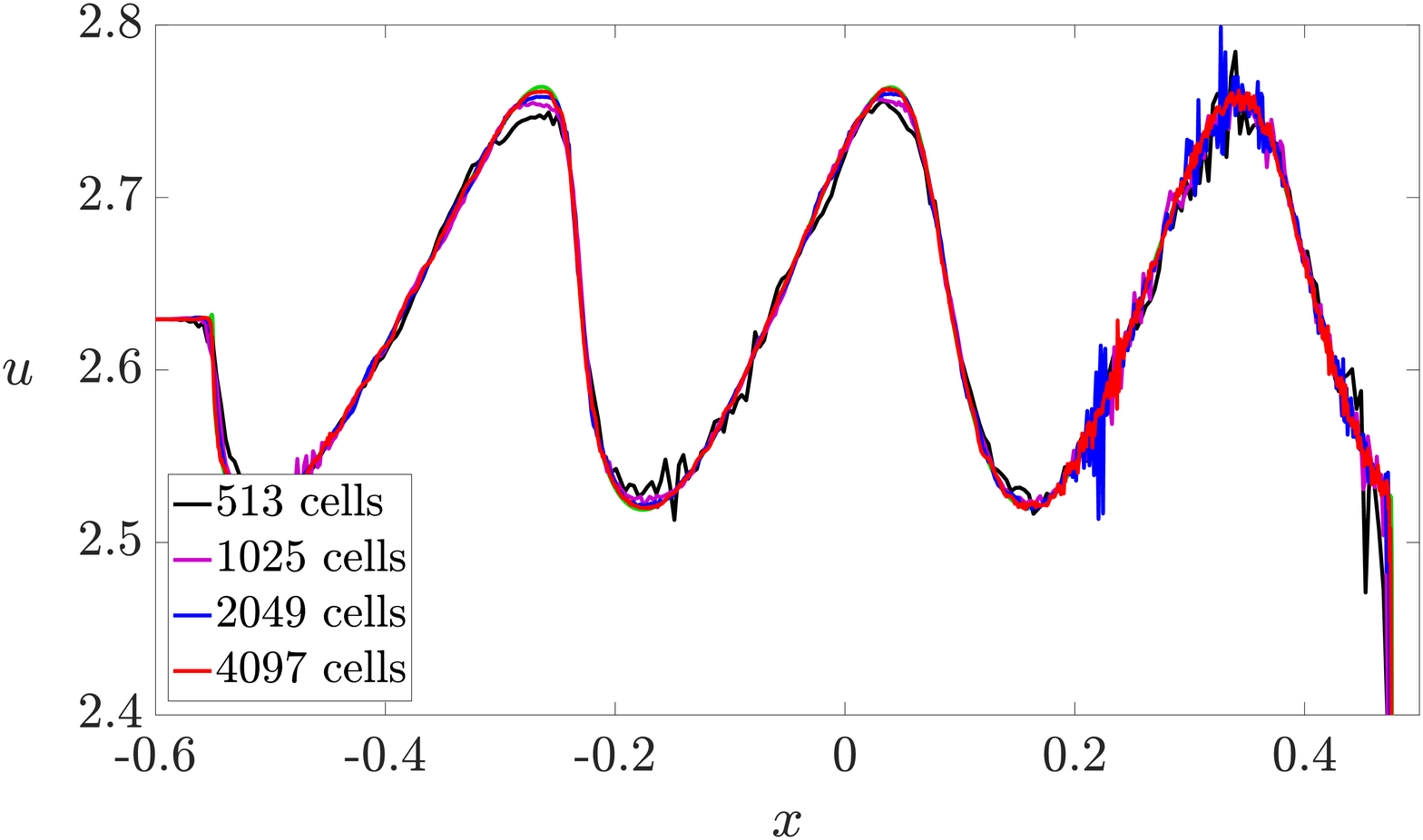}}
\subfigure[WENO-$\Delta^3 u$]{\label{fig:oshu-convergence4}\includegraphics[width=75mm]{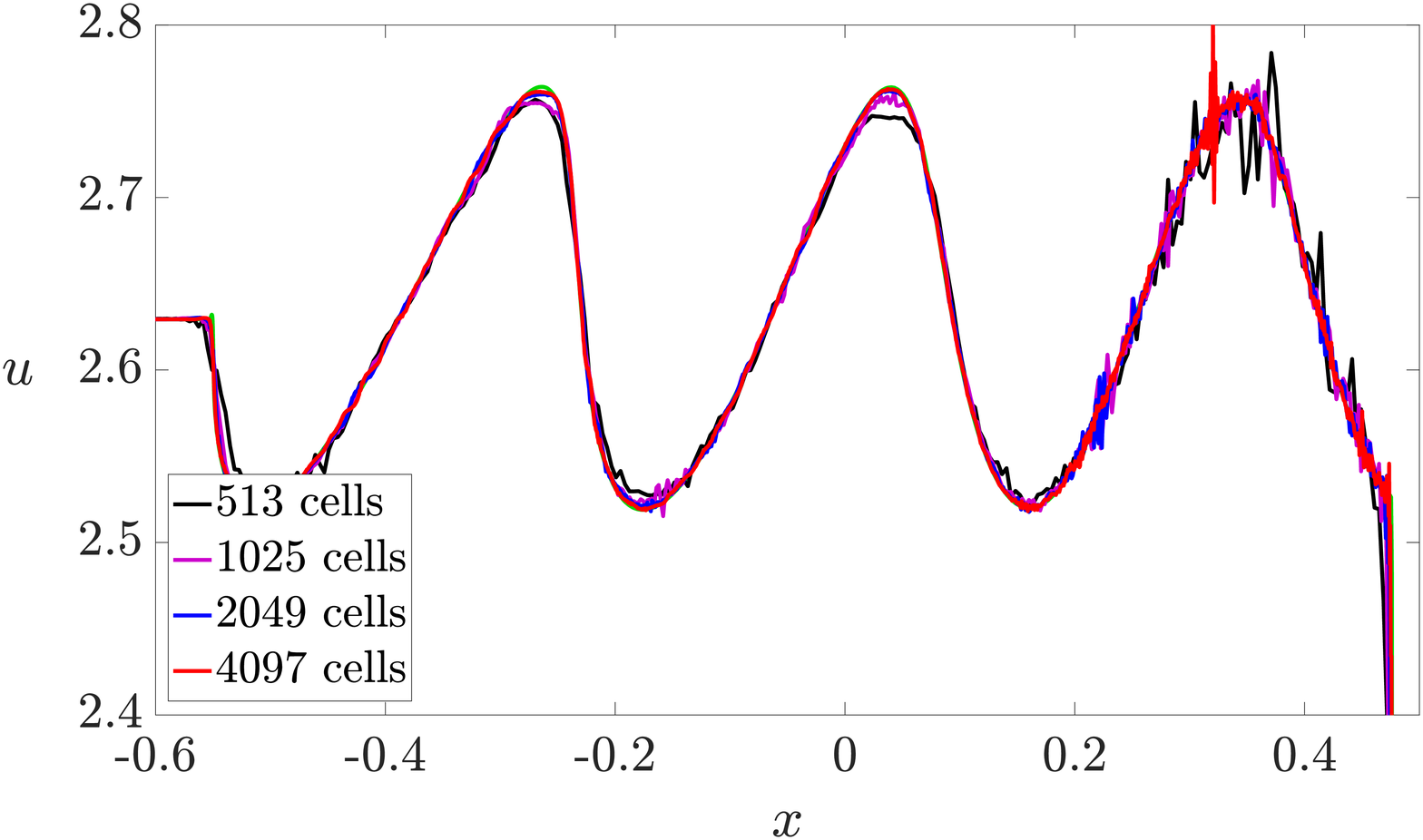}}
\caption{Comparison of the velocity profiles at $t=0.36$ for the Osher-Shu test. The green curve is the 
exact solution.}
\label{fig:hyperviscosity}
\end{figure}

We compare in Fig.{\ref{fig:hyperviscosity}} the WENO-$N$ and WENO-$\Delta^r u$ simulations; 
each subfigure shows the computed velocity, obtained using one of 
the schemes on grids with 513, 1025, 
2049, and 4097 cells, as well as the exact solution. The plots shown are zoomed in on the region behind the 
shock where there is high-frequency noise.

It is clear from these figures that, qualitatively, the WENO-$N$ scheme produces solutions with minimal noise
 that appear to converge to the exact solution. The hyperviscosity schemes, on the other hand, produce 
 solutions with erratic behavior; for instance, despite mesh refinement, 
 the WENO-$\Delta^3 u$ solution on the 4097 cell grid appears 
 much worse than the solutions on the 1025 and 2049 cell grids. Similarly inconsistent convergence 
 behavior can be observed with WENO-$\Delta u$ and WENO-$\Delta^2 u$. This is due to the CFL 
 condition violation. It is interesting to observe, on the other hand, 
 that the WENO-$N$ solutions are not subject to this erratic 
 convergence behavior. This is likely the result of the highly localized (in both space and time) nature 
 of the noise detection. 
 Overall, WENO-$N$ appears to produce noise-free, accurate, and 
 convergent solutions.

Defining the $L^1_t$ norm as
$$
\lVert f \rVert_{L^1_t} = \frac{1}{KM} \sum_{j=1}^{K} \sum_{i=1}^{M} \lvert f(x_i,t_j) \rvert\,,
$$
in Table \ref{table:oshu-noise-error-vel}  
we compute the $L^1$ and $L^1_t $ errors
for  the velocity at time $t = 0.36$ at various mesh refinements. 
Once again, we see that the noise indicator algorithm 
functions as an ``error correcter'', reducing the numerical error through the removal of high-frequency 
noise, while maintaining a relatively high order of accuracy. Among all the schemes considered, WENO-$N$ produces solutions with the smallest errors, 
 providing a quantitative validation of the observations made from Fig.{\ref{fig:hyperviscosity}}.

\begin{table}[H]
\centering
\renewcommand{\arraystretch}{1.0}
\scalebox{0.8}{
\begin{tabular}{|llc|cccc|}
\toprule
\midrule
\multirow{2}{*}{\textbf{Norm}} & \multirow{2}{*}{\textbf{Scheme}} &  & \multicolumn{4}{c|}{\textbf{Cells}}\\

& {}  &   & 513    & 1025   & 2049 & 4097\\
\midrule
 \multirow{10}{*}{\vspace{-5.0em}$\Vert \tilde{u} \rVert_{L^1}$} 
 
 & \multirow{2}{*}{WENO} & Error & 
 $1.003 \times 10^{-2}$  & $5.478 \times 10^{-3}$ & $2.018 \times 10^{-3}$ & $1.258  \times 10^{-3}$\\
				  &  & Order & --     & 0.873   & 1.440 & 0.682 \\[1.125em]

& \multirow{2}{*}{WENO-$\Delta u$} & Error & 
 $1.045 \times 10^{-2}$  & $4.717 \times 10^{-3}$  & $1.990 \times 10^{-3}$  & $9.770  \times 10^{-4}$ \\
				  &  & Order & --     & 1.148   & 1.245 & 1.026 \\[1.125em]

& \multirow{2}{*}{WENO-$\Delta^2 u$} & Error & 
 $1.050 \times 10^{-2}$  & $4.774 \times 10^{-3}$  & $2.459 \times 10^{-3}$  & $1.132  \times 10^{-3}$ \\
				  &  & Order & --     & 1.137   & 0.957 & 1.119 \\[1.125em]
				  
& \multirow{2}{*}{WENO-$\Delta^3 u$} & Error & 
 $1.084 \times 10^{-2}$  & $4.806 \times 10^{-3}$  & $1.981 \times 10^{-3}$  & $1.109  \times 10^{-3}$ \\
				  &  & Order & --     & 1.174   & 1.279 & 0.838 \\[1.125em]
				  
& \multirow{2}{*}{WENO-$N$} & Error & 
 $1.013 \times 10^{-2}$  & $4.432 \times 10^{-3}$  & $1.973 \times 10^{-3}$  & $1.005  \times 10^{-3}$ \\
				  &  & Order & --     & 1.193   & 1.168 & 0.973 \\				  				  				  				  
\midrule
 \multirow{10}{*}{\vspace{-5em}$\lVert \tilde{u} \rVert_{L^1_t}$} 
 
 & \multirow{2}{*}{WENO} & Error & 
 $7.328 \times 10^{-3}$  & $3.223 \times 10^{-3}$ & $1.139 \times 10^{-3}$ & $6.761  \times 10^{-4}$\\
				  &  & Order & --     & 1.185   & 1.501 & 0.752 \\[1.25em]

 & \multirow{2}{*}{WENO-$\Delta u$} & Error & 
 $7.484 \times 10^{-3}$  & $3.348 \times 10^{-3}$ & $1.192 \times 10^{-3}$ & $6.418  \times 10^{-4}$\\
				  &  & Order & --     & 1.161   & 1.490 & 0.893 \\[1.25em]
				  
 & \multirow{2}{*}{WENO-$\Delta^2 u$} & Error & 
 $7.333 \times 10^{-3}$  & $3.316 \times 10^{-3}$ & $1.254 \times 10^{-3}$ & $7.255  \times 10^{-4}$\\
				  &  & Order & --     & 1.145   & 1.403 & 0.789 \\[1.25em]
				  
 & \multirow{2}{*}{WENO-$\Delta^3 u$} & Error & 
 $7.419 \times 10^{-3}$  & $3.340 \times 10^{-3}$ & $1.196 \times 10^{-3}$ & $6.903  \times 10^{-4}$\\
				  &  & Order & --     & 1.151   & 1.482 & 0.793 \\[1.25em]				  

& \multirow{2}{*}{WENO-$N$} & Error & 
 $7.066 \times 10^{-3}$  & $3.004 \times 10^{-3}$  & $1.050 \times 10^{-3}$  & $5.656  \times 10^{-4}$ \\
				  &  & Order & --     & 1.234   & 1.517 & 0.893 \\
\midrule
\bottomrule
\end{tabular}}
\caption{$L^1$ and $L^1_t$ error analysis and convergence tests for the velocity $u$ for the Osher-Shu
problem at time $t=0.36$, with $\tilde{u} = u - u^*$ the difference between the computed
solution $u$ and the ``exact solution'' $u^*$.}
\label{table:oshu-noise-error-vel}
\end{table} 

We note that the numerical error and the order of convergence remains unchanged when
 using the  
density $\rho$ instead of the velocity $u$;
 in particular,  the WENO-N algorithm produces solutions with smaller errors and 
similar rates of convergence as WENO, when errors and accuracy are computed using $\rho$.   
And so, the removal of  high-frequency noise in $u$, in turn, provides a density field that is also
free of high-frequency oscillations (c.f. Remark {\ref{remark-noise}}).

\subsubsection{Stabilizing shock-wall collision for Osher-Shu}
We now turn to the issue of stabilizing shock-wall collision for the Osher-Shu problem. The problem is set up 
as follows: the initial data is \eqref{osher-shu-initial}, the time-step is given by $\Delta t = 5 \times 10^{-4}$ with 
final time $t = 0.63$, and the number of cells is 512, so that the CFL number is 0.6. We impose the
 solid wall boundary conditions \eqref{ghost-node} at the
right boundary $x=1$, and free-flow boundary conditions \eqref{ghost-node-alternate} 
at the left boundary $x=-1$. 
The shock-wave moves to the right
and collides with the wall at $x=1$ 
at time $t \approx 0.5$. Post-collision, there is a large amount of noise present 
in the solution behind the shock-wave, and our aim is to remove the noise while preserving the 
sharpness of the shock front and minimizing the damping of the post-shock low frequency
oscillations. 
\begin{figure}[H]
\centering
\subfigure[$t=0.63$: density]{\label{fig:oshu-after-rho}\includegraphics[width=75mm]{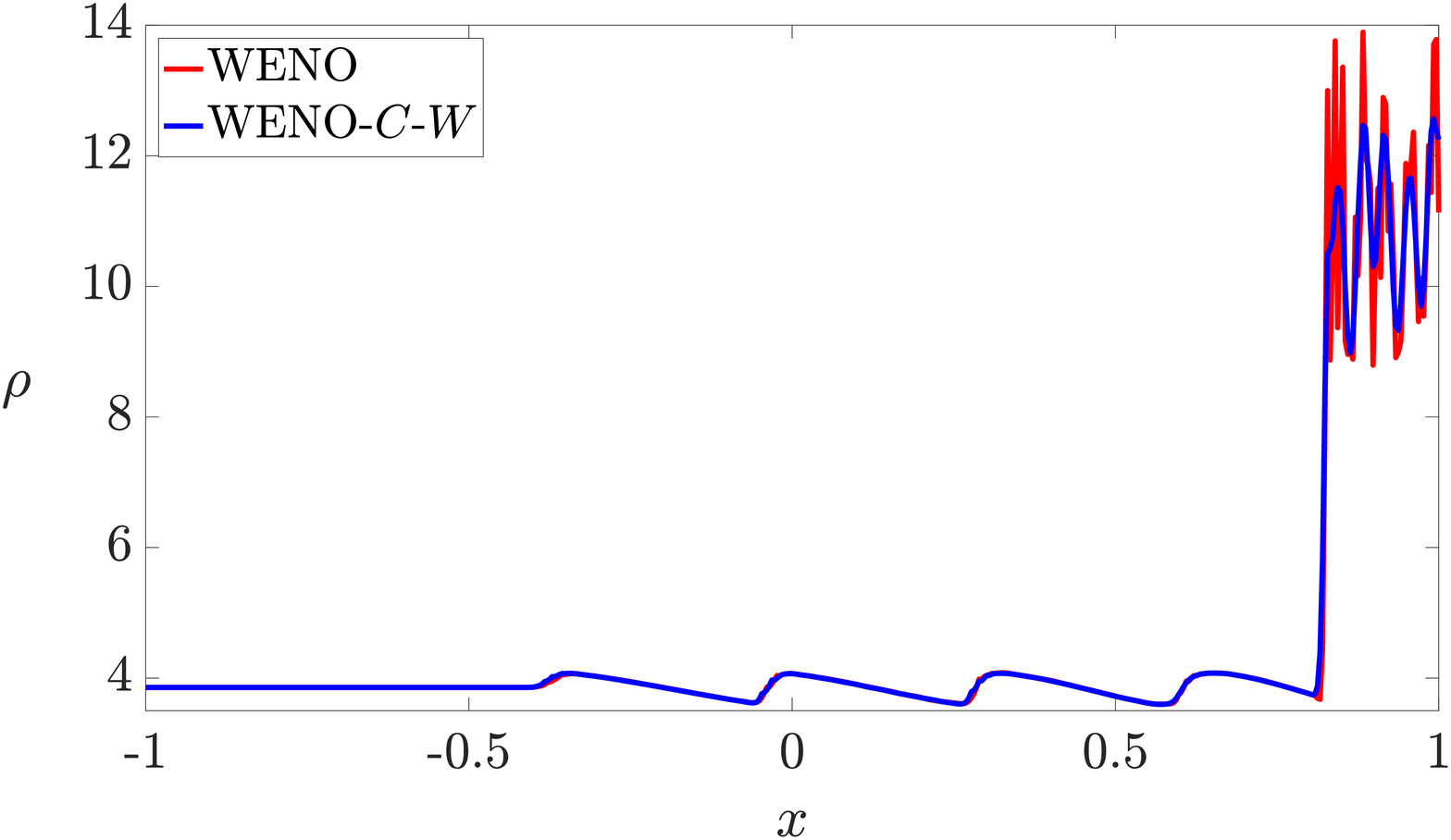}}
\subfigure[$t=0.63$: density, zoomed in]{\label{fig:oshu-after-rho-zoom}\includegraphics[width=75mm]{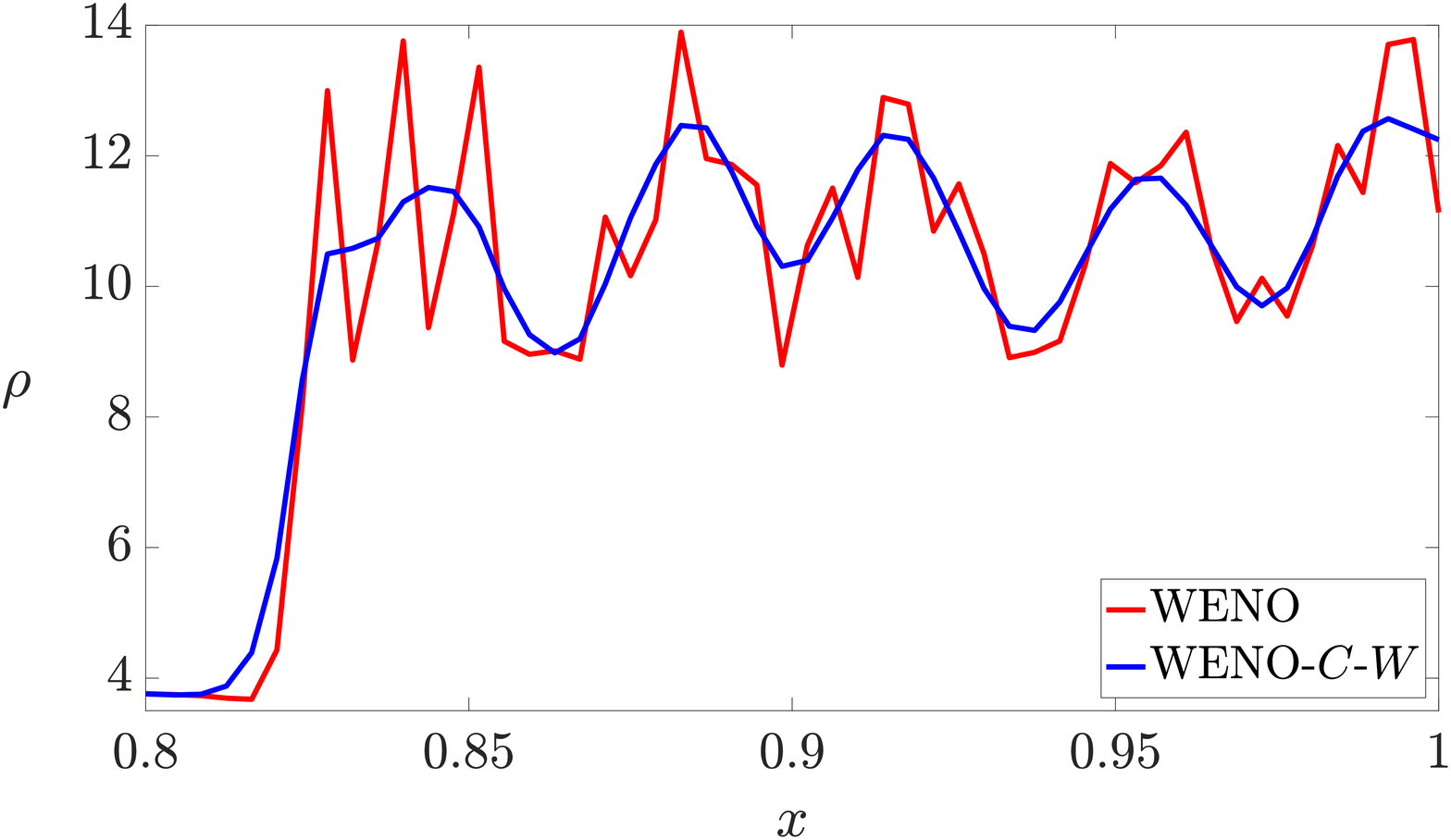}}
\caption{Comparison of WENO vs. WENO-$C$-$W$ for the density just after the shock-wall collision problem for the Osher-Shu problem with 512 cells.}
\label{fig:oshu-collision}
\end{figure}

%
%
%

We employ our WENO-$C$-$W$ scheme and choose the relevant parameters as 
\begin{alignat*}{4}
\beta^u&=1.0, \qquad  \beta^E&=0.0, \qquad   \beta^u_w &= 2.5, \qquad  \beta^E_w &= 0.85 \\
\varepsilon&=1.0, \qquad  \kappa&=5.0, \qquad  \varepsilon_w &=40.0, \qquad  \kappa_w &=4.0 \,.
\end{alignat*}
The results are shown in Fig.\ref{fig:oshu-collision}. 
Post shock-wall collision, WENO produces a noisy solution with high frequency noise
interfering with the sinusoidal  oscillations, while
 WENO-$C$-$W$  produces a solution with a sharp front and  without noise.

\section{Concluding remarks}
In this paper, we have presented three ideas:  the first is a space-time smooth artificial viscosity 
method that is versatile and 
simple to implement; the second is a shock-wall collision scheme that can be used to suppress post-collision 
noise that occurs when a shock-wave collides with a fixed boundary and bounces-back; the third is a wavelet-based noise 
detection and removal scheme that is highly localized and can be used to remove noise present in solutions. 
We have demonstrated the efficacy of the new method on a variety of 1-$D$
test problems with different features, and demonstrated that the solutions produced retain sharp fronts,  correct wave speeds, remain oscillation-free, are not subject to the wall-heating error, and maintain high-order accuracy.

\appendix
\section{The WENO-$|u_x|$ and WENO-Noh schemes}\label{sec:appendix}
For our numerical simulations, we use a variety of combinations of the WENO scheme, the $C$-method, the 
wall $C$-method, and the noise indicator. For the purpose of comparison, we implement two additional
methods. The first is a classical artificial viscosity scheme, WENO-$|u_x|$, and the second is 
WENO-Noh, an artificial viscosity method introduced by Noh \cite{Noh1987}. 
We will employ WENO-Noh primarily as a comparison for the wall $C$-method for shock-wall collision, 
while WENO-$|u_x|$ will serve as a benchmark for the usual $C$-method. 

\subsection{WENO-$|u_x|$: classical artificial viscosity}
This is the classical artificial viscosity scheme, where viscosity is only added to the momentum equation
and the localizing coefficient is given by $|\partial_x u|$. More precisely, we implement the method in the 
following manner:
\begin{subequations}\label{EulerC-classical}
\begin{align}
\partial_t \rho + \partial_x (\rho u) &=0, \label{EulerC-density-classical}\\[0.5em] 
\partial_t (\rho u) + \partial_x (\rho u^2 + p) &=  \partial_x \left( (\Delta x)^2{\beta}^{u} \, \rho \, |\partial_x u| \, \partial_x u \right), \label{EulerC-momentum-classical}\\[0.5em]
\partial_t E + \partial_x (uE + up) &=  0 \,.  \label{EulerC-energy-classical}
\end{align}
\end{subequations}
The time and spatial discretizations are done in as in \S\ref{sec-weno-reconstruction-procedure}. This scheme
will serve primarily as a benchmark for WENO-$C$. 

\subsection{WENO-Noh: an artificial viscosity method of Noh}
This artificial viscosity scheme of Noh \cite{Noh1987} introduces an additional heat conduction term 
to the energy equation, in addition to the usual viscosity term in the momentum equation. For more details, 
we refer the reader to \cite{Noh1987}. We implement the method in the following fashion:
the equations of motion are
\begin{subequations}\label{EulerC-noh}
\begin{align}
\partial_t \rho + \partial_x (\rho u) &=0, \label{EulerC-density-noh}\\[0.5em] 
\partial_t (\rho u) + \partial_x (\rho u^2 + p) &=  \partial_x \left( (\Delta x)^2 {\beta}^{u}_{\operatorname{Noh}} \, \rho \, |\partial_x u| \, \partial_x u \right), \label{EulerC-momentum-noh}\\[0.5em]
\partial_t E + \partial_x (uE + up) &=  \partial_x \left((\Delta x)^2 {\beta}^{E}_{\operatorname{Noh}} \, \rho \, |\partial_x u| \, \partial_x e \right) \,. \label{EulerC-energy-noh}
\end{align}
\end{subequations}
Here, $e= \frac{p}{(\gamma-1) \rho}$ is the specific internal energy of the system. The numerical discretization is then done in an identical fashion to that described in 
\S\ref{sec-weno-reconstruction-procedure}. We will employ this scheme with the aim of fully suppressing 
post-collision noise, even at the expense of a less accurate solution prior to shock-wall collision. 

For readability, we will use Table \ref{table:schemes} to refer to these 
various schemes. 
{\tiny
\begin{table}[H]
\centering
\scalebox{0.8}{
\begin{tabular}{| M{4cm} | M{6cm}|} 
 \hline
 Scheme & Description \\ [0.0em] 
 \hline \hline 
 WENO & standard fifth-order WENO procedure for the usual Euler equations. \\[0.5em] 
\hline
 WENO-$|u_x|$ & WENO scheme with classical artificial viscosity. \\[0.5em]
\hline
 WENO-Noh &  WENO scheme with Noh's artificial viscosity method. \\[0.5em]
 \hline
 WENO-$C$ & WENO scheme with the $C$-method. \\[0.5em] 
 \hline 
 WENO-$C$-$W$ & WENO scheme with the $C$-method and the wall $C$-method outlined in \S\ref{sec:wallvisc}. \\[0.5em] 
 \hline 
 WENO-$N$ & WENO scheme with the noise indicator outlined in \S\ref{sec:noiseind}. \\[0.5em]
 \hline
 WENO-$C$-$N$ & WENO scheme with the $C$-method and the noise indicator. \\[0.5em]
 \hline
 WENO-$W$-$N$ & WENO scheme with the wall $C$-method and the noise indicator. \\[0.5em]
 \hline
 WENO-$C$-$W$-$N$ &  WENO scheme with the $C$-method, the wall $C$-method and the noise indicator. \\[0.5em]
 \hline
  WENO-$\Delta^r u$ &  WENO scheme with linear hyperviscosity \eqref{hyperviscosity}. \\[0.5em]
  \hline
\end{tabular}}
\caption{Various numerical schemes used in the simulations.}
\label{table:schemes}
\end{table}}

\section{Calculation of the exact solution post shock-wall collision}
In this section, we provide details for the calculation of the exact solution to Sod-type problem post shock-wall 
collision. The solution, calculated based on the Rankine-Hugoniot conditions \eqref{RHconditions} 
and the assumption that
the post-shock velocity is identically zero both pre and post shock-wall collision, is valid until the reflected
shock front collides with the contact wave. We assume for simplicity
that the shock front is traveling to the right (so that the shock speed $\dot{\sigma}$ satisfies 
$\dot{\sigma}(t) > 0$ pre shock-wall collision) and collides with, and reflects back off of, the right boundary. We
also assume that the shock $\sigma(t)$ separates two constant states, $\bm{u}_l$ and $\bm{u}_r$, to the 
left and right of the shock, respectively. 

The left states $\bm{u}_l$ and the post-shock velocity $u_r = 0$ are all known; the unknowns are thus the 
post-shock density $\rho_r$, energy $E_r$ (or, equivalently, pressure $p_r$), and 
shock speed $\dot{\sigma}(t)$. The R-H conditions \eqref{RH2} and \eqref{RH1} 
yield
\begin{align}
p_r &= \rho_l u_l^2 + p_l - \dot{\sigma}\rho_l u_l\,, \label{postsol1}\\
\rho_r &= \rho_l - \frac{\rho_l u_l }{\dot{\sigma}}\,, \label{postsol2}
\end{align} 
respectively, so it only remains to calculate the shock speed $\dot{\sigma}$. Substituting \eqref{postsol1} into
\eqref{RH3} and simplifying leads to a quadratic equation for $\dot{\sigma}$,
$$
\dot{\sigma}^2 - \frac{1}{2} (3 - \gamma) u_l \dot{\sigma} - \frac{(\gamma-1)(E_l+p_l)}{\rho_l} = 0\, , 
$$
which has solutions
$$
\dot{\sigma} = \frac{1}{4}(3 - \gamma) u_l \pm \sqrt{  \left(\frac{(3-\gamma)u_l}{4} \right)^2 + \frac{(\gamma-1)(E_l+p_l)}{\rho_l} }\,.
$$
We take the negative root $\dot{\sigma} < 0$, since the shock moves to the left post shock-wall collision:
\begin{equation}\label{postsol3}
\dot{\sigma} = \frac{1}{4}(3 - \gamma) u_l - \sqrt{  \left(\frac{(3-\gamma)u_l}{4} \right)^2 + \frac{(\gamma-1)(E_l+p_l)}{\rho_l} }\, .
\end{equation}

The relations \eqref{postsol1}, \eqref{postsol2}, and \eqref{postsol3} then provide the complete exact solution 
post shock-wall collision, up until the time that the reflected shock front collides with the contact discontinuity.

\section{Comparison of optimized-parameter runs with fixed-parameter runs}\label{sec:appendix2}

In this section, we compare the optimized-parameter runs presented in \S{\ref{sec:simulations}}, 
with a set of runs using fixed parameters. The 
fixed-parameter runs have the 
$C$-equation parameters $\varepsilon$ and $\kappa$ take the fixed value 1, with 
the exception of $\varepsilon_w$, to which we assign the fixed value $\varepsilon_w=50.0$. 
The artificial viscosity parameters $\beta$ are still free to choose, and so vary from problem to problem. 
The particular choices of $\beta$ for each test problem shown in the figures below are listed in the 
corresponding caption. For the initial data, we refer the reader to the relevant section in the main body of the 
paper. 

We present results for the Sod shock-wall collision, Noh, 
LeBlanc shock-wall collision, Peak, and Osher-Shu shock-wall collision
problems. In the figures shown below, Run 1 indicates the solution computed using the optimized set of 
parameters, while Run 2 indicates the solution computed using the fixed set of parameters.

\begin{figure}[H]
\centering
\subfigure[$t=0.20$: velocity, pre-collision]{\label{fig:sod-calibrated-velocity1}\includegraphics[width=75mm]{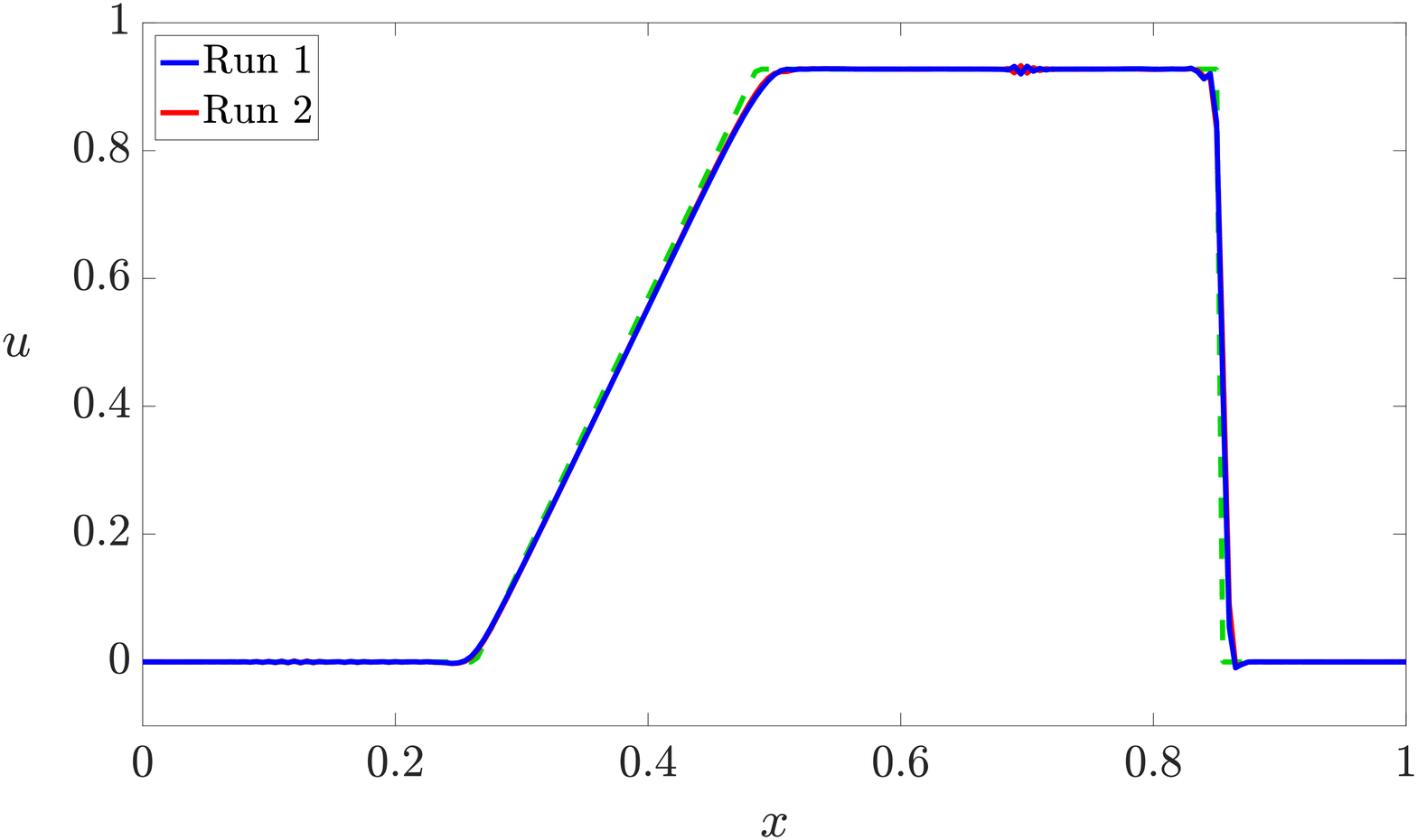}}
\subfigure[$t=0.36$: velocity, post-collision]{\label{fig:sod-calibrated-velocity2}\includegraphics[width=75mm]{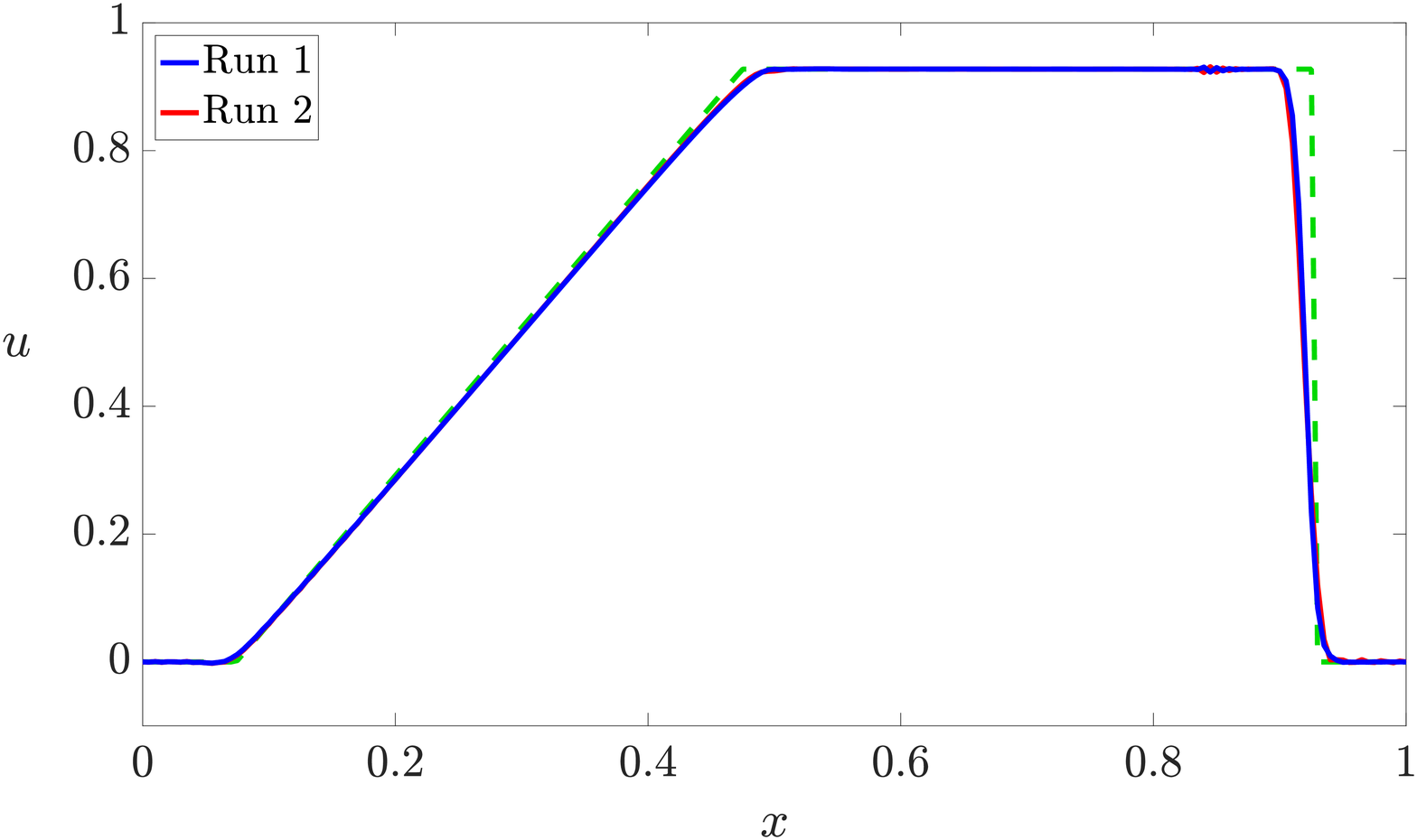}}
\subfigure[$t=0.36$: velocity, post-collision zoom-in]{\label{fig:sod-calibrated-velocity3}\includegraphics[width=75mm]{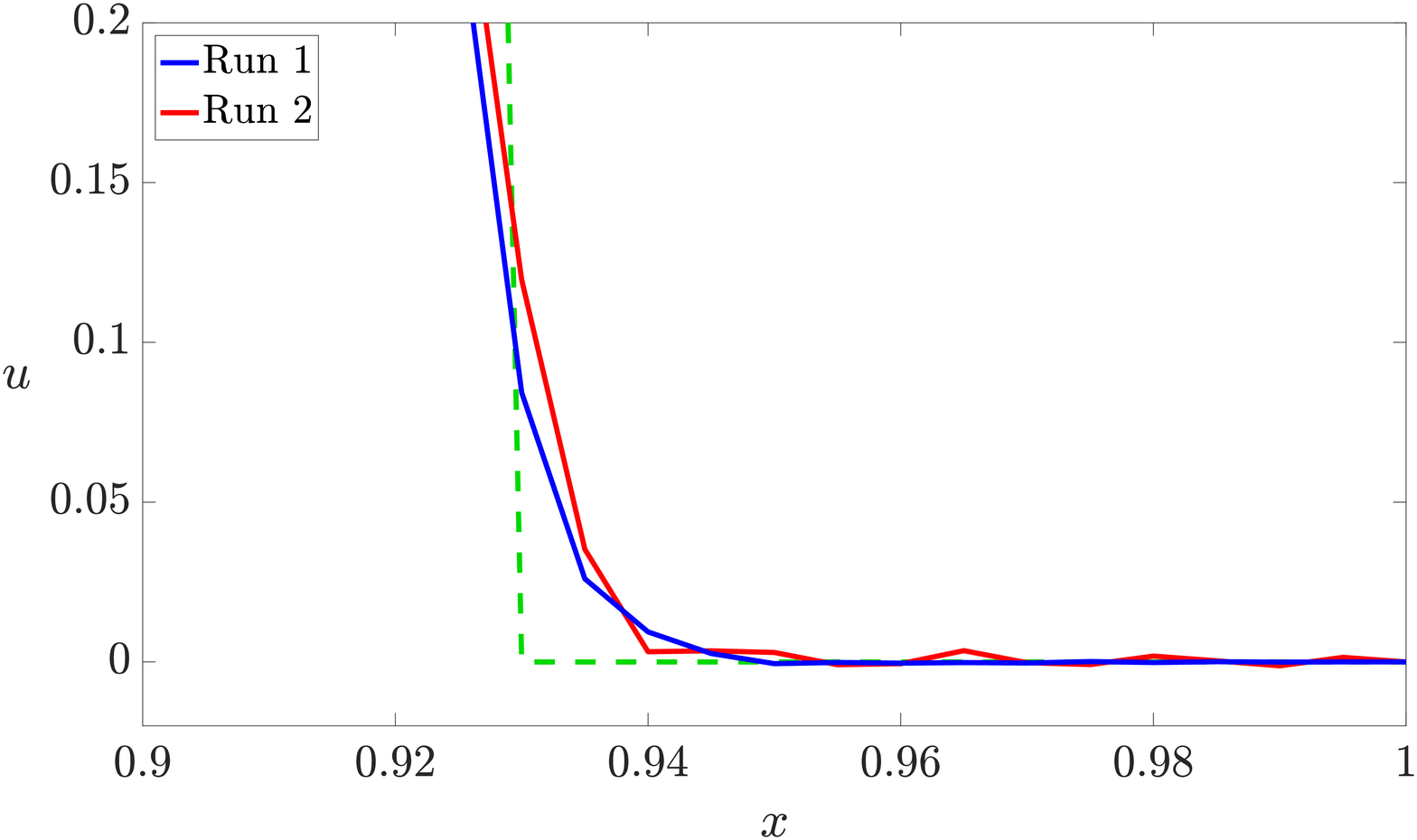}}
\subfigure[$t=0.36$: density, post-collision zoom-in]{\label{fig:sod-calibrated-rho}\includegraphics[width=75mm]{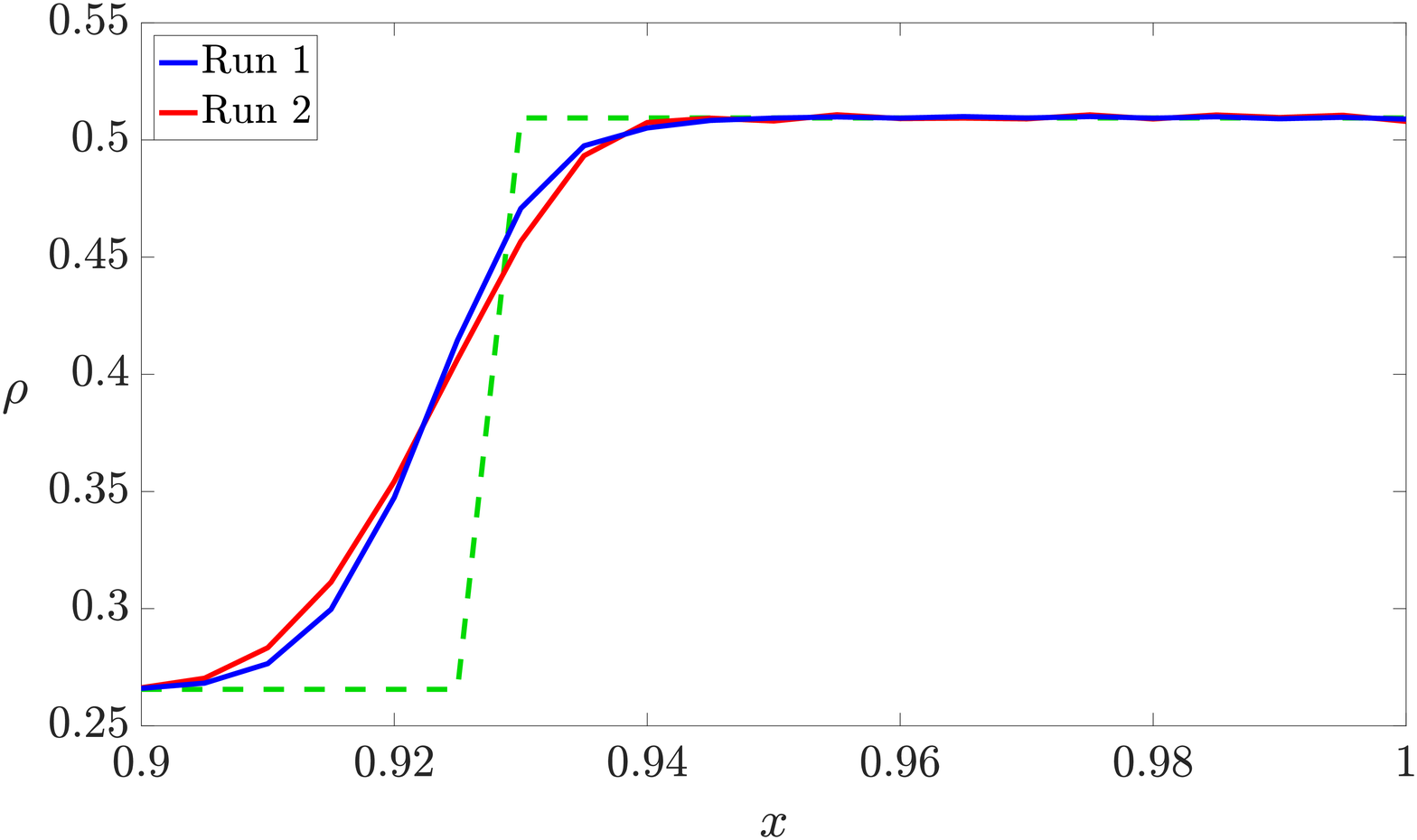}}
\caption{Comparison of the optimized-parameter and fixed-parameter 
WENO-$C$-$W$ runs for the Sod shock tube problem
before and after shock-wall collision. The artificial viscosity parameters for the fixed-parameter Run 2 are 
chosen as $\beta^u=1.0$, $\beta^E=0.0$, $\beta^u_w=5.0$, $\beta^E_w=10.0$.}
\label{fig:sod-calibrated}
\end{figure}  

\begin{figure}[H]
\centering
\subfigure[$t=1.0$: density]{\label{fig:noh-calibrated1}\includegraphics[width=75mm]{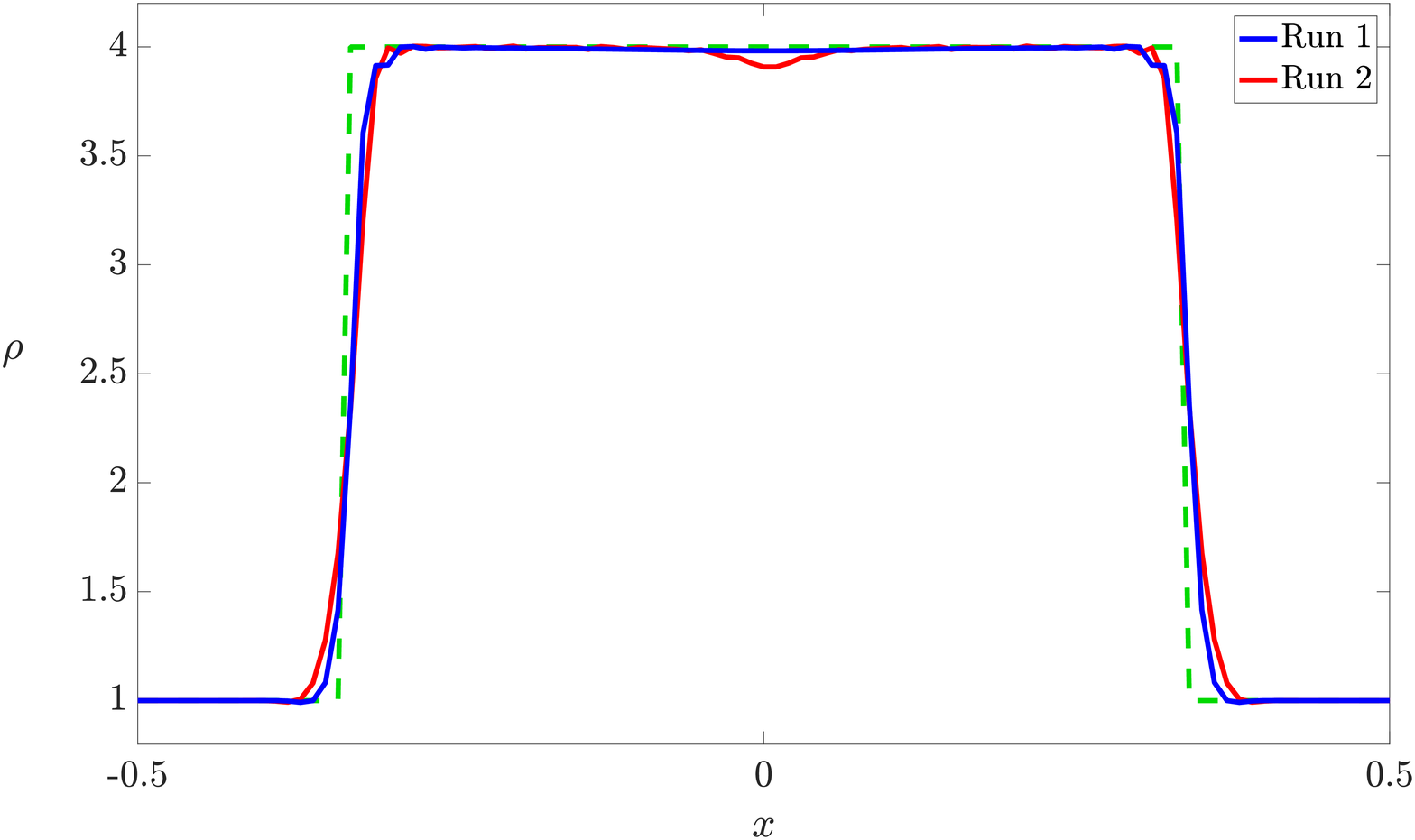}}
\subfigure[$t=1.0$: density zoom-in at shock]{\label{fig:noh-calibrated2}\includegraphics[width=75mm]{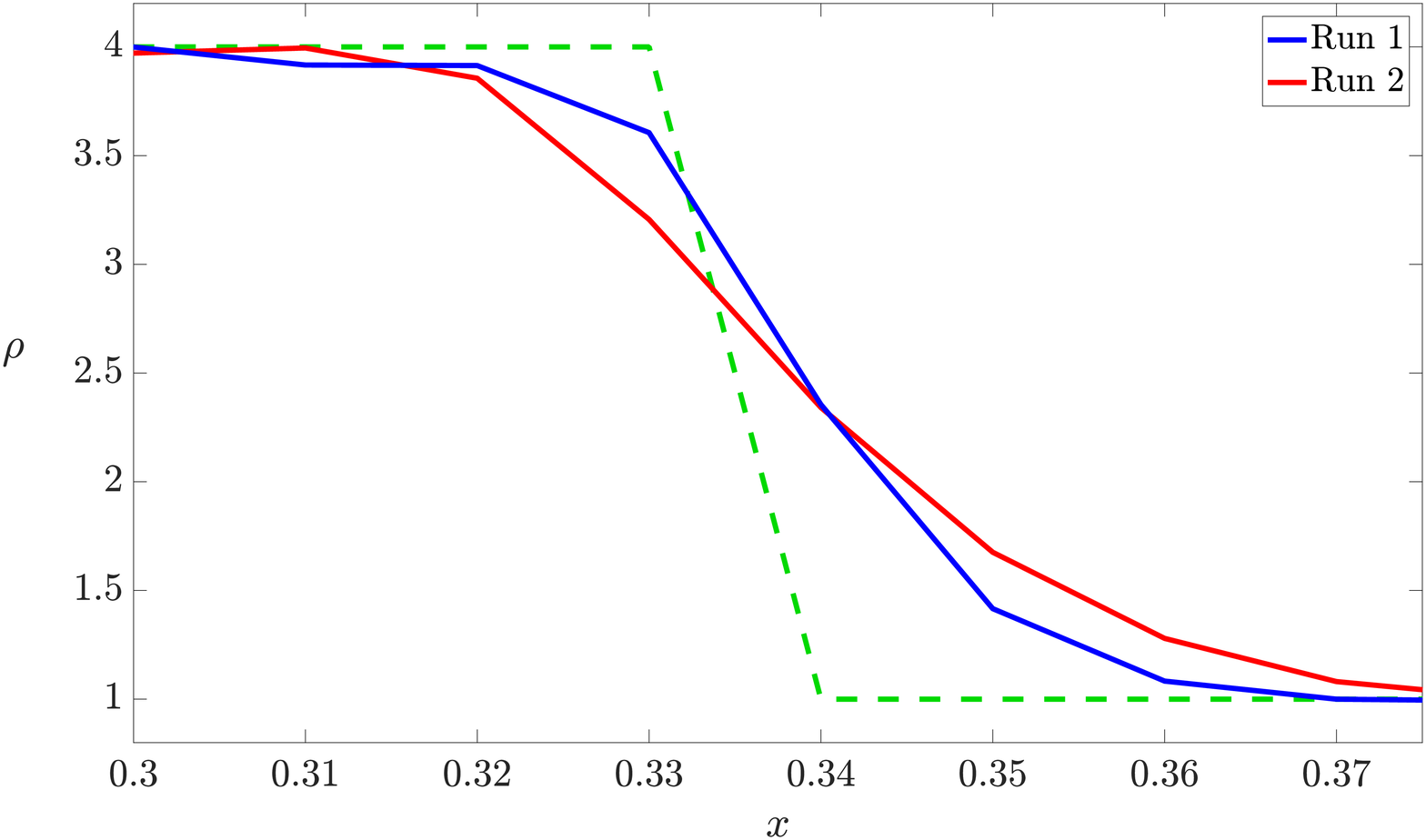}}
\caption{Comparison of the optimized-parameter and fixed-parameter WENO-$C$ runs for the Noh
problem. The artificial viscosity parameters for the fixed-parameter Run 2 are 
chosen as $\beta^u=3.0$, $\beta^E=30.0$.}
\label{fig:noh-calibrated}
\end{figure}

\begin{figure}[H]
\centering
\subfigure[$t=6.0$: internal energy, pre-collision]{\label{fig:leblanc-calibrated-e1}\includegraphics[width=75mm]{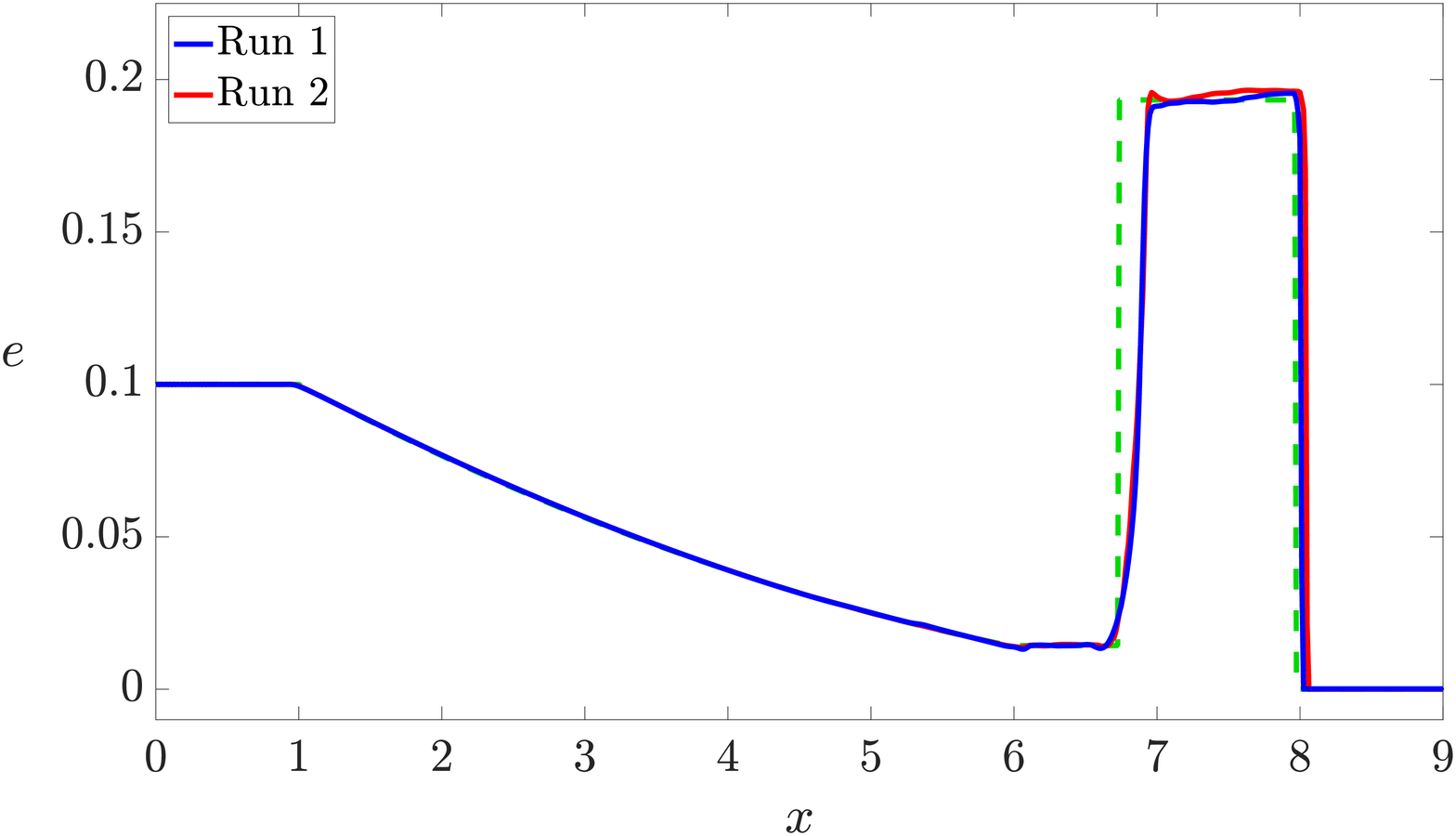}}
\subfigure[$t=8.0$: internal energy, post-collision]{\label{fig:leblanc-calibrated-e2}\includegraphics[width=75mm]{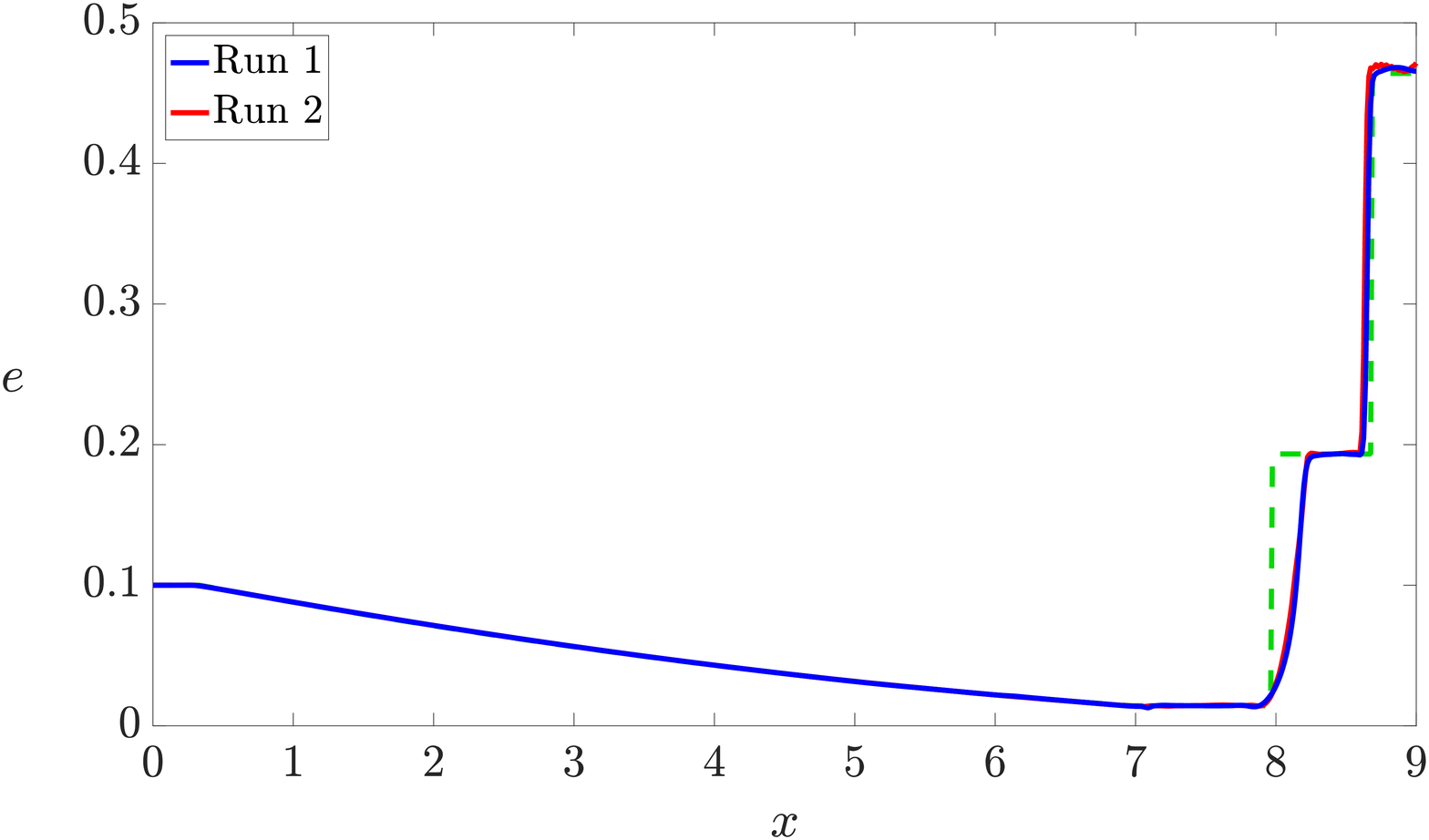}}
\subfigure[$t=8.0$: internal energy, post-collision zoom-in]{\label{fig:leblanc-calibrated-e3}\includegraphics[width=75mm]{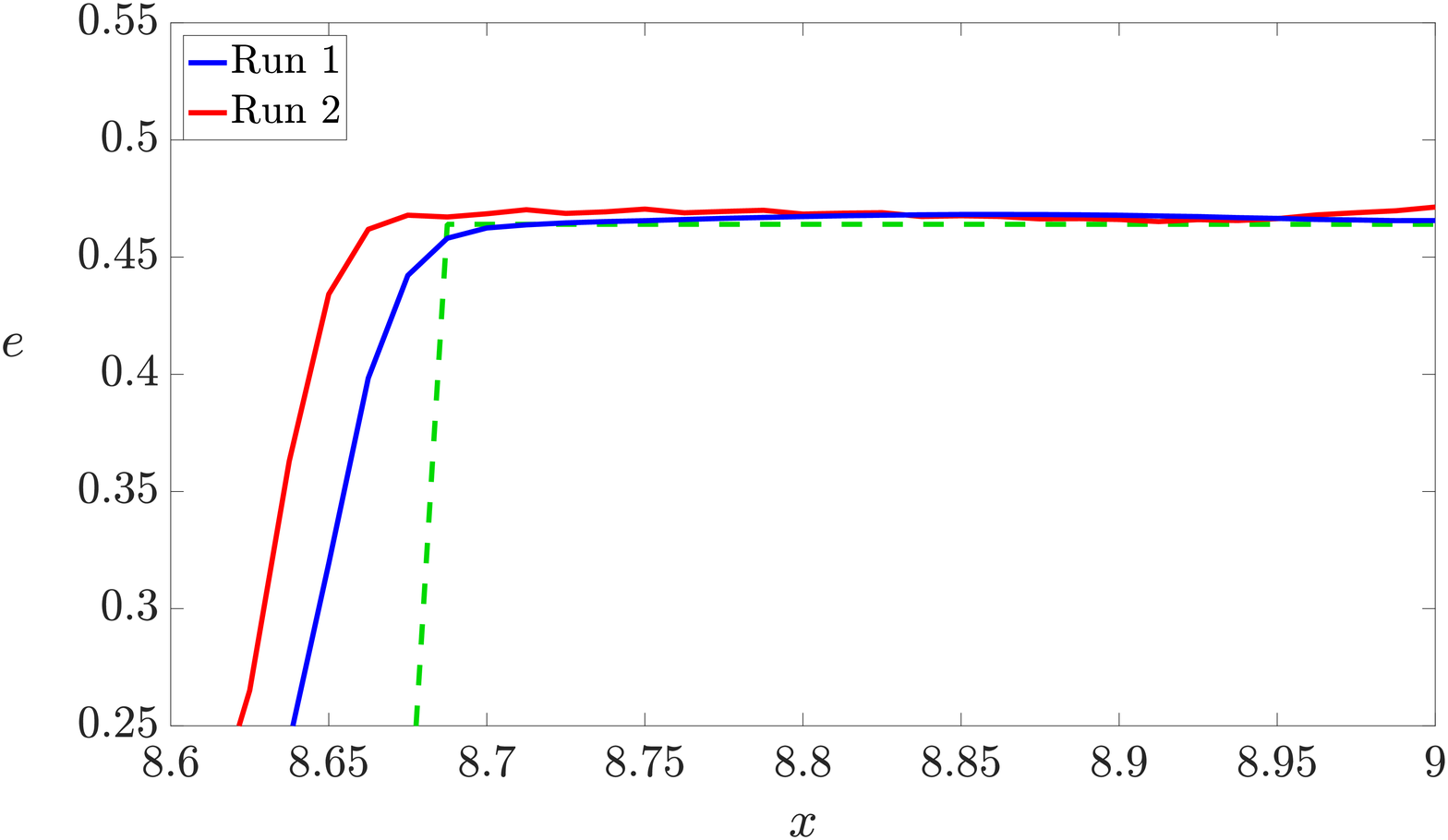}}
\subfigure[$t=8.0$: velocity, post-collision]{\label{fig:leblanc-calibrated-u}\includegraphics[width=75mm]{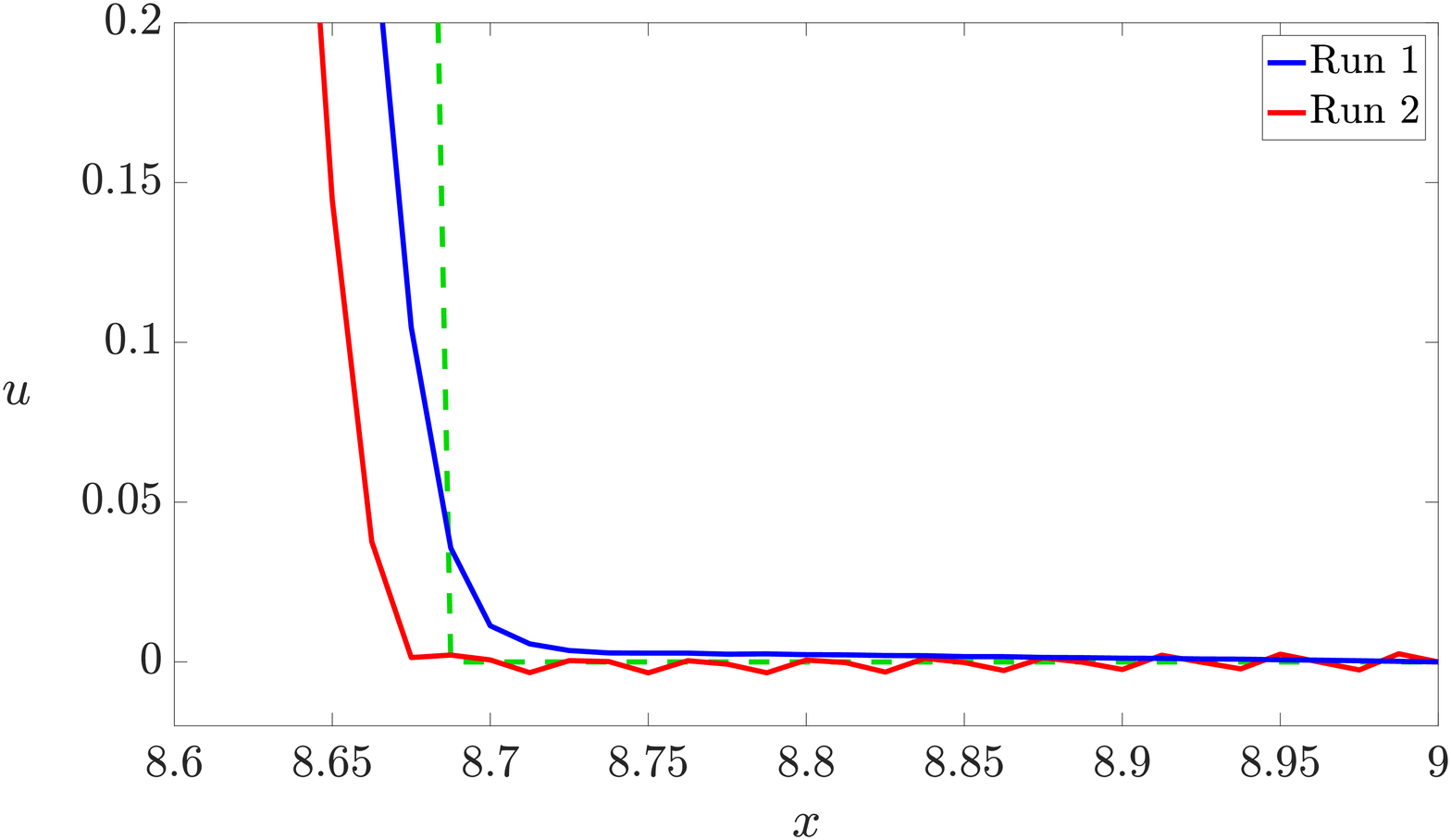}}
\caption{Comparison of the optimized-parameter and fixed-parameter WENO-$C$-$W$ runs for the 
LeBlanc shock tube
problem before and after shock-wall collision. The artificial viscosity parameters for the 
fixed-parameter Run 2 are 
chosen as $\beta^u=0.001$, $\beta^E=0.0$, $\beta^e=0.5$, $\beta^u_w=4.0$, $\beta^E_w=15.0$, 
$\beta^e_w=0.0$.}
\label{fig:leblanc-calibrated}
\end{figure}  

\begin{figure}[H]
\centering
\subfigure[$t=0.0039$: velocity]{\label{fig:peak-calibrated1}\includegraphics[width=75mm]{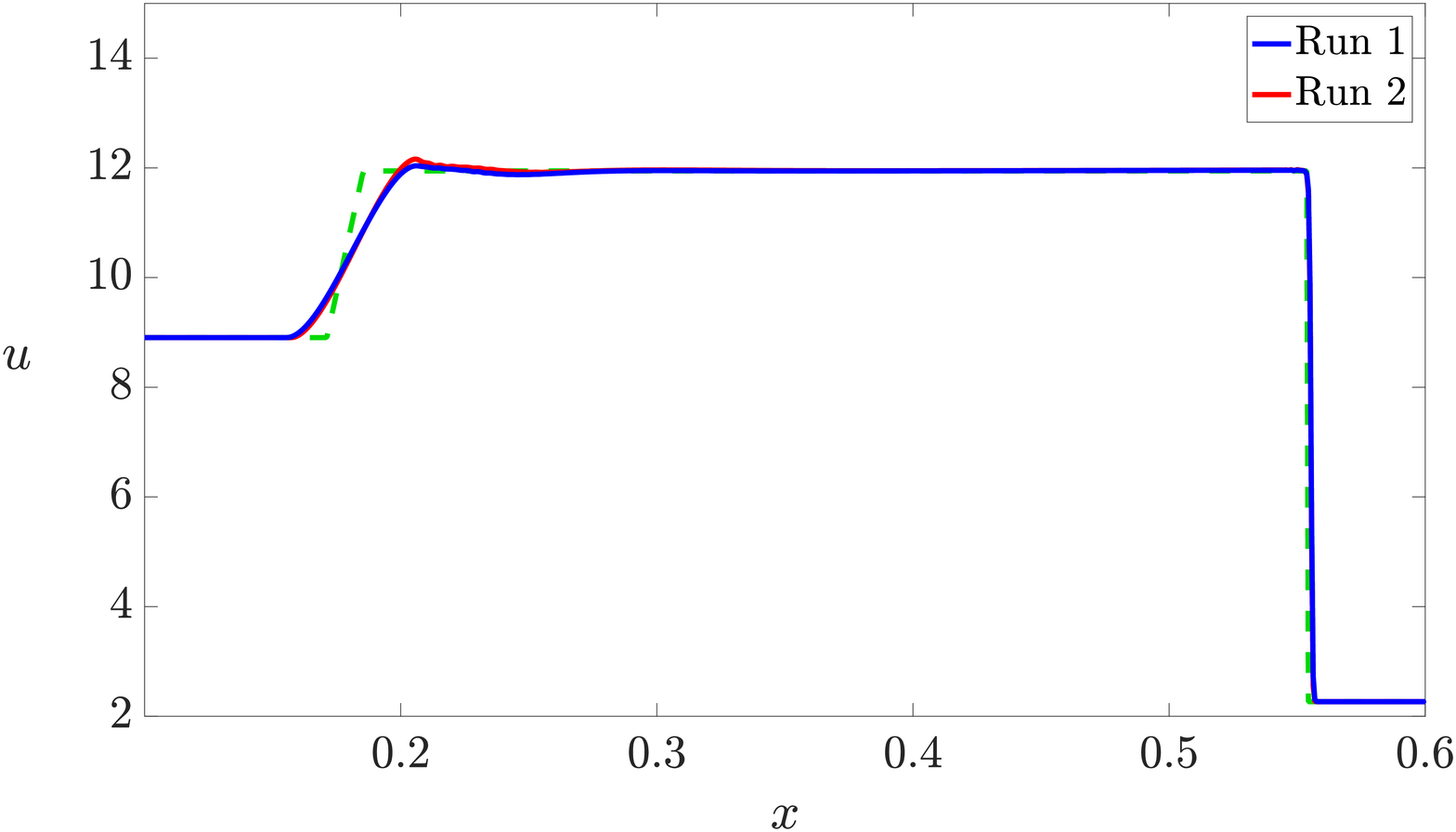}}
\subfigure[$t=0.0039$: velocity zoom-in]{\label{fig:peak-calibrated2}\includegraphics[width=75mm]{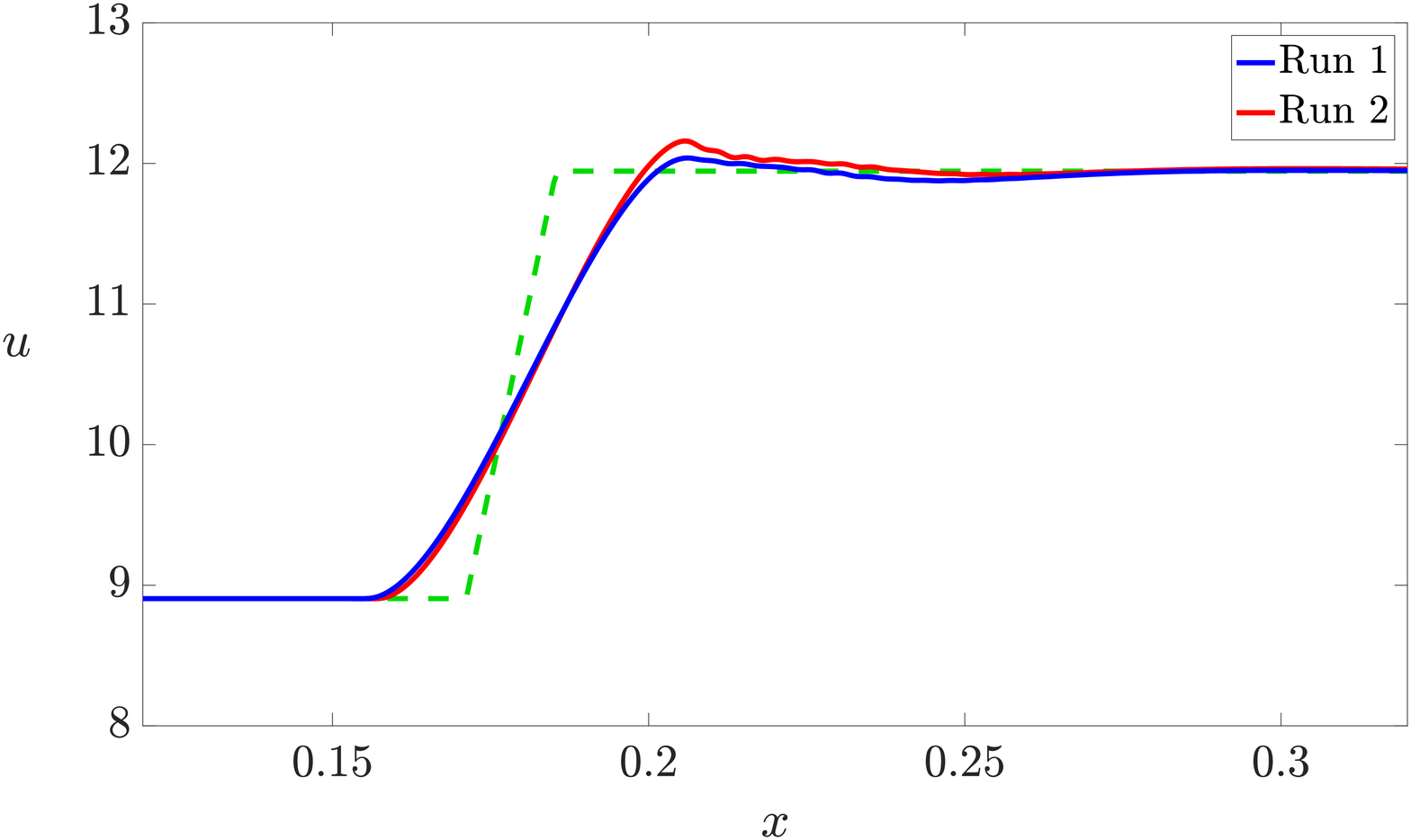}}
\caption{Comparison of the optimized-parameter and fixed-parameter WENO-$C$ runs for the Peak shock tube
problem. The artificial viscosity parameters for the fixed-parameter Run 2 are 
chosen as $\beta^u=1.0$, $\beta^E=0.0$, $\beta^r=10.0$.}
\label{fig:peak-calibrated}
\end{figure}  

\begin{figure}[H]
\centering
\subfigure[$t=0.63$: density, post-collision]{\label{fig:oshu-calibrated1}\includegraphics[width=75mm]{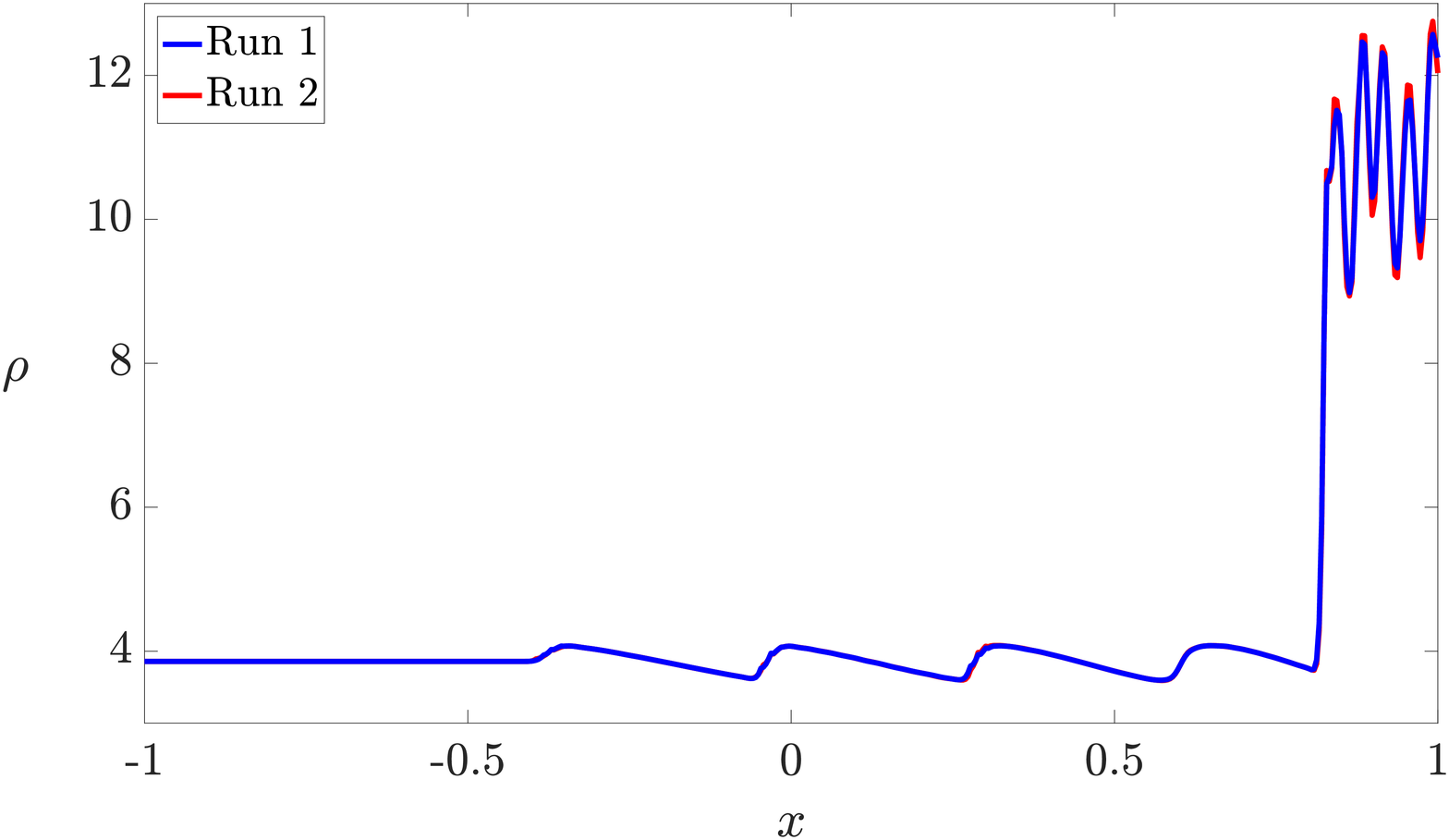}}
\subfigure[$t=0.63$: density, post-collision zoom-in]{\label{fig:oshu-calibrated2}\includegraphics[width=75mm]{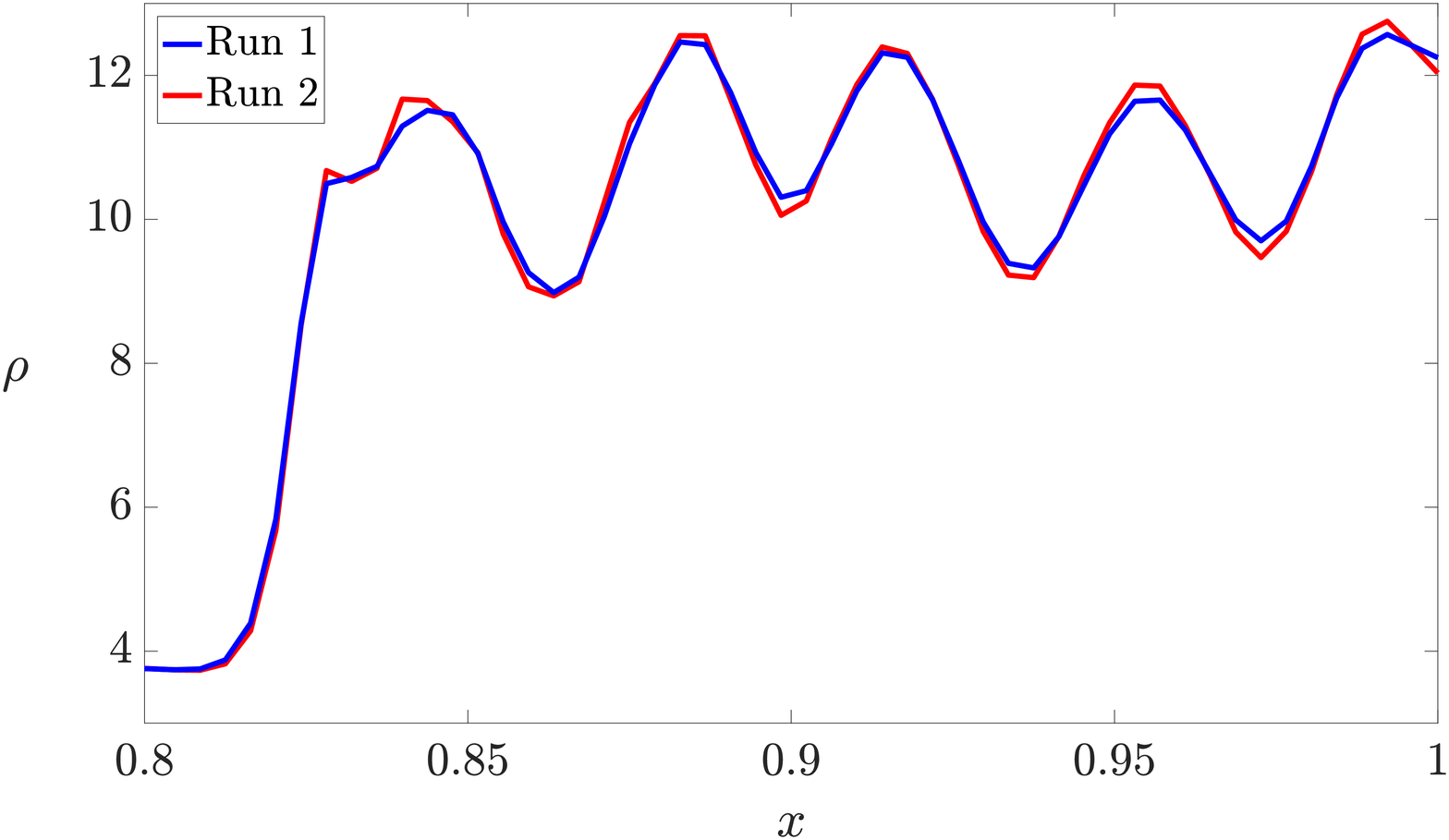}}
\caption{Comparison of the optimized-parameter and fixed-parameter WENO-$C$-$W$ 
runs for the Osher-Shu shock tube
problem after shock-wall collision. The artificial viscosity parameters for the fixed-parameter Run 2 are 
chosen as $\beta^u=1.0$, $\beta^E=0.0$, $\beta^u_w=2.5$, $\beta^E_w=0.85$.}
\label{fig:oshu-calibrated}
\end{figure}

\section*{Acknowledgements} 
Research reported in this publication was supported by the Office of Defense Nuclear Nonproliferation Research and Development and
by the Defense Threat Reduction Agency under Interagency Agreement number HDTRA1825370 (DTRA10027 -- 25370) as work for others.
SS was partially supported by DTRA HDTRA11810022.   

We would like to express our gratitude to the anonymous referees for their numerous 
suggestions that have greatly improved the manuscript.


\begin{thebibliography}{10}

\bibitem{Alpher1954}
R.~A. Alpher and R.~J. Rubin.
\newblock Normal reflection of shock waves from moving boundaries.
\newblock {\em Journal of Applied Physics}, 25(3):395--399, 1954.

\bibitem{BaDa2010}
G.~E. Barter and D.~L. Darmofal.
\newblock Shock capturing with PDE-based artificial viscosity for DGFEM: Part
  I, Formulation.
\newblock {\em Journal of Computational Physics}, 229(5):1810{\textendash}1827,
  2010.
  
  \bibitem{CaCo2004}
A.~W. Cook and W.~H. Cabot.
\newblock A high-wavenumber viscosity for high-resolution numerical methods.
\newblock {\em Journal of Computational Physics}, 195(2):594 -- 601, 2004.

\bibitem{CaCo2004b}
A.~W. Cook and W.~H. Cabot.
\newblock Hyperviscosity for shock-turbulence interactions.
\newblock {\em Journal of Computational Physics}, 203:379--385, 02 2004.

\bibitem{CaShWh1998}
E.~J. Caramana, M.~J. Shashkov, and P.~P. Whalen.
\newblock Formulations of artificial viscosity for multi-dimensional shock wave
  computations.
\newblock {\em Journal of Computational Physics}, 144(1):70 -- 97, 1998.

\bibitem{Coifman1995}
R.~R. Coifman and D.~L. Donoho.
\newblock {\em Translation-Invariant De-Noising}, pages 125--150.
\newblock Springer New York, New York, NY, 1995.

\bibitem{Colella1985104}
P.~Colella.
\newblock {A direct Eulerian MUSCL scheme for gas dynamics}.
\newblock {\em SIAM Journal on Scientific and Statistical Computing},
  6:104--117, 1985.
  
 \bibitem{Colella1984115}
P.~Colella and P.~R. Woodward.
\newblock {The numerical simulation of two-dimensional fluid flow with strong
  shocks}.
\newblock {\em Journal of Computational Physics}, 54:115 -- 173, 1984.

\bibitem{CoWo1984}
P.~Colella and P.~R. Woodward.
\newblock The piecewise parabolic method (PPM) for gas dynamical simulations.
\newblock {\em Journal of Computational Physics}, 54:174 -- 201, 1984.

\bibitem{courant1999supersonic}
R.~Courant and K.O. Friedrichs.
\newblock {\em Supersonic Flow and Shock Waves}.
\newblock Applied Mathematical Sciences. Springer New York, 1999.

\bibitem{DiPerna1983}
R.~J. DiPerna.
\newblock Convergence of the viscosity method for isentropic gas dynamics.
\newblock {\em Communications in Mathematical Physics}, 91(1):1--30, Mar 1983.

\bibitem{DONAT199642}
R.~Donat and A.~Marquina.
\newblock Capturing shock reflections: An improved flux formula.
\newblock {\em Journal of Computational Physics}, 125(1):42 -- 58, 1996.


\bibitem{Farge1992}
M.~Farge.
\newblock Wavelet transforms and their applications to turbulence.
\newblock In {\em Annual review of fluid mechanics, {V}ol.\ 24}, pages
  395--457. Annual Reviews, Palo Alto, CA, 1992.

\bibitem{GeMaDa1966}
R.~A. Gentry, R.~E. Martin, and B.~J. Daly.
\newblock {An Eulerian differencing method for unsteady compressible flow
  problems}.
\newblock {\em J. Computational Physics}, 1:87--118, 1966.

\bibitem{GreenoughRider2004}
J.~A. Greenough and W.~J. Rider.
\newblock A quantitative comparison of numerical methods for the compressible
  euler equations: Fifth-order weno and piecewise-linear godunov.
\newblock {\em J. Comput. Phys.}, 196(1):259--281, May 2004.

\bibitem{Harten1987231}
A.~Harten, B.~Engquist, S.~Osher, and S.~R. Chakravarthy.
\newblock {Uniformly high-order accurate essentially non-oscillatory schemes,
  III}.
\newblock {\em Journal of Computational Physics}, 71:231 -- 303, 1987.

\bibitem{Huynh19951565}
H.~T. Huynh.
\newblock {Accurate upwind methods for the Euler equations}.
\newblock {\em SIAM Journal on Numerical Analysis}, 32:1565--1619, 1995.

\bibitem{Igra1992}
O.~Igra, G.~Ben-Dor, G.~Mazor, and M.~Mond.
\newblock Head-on collision between normal shock waves and a rubber-supported
  plate, a parametric study.
\newblock {\em Shock Waves}, 2(3):189--200, Sep 1992.

\bibitem{JiangShu1996}
G.-S. Jiang and C.-W. Shu.
\newblock Efficient implementation of weighted ENO schemes.
\newblock {\em J. Comput. Phys.}, 126(1):202--228, June 1996.

\bibitem{Landshoff1955}
R.~Landshoff.
\newblock A numerical method for treating fluid flow in the presence of shocks.
\newblock {\em Los Alamos National Laboratory Report, LA-1930}, 1 1955.

\bibitem{Lapidus1967154}
A.~Lapidus.
\newblock {A detached shock calculation by second-order finite differences}.
\newblock {\em Journal of Computational Physics}, 2:154 -- 177, 1967.

\bibitem{MaOs1977}
Andrew Majda and Stanley Osher.
\newblock Propagation of error into regions of smoothness for accurate
  difference approximations to hyperbolic equations.
\newblock {\em Communications on Pure and Applied Mathematics}, 30(6):671--705,
  1977.

\bibitem{VanLeer1979101}
B.~Van Leer.
\newblock {Towards the ultimate conservative difference scheme. V. A
  second-order sequel to Godunov's method}.
\newblock {\em Journal of Computational Physics}, 32:101 -- 136, 1979.

\bibitem{Leveque2002}
R.~J. LeVeque.
\newblock {\em Finite Volume Methods for Hyperbolic Problems}.
\newblock Cambridge Texts in Applied Mathematics. Cambridge University Press,
  2002.

\bibitem{Liska2003995}
R.~Liska and B.~Wendroff.
\newblock {Comparison of several difference schemes on 1D and 2D test problems
  for the Euler equations}.
\newblock {\em SIAM J. Sci. Comput}, 25:995--1017, 2003.

\bibitem{Liu20098872}
W.~Liu, J.~Cheng, and C.-W. Shu.
\newblock {High-order conservative Lagrangian schemes with Lax-Wendroff type
  time discretization for the compressible Euler equations}.
\newblock {\em Journal of Computational Physics}, 228:8872--8891, 2009.

\bibitem{LiuLax1998}
X.-D. Liu and P.~D. Lax.
\newblock Solution of two-dimensional riemann problems of gas dynamics by
  positive schemes.
\newblock {\em SIAM Journal on Scientific Computing}, 19(2):319--340, 1998.

\bibitem{LiuLax2003}
X.-D. Liu and P.~D. Lax.
\newblock Positive schemes for solving multi-dimensional hyperbolic systems of
  conservation laws ii.
\newblock {\em Journal of Computational Physics}, 187(2):428 -- 440, 2003.

\bibitem{LiOsCh1994}
X.-D. Liu, S. Osher, and T. Chan.
\newblock {Weighted essentially non-oscillatory schemes}.
\newblock {\em Journal of Computational Physics}, 115:200--212, 1994.

\bibitem{Loubere2005105}
R.~Loub{\`e}re and M.~Shashkov.
\newblock {A subcell remapping method on staggered polygonal grids for
  arbitrary-Lagrangian-Eulerian methods}.
\newblock {\em Journal of Computational Physics}, 209:105--138, 2005.

\bibitem{Margolin2018}
L.~G. Margolin.
\newblock The reality of artificial viscosity.
\newblock {\em Shock Waves}, Feb 2018.

\bibitem{MaRi2015}
A.~E. Mattsson and W.~J. Rider.
\newblock Artificial viscosity: back to the basics.
\newblock {\em Internat. J. Numer. Methods Fluids}, 77(7):400--417, 2015.

\bibitem{mazor1992head}
G.~Mazor, O.~Igra, G.~Ben-Dor, M.~Mond, H.~Reichenbach, and F. T.~Smith.
\newblock Head-on collision of normal shock waves with a rubber-supported wall.
\newblock {\em Phil. Trans. R. Soc. Lond. A}, 338(1650):237--269, 1992.

\bibitem{Meneveau1991}
C.~Meneveau.
\newblock Analysis of turbulence in the orthonormal wavelet representation.
\newblock {\em J. Fluid Mech.}, 232:469--520, 1991.

\bibitem{Menikoff1993}
R.~Menikoff.
\newblock Errors when shock waves interact due to numerical shock width.
\newblock {\em SIAM J. Sci. Comput.}, 15(5):1227--1242, 1994.

\bibitem{meyer1957}
R.~F. {Meyer}.
\newblock {The impact of a shock wave on a movable wall}.
\newblock {\em Journal of Fluid Mechanics}, 3:309--323, 1957.

\bibitem{Noh1987}
W.~F Noh.
\newblock Errors for calculations of strong shocks using an artificial
  viscosity and an artificial heat flux.
\newblock {\em J. Comput. Phys.}, 72(1):78 -- 120, 1987.

\bibitem{PaPo1988}
T~Passot and A~Pouquet.
\newblock Hyperviscosity for compressible flows using spectral methods.
\newblock {\em Journal of Computational Physics}, 75(2):300 -- 313, 1988.

\bibitem{Quirk1994555}
J.~J. Quirk.
\newblock {A contribution to the great Riemann solver debate}.
\newblock {\em Int. J. Num. Methods Fluids}, 18:555--574, 1994.

\bibitem{ReSeSh2012}
J.~Reisner, J.~Serencsa, and S.~Shkoller.
\newblock A space-time smooth artificial viscosity method for nonlinear
  conservation laws.
\newblock {\em J. Comput. Phys.}, 235:912--933, 2013.

\bibitem{RaReSh2018b}
R.~Ramani, J.~Reisner, and S.~Shkoller.
\newblock A space-time smooth artificial viscosity method with wavelet noise indicator and shock collision scheme, Part 2: the 2-D case.
\newblock {\em Preprint}, 2018.

\bibitem{RaReSh2019}
R.~Ramani, J.~Reisner, and S.~Shkoller.
\newblock A space-time smooth artificial viscosity method for shock-shock and shock-contact collision.
\newblock {\em In preparation}.

\bibitem{Rider2000}
W.~J. Rider.
\newblock Revisiting wall heating.
\newblock {\em J. Comput. Phys.}, 162(2):395 -- 410, 2000.

\bibitem{SchVas2010}
K.~Schneider and O.~V. Vasilyev.
\newblock Wavelet methods in computational fluid dynamics.
\newblock In {\em Annual review of fluid mechanics. {V}ol. 42}, volume~42 of
  {\em Annu. Rev. Fluid Mech.}, pages 473--503. Annual Reviews, Palo Alto, CA,
  2010.
  
 \bibitem{Shu1988439}
C.-W. Shu and S.~Osher.
\newblock {Efficient implementation of essentially non-oscillatory
  shock-capturing schemes}.
\newblock {\em Journal of Computational Physics}, 77:439 -- 471, 1988.

\bibitem{OsherShu1989}
C.-W. Shu and S.~Osher.
\newblock Efficient implementation of essentially nonoscillatory
  shock-capturing schemes. {II}.
\newblock {\em J. Comput. Phys.}, 83(1):32--78, 1989.

\bibitem{Shu2003}
C.-W. Shu.
\newblock High-order finite difference and finite volume weno schemes and
  discontinuous galerkin methods for cfd.
\newblock {\em International Journal of Computational Fluid Dynamics}, 17(2):107--118, 2003.

\bibitem{Toro2009}
E.~F. Toro.
\newblock {\em {Riemann solvers and numerical methods for fluid dynamics}}.
\newblock Springer-Verlay Berling Heidelberg, 2009.

\bibitem{vNRi1950}
J.~Von~Neumann and R.~D. Richtmyer.
\newblock A method for the numerical calculation of hydrodynamic shocks.
\newblock {\em J. Appl. Phys.}, 21:232--237, 1950.

\bibitem{Wilkins1980}
M.~L. Wilkins.
\newblock Use of artificial viscosity in multidimensional fluid dynamic
  calculations.
\newblock {\em Journal of Computational Physics}, 36(3):281 -- 303, 1980.

\end{thebibliography}

\end{document}